\documentclass[aps,prb,reprint,nofootinbib,floatfix,longbibliography]{revtex4-2}

\usepackage{amsmath}
\usepackage{amsfonts}
\usepackage{newtxtext}
\usepackage{newtxmath}
\usepackage{mathtools}
\usepackage[mathcal]{eucal}
\usepackage{bbm}
\usepackage{bm}

\usepackage{tabularray}
\usepackage[table,dvipsnames]{xcolor}
\usepackage[percent]{overpic}
\usepackage{pict2e}
\UseTblrLibrary{booktabs}
\usepackage{array}
\newcolumntype{P}[1]{>{\centering\arraybackslash}p{#1}}
\newcolumntype{M}[1]{>{\centering\arraybackslash}m{#1}}
\definecolor{lgray}{rgb}{.93,.93,.93}

\usepackage[version=4]{mhchem}

\usepackage{enumitem}

\usepackage[bookmarksnumbered,pdfpagelabels=true,plainpages=false,colorlinks=true,linkcolor=blue,citecolor=magenta,urlcolor=blue]{hyperref}
\usepackage[capitalize]{cleveref}
\definecolor{cblue}{RGB}{55,126,184}
\hypersetup{colorlinks=true,linkcolor=cblue,citecolor=cblue,urlcolor=cblue}

\newcommand{\bq}{\bm{q}}
\DeclareMathOperator{\Tr}{Tr}
\DeclareMathOperator{\sign}{sign}
\DeclareMathOperator{\diag}{diag}

\newcommand\Tstrut{\rule{0pt}{2.6ex}}         
\newcommand\Bstrut{\rule[-0.9ex]{0pt}{0pt}} 

\newcommand{\Tpar}{\smash{T_{1\parallel}}}
\newcommand{\Tperp}{\smash{T_{1\perp}}}
\newcommand{\uvec}[1]{\hat{\bm{#1}}}

\newcommand{\Jeig}{J}
\newcommand{\Jmat}{\mathcal{J}}

\newcommand{\irrepvec}{\psi}

\newcommand{\HAFMdual}{HAFM$_{\text{dual}}$}

\newcommand{\Hessian}{\tilde{\mathcal{H}}}
\newcommand{\xyprojector}{\mathcal{P}_{\widetilde{xy}}}

\begin{document}

\title{
    Mapping the Phase Diagram of a Frustrated Magnet: \texorpdfstring{\\}{} 
    Flat Bands and Diabolical Locus from Irrep Collisions on the Pyrochlore Lattice
}
\author{Kristian Tyn Kai Chung}
\affiliation{Max Planck Institute for the Physics of Complex Systems, N\"othnitzer Strasse 38, 01187 Dresden, Germany}

\begin{abstract}
	We map the complete classical phase diagram of the spin Hamiltonian describing pyrochlore rare-earth magnets with all symmetry-allowed nearest-neighbor bond-dependent anisotropic two-spin interactions. 
    We provide a simple derivation of the organization of spins into tensor degrees of freedom describing the multipole moments of a tetrahedron, whose components correspond to irreducible representations (irreps) of the tetrahedral symmetry group $T_d$.
    By parameterizing the Hamiltonian directly in terms of the energies of the individual irreps, we perform an exhaustive search of all possible irrep degeneracies (collisions) which may host stable classical spin liquids arising at classical multi-critical points. 
    We find four one-parameter families of models along which three phases are degenerate, all four of which merge at the Heisenberg antiferromagnetic point and its dual, and we give a complete three-dimensional picture of the phase diagram showing all of the phases and their intersections. 
    The appearance of two copies of a single irrep implies an extra degenerate locus which pierces the phase boundaries at special points, yielding two additional isolated triple points.
    We demonstrate that one-parameter families of Hamiltonians are characterized by a topological invariant describing how the ground state spin configuration winds when adiabatically transported around this degenerate locus, analogous to a diabolical locus in quantum field theory.
    We provide a comprehensive catalog of all flat band degeneracies in the phase diagram and discuss the mechanisms that may allow for or impede the realization of a variety of classical spin liquids described by tensor gauge fields exhibiting nodal line singularities and concomitant multi-fold pinch points.
    Lastly, we provide a list of all cases where three irreps are degenerate above the ground state, which may lead to interesting features in the spin wave spectrum within each ordered phase.  
\end{abstract}

\date{\today}

\maketitle

\tableofcontents

\section{Introduction}

The experimental search for, and theoretical classification of, exotic states of matter such as topological phases and deconfined critical points, along with their associated spectra of fractionalized excitations, is a central aim of modern condensed matter physics.
A core focus is to understand systems whose low-energy physics is not characterized by conventional symmetry breaking with a ``rigid'' order parameter, but instead exhibits emergent liquid-like characteristics due to strong correlations.
A central theme in this effort is that novel physics emerges when multiple conventional ordering channels compete with each other, becoming ``intertwined'' and resulting in physics that is more than the sum of individual parts~\cite{senthilDeconfinedQuantumCritical2004,senthilQuantumCriticalityLandauGinzburgWilson2004,hermeleAlgebraicSpinLiquid2005,balentsCompetingOrdersNonLandauGinzburgWilson2005,senthilCompetingOrdersNonlinear2006,wangDeconfinedQuantumCritical2017,songUnifyingDescriptionCompeting2019,liuCompetingOrdersPyrochlore2019,zouStiefelLiquidsPossible2021}.
One class of systems that have made significant advancements through constant exchange between the experimental and theoretical fronts are frustrated magnets~\cite{lacroixIntroductionFrustratedMagnetism2011,diepFrustratedSpinSystems2020}---specifically, magnetic insulators with antiferromagnetic spin interactions frustrated by triangular motifs, such as occurs on the triangular, kagome, and pyrochlore lattices.
The geometric origin of the frustration makes such lattices highly favorable for supporting spin liquids---strongly correlated paramagnetic states which exhibit emergent deconfined gauge fields~\cite{balentsSpinLiquidsFrustrated2010,lhuillierIntroductionQuantumSpin2011,savaryQuantumSpinLiquids2016,zhouQuantumSpinLiquid2017,wenExperimentalIdentificationQuantum2019,knolleFieldGuideSpin2019,broholmQuantumSpinLiquids2020}.

In three spatial dimensions the pyrochlore lattice is the premier platform for frustrated magnetism, with multiple families of compounds for experimental study~\cite{subramanianOxidePyrochloresReview1983,gardnerMagneticPyrochloreOxides2010,leeFrustratedMagnetismCooperative2010,wiebeFrustrationPressureExotic2015,ghoshBreathingChromiumSpinels2019,reig-i-plessisFrustratedMagnetismFluoride2021} and a solid theoretical understanding of the microscopic physics~\cite{rauFrustratedQuantumRareEarth2019}. 
It hosts two of the most well-known and canonical examples of spin liquids: the pyrochlore Heisenberg anti-ferromagnet (HAFM)~\cite{moessnerLowtemperaturePropertiesClassical1998,henleyPowerlawSpinCorrelations2005,henleyCoulombPhaseFrustrated2010} and the pyrochlore Ising anti-ferromagnet, more commonly known as spin ice~\cite{udagawaSpinIce2021,castelnovoSpinIceFractionalization2012}.
Both of these are known to be \emph{classical} spin liquids, which in the former case refers to the large-spin limit and in the latter refers to the limit where transverse spin-flipping terms are not included in the Hamiltonian.\footnote{
    The latter is especially important because it is one of the only examples where we can derive perturbatively the quantum corrections and map them to a lattice gauge theory owing to the special geometry of the pyrochlore lattice~\cite{hermelePyrochlorePhotonsU12004,bentonSeeingLightExperimental2012,savaryCoulombicQuantumLiquids2012,gingrasQuantumSpinIce2014}, thus demonstrating the emergent gauge fields at the level of the microscopic spin operators, rather than as fluctuations about a mean field ground state~\cite{wenQuantumOrdersSymmetric2002}. 
    The number of references is far too great to list, see Ref.~\cite{udagawaSpinIce2021} for an overview and Refs.~\cite{smithCaseU1piQuantum2022,poreeEvidenceFractionalMatter2025,gaoNeutronScatteringThermodynamic2025,smithExperimentalInsightsQuantum2025} for more recent promising results for the realization of quantum spin ice in Cerium-based pyrochlores.
}
These two models correspond to two distinct limiting cases of the symmetry-allowed spin Hamiltonian: completely isotropic interactions versus the maximally anisotropic limit where spins are constrained to point along their local high-symmetry easy-axes.

Over time the broader problem of considering all symmetry-allowed, bond-dependent, anisotropic spin interactions has become necessary as a plethora of pyrochlore materials have been synthesized and intensively studied~\cite{gardnerMagneticPyrochloreOxides2010}.
Spatial anisotropies are generically present in these magnets because the ``spins'' are actually crystal field doublets of $4\!f$ rare earth ions~\cite{bertinCrystalElectricField2012}, for which strong spin-orbit coupling makes the interactions sensitive to the local crystalline symmetries~\cite{rauFrustratedQuantumRareEarth2019}.
Such bond-dependent anisotropies intermediate between the isotropic and easy-axis limits result in new types of frustration which can yield interesting novel spin liquids~\cite{rauSpinOrbitPhysicsGiving2016,essafiGenericNearestneighborKagome2017,maksimovAnisotropicExchangeMagnetsTriangular2019}, most famously in the exactly solved Kitaev Honeycomb model~\cite{kitaevAnyonsExactlySolved2006,hermannsPhysicsKitaevModel2018,takagiConceptRealizationKitaev2019}. 
Indeed, within the phase diagram of the microscopic pyrochlore pseudo-spin-1/2 Hamiltonian with four symmetry-allowed nearest-neighbor spin couplings, a number of classical spin liquids beyond the HAFM and spin ice models have been discovered~\cite{taillefumierCompetingSpinLiquids2017,franciniHigherRankSpinLiquids2025,bentonSpinliquidPinchlineSingularities2016,yanRank2U1Spin2020,lozano-gomezCompetingGaugeFields2024}, including some exhibiting emergent \emph{tensor} gauge fields~\cite{bentonSpinliquidPinchlineSingularities2016,yanRank2U1Spin2020,lozano-gomezCompetingGaugeFields2024}, which are expected to describe fractonic spin liquids~\cite{pretkoSubdimensionalParticleStructure2017,premPinchPointSingularities2018}.

Despite many studies, however, a comprehensive picture of the entire phase diagram and how the various models studied fit together has not yet appeared. 
Given the maturity of research on pyrochlore magnetism, recent advances in the understanding of classical spin liquids~\cite{bentonTopologicalRouteNew2021,yanClassificationClassicalSpin2024,yanClassificationClassicalSpin2024a,fangClassificationClassicalSpin2024}, along with many new avenues for experimental research~\cite{ghoshBreathingChromiumSpinels2019,reig-i-plessisFrustratedMagnetismFluoride2021,nutakkiClassicalHeisenbergModel2023}, it is warranted to give a complete account of the phase diagram and the various places in it where spin liquidity exists or may be present, how the variety of known examples fit together, and how the tensorial nature of these spin liquids arises.
Whereas previous studies have looked at either two-dimensional cross sections of the phase diagram~\cite{wongGroundStatePhase2013,javanparastOrderbydisorderCriticalityXY2015,yanTheoryMultiplephaseCompetition2017,rauFrustratedQuantumRareEarth2019,weiExactDiagonalization16site2023} or certain 1-parameter models~\cite{taillefumierCompetingSpinLiquids2017,noculakClassicalQuantumPhases2023,yanRank2U1Spin2020}, it is highly desirable to have a complete picture that plainly visualizes how such cuts of the phase diagram fit together, along with how the variety of spin liquids coincide with special degeneracies of the phase diagram. 
Additionally, recent work in the study of so-called `diabolical loci''~\cite{hsinBerryPhaseQuantum2020} and ``unnecessary critical points''~\cite{biAdventureTopologicalPhase2019,jianGenericUnnecessaryQuantum2020,prakashMultiversalityUnnecessaryCriticality2023,prakashClassicalOriginsLandauincompatible2025} has emphasized that phase diagrams may contain non-trivial structures other than phase boundaries, and it would be interesting to know whether such structures exist in the phase diagrams of frustrated magnets.
These are the tasks we take up in this paper.

\begin{figure}[t]
    \centering
    \includegraphics[width=\columnwidth]{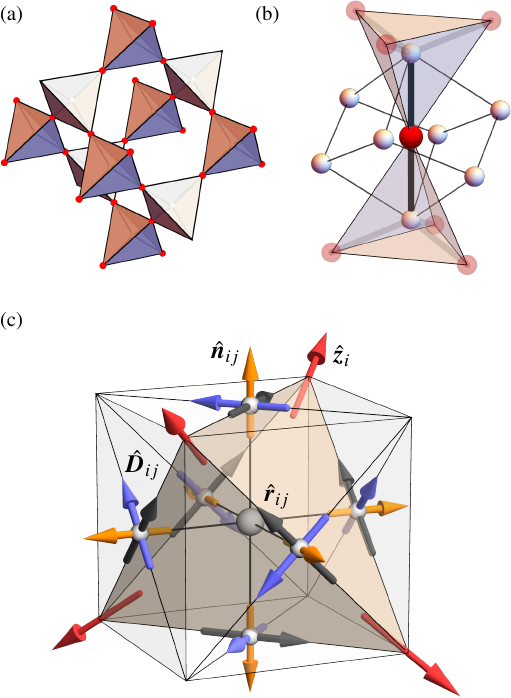}
    \caption{
        (a) An illustration of the pyrochlore lattice of corner sharing tetrahedra. (b) Each rare earth cation (red) sits in a cage of oxygen anions (white) with local symmetry group $D_{3d}$. (c) A single tetrahedron inscribed in a cube, with symmetry group $T_d$. The bond-dependent anisotropy vectors which appear in the symmetry-allowed spin-spin interactions in \cref{eq:H_named_interactions} are illustrated:
        the local three-fold easy-axis $\uvec{z}_i$ at each corner of the tetrahedron is indicated by a red arrow; 
        the nearest-neighbor separation vector $\uvec{r}_{ij}$ (gray arrows) point along the edges of the tetrahedron; 
        the normal vectors $\uvec{n}_{ij}$ (orange arrows) point out of the cube at the midpoint of each tetrahedron edge; 
        the DM vectors $\smash{\uvec{D}_{ij}}$ (light blue arrows) then lie in the cube faces pointing along the face diagonals orthogonal to the edges of the tetrahedron (thin black lines). 
        Since $\smash{\cramped{\uvec{r}_{ij} = -\uvec{r}_{ji}}}$, $\smash{\cramped{\uvec{D}_{ij} = -\uvec{D}_{ji}}}$; here we have made a particular choice of $\uvec{r}_{ij}$. 
        The total interaction $\smash{\cramped{\uvec{D}_{ij}\cdot  (\bm{S}_i \times \bm{S}_j}})$ does not depend on the order of $i,j$. 
        }
    \label{fig:tetrahedron_vectors}
\end{figure}

A summary of the layout and core results are as follows. 
In \cref{sec:symmetry_classification} we introduce the Hamiltonian and its various parameterizations, where \cref{fig:tetrahedron_vectors} illustrates the bond-dependent nature of the anisotropic interactions. 
We then explain the origin of the tensorial nature of pyrochlore spin liquids by deriving the tensor multipole decomposition of a single tetrahedron (\cref{tab:multipoles}) and exposing how it corresponds to an analysis of the irreducible representations (irreps) of the tetrahedral symmetry group, where each of the four distinct irreps corresponds to one ground state phase illustrated in \cref{fig:ground_states}. 
We pay particular attention to the ``canting'' of spins due to the continuous mixing of two copies of the $T_1$ irrep appearing in the symmetry decomposition, illustrated in \cref{fig:T1_mixing}.
In \cref{sec:mapping_the_phase_diagram} we then take up the issue of parameterizing the Hamiltonian in a way that can be used to expose all of the possible ground state degeneracies, which we accomplish by inverting the relation between the irrep energies and the exchange parameters, which allows the irrep energies to be tuned independently.
This enables us to give the complete classification of all of the multi-critical loci where ground state irreps collide, i.e. different phases become degenerate and thus compete and potentially intertwine to produce spin liquids, as evidenced by the presence of flat bands in the interaction matrix. 
In particular, we show that there is one point in the phase diagram where all four phases become degenerate---the maximally symmetric HAFM. It lies at the meeting point of four \emph{triple lines}, listed in \cref{tab:high_degeneracies}.
We provide a complete picture of the full phase diagram in \cref{fig:full-phase-diagram}, which shows all of the phases, phase boundaries, and the four triple lines which merge at the HAFM point. 
In the process we uncover a locus along which the two copies of the repeated $T_1$ irrep are degenerate, and in \cref{sec:canting_cycles} we show that this special locus defines a topological invariant for 1-parameter families of models which encircle it, analogous to a diabolical locus or conical intersection.
Finally, in \cref{sec:flat_bands} we give a full catalog of all degenerate combinations which yield flat band degeneracies in \cref{tab:flat_bands_phases,tab:flat_bands_phase_boundaries,tab:flat_bands_triple_lines}, along with the spin structure factors computed in the self-consistent Gaussian approximation in \cref{fig:structure_factors_triple_lines,fig:structure_factors_special}, and discuss the connection between pinch lines and fourfold pinch points in tensor spin liquids, illustrated in \cref{fig:fourfold_pinchpoint3d}.
We end in \cref{sec:triple_ferro} with a discussion of all triple degeneracies that occur in excited states, shown in \cref{fig:triple_lines_AFM_FM}, which give the phase diagram some further interesting structure and may have interesting imprints on the excitations above the ground state within each phase.

\section{Symmetry Classification of Ground States}
\label{sec:symmetry_classification}

For rare earth pyrochlore magnetic insulators with the familiar \ce{R2M2O7} structure, the magnetic $f$-electron rare earth (R) ions form a pyrochlore lattice of corner-sharing tetrahedra, shown in \cref{fig:tetrahedron_vectors}(a), with the transition metal (M) ions forming a second interwoven pyrochlore lattice (not shown)~\cite{rauFrustratedQuantumRareEarth2019,subramanianOxidePyrochloresReview1983,gardnerMagneticPyrochloreOxides2010}. 
Each rare earth ion sits in the center of a distorted cube of oxygen atoms shown in \cref{fig:tetrahedron_vectors}(b), which generate local crystal fields that split the electronic ground state degeneracy of the central ion.
In most such compounds this results in a crystal field ground state doublet typically protected by a gap of order $100$K~\cite{bertinCrystalElectricField2012}, and the full crystal field Hamiltonian can be projected into this doublet Hilbert space to obtain an effective low-energy pseudo-spin-1/2 Hamiltonian.
There are three allowed types of doublets---Kramers effective-spin-1/2; Kramers dipolar-octupolar; and non-Kramers~\cite{rauFrustrationAnisotropicExchange2018,rauFrustratedQuantumRareEarth2019}. 
For the effective spin-1/2 doublet the pseudo-spin operators transform as ordinary dipoles, and its most general Hamiltonian is our main concern in this work. 
For the non-Kramers doublet the local-$z$ operators transform as dipoles while the local-$xy$ operators transform as quadrupoles, so they cannot couple, but its Hamiltonian is just a restricted version of that of the effective spin-1/2 Hamiltonian with one coupling set to zero. 
The dipolar-octupolar case has a different Hamiltonian which reduces to a simple XYZ-model, and we will not consider it here.\footnote{
    The dipolar-octupolar classical phase diagram can be parameterized as the surface of a cube, with three copies of the spin ice phase and three copies of the all-in-all-out phase on the six faces~\cite{huangQuantumSpinIces2014}.
}

We consider the most general nearest-neighbor spin model on the pyrochlore lattice, 
\addtocounter{equation}{1} 
\begin{equation}
	H = \frac{1}{2}\sum_{ i,j } \sum_{\alpha,\beta\in \{x,y,z\}} S_i^\alpha \Jmat_{ij}^{\alpha\beta} S_j^\beta,
    \tag{\theequation a}
	\label{eq:H_generic}
\end{equation}
where the first sum is over sites $i$ and $j$ of the pyrochlore lattice and $\alpha,\beta$ index the three components of a spin. 
The matrix $\Jmat$ contains all symmetry-allowed nearest-neighbor spin-spin interactions, and we refer to it as the interaction matrix, restricting to only nearest-neighbor interactions.  
The allowed couplings can be deduced by symmetry by considering all spatial symmetries which swap two spins at sites $i$ and $j$, and demanding that $\sum_{\alpha\beta}S_i^\alpha \Jmat_{ij}^{\alpha\beta}S_j^\beta$ be invariant. 

\subsection{Parameterizations of Anisotropic Interactions}

The Hamiltonian can be parameterized either in a global or a local basis, each of which can be useful to identify special parameter points~\cite{yanTheoryMultiplephaseCompetition2017,rauFrustratedQuantumRareEarth2019,rossQuantumExcitationsQuantum2011}. 
If we take $\alpha,\beta$ in \cref{eq:H_generic} to refer to the global frame Cartesian components of a spin, with $x$, $y$, and $z$ axes along the cubic symmetry axes, then for two spins in the $x,y$ plane the interaction matrix has four independent components,
\begin{equation}
	\Jmat_{ij}^{\alpha\beta} 
	= 
	\begin{pmatrix*}[c]
	 J_1 & J_3 & -J_4 \\
	 J_3 & J_1 & -J_4 \\
	 J_4 & J_4 & J_2
	\end{pmatrix*}^{\alpha\beta}.
    \tag{\theequation b}
	\label{eq:Jijab}
\end{equation}
Since every nearest-neighbor bond is symmetry-equivalent, there are four free parameters in the Hamiltonian.

These symmetry-allowed interactions can equivalently be parameterized in a coordinate-free notation as ($\mathrm{T}$ denotes transpose)
\begin{align}
	H = \sum_{\langle ij \rangle} &\Big[
	J_{\text{Heis}} \bm{S}_i \cdot \bm{S}_j 
	+ 
    J_{\text{PD}} 
    \bm{S}_i [\uvec{r}_{ij} \uvec{r}_{ij}^\mathrm{T}] \bm{S}_j
    + 
    J_{\text{K}}
    \bm{S}_i [\uvec{n}_{ij} \uvec{n}_{ij}^\mathrm{T}] \bm{S}_j
    \nonumber\\
	&+ J_{\text{DM}} \bm{D}_{ij} \cdot (\bm{S}_i \times \bm{S}_j)\Big] 
	 +\frac{J_{\text{SIA}}}{2}  \sum_i  \vert \bm{S}_i \cdot \uvec{z}_i\vert^2,
	 \label{eq:H_named_interactions}
\end{align}
where we refer to the four interactions as: isotropic Heisenberg (Heis), pseudo-dipolar (PD), Kitaev-like (K), Dzyaloshinskii-Moriya (DM), and we also include a single-ion anisotropy (SIA).\footnote{
    While single-ion anisotropy is irrelevant for spin-1/2 operators, it will be important for us to include since it is a symmetry-allowed parameter required to match the number of parameters in the Hamiltonian with the number of irreducible representations of the point symmetry group, thus allowing us to invert the relation between irrep energies and exchange parameters in \cref{sec:model_space_parameterization}. 
} 
The geometry of these bond-dependent anisotropic interactions is illustrated in \cref{fig:tetrahedron_vectors}, which shows a single tetrahedron of the pyrochlore lattice inscribed in a cube: $\smash{\cramped{\uvec{r}_{ij} \propto \bm{r}_j - \bm{r}_i}}$ point from site~$i$ to site~$j$ (gray arrows), where $\bm{r}_i$ is the position of site~$i$; $\uvec{n}_{ij}$ are the normal vectors to the cube faces (orange arrows); and $\smash{\cramped{\uvec{D}_{ij}= \uvec{n}_{ij}\times \uvec{r}_{ij}}}$ are the DM vectors (blue arrows).
We call the third term Kitaev-like because it involves only $x$, $y$, or $z$ spin interactions on each edge---a $z$-$z$ interaction if $\bm{r}_{ij}$ is in the $x,y$ plane and the others by cyclic permutation. 
The pseudo-dipolar term is so called because it appears when truncating a dipole-dipole interaction at nearest-neighbor.
In the single-ion anisotropy, $\uvec{z}_i$ is the local easy-axis direction.

In addition to the global frame, one can utilize a local symmetry-adapted basis, with a separate frame defined for each spin. The standard definitions take the local $\uvec{z}_i$ along the 3-fold easy-axis, $\hat{\bm{y}}_i$ along a 2-fold axis, and $\hat{\bm{x}}_i = \hat{\bm{y}}_i \times \uvec{z}_i$ lying in a mirror plane. 
An explicit choice of local coordinates is given in \cref{apx:conventions}.
In terms of the local frame $S_i^x$, $S_i^y$, and $S_i^z$ components, defining $S_i^\pm = S_i^x \pm i S_i^y$, the Hamiltonian is given by~\cite{rossQuantumExcitationsQuantum2011,rauFrustratedQuantumRareEarth2019}
\begin{align}
	&H 
	= 
	\frac{J_{\text{SIA}}}{2} 
    \sum_i (S_i^z)^2 
    + 
	\sum_{\langle ij \rangle} 
    \Big[ 
    J_{zz} S_i^z S_j^z 
	- 
    J_{\pm} (S_i^+ S_j^- + S_i^- S_j^+)
 	\nonumber 
    \\
	&+ 
    J_{\pm\pm} (\gamma_{ij} S_i^+ S_j^+ + \gamma_{ij}^* S_i^- S_j^-)
    + 
    J_{z\pm} \left(\zeta_{ij} (S_i^+ S_j^z + S_i^z S_j^+) + \text{h.c.}\right)
    \!\!
	\Big] ,
	\label{eq:H_local}
\end{align}
where $\gamma_{ij}$ are cube roots of unity and $\zeta_{ij} = -\gamma_{ij}^*$. 
For non-Kramers doublets, the Hamiltonian has $J_{z\pm} = 0$. 
The global and local parameterizations correspond to a change of basis for $\Jmat$, and are related by a linear map
\addtocounter{equation}{1} 
\begin{equation}
    \begin{pmatrix*}[l]
        J_{zz} \\[1ex] J_{\pm\pm} \\[1ex] J_{z\pm} \\[1ex] J_{\pm}
    \end{pmatrix*}
    = 
   \begin{pmatrix*}[r]
      -\frac{2}{3} & \frac{1}{3} & -\frac{2}{3} & -\frac{4}{3} \\[1ex]
     \frac{1}{6} & \frac{1}{6} & -\frac{1}{3} & \frac{1}{3} \\[1ex]
     \frac{1}{3 \sqrt{2}} & \frac{1}{3 \sqrt{2}} & \frac{1}{3 \sqrt{2}} & -\frac{1}{3 \sqrt{2}} \\[1ex]
     \frac{1}{3} & -\frac{1}{6} & -\frac{1}{6} & -\frac{1}{3} 
    \end{pmatrix*}
    \begin{pmatrix*}[l]
        J_1 \\[1ex] J_2 \\[1ex] J_3 \\[1ex] J_4
    \end{pmatrix*},
    \tag{\theequation a}
    \label{eq:JloctoJ14}
\end{equation}
with inverse map
\begin{equation}
    \begin{pmatrix*}[l]
        J_1 \\[1ex] J_2 \\[1ex] J_3 \\[1ex] J_4
    \end{pmatrix*}
    = 
   \begin{pmatrix*}[r]
         -\frac{1}{3} & \frac{2}{3} & \frac{2 \sqrt{2}}{3} & \frac{4}{3} \\
         \frac{1}{3} & \frac{4}{3} & \frac{4 \sqrt{2}}{3}  & -\frac{4}{3} \\
         -\frac{1}{3} & -\frac{4}{3} & \frac{2 \sqrt{2}}{3} & -\frac{2}{3} \\
         -\frac{1}{3} & \frac{2}{3} & -\frac{\sqrt{2}}{3} & -\frac{2}{3}
    \end{pmatrix*}
    \begin{pmatrix*}[l]
        J_{zz} \\[1ex] J_{\pm\pm} \\[1ex] J_{z\pm} \\[1ex] J_{\pm}
    \end{pmatrix*}.
    \tag{\theequation b}
    \label{eq:J14toJloc}
\end{equation}
The relations between the three parameterizations \cref{eq:Jijab}, \cref{eq:H_named_interactions}, and \cref{eq:H_local} are given in \cref{apx:parameter_maps}.
The local basis parameters in \cref{eq:H_local} will play a key role throughout this work, as they expose a number of important symmetries and dualities.
In particular, the Hamiltonian has a duality by applying a $\pi$ rotation to each spin about its local $\uvec{z}_i$ axis and changing the sign of $J_{z\pm}$.
Furthermore, when $J_{z\pm} = 0$, there is an additional duality applying a $\pi/2$ rotation to each spin about its local axis and switching the sign of $J_{\pm\pm}$.

\begin{figure*}
    \centering
    \includegraphics[width=\textwidth]{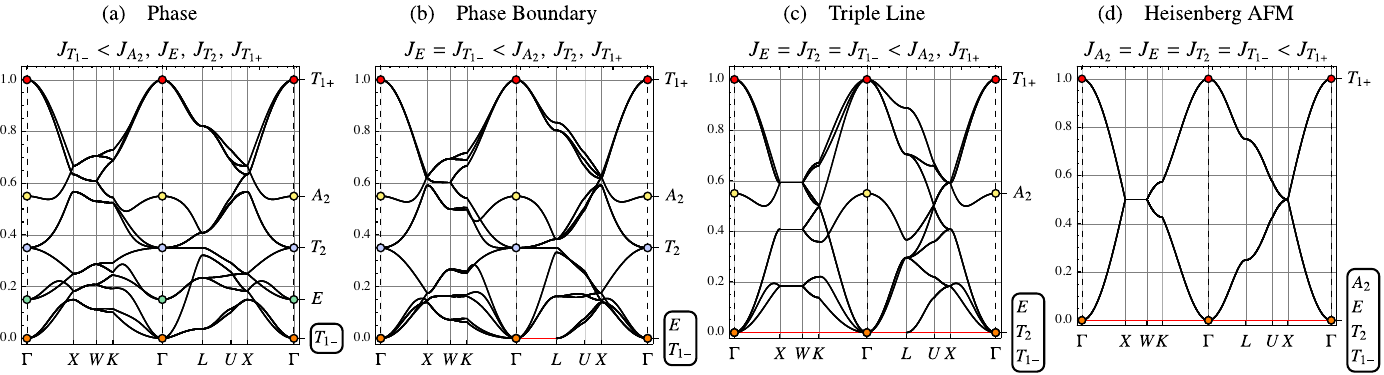}
    \caption{
    Tuning the symmetry-allowed couplings in a single tetrahedron allows for the tuning of the $\bm{q}=\bm{0}$ eigenvalues of the interaction matrix, labeled by irreducible representations of the point group $T_d$. 
    (a) A phase is the locus of parameters where a single irrep has the lowest energy. 
    (b) A phase boundary is the locus where two irreps have the lowest energy. (c) A triple line is the locus where three irreps have the lowest energy. 
    (d) There is a point in the phase diagram where all four phases become degenerate, the Heisenberg anti-ferromagnet (HAFM). 
    Increasing the degeneracy of $\bm{q}=\bm{0}$ eigenvalues leads to increased degeneracies away from $\bm{q}=\bm{0}$ in the form of flat lines, flat planes, and flat bands, indicated in red. 
    These flat degeneracies are tabulated in \cref{tab:flat_bands_phase_boundaries,tab:flat_bands_phases,tab:flat_bands_triple_lines}.
    A visualization of the Brillouin zone and the path followed in this figure is given in \cref{fig:FBZ_path}.}
    \label{fig:bands}
\end{figure*}

In this work we will set $\cramped{J_{\text{SIA}} = 0}$, as is appropriate for effective spin-1/2 systems where $(S_i^z)^2$ is a constant multiple of the identity.
However, for the purpose of the symmetry irreducible representation analysis we will employ, however, it is important that we keep $J_{\text{SIA}}$ as a formal parameter. 
The symmetry analysis does not know about the underlying spin quantum number, only the vector transformation properties of the spin operators.
Since= single ion anisotropy us a symmetry-allowed  coupling, its inclusion matches the number of parameters in the Hamiltonian with the number of tunable eigenvalues of the interaction matrix.
This matching is necessary to invert their relation and re-parameterize the phase diagram in terms of the eigenvalues, which is key to our mapping of the phase diagram. 
One can consider turning on single-ion anisotropy adiabatically starting from the phase diagram we present here, which will drive the system towards either the easy-plane or easy-axis configurations.

In this work we will treat the spins as classical fixed-length vectors, with spin length normalized to unity, $\vert \bm{S}_i\vert^2 = 1$.
There are two ways to justify this classical approximation.
The standard view is that it arises in the large-$S$ limit of the quantum model.
An alternative perspective is that it is the restriction of the quantum $S=1/2$ problem to the variational manifold of unentangled product states, whose local degrees of freedom are SU(2) spin coherent states on the Bloch sphere, $\mathbbm{C}P^1\simeq S^2$.
While these two perspectives give the same interpretation---unentangled local dipole moments---the latter can be generalized to different spin quantum number.
In particular, the approach can be modified to higher spin quantum number to account for multipolar degrees of freedom using SU($2S+1$) coherent states~\cite{remundSemiclassicalSimulationSpin12022,zhangClassicalSpinDynamics2021,chungGeometricallyFrustratedQuadrupoles2025}, particularly relevant in the case of spin-1 pyrochlores~\cite{zhangDynamicalStructureFactor2019,plumbContinuumQuantumFluctuations2019,hagymasiPhaseDiagramAntiferromagnetic2024}.

\subsection{Band Structure and Degeneracies}

The classical phase diagram of this Hamiltonian can be mapped out by studying a single tetrahedron, because the Hamiltonian decomposes into identical copies on each tetrahedron $t$,
\begin{equation}
	H = \frac{1}{2}\sum_{t} \sum_{\mu,\nu=1}^4 \sum_{\alpha,\beta} S_{t,\mu}^\alpha \Jmat_{\mu\nu}^{\alpha\beta} S_{t,\nu}^\beta,
	\label{eq:H_tetrahedra}
\end{equation}
where $\bm{S}_{t,\mu}$ is the spin on tetrahedron $t$ and sublattice $\mu$. 
The interaction matrix $\Jmat$ is the same on every tetrahedron, and the factor of 1/2 accounts for the double-counting of each nearest-neighbor pair in the sum over $\mu$ and $\nu$.
Since the Hamiltonian is the same on every tetrahedron, by studying the ground states of a single tetrahedron the ground states of the entire system can be determined by ``lego-brick-rules'', i.e. by attaching tetrahedra together at their corners and matching the corresponding spin~\cite{yanTheoryMultiplephaseCompetition2017}. 

Another way to say this is to note that each FCC primitive unit cell contains four spins, and the possible \textit{uniform} ground states, i.e. those that repeat in every FCC unit cell, must have the same spin configuration on every tetrahedron. 
Thus by solving the single-tetrahedron ground states, one can construct all the zero-wavevector ground states for the entire lattice. 
Because all of the interactions are internal to each tetrahedron, it follows that every single-tetrahedron ground state gives a uniform ground state for the entire system. 

This can also be understood by diagonalizing the quadratic form in \cref{eq:H_generic} via Fourier transform. 
Since $\Jmat$ commutes with the magnetic space group symmetries, it is translationally invariant and thus block-diagonalized into $12\times 12$ blocks in the Fourier basis (corresponding to four 3-component spins per unit cell), each block labeled by a crystal momentum wavevector $\bm{q}$. 
Diagonalizing therefore yields twelve bands of eigenvalues in reciprocal space, $\Jeig_n(\bq)$, where $n=1,\cdots 12$, each corresponding to a normalized eigenvector $\smash{\irrepvec_{\bq,n}}$ and a corresponding normal mode
\begin{equation}
	S_{\bq,n} = \sum_{i}\sum_{\alpha} [\irrepvec_{\bq,n}]_{i}^{\alpha} S_i^\alpha,
\end{equation}
so that the Hamiltonian is diagonalized into a sum of decoupled quadratic modes 
\begin{equation}
	H = \frac{1}{2} \sum_{\bq}\sum_{n=1}^{12} \Jeig_n(\bq) \vert S_{\bq,n}\vert^2.
    \label{eq:H_bands}
\end{equation}
In principle, the minimum eigenvalue normal mode determines the ground state~\cite{luttingerTheoryDipoleInteraction1946,litvinLuttingerTiszaMethod1974}.\footnote{
    It may occur that one cannot use the corresponding eigenvectors to construct a state satisfying the spin-length constraint $\vert\bm{S}_i\vert^2 = 1$ on every site, in which case the ground state will necessarily include contributions from some higher-energy normal modes. This does not occur in the pyrochlore model we are considering here.
    }
With only nearest-neighbor interactions on the pyrochlore lattice, the minimum eigenvalue of the interaction matrix is always found at the Brillouin zone center. This is a direct consequence of the corner-sharing tetrahedral geometry: minimizing the energy of each tetrahedron simultaneously minimizes the energy of the full lattice, leaving no energetic incentive for spatially modulated order. As a result, the nearest-neighbor model supports only $\bq=0$ ordered phases (or disordered phases). By contrast, further-neighbor interactions introduce additional $\bq$-dependence into the interaction matrix, producing higher-harmonic modulations of the bands that can shift the minimum to finite wavevector and thereby stabilize $\bq\neq0$ order.

The $\cramped{\bq=\bm{0}}$ block of the interaction matrix commutes with a 12-dimensional representation of the tetrahedral symmetry group $\cramped{T_d\subseteq O(3)}$.\footnote{
    In full generality one could consider how the entire system transforms under the full space group. Since we only consider nearest-neighbor interactions within a unit cell, we restrict our attention to the point group $\mathrm{O}_h$, which is the symmetry of a single primitive unit cell containing two tetrahedra. Furthermore, since we do not consider inversion-breaking breathing anisotropy, we can restrict our attention to the inversion-even subgroup $T_d \simeq \mathrm{O}_h/\mathbbm{Z}_2$, or equivalently to the inversion-even (\textit{gerade}) irreps of $\mathrm{O}_h$. 
} 
This representation decomposes into a collection of irreducible representations (irreps), and the dimensions of the irreps appearing in the decomposition correspond to the degeneracies of the eigenvalues at $\cramped{\bq=\bm{0}}$, i.e. the symmetry protects a set of band touchings at the zone center, which we discuss in detail in \cref{sec:irreps_symmetries_multipoles}.
Classical spin liquids arise when there is at least one flat band at the bottom of the spectrum~\cite{yanClassificationClassicalSpin2024,yanClassificationClassicalSpin2024a,davierCombinedApproachAnalyze2023,henleyPowerlawSpinCorrelations2005,isakovDipolarSpinCorrelations2004}, which will be discussed in detail in \cref{sec:flat_bands}.
This generally occurs when multiple ground state irreps collide, with more irreps giving more degeneracy and more potential flat bands.

This is illustrated in \cref{fig:bands}.
There are five distinct eigenvalues of the interaction matrix at the $\Gamma$ point, each labeled by an irreducible representation (irrep) of the point group $T_d$ and discussed in detail in \cref{sec:irreps_symmetries_multipoles}, the degeneracy of the bands at the zone center corresponding to the dimension of the corresponding irrep.
\Cref{fig:bands}(a) shows the band structure at a \emph{generic} point in the phase diagram, where all five eigenvalues are distinct, and a quadratic band minimum at $\bq = \bm{0}$ (the $\Gamma$ point). 
Correspondingly, the system exhibits a unique uniform ground state (up to symmetry).
\Cref{fig:bands}(b) shows the band structure at a point on a phase boundary between two phases, where two different irreps are degenerate at zero energy.
This results in a flat line of zero-energy eigenvalues along the $(hhh)$ high-symmetry line ($\Gamma \to L$, marked in red). 
\Cref{fig:bands}(c) shows the band structure on a triple line where three phases are degenerate, with three degenerate zero-energy irreps, exhibiting completely flat bands throughout the entire Brillouin zone.
Lastly, \cref{fig:bands}(d) shows the band structure at the maximally isotropic Heisenberg antiferromagnet point, where four eigenvalues are degenerate at zero energy.
As we will show, at this special point all four phases meet, and the system exhibits the maximum zero-energy degeneracy, with six of twelve bands completely flat.

\begin{figure*}[t]
    \includegraphics[width=\textwidth]{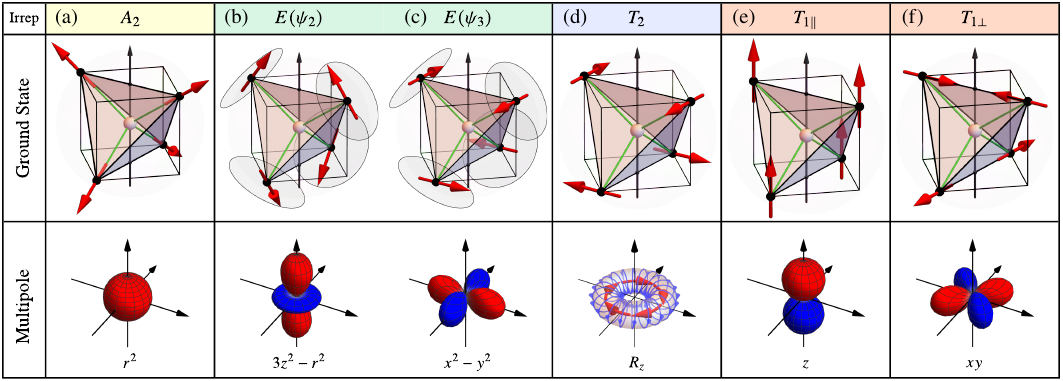}
	\caption{Ground states of a single tetrahedron, each corresponding to an irrep of $T_d$. Below each configuration is shown the corresponding magnetic multipole moment to which this irrep corresponds, along with labels of their symmetries ($R_z$ indicates a rotation, i.e. angular momentum pseudovector), cf.~\cref{tab:multipoles}. Black arrows indicate cubic axes. The $T_1$ and $T_2$ irreps each have two additional components given by the same configuration but along the other two Cartesian axes. Common names from left to right: (a) All-in-all-out (AIAO, $A_2$), (b,c) $\Gamma_5$ ($E$, split into $\psi_2$ and $\psi_3$ components), (d) Palmer-Chalker (PC, $T_2$), (e) Collinear ferromagnet ($T_{1,\parallel}$). The last configuration (f) may be called a biaxial antiferromagnet ($T_{1,\perp}$), since it corresponds to a biaxial quadrupole configuration. Linear combinations of the two $T_1$ irreps are splayed (or canted) ferromagnet (SFM) configurations, cf. \cref{fig:T1_mixing}. Here we show a choice of basis corresponding to a collinear ferromagnet and a coplanar anti-ferromagnet, which were derived from the multipole decomposition of a tetrahedron. Colors for each irrep match colors used in other figures in the rest of the paper.
    }
	\label{fig:ground_states}
\end{figure*}

\subsection{Irreducible Representations and Multipole Decomposition}
\label{sec:irreps_symmetries_multipoles}

While the irreducible representation analysis has been known and utilized to study the phase diagram~\cite{yanTheoryMultiplephaseCompetition2017,pandeyAnalyticalTheoryPyrochlore2020}, we present here an intuitive derivation that directly relates it to the multipole moments of a tetrahedron. 
Consider four spins on the corners of a single tetrahedron, with a total of twelve components $S_\mu^\alpha$. 
Under the action of $T_d$, the corners of the tetrahedron are permuted while the spins rotate as angular momenta (i.e. pseudo-vectors). 
We thus have a 12-dimensional representation of $T_d$, where each $g \in T_d$ is represented by a $12\times 12$ matrix $\rho(g)$, such that after a rotation the spin components are given by
\begin{equation}
	[\rho(g)S]_\mu^\alpha = \sum_{\nu,\beta} [\rho(g)]_{\mu\nu}^{\alpha\beta} S_\nu^\beta. 
\end{equation}
This is a tensor product representation of a four-dimensional permutation representation and a 3-dimensional pseudo-vector representation, i.e. it can be decomposed as
\begin{equation}
	[\rho(g)]_{\mu\nu}^{\alpha\beta} = [\pi(g)]_{\mu\nu} [\bar{R}(g)]^{\alpha\beta},
	\label{eq:rep_matrix_pi_R}
\end{equation}
where $\pi(g)$ is a permutation matrix and $\bar{R}(g)$ is a pseudo-vector rotation matrix, $\bar{R}(g) \equiv \det[R(g)]R(g)$ with $R(g)$ an ordinary rotation matrix.\footnote{
    For example, a reflection acts on a polar vector as $-1$ (inversion) times a $\pi$ rotation in the reflection plane, while on a pseudovector (axial vector) it acts only as a $\pi$ rotation. 
    }

The fact that this is a tensor product representation allows us to straightforwardly deduce the structure of the irrep decomposition and construct the corresponding invariant linear combinations of spin components. 
First, we note that $T_d$ has five irreducible representations, which are commonly denoted $A_1$, $A_2$, $E$, $T_1$, and $T_2$, and which we also denote by $\bm{1}$, $\bm{\bar{1}}$, $\bm{2}$, $\bm{\bar{3}}$, and $\bm{3}$, respectively. 
The number indicates the dimension of the irrep while the bar indicates that it picks up an additional sign under inversions---${\bm{1}}$ is a scalar representation, $\bar{\bm{1}}$ is a pseudoscalar representation, ${\bm{3}}$ is a vector representation, $\bar{\bm{3}}$ is a pseudovector representation.
In particular, the $T_1$ or $\bm{\bar{3}}$ irrep acts as pseudovector rotations, while the $T_2$ or $\bm{3}$ irrep acts as vector rotations.  
Let $\bm{12}$ denote the full twelve-dimensional reducible representation $\rho$, and let $\bm{4}$ denote the permutation representation, then starting from \cref{eq:rep_matrix_pi_R} we have
\begin{align}
	\bm{12} &= \bm{4} \otimes \bm{\bar{3}}  \nonumber \\
	&= (\bm{1} \oplus \bm{3})\otimes \bm{\bar{3}} \nonumber \\
	&= (\bm{1}\otimes\bm{\bar{3}}) \oplus (\bm{3}\otimes \bm{\bar{3}}).
	\label{eq:irreps12_4_3}
\end{align}
Here we have utilized a simple fact: every permutation representation always contains into a single copy of the trivial representation $\bm{1}$ plus a remainder. 
Furthermore, the permutation representation is faithful, so the remainder must be the faithful $\bm{3}$ irrep.\footnote{
    Faithful means that no two group elements map to the same matrix in the representation.
    }
We can thus anticipate that the twelve dimensional representation can be decomposed into a 3-component pseudovector which is invariant under permuting the spins ($\bm{1}\otimes\bm{3}$), and the remaining nine spin components can be grouped into a $3\times 3$ tensor
This tensor will naturally decompose into its scalar trace, its three anti-symmetric components, and its five symmetric components,\footnote{
    The trace corresponds to the $\bm{\bar{1}}$ representation since one index of the tensor picks up an extra sign under inversions. 
    Similarly, the anti-symmetric part of a rank-2 tensor would normally act as a pseudovector, but the extra sign turns it into a proper vector. 
    We put the bar on the $\bm{\bar{5}}$ to indicate that this piece also gets the extra sign. 
    }
\begin{equation}
	\bm{12} = \bm{\bar{3}} \oplus (\bm{\bar{1}} \oplus \bm{3} \oplus \bm{\bar{5}}),
	\label{eq:irreps12_3_1_3_5}
\end{equation}
Since $T_d$ only has irreps up to three dimensions, the $\bm{5}$ irrep must decompose further into $\bm{2}\oplus \bar{\bm{3}}$ corresponding to the trace-free diagonal and symmetric off-diagonal tensor components. 
In sum, we have the final decomposition
\begin{align}
	\bm{12} &= \bm{\bar{3}} \oplus (\bm{\bar{1}} \oplus \bm{3} \oplus( \bm{2} \oplus \bm{\bar{3}})) \nonumber \\
	&\equiv \Tpar \oplus (A_2 \oplus T_2 \oplus (E \oplus \Tperp)),
	\label{eq:irreps_all}
\end{align}
where we have given two distinct labels to the two copies of the $T_1$ ($\bar{\bm{3}}$) irrep.
Each of the irreps corresponds to a ground state of the tetrahedron. 
Each irrep $I$ corresponds to a linear combination of the four spins with the general form
\begin{equation}
    m_{I}^{a} = \frac{1}{2}\sum_\mu (\uvec{\irrepvec}_I^a)_\mu \cdot \bm{S}_\mu,
    \label{eq:ordering_vectors}
\end{equation}
where $a$ indexes the components of the irrep. 
Each corresponds to a distinct set of ground states of the tetrahedron where the spins are aligned along the local vectors $(\uvec{\irrepvec}_I^a)_\mu$, which are summarized in \cref{tab:multipoles}.

\subsubsection{Multipole Decomposition}

Here we give these irreps physical meaning as magnetic multipole moments of the tetrahedron. 
Thinking of the four spins as an array of dipoles, we construct the net dipole moment (rank-1) and the net rank-2 moment tensor as 
\begin{equation}
    \mathcal{D}^\alpha \propto \sum_{i \in t} S_i^\alpha \qquad \mathcal{B}^{\alpha\beta} \propto \sum_{i\in t} r_i^\alpha S_i^\beta,
    \label{eq:tensor_moments}
\end{equation}
where $\bm{r}_i$ is the position of the $i$'th spin relative to the center of the tetrahedron. 
These correspond to the $\bar{\bm{3}}$ and $(\bm{3}\otimes \bar{\bm{3}}$ terms in the decomposition. 
The rank-2 moment decomposes into its trace---the magnetic monopole moment ($\bar{\bm{1}}$); its antisymmetric part---the magnetic toroidal dipole moment ($\bm{3}$); and its symmetric trace-free part----the magnetic quadrupole moment ($\bar{\bm{5}}$).
For a system with full $O(3)$ symmetry, the 5\nobreakdash-component quadrupole tensor would be irreducible, but when restricting to the subgroup of cubic symmetries it decomposes into~$\cramped{\bm{5} = \bm{2} \oplus \bm{3}}$.

$\bm{T_{1\parallel}}$ \textbf{\emph{Dipole Moment}}---The obvious normal mode is the net magnetic dipole moment of a single tetrahedron, which we normalize as
\begin{equation}
	\mathcal{D}^\alpha \equiv m_{\Tpar}^\alpha = \frac{1}{2}\sum_{\mu} S_\mu^\alpha = \frac{1}{2}\sum_\mu \uvec{e}^\alpha \cdot \bm{S}_\mu,
	\label{eq:irreps_T1_par}
\end{equation}
where we defined $\uvec{e}^\alpha$ as the unit vector pointing along the global Cartesian axis $\alpha \in \{x,y,z\}$.
We use the symbol $\mathcal{D}$ to denote that this corresponds to the net moment of the tetrahedron. 
This quantity is invariant under permutations of the four sublattices, so corresponds to the factor $(\bm{1}\otimes\bm{\bar{3}})$ in \cref{eq:irreps12_4_3}, or $\Tpar$ in \cref{eq:irreps_all}.
As a sum of spins it naturally transforms as a pseudovector.
This normal mode is saturated when the four spins are collinear, corresponding to a set of ferromagnetic ground states shown in \cref{fig:ground_states}(e).

The remaining degrees of freedom are packaged into the rank-2 moment tensor transforming in the $(\bm{3}\otimes \bar{\bm{3}})$ representation, which we normalize as
\begin{equation}
	\bm{\mathcal{B}} = \frac{\sqrt{3}}{2}\sum_\mu \uvec{z}_\mu \otimes \bm{S}_\mu, \quad \mathcal{B}^{\alpha\beta} =\frac{\sqrt{3}}{2} \sum_{\mu} \hat{z}_\mu^{\alpha} S_\mu^{\beta},
	\label{eq:irreps_B}
\end{equation}
where we have replaced the displacement from the center of the tetrahedron $\bm{r}_i$ in \cref{eq:tensor_moments} with their unit vector equivalents, which are the same as the four easy-axis unit vectors $\hat{\bm{z}}_\mu$ (red arrows in \cref{fig:tetrahedron_vectors}).
The remaining normal modes are extracted by decomposing this tensor into its trace, anti-symmetric, and symmetric components. 

\begin{table}[t]
    \centering
    \includegraphics[width=\columnwidth]{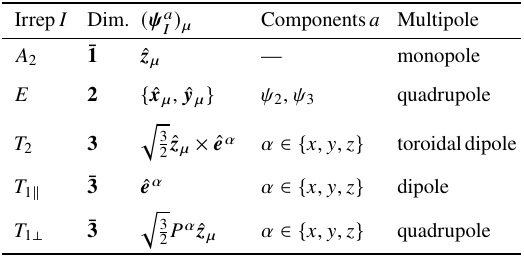}
    \caption{The twelve components of the four spins (magnetic dipoles) on the corners of a single tetrahedron decompose into irreducible representations of the tetrahedral symmetry group $T_d$. Each corresponds to a multipole moment of the tetrahedral spin configuration, where the SO(3) quadrupole moment is split into diagonal and off-diagonal components due to reduction to a cubic symmetry subgroup. Each irrep $I$ has a corresponding set of local unit vectors denoted $(\uvec{\irrepvec}_I^a)_\mu$, where $a$ indexes the components of the irrep and $\mu$ indexes the four corners of the tetrahedron, defining the normal modes and ground states. These are derived from the multipole decomposition of a single tetrahedron, and correspond to the $\bm{q}=\bm{0}$ eigenvectors of the interaction matrix. The vectors $\uvec{e}^\alpha\in\{\uvec{x},\uvec{y},\uvec{z}\}$ are the Cartesian cubic axes along the [001] directions, while the $\uvec{z}_\mu$ lie along local three-fold easy-axes along the [111] directions, cf. \cref{fig:tetrahedron_vectors}. Definitions of the vectors $\uvec{x}_\mu$ and $\uvec{y}_\mu$ are given in \cref{apx:conventions}. The matrices $P^{\alpha}$ are act as a projector orthogonal to $\uvec{e}^\alpha$ times a reflection swapping the remaining two cubic axes.
    }
    \label{tab:multipoles}
\end{table}

$\bm{A_2}$ \textbf{\emph{Monopole Moment}}---The simplest irrep is the psuedo-scalar $A_2$ irrep, extracted from the trace of~$\mathcal{B}$,\footnote{
    Note that each $\bm{m}_I$ is defined with a normalization factor so that \cref{eq:ordering_vectors} is satisfied, which ensures that each is a linear combinations of spins, $\sum_{\mu,\alpha} c_{\mu}^{\alpha} S_\mu^\alpha$, where the coefficients $c_\mu^\alpha$ form a unit-length twelve-component vector.
}
\begin{equation}
	\sqrt{3} m_{A_2} 
	:= 
	\sum_\alpha \mathcal{B}^{\alpha\alpha} 
	= 
	\frac{\sqrt{3}}{2}\sum_\mu \uvec{z}_\mu \cdot \bm{S}_\mu.
    \label{eq:irrep_A2}
\end{equation}
This normal mode is saturated when the spins are ``all-out'' or ``all-in'' along their local easy-axes. 
Such an all-in-all-out configurations is shown in \cref{fig:ground_states}(a). 
This scalar changes sign under inversion since the spins transform as angular momenta, thus transforming as the $A_2$ irrep.
It is natural to identify as the magnetic monopole moment or ``magnetic charge'' of the spin configuration, and indeed it behaves as such in real materials~\cite{castelnovoMagneticMonopolesSpin2008}.

$\bm{T_2}$ \textbf{\emph{Toroidal Dipole Moment}}---Next we can extract the anti-symmetric part, 
\begin{equation}
	\sqrt{2} m_{T_2}^\alpha 
    := 
    \sum_{\beta,\gamma} \epsilon^{\alpha\beta\gamma} \mathcal{B}^{\beta\gamma} 
    = 
    \frac{\sqrt{3}}{2}\sum_\mu (\uvec{z}_\mu \times \uvec{e}^\alpha)\cdot \bm{S}_\mu.
    \label{eq:irrep_T2}
\end{equation}
First, note that this contains no contributions from the easy-axis components which are killed by the cross product, thus it corresponds to a set of ground states with spins orthogonal to their easy-axes.
This normal mode is saturated when the spins are coplanar and form a chiral configuration relative to one of the Cartesian axes, known as the Palmer-Chalker ground states~\cite{palmerOrderInducedDipolar2000}. 
There are six such ground states, corresponding to two chiralities about each of the three Cartesian cubic axes, one is shown in \cref{fig:ground_states}(d).

$\bm{T_{1\perp}}$ \textbf{\emph{Quadrupole Moment (off-diagonal)}}---Lastly we have the symmetric part of the tensor, the five-component quadrupole moment, which separates under cubic symmetries into its diagonal-trace-free and off-diagonal parts.\footnote{
    This decomposition is widely known as $e_g \oplus t_{2g}$ in the context of the larger cubic point symmetry group $O_h$.
}  
The symmetric off-diagonal components of the tensor can be extracted by defining 
\begin{equation}
    \sqrt{2} m_{\Tperp}^\alpha  
    = 
    \sum_{\beta,\gamma} \vert \epsilon^{\alpha\gamma\beta}\vert \mathcal{B}^{\gamma\beta} 
    = 
    \frac{\sqrt{3}}{2}\sum_\mu \sum_{\beta,\gamma} \vert \epsilon^{\alpha\gamma\beta}\vert \hat{z}_\mu^\gamma S_\mu^\beta,
    \label{eq:T1_perp}
\end{equation}
the corresponding normal mode vectors lie in the Cartesian plane orthogonal to the $\alpha$ axis with two pointing towards the central axis and the other two pointing away, shown in \cref{fig:ground_states}(f). 
To see this, note that for each $\alpha$ the matrix $P^{\alpha}$ with components $P_{\beta\gamma}^{\alpha} = \vert \epsilon^{\alpha\beta\gamma} \vert$ acts as (\textit{i}) a projector orthogonal to $\uvec{e}^\alpha$ and (\textit{ii}) a reflection swapping the remaining two Cartesian axes. 
It projects each $\uvec{z}_\mu$ into one Cartesian plane, where they all point away from the center of the tetrahedron, then two of them on opposite corners are mirrored.

$\bm{E}$ \textbf{\emph{Quadrupole Moment (diagonal)}}---Finally, the diagonal-trace-free part.
This is the hardest to represent in a simple geometric form, but it can be expressed as the three components of $\mathcal{B}^{\alpha\alpha} - (1/3) \Tr\mathcal{B}$.
By construction only two of the components are linearly independent since summing over $\alpha$ yields zero. 
However, there is a freedom to distribute the trace subtraction over the three elements. 
It is useful therefore to define a 3-component quantity which contains the diagonal components, combining the $A_2$ and $E$ irreps together,
\begin{equation}
    \mathcal{Q}^{\alpha} = \mathcal{B}^{\alpha\alpha}
\end{equation}
such that the trace is given by
\begin{equation}
    \mathcal{Q}_{A_2} \equiv \sqrt{3}\, m_{A_2} := \Tr\mathcal{B} = (1,1,1)\cdot \bm{\mathcal{Q}}.
\end{equation}
We can then separate the trace-free $E$ components by choosing two basis vectors orthogonal to $(1,1,1)$, defining
\addtocounter{equation}{1} 
\begin{align}
    \mathcal{Q}_{\psi_2} \equiv\sqrt{3} \, m_{E}^{\psi_2} &= \sqrt{3}\,\uvec{x}_1 \cdot \bm{\mathcal{Q}}
    \,\,\, \text{with} \,\,\, 
    \uvec{x}_1=\frac{1}{\sqrt{6}}(1, 1, -2), 
    \tag{\theequation a}
    \\ 
    \mathcal{Q}_{\psi_3} \equiv \sqrt{3} \, m_{E}^{\psi_3} &= \sqrt{3}\,\uvec{y}_1 \cdot \bm{\mathcal{Q}}
    \,\,\, \text{with} \,\,\, 
    \uvec{y}_1 = \frac{1}{\sqrt{2}}(1,-1,0),
    \tag{\theequation b}
\end{align}
which are then packaged into the 2-component $E$ normal mode,
\begin{equation}
    \bm{m}_E^a = 
        \begin{pmatrix}
            m_E^{\psi_2} \\[4pt] m_E^{\psi_3}
        \end{pmatrix}^a.
    \label{eq:irrep_E}
\end{equation}
The corresponding ordering vectors are harder to write down than the previous ones, but they can be obtained by noting that  $\uvec{x}_1$ and $\uvec{y}_1$ form an orthonormal basis with $\uvec{z}_1$ on one corner of the tetrahedron, which can be extended to an orthonormal basis $\{\uvec{x}_\mu,\uvec{y}_\mu,\uvec{z}_\mu
\}$ on each of the four corners of the tetrahedron using the fourfold improper rotation symmetry (ninety degree rotations about a Cartesian axis followed by mirroring through the orthogonal plane), given explicitly in \cref{apx:conventions}.
These correspond to the so-called $\psi_2$ and $\psi_3$ configurations, shown in \cref{fig:ground_states}(b,c), where in the $\psi_2$ configurations spins lie in local mirror planes, while in the $\psi_3$ configurations spins lie along a local 2-fold rotation axis.

\subsubsection{Single Tetrahedron Irrep Energies}

Inverting these definitions, we can re-package the irrep normal modes $m_I^a$ into the vector and tensor degrees of freedom as
\addtocounter{equation}{1} 
\begin{align*}
    \bm{\mathcal{D}} &= \begin{pmatrix}
        m_{\Tpar}^x \\[1pt]
        m_{\Tpar}^y \\[1pt]
        m_{\Tpar}^z
    \end{pmatrix},
    \tag{\theequation a}
    \label{eq:tensor_M}
    \\[1ex]
    \bm{\mathcal{B}} &= \frac{1}{\sqrt{2}}\begin{pmatrix}
        \sqrt{2}\, \mathcal{Q}^x & m_{T_{1\perp}}^z + m_{T_2}^z & m_{\Tperp}^y - m_{T_2}^y 
        \\[4pt]
        m_{\Tperp}^z - m_{T_2}^z & \sqrt{2}\, \mathcal{Q}^y  & m_{\Tperp}^x + m_{T_2}^x 
        \\[4pt]
        m_{\Tperp}^y + m_{T_2}^y & m_{\Tperp}^x - m_{T_2}^x & \sqrt{2}\, \mathcal{Q}^z
    \end{pmatrix}, 
    \tag{\theequation b}
    \label{eq:tensor_B}
    \\[4pt]
    \bm{\mathcal{Q}} &= 
        \frac{1}{\sqrt{3}}
        \begin{pmatrix}
            m_{A_2} + \frac{1}{\sqrt{2}} m_E^{\psi_2} + \sqrt{\frac{3}{2}} m_E^{\psi_3} 
            \\[1pt]
            m_{A_2} + \frac{1}{\sqrt{2}} m_E^{\psi_2} - \sqrt{\frac{3}{2}} m_E^{\psi_3} 
            \\[1pt]
            m_{A_2} - \sqrt{2} m_E^{\psi_2}
    \end{pmatrix}.
    \tag{\theequation c}
    \label{eq:tensor_Q}
\end{align*}
Thus we have demonstrated how the irreducible representation normal modes can be conveniently packaged into a tensors which encode the tetrahedral multipole moments.
We can compute the energies of the different irreps, i.e. the $\bq=\bm{0}$ eigenvalues in \cref{eq:H_bands}, as\footnote{
    The prefactor normalizes the $(\uvec{\irrepvec}_I^a)_\mu$ when treated as a single 12-component vector. Note that since these are defined as unit vectors on each sublattice, it follows that $\sum_{\mu} \vert (\uvec{\irrepvec}_I^a)_\mu\vert^2 = 2$.
}
\begin{equation}
    \Jeig_I^a 
    = 
    \frac{1}{4}
    \sum_{\mu>\nu}\sum_{\alpha\beta} 
    (\hat{\irrepvec}_I^a)_\mu^\alpha 
    \Jmat_{\mu\nu}^{\alpha\beta} 
    (\hat{\irrepvec}_I^a)_\nu^\beta.
    \label{eq:irrep_energies}
\end{equation} 
For $I\neq T_1$, these are the components of the $\bm{q}=\bm{0}$ eigenvectors of $\Jmat$ on a single tetrahedron, and the $\Jeig_I^a$ are the corresponding eigenvalues.
Symmetry guarantees that eigenvalues corresponding to the same irrep $I$ are degenerate, i.e. the $\smash{J_I^a\equiv J_I}$ are independent of $a$. These values are listed in \cref{tab:irrep_energies}. 
The single-tetrahedron Hamiltonian can be expressed as
\begin{equation}
    H = \sum_{I \in \{A_2, E, T_2, \Tpar, \Tperp\}} J_I \vert \bm{m}_I\vert^2 
    \, + \,\,
    2J_{\Tpar\cdot\Tperp} \bm{m}_{\Tpar}\cdot \bm{m}_{\Tperp}
\end{equation}
where we included a symmetry-allowed cross-term between the two $T_1$ irreps, which we treat in the next section. 
The tetrahedron energy can also be expressed in terms of the multipole moments by using the following identities
\addtocounter{equation}{1} 
\begin{align}
    \vert\bm{\mathcal{D}}\vert^2 &= \vert\bm{m}_{\Tpar}\vert^2 
    \tag{\theequation a}
    \\
    \frac{1}{4}\Tr[(\bm{\mathcal{B}}+\bm{\mathcal{B}}^T)^2] &= m_{A_2}^2 + \vert\bm{m}_E\vert^2 + \vert\bm{m}_{\Tperp}\vert^2,
    \tag{\theequation b}
    \\
    \frac{1}{4}\Tr[(\bm{\mathcal{B}}-\bm{\mathcal{B}}^T)^2] &= -\vert\bm{m}_{T_2}\vert^2,
    \tag{\theequation c}
    \\
    \frac{1}{3}\Tr[\bm{\mathcal{B}}]^2 &= m_{A_2}^2,
    \tag{\theequation d}
    \\
    \vert \bm{\mathcal{Q}}\vert^2 &= m_{A_2}^2 + \vert\bm{m}_E\vert^2.
    \tag{\theequation e}
\end{align}
The last term is only allowed due to the reduction from SO(3) to cubic rotational symmetries, which splits the energies of the $E$ and $\Tperp$ irreps. 
Note that we have the sum rule
\begin{equation}
    \Tr[\bm{\mathcal{B}}\bm{\mathcal{B}}^T] + \vert\bm{\mathcal{D}}\vert^2 = \sum_I \vert\bm{m}_I\vert^2 = \sum_\mu \vert\bm{S}_\mu\vert^2 = \text{const}.
\end{equation}
The cubic symmetry-allowed coupling between the $\smash{\Tperp}$ and $\smash{\Tpar}$ can be written as
\begin{equation}
    \bm{m}_{\Tpar}\cdot\bm{m}_{\Tperp} = \frac{1}{\sqrt{2}}\sum_{\alpha\beta\gamma} \vert\epsilon^{\alpha\beta\gamma}\vert \mathcal{D}^\alpha \mathcal{B}^{\beta\gamma}.
\end{equation}
We discuss decoupling the two $T_1$ irreps in the next section.

\begin{figure}[t]
    \centering
    \includegraphics[width=\columnwidth]{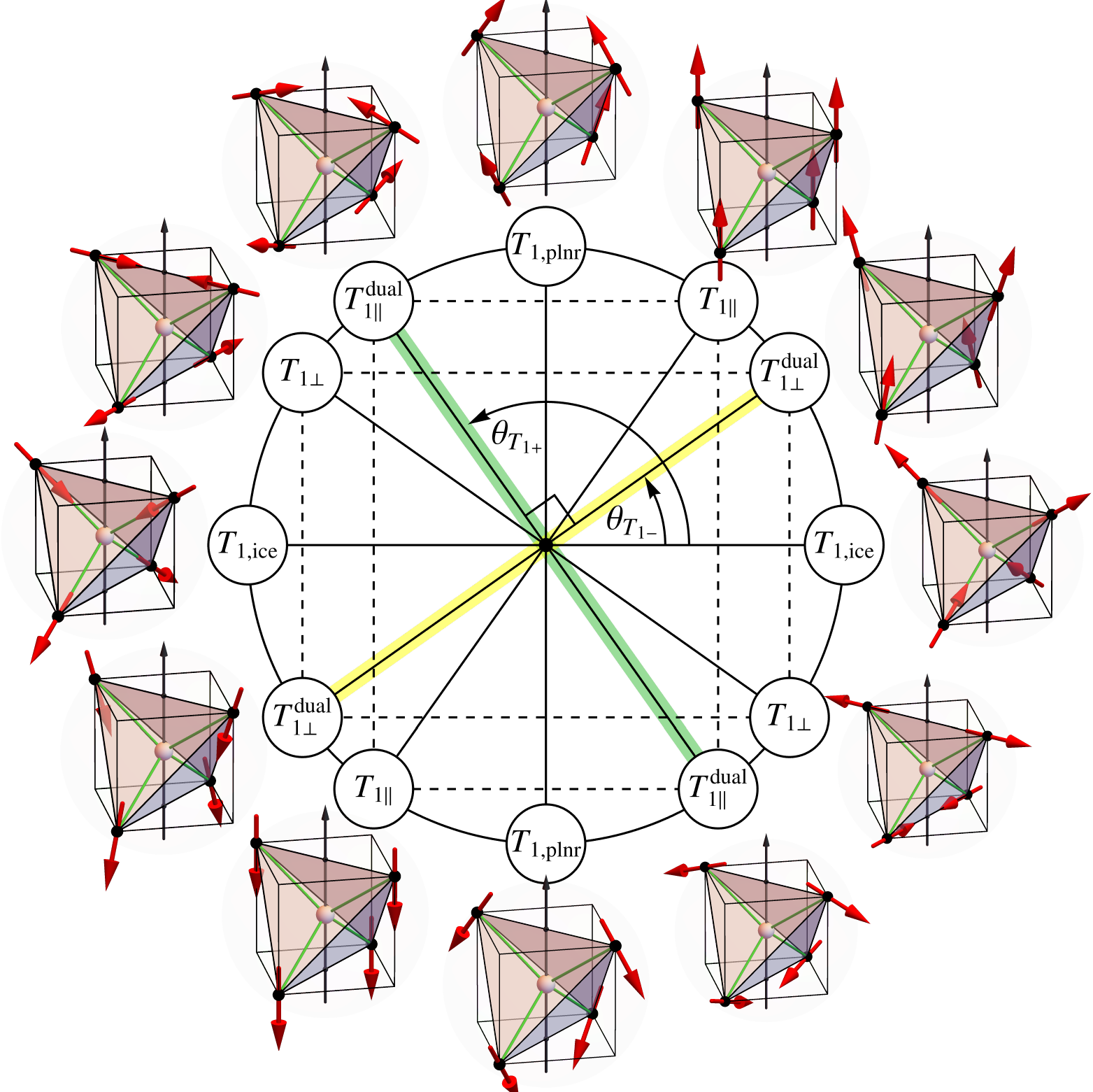}
    \caption{The mixing of the two $T_1$ irreps induces canting of the spins. The canting angle relative to the local easy-axes $\theta_c$, which decouples the two $T_1$ irreps, is defined modulo $\pi/2$ (\textit{cf}.~\cref{eq:canting_angle_local}). Defining $\theta_{T_{1-}}$ as the canting angle of the lower-energy $T_1$ irrep, then $\theta_{T_{1+}} = \theta_{T_{1-}}+\pi/2$ is the canting angle of the higher-energy $T_1$ irrep, and these two angles are defined modulo $\pi$. There are some special canting angles which we note here: $T_{1,\text{ice}}$ corresponds to spins aligned in 2-in 2-out fashion along the easy-axes, whose partner $\smash{T_{1,\text{planar}}}$ is ferromagnetic with spins orthogonal to the easy-axes. $T_{1\parallel}$ is the collinear spin configuration along the global Cartesian axes ($\smash{\theta_{\Tpar}=\cos^{-1}(1/\sqrt{3})}$), whose partner $\smash{T_{1\perp}}$ has all spins lying in a plane, cf. \cref{fig:ground_states}(e,f). 
    Each angle $\theta$ has a dual angle $-\theta$  indicated by the dashed lines, corresponding to rotating all spins by $\pi$ about their local easy-axes, which is the dual irrep when the sign of $J_{z\pm}$ is flipped. Note also that sending $\theta\to\theta+\pi$ corresponds to flipping all spins.}
    \label{fig:T1_mixing}
\end{figure}

\subsection{\texorpdfstring{$T_1$}{T1} Mixing and Canting Angle}

Since there are two $T_1$ irreps they can have a symmetry-allowed coupling in the Hamiltonian and can therefore mix. 
The $\Tpar$ and $\Tperp$ irrep order parameters derived from the multipole decomposition we will refer to as the global $T_1$ irreps, since they involve the components of spins relative to the global Cartesian axes. 
Their coupling in the Hamiltonian is of the form
\begin{equation}
    H_{\Tpar\oplus\Tperp} = 
    \begin{pmatrix}
	   \bm{m}_{\Tpar} \\ \bm{m}_{\Tperp}
	\end{pmatrix}^T
    \begin{pmatrix}
	   J_{\Tpar} & J_{\Tpar\cdot\Tperp} \\ J_{\Tpar\cdot\Tperp} & J_{\Tperp}
	\end{pmatrix}
    \begin{pmatrix}
	   \bm{m}_{\Tpar} \\ \bm{m}_{\Tperp}
	\end{pmatrix},
    \label{eq:T1_quadratic_form}
\end{equation}
where the cross term is controlled by the global $J_3$ parameter in \cref{eq:Jijab} and the single ion anisotropy,
\begin{equation}
    J_{\Tpar\cdot\Tperp} = \frac{\sqrt{2}}{3}(J_{\text{SIA}} + 3J_3),
    \label{eq:T1_mixing_J3}
\end{equation}
which prefer for spins to cant away from the global Cartesian axes and towards the local three-fold axes.
The amount of canting is determined by eliminating the coupling between the two $T_1$ irreps, which is achieved by defining rotated $T_1$ modes
\begin{equation}
	\begin{pmatrix}
	\bm{m}_{T_{1\phi}}\\
	\bm{m}_{T_{1\phi'}}
	\end{pmatrix}
	=
	\begin{pmatrix}
	\cos\phi & \sin \phi \\ -\sin\phi & \cos\phi
	\end{pmatrix}
	\begin{pmatrix}
	\bm{m}_{\Tpar} \\ \bm{m}_{\Tperp}
	\end{pmatrix}.
     \label{eq:canting_irreps}
\end{equation}
These rotated $T_1$ modes have a simple interpretation: each spin cants in one of the mirror planes according to
\begin{equation}
    {m}^{\alpha}_{T_{1\phi}} 
    = 
    \sum_\mu 
    \underbrace{
        (
        \cos(\phi) \uvec{e}^\alpha 
        + 
        \sin(\phi) (\uvec{\irrepvec}_{T_{1,\perp}}^{\alpha})_\mu)
    }_{
        (\displaystyle \uvec{\irrepvec}_{T_{1\phi}}^\alpha)_\mu
    } 
    \cdot \bm{S}_\mu .
    \label{eq:canted_irrep}
\end{equation}
and similarly for $T_{1\phi'}$. 
The critical angle $\phi_{c}$ for which the cross-term between the two $T_1$ irreps vanishes is given in terms of the global irrep parameters by solving the equation
\begin{equation}
    \tan(2\phi_{c}) = \frac{2J_{\Tpar\cdot\Tperp}}{J_{\Tperp}-J_{\Tpar}}.
    \label{eq:canting_angle_global}
\end{equation}
Note that $\phi_c$ is only defined modulo $\pi/2$, i.e. it does not tell us whether the $T_1$ irrep with angle $\phi_c$ or $\phi_c + \pi/2$ has lower energy. 
Diagonalizing the matrix in \cref{eq:T1_quadratic_form} yields the energies of the decoupled $T_1$ normal modes
\begin{equation}
    J_{T_{1\pm}} = \frac{1}{2}\left( J_{\Tpar}+J_{\Tperp} \pm \sqrt{(J_{\Tpar}-J_{\Tperp})^2 + 4 {J_{\Tpar\cdot\Tperp}}^2}\,\right),
\end{equation}
where the $\pm$ indicates the higher or lower energy eigenvalue.
For solutions of \cref{eq:canting_angle_global} in the range $\phi_{c}\in(-\pi/4,\pi/4)$, if $J_{\Tpar}< J_{\Tperp}$ then $T_{1-}=T_{1\phi_{c}}$, while the opposite inequality implies $T_{1-} = T_{1\phi_c'}$.

\subsubsection{Easy-Axis Limit of \texorpdfstring{$T_1$}{T1} Irreps}

It is through canting and mixing the two $T_1$ irreps that one connects the global irreps to the local symmetry-adapted description in \cref{eq:H_local}. 
Writing the canting angle equation \cref{eq:canting_angle_global} in terms of the local couplings,
\begin{equation}
    \tan(2\phi_{c}) = 2\sqrt{2} - \frac{72 J_{z\pm}}{2J_{\pm} + 4 J_{\pm\pm} + J_{zz} + 16\sqrt{2} J_{z\pm} - J_{\text{SIA}}},
\end{equation}
exposes a special critical angle---when $J_{z\pm}=0$ the canting angle is given by
\begin{equation}
    \phi_0 \equiv \tan^{-1}(2\sqrt{2})/2 = \cos^{-1}(\sqrt{2/3}).
\end{equation}
At this special canting angle, the local vectors in \cref{eq:canted_irrep} are aligned along or orthogonal to the local three-fold axis $\uvec{z}_\mu$.
This occurs precisely when $J_{z\pm}=0$, i.e. when the local $z$-components of the spin are completely decoupled from the local transverse components.
We refer to the two corresponding irreps as
\begin{equation}
    T_{1\phi_0} \equiv T_{1,\text{planar}}, \quad T_{1\phi_0'}\equiv T_{1,\text{ice}},
\end{equation}
using the notation of Ref.~\cite{yanTheoryMultiplephaseCompetition2017}.
The $T_{1,\text{ice}}$ irrep corresponds to the ground states of spin ice, with two spins pointing ``in'' and two spins pointing ``out'' of the tetrahedron.
The $T_{1,\text{planar}}$ irrep corresponds to ferromagnetic configurations with spins in their local easy-planes.
It is very useful to express the canting angle relative to the $[111]$ local easy-axes, rather than the $[001]$ global Cartesian axes, by defining 
\begin{equation}
    \theta_c = \phi_c - \phi_0.
\end{equation} 
From hereon we will refer to $\theta_c$ as \emph{the} canting angle, and $\phi_c$ will not be used.
Relative to the [111] axis, the analog of \cref{eq:canting_angle_global} is
\begin{equation}
    \tan(2\theta_c) = 
    \frac{
        2J_{T_{1,\text{ice}}\cdot T_{1,\text{planar}}}
    }{
        J_{T_{1,\text{ice}}} \! - J_{T_{1,\text{planar}}}
    } = \frac{8J_{z\pm}}{2J_{\pm}+4J_{\pm\pm}+J_{zz}-J_{\text{SIA}}}.
    \label{eq:canting_angle_local}
\end{equation}
The corresponding couplings are listed in \cref{tab:irrep_energies}.
From this equation it is evident that switching the sign of $J_{z\pm}$ corresponds to switching the sign of $\theta_c$.

This is all summarized in \cref{fig:T1_mixing}, which shows how the spin configuration of the $T_1$ irreps evolves as the canting angle rotates. 
To each configuration there is a corresponding dual configuration with the opposite sign of the canting angle, corresponding to a $\pi$ rotation of each spin about its local easy-axis. 
For a given set of exchange parameters, one of the two $T_1$ irreps will have a lower energy; we denote the corresponding angle by $\theta_{T_{1-}}$, and the higher energy irrep has angle $\theta_{T_{1+}} = \theta_{T_{1-}} + \pi/2$. 
These angles are defined modulo $\pi$, since a $\pi$ rotation returns to the same $T_1$ irrep with the spin directions reversed, i.e. the time-reversed state.

\begin{table*}[t]
    \centering
    \includegraphics[width=\textwidth]{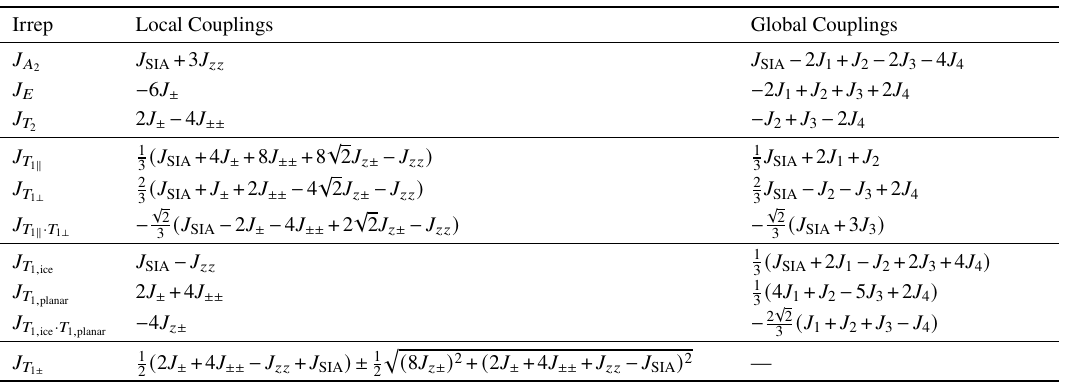}
    \caption{Irrep energies, \cref{eq:irrep_energies}, and $T_1$ couplings, \cref{eq:T1_quadratic_form}, in terms of the global and local basis interaction parameters.}
    \label{tab:irrep_energies}
\end{table*}

\section{Irrep Parameterization of the Model Space}
\label{sec:model_space_parameterization}

In the remainder of this paper we will elucidate the structure of the full classical phase diagram of the nearest-neighbor spin-1/2 model on the pyrochlore lattice, focusing especially on the locations of the phase boundaries and the triple and quadruple points where three or four phases become degenerate, respectively. 
The irrep eigenvalues are given in \cref{tab:irrep_energies}.
For a given set of parameters, the classical ground state is given by finding which irrep takes the minimum eigenvalue and aligning the spins along the associated ordering vectors in \cref{eq:ordering_vectors}.
For classical fixed-length vector spins this is the minimal energy state, and away from phase boundaries this should be an accurate product state ansatz for the quantum ground state.
As we will see in detail, there are effectively four phases, one for each irrep if we don't distinguish between the two $T_1$ irreps.
Each phase is the locus of parameter space within which one irrep has the lowest energy and the others are separated by a gap. 
In order to map the phase diagram it is crucial to identify the locus of parameters along which multiple ground states are degenerate, i.e. phase boundaries, triple points, and quadruple points.

\subsection{Irrep Parameterization and \texorpdfstring{$J_{z\pm}$}{Jzpm} Duality}
\label{sec:jzpm_duality}

Our strategy is to parameterize the phase diagram not in terms of the basis-dependent spin-spin couplings, but rather in terms of the basis-independent irrep energies $J_I$. 
To achieve this, we first invert the relations in \cref{tab:irrep_energies} in order to obtain the couplings as functions of the irrep eigenvalues. 
The resulting equations are
\begin{subequations}
\label{eq:Jirrep_Jloc}
\begin{align}
	J_{zz} &= \frac{1}{12}[3J_{A_2} - 2J_E - 3J_{T_2} - 3(J_{T_{1+}}+ J_{T_{1-}})]
    \\
	J_{\pm} &= -\frac{J_E}{6} 
    \\
	J_{\pm\pm} &= -\frac{1}{12}(J_E + 3 J_{T_2}) 
    \\
	\vert J_{z\pm}\vert &= \frac{1}{12} \sqrt{-(2J_E + 3 J_{T_2} + 3J_{T_{1+}})(2J_E + 3 J_{T_2} + 3J_{T_{1-}})}  \label{eq:Jzpm}
    \\
	J_{\text{SIA}} &= \frac{1}{4}[J_{A_2} + 2J_E + 3J_{T_2} + 3(J_{T_{1-}} + J_{T_{1+}})] \label{eq:JSIA}
\end{align}
\end{subequations}
For any model with a given sign of $J_{z\pm}$, there is another model with the opposite sign of $J_{z\pm}$ with the same irrep energies, because only its square appears in the irrep energies $J_{T_{1\pm}}$ in \cref{tab:irrep_energies}.

This corresponds to a duality transformation of the Hamiltonian: applying a $\pi$ rotation of each spin about its local $\uvec{z}_i$ axis sends $S_i^\pm \to -S_i^\pm$, changing the sign of the $S^z S^{\pm}$ term in \cref{eq:H_local} while leaving the rest of the terms invariant. 
This sign change can then be cancelled by changing the sign of $J_{z\pm}$. 
Furthermore, such a rotation maps each irrep back to itself: $A_2$ is clearly invariant under this transformation; for the easy-plane $E$ and $T_2$ configurations this sends a spin configuration to its time-reversed configuration; and $T_1$ configurations are turned into other $T_1$ configurations. As such, reversing the sign of $J_{z\pm}$ relates two Hamiltonians in the same phase. 
This is clear from \cref{tab:irrep_energies} since the energies of the irreps are unchanged.

It is convenient to parameterize the splitting between the two $T_1$ irreps, 
\begin{equation}
    J_{T_{1\pm}} \equiv J_{T_1} \pm J_{\delta T_1} 
    \quad 
    \text{with}
    \quad 
    J_{\delta T_1}\geq 0 ,
    \label{eq:T1_splitting}
\end{equation}
in order to write \cref{eq:Jzpm} in the form
\begin{equation}
    \vert J_{z\pm}\vert = \frac{1}{12} \sqrt{ (3J_{\delta T_1})^2 - (2J_{E} + 3J_{T_2} + 3J_{T_1})^2}.
    \label{eq:Jzpm_T1_splitting}
\end{equation}
In this form we can see the close relation between $J_{z\pm}$ and the splitting of the two $T_1$ irreps.
In particular, there is a constraint on the possible eigenvalues,
\begin{equation}
    3J_{\delta T_1} \geq \vert 2J_{E} + 3J_{T_2} + 3J_{T_1} \vert,
    \label{eq:T1_inequality}
\end{equation}
in order that $J_{z\pm}$ is real.
It is evident that if the two $T_1$ irreps are degenerate ($J_{\delta T_1}=0$) then $J_{z\pm}$ must be zero, saturating the inequality.
\Cref{eq:Jzpm_T1_splitting} suggests that it is convenient to define a dimensionless parameter 
\begin{equation}
    \alpha \equiv \sign(J_{z\pm}) \frac{(2J_{E} + 3J_{T_2} + 3J_{T_1})}{3 J_{\delta T_1}},
    \label{eq:alpha}
\end{equation}
so that
\begin{equation}
    \vert J_{z\pm} \vert = \frac{J_{\delta T_1}}{4}\sqrt{1 - \alpha^2},
\end{equation}
such that the inequality \cref{eq:T1_inequality} is equivalent to $\vert\alpha\vert \leq 1$. 
The sign prefactor in \cref{eq:alpha} has been chosen because it allows us to interpret the canting formula \cref{eq:canting_angle_local} geometrically---it simplifies to
\begin{equation}
    \tan(2\theta_c) = \frac{\sqrt{1-\alpha^2}}{\alpha} = \tan\left(\cos^{-1}(\alpha)\right), 
\end{equation}
from which we obtain the convenient formula
\begin{equation}
    \theta_c = \frac{1}{2} \cos^{-1}(\alpha) = \cos^{-1}\left(\sqrt{\frac{\alpha+1}{2}}\right) \mod \pi/2.
    \label{eq:canting_angle_alpha}
\end{equation}

\begin{table*}[t]
	\centering
    \begin{tabular}{ 
    M{.15\textwidth} 
    M{.10\textwidth}  
    p{.53\textwidth}  
    M{.12\textwidth}
    M{.06\textwidth}}
	\toprule
	Degenerate & Tuned & \hspace{23ex} Parameterization & Range & Refs.
    \\
	Irreps &  Irreps &  & 
	\\
    \midrule
		$A_2\oplus E \oplus T_2 \oplus T_{1\perp}$ 
		& 
		$T_{1\parallel}$ 
		&
        \begin{tabular}{p{.25\columnwidth}p{.25\columnwidth}p{.25\columnwidth} p{.25\columnwidth}} 
            $J_{zz} = -\frac{1}{12}\Delta_{T_{1+}}$
            &
    		$J_{\pm} = \frac{1}{24}\Delta_{T_{1+}}$
            &
    		$J_{\pm\pm} = \frac{1}{12}\Delta_{T_{1+}}$
            &
    		$\vert J_{z\pm} \vert = \frac{1}{6\sqrt{2}}\Delta_{T_{1+}}$
        \end{tabular}
        &
        ---
        &
        \cite{moessnerLowtemperaturePropertiesClassical1998,henleyPowerlawSpinCorrelations2005}
	\\
    \midrule
		$E \oplus T_2 \oplus T_{1-}$ 
		& 
		$A_2\oplus T_{1+}$ 
		&
        \begin{tabular}{p{.4\columnwidth} p{.66\columnwidth}}
            $J_{zz} = \frac{1}{36}(11\Delta_{A_2} - 3 \Delta_{T_{1+}})$ 
        	& 
            $J_{\pm} = \frac{1}{72}(\Delta_{A_2} + 3 \Delta_{T_{1+}}) $
            \\[.7ex]
            $J_{\pm\pm} = \frac{1}{36}(\Delta_{A_2} + 3 \Delta_{T_{1+}})$ 
            &
        	$\vert J_{z\pm} \vert = \frac{1}{36}\sqrt{(3 \Delta_{T_{1+}}-2\Delta_{A_2})(2\Delta_{A_2} + 6 \Delta_{T_{1+}})}$
        \end{tabular}
        &
        $-3\leq {\displaystyle  \frac{\Delta_{A_2}}{\Delta_{T_{1+}}}} \leq \frac{3}{2}$
        &
        \cite{bentonSpinliquidPinchlineSingularities2016}
    \\[4ex]
		$A_2\oplus T_2 \oplus T_{1-}$ 
		& 
		$E \oplus T_{1+}$ 
		& 
        \begin{tabular}{p{.4\columnwidth} p{.66\columnwidth}}
    		$J_{zz} = -\frac{1}{36}(2\Delta_E + 3 \Delta_{T_{1+}}) $
    		&
            $J_{\pm} = \frac{1}{72}(-10\Delta_{E} + 3 \Delta_{T_{1+}}) $
    	    \\[.7ex]
            $J_{\pm\pm} = \frac{1}{36}(-\Delta_{E} + 3 \Delta_{T_{1+}})$
            &
    		$\vert J_{z\pm} \vert = \frac{1}{36}\sqrt{(6 \Delta_{T_{1+}}-2\Delta_{E})(3 \Delta_{T_{1+}} + 2\Delta_E)}$
        \end{tabular}
        &
        $-\frac{3}{2}\leq \displaystyle\frac{\Delta_{E}}{\Delta_{T_{1+}}} \leq 3$
        &
        \cite{franciniHigherRankSpinLiquids2025}
    \\[4ex]
		$A_2\oplus E \oplus T_{1-}$ 
		& 
		$T_2 \oplus T_{1+}$ 
		& 
        \begin{tabular}{p{.4\columnwidth} p{.66\columnwidth}}
    		$J_{zz} = -\frac{1}{12}(\Delta_{T_2} + \Delta_{T_{1+}})$
            &
            $J_{\pm} = \frac{1}{24}(\Delta_{T_2} + \Delta_{T_{1+}})$
    	    \\[.7ex]
            $J_{\pm\pm} = \frac{1}{12}(\Delta_{T_{1+}}-2\Delta_{T_2}) $
            &
    		$\vert J_{z\pm} \vert = \frac{1}{12}\sqrt{(2\Delta_{T_{1+}}-\Delta_{T_2})(\Delta_{T_{1+}} + \Delta_{T_2})}$
        \end{tabular}
        &
        $-1\leq \displaystyle\frac{\Delta_{T_2}}{\Delta_{T_{1+}}} \leq 2$
        &
        \cite{franciniExactNematicMixed2025}
    \\[4ex]
  		$A_2\oplus E \oplus T_2 $ 
		& 
		$T_{1-}\oplus T_{1+}$ 
		& 
        \begin{tabular}{p{.22\columnwidth}p{.185\columnwidth}p{.21\columnwidth} p{.42\columnwidth}} 
            $J_{zz} = -\frac{1}{6}\Delta_{T_1} $
            &
            $J_{\pm} = \frac{1}{12}\Delta_{T_1}$ 
            &
            $J_{\pm\pm} = \frac{1}{6}\Delta_{T_1} $
            &
            $\vert J_{z\pm} \vert = \frac{1}{12}\sqrt{(3\Delta_{\delta T_1})^2-\Delta_{T_1}^2}$
        \end{tabular}
        &
        $-3 \leq {\displaystyle \frac{\Delta_{T_1}}{\Delta_{\delta T_1}}} \leq 3$
	\\[2.5ex]
    \midrule
		$T_2 \oplus T_{1,\text{ice}} \oplus T_{1,\text{plnr}}$ 
		& 
		$E \oplus A_2$ 
		& 
        \begin{tabular}{p{.25\columnwidth}p{.25\columnwidth}p{.25\columnwidth} p{.25\columnwidth}} 
            $J_{zz} = \frac{1}{4}\Delta_{A_2}$
            &
            $J_{\pm} = -\frac{1}{8}\Delta_{A_2}$
            &
            $J_{\pm\pm} = 0$
            &
            $J_{z\pm} = 0$
        \end{tabular}
        &
        $\Delta_E = \Delta_{A_2}$
        &
        \cite{taillefumierCompetingSpinLiquids2017}
	\\[1ex]
		$\phantom{A_2}\mathllap{E} \oplus T_{1,\text{ice}} \oplus T_{1,\text{plnr}}$ 
		& 
		$T_2\oplus A_2$ 
		& 
        \begin{tabular}{p{.25\columnwidth}p{.25\columnwidth}p{.25\columnwidth} p{.25\columnwidth}} 
            $J_{zz} = \frac{1}{4}\Delta_{A_2}$
            &
            $J_{\pm} = \frac{1}{24}\Delta_{A_2}$
            &
            $J_{\pm\pm} = -\frac{1}{12}\Delta_{A_2}$
            &
            $J_{z\pm} = 0$
        \end{tabular}
        &
        $\Delta_{T_2} = \frac{2}{3}\Delta_{A_2}$
        &
        \cite{lozano-gomezCompetingGaugeFields2024}
	\\[1ex]
		$A_2 \oplus T_{1,\text{ice}} \oplus T_{1,\text{plnr}}$ 
		& 
		$E\oplus T_2$ 
		& 
		\hspace{.22\textwidth} --- 
        & 
        ---
        &
	\\
	\bottomrule
    \end{tabular}
	\caption{List of fourfold and threefold degenerate models. There are two parameter combinations (up to an overall scale) yielding fourfold degeneracy, both of which have $T_{1\parallel}$ gapped: the Heisenberg antiferromagnet (HAFM) (with $J_{z\pm}>0$), and its $J_{z\pm}$-dual (with $J_{z\pm}<0$), denoted \HAFMdual.
    By gapping one of the four degenerate irreps, we obtain four lines (up to a scale) along which three phases meet, which go from the HAFM, to $J_{z\pm} = 0$, then return with the opposite sign of $J_{z\pm}$ to the \HAFMdual. 
    For the last line we used the $T_1$ splitting parameters defined in \cref{eq:T1_splitting_relative_Delta}.
    The remaining three combinations of three degenerate irreps have both $T_1$ irreps degenerate, which is only possible when $J_{z\pm}=0$. 
    Two of these combinations are possible, corresponding to isolated triple points (up to a scale) which do not lie on one of the four triple lines.
    There is one other possible combination of three degenerate and two gapped irreps, but it does not admit any solutions with real $J_{z\pm}$, because $A_2$ is the charge irrep for $T_{1\text{ice}}$, and their ground states are mutually incompatible (cf. \cref{sec:charge-flux-intertwined}).
    We have given references for cases where examples have appeared previously in the literature (cf. \cref{sec:pyro_spin_liquids}).
    Flat band degeneracies for all cases in this table are listed in \cref{tab:flat_bands_triple_lines}.
    Note that reversing the signs of all couplings produces another set of models with triple degeneracies higher in the interaction matrix spectrum, cf. \cref{fig:triple_lines_AFM_FM}.
    }
	\label{tab:high_degeneracies}
\end{table*}

\subsection{Parameter Space vs. Model Space}
\label{sec:parameter_space_vs_model_space}

Any real choice of the five symmetry-allowed nearest-neighbor spin-spin coupling parameters in \cref{eq:H_named_interactions} or \cref{eq:H_local}, each a real number, defines an interaction matrix $\Jmat$ and thus a Hamiltonian.
We call this $\mathbbm{R}^5$ five-dimensional space the \emph{parameter space} ($\mathbbm{R}^4$ if we set $J_{\text{SIA}}=0$).
Within this space is a special point where all couplings are zero, corresponding to the trivial Hamiltonian $H=0$.
Many of the remaining non-trivial parameter sets are physically equivalent up to an overall energy scale.
Ideally one would like to parameterize the space of Hamiltonians modulo such rescalings, which we call the \emph{model space}. 

One way to parameterize the model space is to simply take a unit sphere in the parameter space, e.g. using the local couplings in \cref{eq:H_local} (setting $J_{\text{SIA}}=0$)
\begin{equation}
    J_{zz}^2 + J_{\pm\pm}^2 + J_{z\pm}^2 + J_{\pm}^2 = 1.
    \label{eq:unit_sphere}
\end{equation}
Of course one could alternatively take the unit sphere in the basis $J_1\cdots J_4$ in \cref{eq:H_generic}, or those in \cref{eq:H_named_interactions}. 
The unit spheres in these spaces do not coincide, however, since the transformation relating these bases of the parameter space, \cref{eq:JloctoJ14,eq:J14toJloc}, are not orthogonal. 
The parameter space does not have a canonical metric to measure distances between parameter sets or to define orthogonality, so such a choice of sphere is arbitrary. 
In any case such a choice does not yield a useful parameterization of the space of models.
In this section we provide a parameterization by directly manipulating the irrep energies.

In full generality, a model is an equivalence class of parameters, where two parameter sets are equivalent if the band structure of the interaction matrix $\Jmat$ eigenvalues (i.e. the $\Jeig_n(\bq)$ in \cref{eq:H_bands}) differ by either a shift or a re-scaling, i.e. a changing of the total bandwidth of the band structure of $\Jmat$ without changing the band structure. 
For nearest-neighbor interactions the band structure of $\Jmat$ is completely determined by the $\bq=\bm{0}$ eigenvalues, i.e. the irrep eigenvalues $J_I$.
Therefore, a set of irrep eigenvalues determines an interaction matrix $\Jmat$, with the caveats that ($\textit{i}$) each allowed set of irrep eigenvalues corresponds to a pair of models with opposite signs of $J_{z\pm}$, and ($\textit{ii}$) the inequality \cref{eq:T1_inequality} must be satisfied for $J_{z\pm}$ to be real.
Consider then the 5-dimensional space of real irrep eigenvalues $J_I$. 
We define an equivalence relation under affine rescalings
\begin{equation}
    J_I \sim J_I' \quad \text{if}\quad J_I' = a J_I + b
    \label{eq:equivalence_of_parameters}
\end{equation}
for some real numbers $a$ and $b$ with $a>0$.
An equivalence class $[J_I]$ defines a model, and the space of equivalence classes is  the model space, which has the topology of a 3-sphere.

\subsection{Parameterizing the Model Space with \texorpdfstring{$J_{\text{SIA}}=0$}{JSIA=0}}

In order to explore the model space, we require a parameterization which selects a single representative of each equivalence class of parameter sets, i.e. one which effectively fixes $a$ and $b$ in \cref{eq:equivalence_of_parameters}.
To do so, we first choose an arbitrary energy scale $J_0$ and measure each irrep energy relative to it, defining
\begin{equation}
    J_I = J_0 + \Delta_I.
    \label{eq:irrep_delta}
\end{equation}
To fix $b$, we restrict the allowed parameter sets $\Delta_I$ to those for which $\mathrm{min}(\Delta_I) = 0$.
In other words, we fix the energy scale to be $J_0 = \mathrm{min}(J_I)$, i.e. the ground state energy. 
In order to fix $a$ we have to fix the overall scale.
Here many choices are possible, but the most obvious is to set $\mathrm{max}(\Delta_I) = 1$.
In other words, we fix the total bandwidth of $\Jmat$ to be unity, which is sensible because it is the natural energy scale in the Hamiltonian.

We note with this choice of scale, the $L^2$ norm on the space of real eigenvalues induces a dimensionless metric on the model space, which can (in principle) be used to define how far apart two models are from each other. 
While the parameter space does not have a canonical metric, this would be the natural candidate for a ``useful'' metric---it measures how far apart two parameter sets are while holding the total bandwidth constant, i.e. modulo overall rescaling of the Hamiltonian.
This may be a useful definition for quantitatively assessing how close a compound is to a phase boundary or a spin liquid~\cite{scheieDynamicalScalingSignature2022}.

For the purposes of this paper, we will set $J_{\text{SIA}} = 0$, as is physically appropriate for spin-1/2 systems. 
This leads to helpful simplifications, but the analysis could be extended to include non-zero $J_{\text{SIA}}$.
Setting \cref{eq:JSIA} equal to zero and solving for $J_0$, we obtain 
\begin{equation}
	J_0 = -\frac{1}{12}\left[\Delta_{A_2} + 2 \Delta_{E} + 3\Delta_{T_2} + 3(\Delta_{T_{1-}} + \Delta_{T_{1+}})\right] .
\end{equation}
Substituting back into the remaining four equations, we obtain
\begin{subequations}
\label{eq:Delta_equations}
\begin{align}
	J_{zz} &= \frac{1}{36}
		[
		11 \Delta_{A_2} - 
		2  \Delta_E - 
        3  \Delta_{T_2} - 
		3 (\Delta_{T_{1-}} +
		   \Delta_{T_{1+}})
		] ,
	\\
	J_{\pm} &= \frac{1}{72}
		[
 		   \Delta_{A_2} - 
		10 \Delta_E + 
        3  \Delta_{T_2} + 
		3 (\Delta_{T_{1-}} +
		   \Delta_{T_{1+}})
		] ,
	\\
	J_{\pm\pm} &= \frac{1}{36}
		(
		   \Delta_{A_2} - 
		   \Delta_E -
        6  \Delta_{T_2} + 
		3 (\Delta_{T_{1-}} +
		   \Delta_{T_{1+}})
		) ,
	\\
	\vert J_{z\pm}\vert &= \frac{1}{36}\sqrt{(9\Delta_{\delta T_1})^2 - [2\Delta_{A_2} - 2\Delta_E -3\Delta_{T_2} + 3\Delta_{T_1}]^2} \label{eq:Jzpm_Deltas}
\end{align}
\end{subequations}
where in \cref{eq:Jzpm_Deltas} we have defined for convenience the equivalent of \cref{eq:T1_splitting}, the average and the splitting between the two $T_1$ irreps, $\Delta_{T_{1\pm}} = \Delta_{T_1} \pm \Delta_{\delta T_1}$, or inversely
\begin{equation}
    \Delta_{T_1} \equiv
    \frac{1}{2}(\Delta_{T_{1+}}+\Delta_{T_{1-}}),
    \quad
    \Delta_{\delta T_1} 
    \equiv 
    \frac{1}{2} (\Delta_{T_{1+}}-\Delta_{T_{1-}}) \geq 0 .
    \label{eq:T1_splitting_relative_Delta}
\end{equation}
The full model space is parameterized by tuning the $\Delta_I$ subject to the constraints that
\begin{enumerate}
    \item $\min \Delta_I = 0$,
    \item $\max \Delta_I = 1$,
    \item $\Delta_{T_1+} \geq \Delta_{T_{1-}}$, 
    \item $J_{z\pm}$ is real, i.e. \cref{eq:T1_inequality} is satisfied.
\end{enumerate}
The model space is 3-dimensional, since at a generic point with one $\Delta_I$ set to zero and one set to unity there are three free parameters to vary.
A phase is then the 3-dimensional locus of parameters where one of the $\Delta_I$ is zero, and the rest are positive with the max fixed to unity.
While there are five $\Delta_I$ parameters, we will see there are only four distinct phases because the two $T_1$ irreps mix continuously.
Phase boundaries are two-dimensional and occur when two $\Delta_I$ are zero while the remainder are positive. 
Triple points are one-dimensional lines along which three $\Delta_I$ are zero, and the remaining one is allowed to vary.
Lastly, it is possible to have a quadruple point, an isolated point where four phases are degenerate.

\begin{figure}
    \includegraphics[width=\columnwidth]{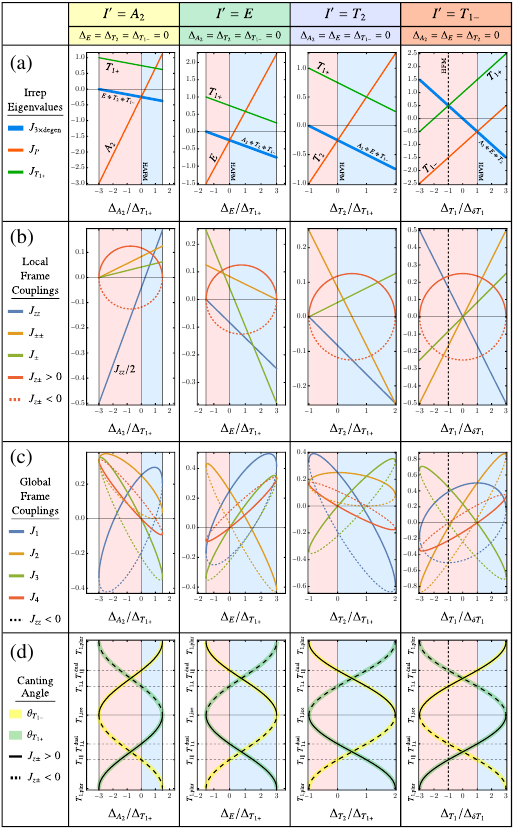}
	\caption{
        Variation of (a) irrep energies, (b) local couplings, (c) global couplings, and (d) canting angle on the four triply-degenerate lines in the phase diagram listed in \cref{tab:high_degeneracies}, characterized by tuning one of the irreps $I' \in\{A_2, E, T_2, T_1\}$ while keeping the other three degenerate.
        Columns are colored according to the irrep colors used in \cref{fig:jzpm0,fig:full-phase-diagram,fig:phase_diagram_HDM,fig:triple_lines_AFM_FM}.
        (a) Irrep eigenvalues $J_I$, with the triply degenerate eigenvalues denoted by a thick blue line and the tuned eigenvalue denoted by a red lin. 
        (b) Local couplings for the parameterization of the Hamiltonian \cref{eq:H_local} computed from \cref{eq:Delta_equations}. 
        For $I'=A_2$ (first column) we have scaled $J_{zz}$ by a factor of 2.
        (c) Global couplings in \cref{eq:Jijab}, with $J_{z\pm}<0$ indicated by dashed lines. 
        (d) Evolution of the canting angles $\theta_{T_{1-}}$ (yellow) and $\theta_{T_{1+}}$ (green).
        Special angles are labeled according to \cref{fig:T1_mixing}, and $J_{z\pm}<0$ indicated by the dashed lines.
        The intersections of three phases, where three ground states are degenerate, occurs for $\Delta_{I'}/\Delta_{T_{1+}} > 0$ and $\Delta_{T_{1+}}>0$ (blue region on the right side of each plot, cf. \cref{fig:full-phase-diagram}). 
        For $\Delta_{I'}<0$ the irrep $I'$ is the ground state but the other three are degenerate at higher energy (red region on the left side of each plot, cf. \cref{fig:triple_lines_AFM_FM}(a)). 
        Each of these triple lines forms a closed circle in the phase diagram (cf. \cref{fig:triple_lines_AFM_FM}(a)).
        The triple line $I'=T_{1-}$ (last column) intersects the Heisenberg ferromagnet (HFM) and its $J_{z\pm}$-dual, indicated by a black vertical line (cf. \cref{fig:phase_diagram_HDM,fig:triple_lines_AFM_FM}). 
	} 
	\label{fig:3-fold-lines-params}
\end{figure}

Note that while the model space has $0\leq \Delta_I \leq 1$, we will also allow the $\Delta_I$ to vary outside this range in order to parameterize collections of models with interesting properties (i.e. irrep degeneracies) and to identify special points in the phase diagram (cf. \cref{fig:3-fold-lines-params}).
Such parameter sets can always be re-scaled uniquely back to the model space.

\section{Mapping the Phase Diagram}
\label{sec:mapping_the_phase_diagram}

The parameterization of \cref{eq:Delta_equations} allows us to efficiently explore the parameter space in a way that makes the phase boundaries and special subspaces manifest, while decoupling the rescaling dimension. 
We will visualize the phase diagram by stereographically projecting the 3-sphere \cref{eq:unit_sphere} into 3-dimensional space using the mapping
\begin{equation}
    (J_{zz}, J_{\pm\pm}, J_{z\pm}, J_{\pm}) \mapsto (X,Y,Z)\equiv \left( \frac{J_{zz}}{1-J_{\pm}},\frac{J_{\pm\pm}}{1-J_{\pm}},\frac{J_{z\pm}}{1-J_{\pm}}\right).
    \label{eq:stereo_projection}
\end{equation}
In particular, this 3-sphere has a 2-sphere subspace on which $J_{z\pm}=0$, dividing into into two hemispheres. 
Since the phases are dual under changing the sign of $J_{z\pm}$, the phase diagram in these two hemispheres is reflected through the $J_{z\pm}=0$ ``equator'', i.e. the ``northern'' and ``southern'' hemispheres of the phase diagram are mirror images of each other. 
For completeness, the inverse mapping of \cref{eq:stereo_projection} is given by
\begin{equation}
    (X,Y,Z)\mapsto \frac{1}{R^2+1}\left(2X,2Y,2Z, R^2-1\right),
\end{equation}
where $R^2 = X^2+Y^2+Z^2$.

\subsection{Quadruple Points: Heisenberg Antiferromagnet}

To map the structure of the phase diagram, we begin with the highest-degeneracy four-fold degenerate quadruple points.
These can be searched for manually by setting all but one $\Delta_I$ to zero in \cref{eq:Delta_equations} and checking whether solutions exist with real $J_{z\pm}$. 
This immediately rules out the possibility of fourfold degeneracies with only $A_2$, $E$, or $T_2$ gapped.
This leaves only one possibility: $\Delta_{T_{1+}}>0$ and all others are zero. 
This corresponds to the Heisenberg antiferromagnet (HAFM) point, with Hamiltonian
\begin{equation}
	H_{\text{HAFM}} 
    = 
    \frac{\Delta_{T_{1+}}}{4}\sum_{\langle ij \rangle} \bm{S}_i \cdot \bm{S}_j 
    =
    E_0 + \Delta_{T_{1+}} \sum_t \vert\bm{m}_{\Tpar}\vert^2
    ,
    \label{eq:HAFM}
\end{equation}
which has $J_{z\pm}>0$.
The HAFM has a massively degenerate classical ground state manifold characterized by the zero-net-moment constraint $\bm{m}_{\Tpar} = \bm{0}$ on every tetrahedron, corresponding to the large degeneracy of the irreps.
This is a prototypical example of a classical spin liquid, and has been extensively studied~\cite{reimersAbsenceLongrangeOrder1992,canalsPyrochloreAntiferromagnetThreeDimensional1998,canalsSpinliquidPhasePyrochlore2001,moessnerLowtemperaturePropertiesClassical1998,isakovDipolarSpinCorrelations2004,henleyPowerlawSpinCorrelations2005,henleyCoulombPhaseFrustrated2010,conlonAbsentPinchPoints2010,iqbalQuantumClassicalPhases2019}.
Since every model determined by a set of eigenvalues with $J_{z\pm}\neq 0$ has a dual with the opposite sign of $J_{z\pm}$, there is a second fourfold-degenerate point in the phase diagram which we will refer to as \HAFMdual~\cite{rauFrustratedQuantumRareEarth2019}.
The Hamiltonian at this point can be written as 
\begin{equation}
	H_{\text{\HAFMdual}} 
    =
    E_0 + \Delta_{T_{1+}} \sum_t \vert\bm{m}_{T_{1\parallel}^{\text{dual}}}\vert^2,
\end{equation}
where the gapped $\smash{T_{1\parallel}^{\text{dual}}}$ spin configuration, as well as the zero-energy $\smash{T_{1\perp}^{\text{dual}}}$, are shown in \cref{fig:T1_mixing}.

\subsection{Triple Lines and Points}

Starting from the HAFM points where all four phases are degenerate, $\Delta_{A_2} = \Delta_{E} = \Delta_{T_2} = \Delta_{T_{1-}} = 0$, one can lift the degeneracy of one of the four phases, thus tuning along a line where three phases are degenerate. 
The resulting parameter sets for these four triply degenerate lines are given in \cref{tab:high_degeneracies}.
For the ones where the gapped irrep is $I'\in\{A_2,E,T_2\}$, the line can be parameterized by the ratio $\Delta_{I'}/\Delta_{T_{1+}}$. 
The fourth line where both $T_1$ irreps are gapped is better parameterized by the average and splitting of the two $T_1$ irreps, \cref{eq:T1_splitting_relative_Delta}.

The point $\Delta_{I'} = 0$ corresponds to the two HAFM points, from which one can tune $\Delta_{I'}/\Delta_{T_{1+}}>0$ continuously until it reaches a maximum allowed value at which $J_{z\pm} = 0$. 
Each line then continues by reversing the sign of $J_{z\pm}$ and tuning $\Delta_{I'}/\Delta_{T_{1+}}$ back to zero, ending at the other HAFM point.
\Cref{fig:3-fold-lines-params} shows the evolution of (a) the local exchange couplings, (b) the global exchange couplings, (c) the irrep energies, and (d) the canting angle along these lines. 
The right side of each plot with blue background is the region where the triply degenerate irreps are the ground state. 
The left side, with red background, has the triply degenerate irreps in the excited states, and is discussed further in \cref{sec:triple_ferro}.

There are three other possible combinations of three degenerate ground state irreps in which $\Delta_{T_{1-}}=\Delta_{T_{1+}}=0$ along with one other irrep.
According to \cref{eq:Jzpm_T1_splitting} this is only possible when $J_{z\pm}=0$.
Two such cases are possible, when $\Delta_{E}=0$ or $\Delta_{T_2}=0$, corresponding to isolated triple points in the phase diagram. 
The third case with $\Delta_{A_2}=0$ has no solutions.
This exhausts all possibilities for three irreps to be degenerate in the ground state.

\begin{figure*}[t]
    \centering
    \includegraphics[width=\textwidth]{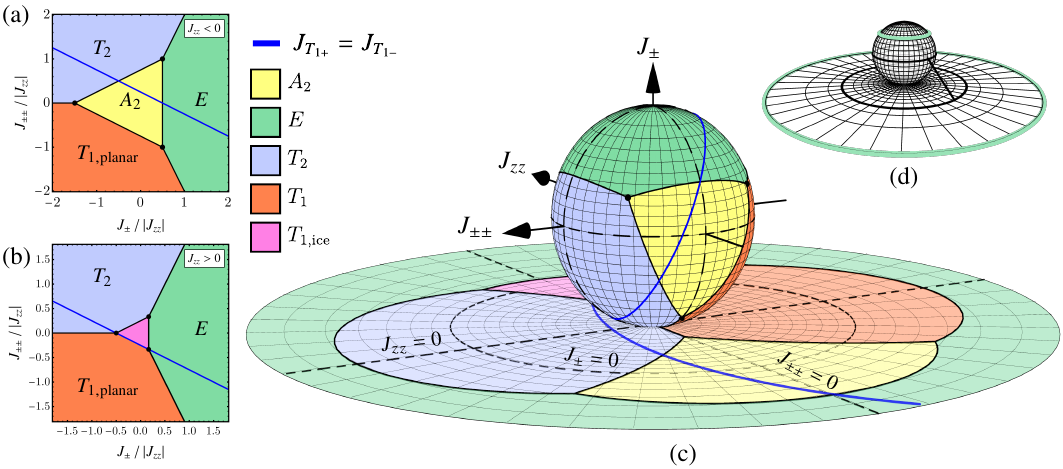}
    \caption{The phase diagram restricted to $J_{z\pm}=0$. (a) Fixing $J_{zz}<0$ and varying $J_{\pm}$ and $J_{\pm\pm}$, three triple-points are indicated by black dots. The blue line indicates the locus on which the two $T_1$ irreps are degenerate defined by \cref{eq:T1_degeneracy}, i.e. along this line all canting angles in \cref{fig:T1_mixing} are degenerate. (b) Fixing $J_{zz}>0$, again showing three triple points. The degenerate $T_1$ line (blue) separates the $T_{1,\text{planar}}$ from the $T_{1,\text{ice}}$ ground states, though they are continuously connected (i.e. in the same phase) out of the $J_{z\pm}=0$ phase. The intersection of this line with the phase boundary between the $T_1$ and either $T_2$ or $E$ phases yields the isolated triple points. (c) The two phase diagrams at left can be joined together by considering a unit 2-sphere in the 3-dimensional parameter space of $J_{zz}$, $J_{\pm}$, and $J_{\pm\pm}$, shown here. 
    The $T_{1,\text{ice}}$ phase is on the back of the sphere.
    In order to visualize the entire surface of the sphere we perform a stereographic projection (illustrated in (d) inset) onto the flat plane. Here we show a disk corresponding to $J_{\pm} \leq 0.53$. The excluded region corresponding to the northern cap of the sphere, see inset (d), is always in the $E$ phase.
    }
    \label{fig:jzpm0}
\end{figure*}

\subsection{The \texorpdfstring{$J_{z\pm} = 0$}{Jzpm=0} Plane}

Since the model space is a 3-sphere, the $J_{z\pm} = 0$ locus may be thought of as the ``equator'', with the topology of a 2-sphere. 
It divides the 3-sphere into northern and southern hemispheres which are dual to each other under changing the sign of $J_{z\pm}$ and performing a $\pi$ spin rotation about the local axes.
Within the $J_{z\pm}=0$ plane this spin-rotation duality is promoted to a symmetry.
Furthermore, within this subspace there is an additional duality: performing a $\pi/2$ rotation of each spin about the local~$\uvec{z}_i$ sends $\cramped{S_i^{\pm} \to \mp i S_i^{\pm}}$, which can be compensated by switching the sign of $J_{\pm\pm}$~\cite{rauFrustratedQuantumRareEarth2019}. 
Unlike the duality switching the sign of $J_{z\pm}$ discussed in \cref{sec:jzpm_duality}, which relates two parameter sets in the same phase, reversing the sign of $J_{\pm\pm}$ relates different phases since it changes the irrep energies in \cref{tab:irrep_energies}.
Referring to the ground states in \cref{fig:ground_states} and \cref{fig:T1_mixing}, the $J_{\pm\pm}$ duality switches $\psi_2\leftrightarrow \psi_3$ and $T_2\leftrightarrow T_{1,\text{planar}}$.
This duality ensures that the phase boundaries will be symmetric across $J_{\pm\pm}=0$ within the $J_{z\pm}=0$ subspace, and it is promoted to a symmetry when $J_{\pm\pm}=J_{z\pm}=0$.

Within the $J_{z\pm}=0$ subspace, the relevant couplings are $J_{zz}$, $J_{\pm}$, and $J_{\pm\pm}$, meaning that there is no direct coupling between the local easy-axis and easy-plane spin components.
The ground states are therefore either easy-axis Ising orders (ferro- or antiferromagnetic), or easy-plane XY orders.
The Ising ferromagnetic order occurs for dominant $\cramped{J_{zz} \ll 0}$ and corresponds to the $A_2$ all-in-all-out order. 
The Ising anti-ferromagnetic state occurs in the opposite limit $\cramped{J_{zz} \gg 0}$, corresponding to the $T_{1,\text{ice}}$ ground states.
For dominant $\cramped{J_{\pm}\gg 0}$ the $E$ irrep is the ground state.
Further selection of the energetically degenerate $\psi_2$ and $\psi_3$ configurations occurs due to order-by-disorder induced by asymmetric fluctuations about the ground state, discussed further in \cref{sec:flat_bands}.
Lastly, large $\cramped{J_{\pm\pm}\gg 0}$ selects the $T_2$ order, while large $\cramped{J_{\pm\pm}\ll 0}$ selects the $T_{1,\text{planar}}$ configurations. 
Since $J_{zz}=0$ is a degenerate case where the Hamiltonian becomes a pure XY model, it is standard to use $J_{zz}$ as the energy scale, considering two separate cases depending on the sign of $J_{zz}$, while varying the ratios $\cramped{J_{\pm}/\vert J_{zz}\vert}$ and $\cramped{J_{\pm\pm}/\vert J_{zz}\vert}$. 
This yields the two phase diagrams in \cref{fig:jzpm0}(a) and (b).

\begin{figure*}[th]
    \includegraphics[width=\textwidth]{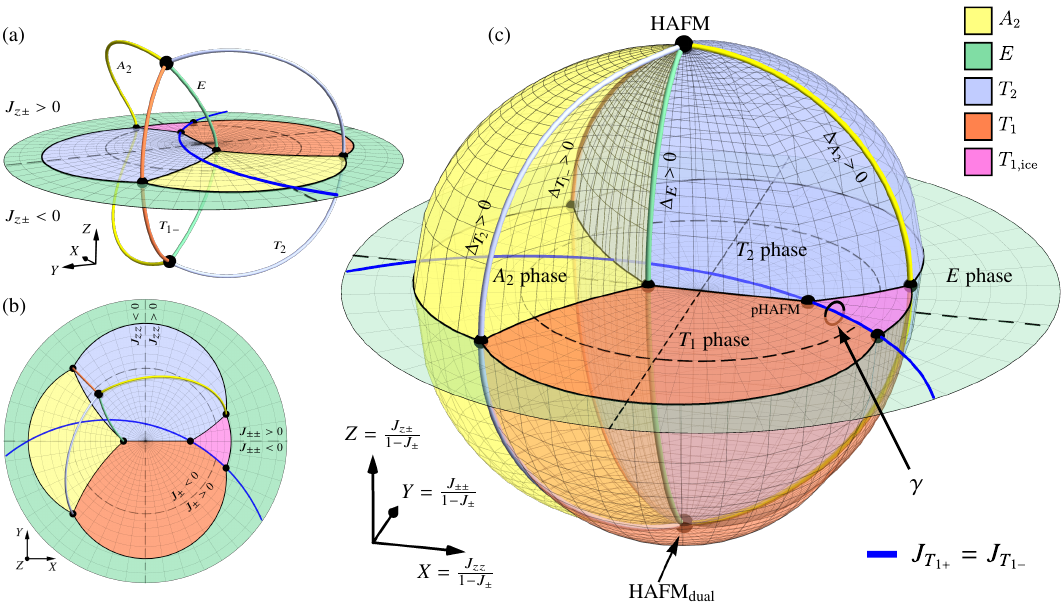}
	\caption{The full phase diagram, including the four 3-fold degenerate lines which run from the HAFM to the \HAFMdual models, crossing through the plane with $J_{z\pm} = 0$. (a) Stereographically projected triple lines listed in \cref{tab:high_degeneracies}, which pass through the $J_{z\pm}=0$ plane.\footnote{These lines can be parameterized as in \cref{eq:triple_line_parameterization}, then normalized to the unit sphere \cref{eq:unit_sphere}, then mapped via \cref{eq:stereo_projection}.} Phases in the plane are colored the same as in \cref{fig:jzpm0}, and the triply-degenerate lines are colored according to which of the four irreps is gapped along with $T_{1+}$. Above (below) the plane corresponds to $J_{z\pm}$ positive (negative). 
    (b) A top-down view of (a), with the $J_{zz}=0$, $J_{\pm}=0$, and $J_{\pm\pm}=0$ lines indicated. 
    (c) A ``cartoon'' version of the projection where we have ``straightened out'' the lines and phase boundaries to emphasize the topology of the phase diagram. The $J_{z\pm}=0$ plane is shown along with the four triple lines as in (a) and (b), along with the phase boundaries. Each phase has three boundaries, one with each of the other phases. The exterior region of the diagram is in the $E$ phase. 
    The $A_2$ and $T_2$ phases are enclosed by the yellow and blue phase boundaries, respectively. When turning on $J_{z\pm}\neq 0$ there is only a single contiguous $T_1$ phase, indicated by the red phase boundary. The $J_{z\pm}>0$ portion of the $T_1\oplus E$ phase boundary is not shown in order to expose the interior region of the $T_1$ phase and its phase boundaries with the $A_2$ and $T_2$ phases. An undistorted version of the phase boundaries according to the stereographic projection \cref{eq:stereo_projection} is shown in \cref{fig:phase_diagram_HDM}.
    }
	\label{fig:full-phase-diagram}
\end{figure*}

\subsubsection{Line of \texorpdfstring{$T_1$}{T1} Degeneracy}

Within these two diagrams one can clearly distinguish six triple points where three phases are degenerate, corresponding to the four lines and two special points that we identified in \cref{tab:high_degeneracies}.
The two special points occur when both $T_1$ irreps are degenerate, at the ends of the line separating the $T_{1,\text{planar}}$ and $T_{1,\text{ice}}$ regions of the phase diagram. 
As we saw from \cref{eq:Jzpm_T1_splitting}, the degeneracy of the two $T_1$ irreps is only possible when $J_{z\pm} = 0$, and meaning that there is \emph{no} phase boundary between two distinct $T_1$ phases when $J_{z\pm} \neq 0$, i.e. there is only a single $T_1$ phase. 
It is therefore interesting to identify the locus on which the two $T_1$ irreps are degenerate, i.e. all canting angles in \cref{fig:T1_mixing} are degenerate, which is contained within the $J_{z\pm} = 0$ subspace.
From \cref{tab:irrep_energies} (setting $J_{\text{SIA}}=0$) this occurs when the square root which splits to the two $T_1$ energies is zero, i.e. when
\begin{equation} 
    2J_{\pm} + 4J_{\pm\pm} + J_{zz} = 0 
    \quad 
    \text{and} 
    \quad J_{z\pm} = 0 .
    \label{eq:T1_degeneracy}
\end{equation}
The solutions of this equation can be parameterized by a single angle $\vartheta$ as
\begin{equation}
    \begin{pmatrix}
        J_{zz} \\ J_{\pm\pm} \\ J_{z\pm} \\ J_{\pm}
    \end{pmatrix}
    = 
    \mathcal{N}(\vartheta) 
    \begin{pmatrix}
    2 \cos(\vartheta) - 4 \sin(\vartheta)\\
    5 \sin(\vartheta)\\
    0\\
    -\cos(\vartheta) - 8\sin(\vartheta)
    \end{pmatrix}
\end{equation}
where $\mathcal{N}(\vartheta)>0$ is an arbitrary overall rescaling of the energy.
These solutions are shown by the blue line in \cref{fig:jzpm0}(a,b).
Along this line all canted $T_1$ configurations in \cref{fig:T1_mixing} are degenerate.
Indeed, part of this line forms the boundary between the $T_{1,\text{ice}}$ (pink) and $T_{1,\text{planar}}$ (red) portions of the phase diagram, clearly shown in \cref{fig:jzpm0}(b).

\subsubsection{Stereographic Projection}

In order to unify \cref{fig:jzpm0}(a) and (b) into a single two-dimensional phase diagram, we consider the unit sphere in the space spanned by $J_{\pm\pm}$, $J_{zz}$, and $J_{\pm}$. 
This is the ``equator'' of the 3-sphere defined by \cref{eq:unit_sphere}.
This 2-sphere is shown in \cref{fig:jzpm0}(c), with the phases colored the same as in \cref{fig:jzpm0}(a,b). 
In this figure the $T_{1,\text{ice}}$ region is on the back of the sphere and not visible. 
In order to visualize the entire surface of the sphere, we use the stereographic map defined by \cref{eq:stereo_projection} to project the surface of the sphere onto a plane, as illustrated in \cref{fig:jzpm0}(d). 
We have chosen the azimuthal axis to be the $J_{\pm}$ axis, which becomes the radial direction on the projected surface.
The top of the sphere is in the $E$ phase, so that after projecting onto a plane all of the phase boundaries appear near the origin,  surrounded by the large-$J_{\pm}$ $E$ phase, which extends to infinity.

\subsection{Big Picture: The Full Phase Diagram}

We now have all the pieces to put together the full phase diagram using the stereographic projection \cref{eq:stereo_projection} from the 3-sphere to 3-dimensional euclidean space. 
Using the projected $J_{z\pm}=0$ plane from \cref{fig:jzpm0} as our baseline, turning on $J_{z\pm}$ takes us into the third (vertical) dimension. 
First, \cref{fig:full-phase-diagram}(a,b) adds to the $J_{z\pm}=0$ plane the four triple lines, along each of which three irreps are degenerate and one is gapped.
They are labeled and colored according to which irrep is gapped along that line with \cref{fig:full-phase-diagram}(a) showing an edge-one perspective and (b) showing a top-down view.
All four meet at the two HAFM points above and below the plane, with the region above (below) this plane corresponding to $J_{z\pm}$ positive (negative). 
The HAFM sits above the plane while its dual sits below the plane.

In order to visualize the phase boundaries in this three-dimensional space in a static two-dimensional image, it is convenient to deform the phase diagram in order to ``straighten out'' the four triple lines.
We thus provide the topologically equivalent and simpler to visualize picture shown in \cref{fig:full-phase-diagram}(c). 
By choosing the ``point at infinity'' to be deep in the $E$ phase, all of the phase boundaries are located near the origin, and the exterior region is all in the $E$ phase.
We could have chosen the point at infinity to be deep within any of the four phases, and obtained a qualitatively similar phase diagram with the other three phases located near the origin.
We have colored the phase boundaries surrounding the $A_2$ phase yellow, those surrounding the $T_2$ phase blue, and those surrounding the $T_1$ phase red. 
Gridlines on the phase boundaries are added as guides to the eye to emphasize the 3-dimensionality of the surfaces, but do not have any special meaning. 
We have excluded the $J_{z\pm}>0$ phase boundary between the $T_1$ phase and the $E$ phase in order to show the interior region of the $T_1$ phase and expose its phase boundaries with the $A_2$ and $T_2$ phases along with their triple line. 
Each of the four phases is enclosed by three phase boundaries where it touches one of the other phases.
The $J_{z\pm}\geq 0$ region of each phase is topologically equivalent to a tetrahedron, and the same for the $J_{z\pm}\leq 0$ region, thus each phase has the topology of two tetrahedra glued along a triangular face.
\Cref{fig:full-phase-diagram} constitutes a primary result of this work: a complete picture of the pyrochlore phase diagram showing all of the phases, phase boundaries, triple lines, and quadruple points.

\section{Topological Canting Cycles and Diabolical Locus}
\label{sec:canting_cycles}

In this section we investigate the locus on which the two $T_1$ irreps are degenerate, defined by \cref{eq:T1_degeneracy}. 
Whereas the energetic crossing of two different irreps in the ground state implies a phase transition, there is only one $T_1$ phase in which the canting angle varies continuously, so the crossing of the $T_1$ irreps is not a phase transition. 
This is a type of level repulsion: in order to make the square root zero in the last row of \cref{tab:irrep_energies}, we have to tune two separate parameter sets to be zero.
Thus this locus is analogous to a conical intersection or diabolical point in the spectrum of a quantum system~\cite{berryDiabolicalPointsSpectra1997}, but in the space of parameters, known in quantum field theory as a diabolical locus~\cite{hsinBerryPhaseQuantum2020}. 
In \cref{fig:full-phase-diagram} one can see that the locus where the two $T_1$ irreps are degenerate (blue arc in the $J_{z\pm}=0$ plane) ``pierces'' each of the $T_1\oplus T_2$ and $T_1 \oplus E$ phase boundaries once, giving rise to the two isolated triple points in the phase diagram. 
It passes within the $T_1$ phase as a special line of degeneracy along which all canting angles are degenerate in the ground state, but it does not form a phase boundary. 
We have indicated with a small black loop a path surrounding the line of $T_1$ degeneracy entirely contained within the $T_1$ phase. 
Moving along such a loop, one $T_1$ irrep is the unique ground state and the other is gapped, while the canting angle varies continuously.
In other words the eigenvectors of the interaction matrix rotate in the $T_1$ subspace, going around the cycle shown in \cref{fig:T1_mixing}.
Starting from a point with $T_{1-}=T_{1,\text{ice}}$ ($\theta_{T_{1-}}=0)$ and $T_{1+} = T_{1,\text{planar}}$, going halfway around the loop the canting angle continuously rotates by $\pi/4$ until the two $T_1$ irreps have swapped, then continuing around the ground state returns to $T_{1,\text{ice}}$ but with all spins reversed, i.e. a $\pi$ rotation in \cref{fig:T1_mixing}.
Going around the loop twice returns to the original ground state.
Thus there is a topological winding number associated to paths in parameter space which enclose this locus, and we can identify it as a source of a sort of Berry curvature~\cite{berryDiabolicalPointsSpectra1997}.

\subsection{Two Families of Hamiltonians}
\label{sec:two_families}

\begin{figure}[t]
    \centering
    \includegraphics[width=\columnwidth]{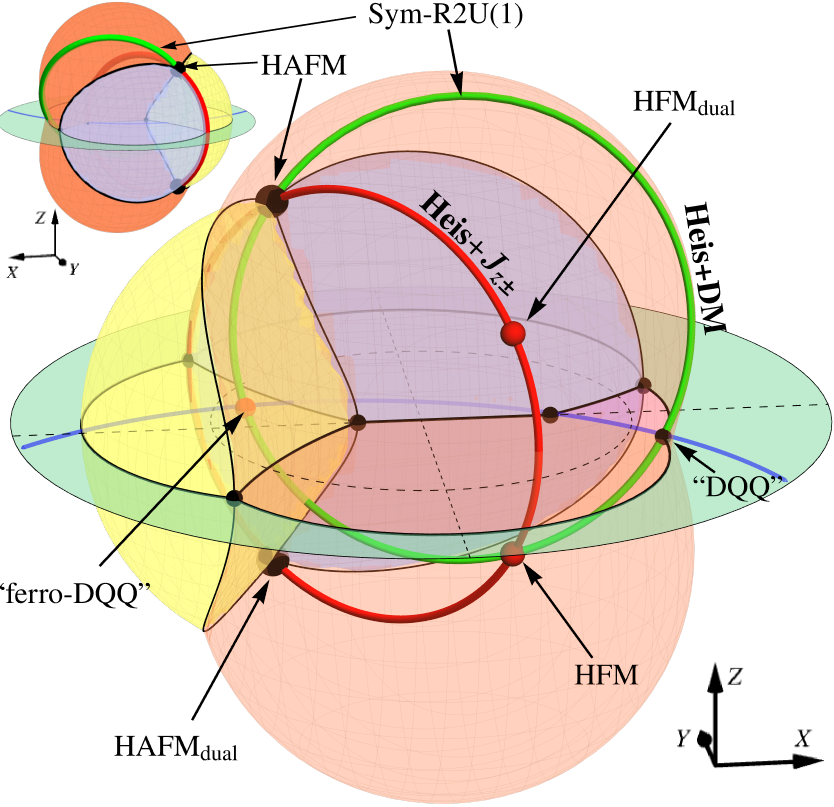}
    \caption{Here we show the ``true'' 3D stereographically projected phase diagram from \cref{fig:full-phase-diagram}(a,b) with the actual phase boundaries indicated, whereas \cref{fig:full-phase-diagram}(c) showed a deformed version that emphasized the topology. 
    The inset (top right) shows the same from the back side.
    Here we have colored the triple lines black, along with the intersections of the phase boundaries with the $J_{z\pm}=0$ plane.
    The red circle is the Heisenberg-plus-$J_{z\pm}$ family, which corresponds to the triple line along which $A_2$, $E$, and $T_2$ are degenerate (last column of \cref{fig:3-fold-lines-params}), which contains the Heisenberg anti-ferromagnet (HAFM), the Heisenberg ferromagnet (HFM), and both of their $J_{z\pm}$-duals. 
    The green circle is the Heisenberg-plus-Dzyloshinskii-Moriya (H+DM) family~\cite{yanRank2U1Spin2020,noculakClassicalQuantumPhases2023,hickeyOrderbydisorderQuantumZeropoint2025}, which passes through the HAFM, the HFM, the ``DQQ'' isolated triple point and its ferromagnetic counterpart. Because this family has $J_3 = 0$, it lies on a surface of constant canting where the $T_1$ irreps decompose as $\Tpar\oplus\Tperp$. This surface can be parameterized as a sphere in the $J_1$-$J_2$-$J_4$ space, corresponding to $\alpha = 1/3$ in \cref{fig:canting_foliation}. Along an arc from the DQQ point to the HAFM the H+DM model runs along the intersection of this constant-canting surface and the $E\oplus T_1$ phase boundary (red) where the ground state irreps are $E\oplus \Tperp$ which form a symmetric trace-free tensor, which was studied in Ref.~\cite{yanRank2U1Spin2020}.
    }
    \label{fig:phase_diagram_HDM}
\end{figure}

To investigate the canting behavior in more detail, as well as to illustrate the utility of having a complete phase diagram, we consider two particular one-parameter models---one which has constant canting angle and one on which the canting angle winds---and show how they fit in and are related in the phase diagram. 
\Cref{fig:phase_diagram_HDM} shows the same phase diagram as \cref{fig:full-phase-diagram}(a,b), but with the ``true'' phase boundaries computed from the stereographic projection, rather than the deformed version shown in \cref{fig:full-phase-diagram}(c). 
Note in particular that the $T_1$ phase appears larger than the $A_2$ and $T_2$ phases---while the $T_1$ and $T_2$ phases are symmetric in the $J_{z\pm}=0$ plane, as seen in \cref{fig:full-phase-diagram}(b), turning on $J_{z\pm}\neq 0$ promotes the $T_1$ phase (which is the ground state in the large-$\vert J_{z\pm}\vert$ limit).

The first family is what we will call the Heisenberg-plus-$J_{z\pm}$ model, by tuning both $J_{\text{Heis}}$ from \cref{eq:H_named_interactions} and $J_{z\pm}$ from \cref{eq:H_local}.
Up to an overall scale, this is a 1-parameter family of Hamiltonians, which can be parameterized as $J_{\text{Heis}}=\cos\vartheta$ and $J_{z\pm}=\sin\vartheta$. 
This family is interesting because it corresponds to the triple line with $A_2$, $E$ and $T_2$ degenerate, while both $T_1$ irreps are tuned---the last triple line in \cref{tab:high_degeneracies} and the last column in \cref{fig:3-fold-lines-params}. 
It is shown by the red circle in \cref{fig:phase_diagram_HDM}, which passes through both the Heisenberg ferromagnet (HFM) and anti-ferromagnet (HAFM) points, along with both of their duals. 
Part of this line lies on the triple intersection of the $A_2$, $E$, and $T_2$ phases, while the remainder passes through the $T_1$ phase. 
Since this family encloses the line where both $T_1$ irreps are degenerate, the canting angle winds by $\pi$ as we go around it while the two $T_1$ irrep energies never cross, as seen in the last column of \cref{fig:3-fold-lines-params}(c) and (d).

The second family is the Heisenberg-plus-Dzyloshinskii-Moriya (H+DM) line~\cite{elhajalOrderingPyrochloreCompounds2004,elhajalOrderingPyrochloreAntiferromagnet2005,canalsIsinglikeOrderDisorder2008,noculakClassicalQuantumPhases2023}, with Hamiltonian
\begin{equation}
    H_{\text{H}+J_{z\pm}} = \sum_{\langle ij \rangle} [J_{\text{Heis}} \bm{S}_i \cdot \bm{S}_j + J_{\text{DM}} \bm{D}_{ij} \cdot \bm{S}_i \times \bm{S}_j],
\end{equation}
which was recently studied in detail in Ref.~\cite{noculakClassicalQuantumPhases2023}.
This is a 1-parameter family of models up to a scale, which can be parameterized by $J_{\text{Heis}}=\cos\vartheta$ and $J_{\text{DM}}=\sin\vartheta$. 
This family is experimentally interesting because the DM interaction may be the leading anisotropic correction to the Heisenberg interaction in transition-metal pyrochlore insulators~\cite{noculakClassicalQuantumPhases2023}.
The resulting family is shown by the green circle in \cref{fig:phase_diagram_HDM}. 
Starting from the HAFM, it enters the $A_2$ phase, passes through the $A_2\oplus T_1$ phase boundary into the $T_1$ phase, meets the HFM point, then meets the isolated triple point on the $T_1\oplus E$ phase boundary (studied in more detail in Ref.~\cite{lozano-gomezCompetingGaugeFields2024} where it was dubbed the dipolar-quadrupolar-quadrupolar or ``DQQ'' point), then runs along the $T_1\oplus E$ phase boundary until it meets the HAFM again.
This family is particularly interesting because it has $J_3=0$, meaning that the canting angle is constant and the $T_1$ irreps decouple into $\Tpar$ and $\Tperp$. 
The two $T_1$ irreps cross each other at $J_{z\pm}=0$, where the family intersections the 1-dimensional locus where the two $T_1$ irreps are degenerate (blue line in \cref{fig:phase_diagram_HDM}). 
On the side with $J_{z\pm}>0$ we have $\Delta_{\Tpar}<\Delta_{\Tperp}$, while on the side with $J_{z\pm}<0$ we have $\Delta_{\Tpar}>\Delta_{\Tperp}$.

\begin{figure*}[t]
    \centering
    \includegraphics[width=\textwidth]{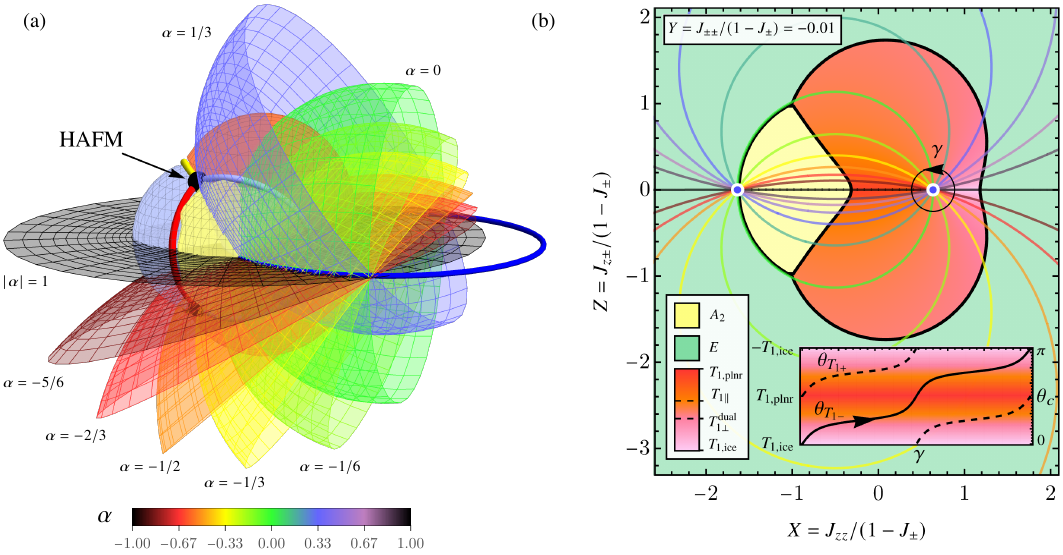}
    \caption{Illustration of how the phase diagram is foliated into surfaces of constant canting angle, which are spheres that intersect along the locus where the two $T_1$ irreps are degenerate. Here we parameterize the canting angle using the parameter $\alpha\in[-1,1]$ using \cref{eq:canting_angle_alpha}. (a) The 3D stereographically projected phase diagram from \cref{fig:full-phase-diagram}(a,b), along with the full circular locus of $T_1$ degeneracy (blue circle). For various canting angles we show a portion of the corresponding surface of constant canting (cf. the color legend below). Each such surface is a sphere, which we have cut open for visualization. All of these surfaces intersect on the degenerate $T_1$ locus, where all canting angles have the same energy. 
    Note that the $\alpha=1/3$ surface corresponds to $\Tpar\oplus\Tperp$ decoupling, while $\vert\alpha\vert=1$ corresponds to $T_{1,\text{ice}}\oplus T_{1,\text{planar}}$, whose corresponding sphere is the $J_{z\pm}=0$ plane.
    (b) A 2D cross-section of (a) which intersects the degenerate $T_1$ locus at two points (blue circles with white borders). 
    Within the $T_1$ phase the ground state canting angle $\theta_{T_{1-}}$ evolves continuously (pink-to-red gradient). 
    A black loop with arrowhead labeled $\gamma$ indicates a path in the parameter space that encloses the degenerate $T_1$ locus within the $T_1$ phase.
    This loop can be compared with the one labeled $\gamma$ in \cref{fig:full-phase-diagram}.
    Adiabatically tuning the system along this path, the ground state evolves halfway around \cref{fig:T1_mixing}.
    The inset demonstrates the evolution of the canting angle along this path, starting at $T_{1,\text{ice}}$ and ending at $-T_{1,\text{ice}}$.
    Two cycles are required to return to the same state. 
    }
    \label{fig:canting_foliation}
\end{figure*}

\subsection{Surfaces of Constant Canting}

Since the condition for the decoupling into $\Tpar$ and $\Tperp$ to occur is that $J_3=0$, the one-dimensional H+DM family can be extended to a two-dimensional family on which the canting angle is constant, by additionally tuning the $J_2$ coupling in \cref{eq:Jijab} (or the Kitaev-like coupling in \cref{eq:H_named_interactions}). 
This family can be parameterized by the unit sphere in the $J_1$-$J_2$-$J_4$ parameter space, meaning that it is a 2-sphere in the phase diagram which intersects the circular $T_1$-degenerate locus (only part of which is shown in \cref{fig:full-phase-diagram,fig:phase_diagram_HDM} by the blue arc). 
On the $J_{z\pm}$ positive (negative) side of this sphere the lower-energy $T_1$ irrep is $\Tpar$ ($\Tperp$).

For any canting angle $\theta$ there is a corresponding 2-sphere on which the canting angle is constant, given by the intersection of the 3-sphere \cref{eq:unit_sphere} and the linear subspace satisfying \cref{eq:canting_angle_local}, written as
\begin{equation}
    \tan(2\theta)(2J_{\pm} +4J_{\pm\pm} + J_{zz}) - 8J_{z\pm} = 0.
\end{equation}
Thus the entire 3-sphere phase diagram is foliated into 2-sphere subspaces which intersect each other on the locus where the two $T_1$ irreps are degenerate defined by \cref{eq:T1_degeneracy}.
Along any path on such a surface that crosses this line the energies of the two $T_1$ irreps cross each other.
In particular, the $J_{z\pm}=0$ 2-sphere in \cref{fig:jzpm0} is the constant-canting surface with $\theta=0$, and the blue line divides it into two halves on which $T_{1-}$ is either $T_{1,\text{ice}}$ or $T_{1,\text{planar}}$.

This foliation of the phase diagram is illustrated in \cref{fig:canting_foliation}. 
We use the convenient parameter $\alpha\in[-1,1]$ defined in \cref{eq:alpha} to parameterize the canting angle as in \cref{eq:canting_angle_alpha}. 
Note that $\alpha=1/3$ corresponds to the decomposition $\Tpar\oplus \Tperp$ (occurring when $J_3 = 0$), and $\vert\alpha\vert=1$ corresponds to $T_{1,\text{ice}}\oplus T_{1,\text{planar}}$ (occurring when $J_{z\pm}=0$).
\Cref{fig:canting_foliation}(a) shows a collection of constant-canting 2-spheres for different values of $\alpha$, each of which is only partially shown to make the foliation structure visible.
These surfaces are the ``leaves'' of the foliation. Since the leaves at $\alpha=\pm 1$ coincide, $\alpha$ may be regarded as a periodic parameter. 
Note that \emph{all} of these constant-canting surfaces intersect along the circle where the two $T_1$ irreps become degenerate, indicated in blue. 
Indeed the equations defining this locus, \cref{eq:T1_degeneracy}, can be equivalently written as 
\begin{equation}
    J_3 = 0 \quad \text{and} \quad J_{z\pm} = 0,
\end{equation}
which defines the intersection of the $\vert \alpha\vert = 1$ and $\alpha = 1/3$ surfaces.
Furthermore, this locus divides each constant-canting 2-sphere into two regions, on each of which one of the two $T_1$ irreps has the lower energy.
\Cref{fig:canting_foliation}(b) shows a cross section of \cref{fig:canting_foliation}(a), with various constant-canting surfaces indicated, again demonstrating how the phase diagram is foliated.

\subsection{``Diabolical'' Loci from Duplicate Irreps}

This foliation also reveals an unexpected topological feature of the $T_1$ phase. Although the ground state evolves continuously along any path within the phase, loops that link the $T_1$-degenerate locus are not topologically trivial. 
Consider adiabatically varying the Hamiltonian parameters around a closed loop that links the $T_1$-degenerate locus while remaining within the $T_1$ phase, such as the path $\gamma$ shown in \cref{fig:full-phase-diagram}(c). 
Along this loop the ground state is continuously transported through the family of $T_1$ states shown in \cref{fig:T1_mixing}. 
Rather than returning to its initial spin configuration after one circuit, the canting angle $\theta_{T_{1-}}$ advances by only $\pi$, as illustrated in the inset of \cref{fig:full-phase-diagram}(b). 
Consequently, the system returns to the time-reversed configuration (i.e. with all spins reversed), and only after a second circuit is the original spin configuration recovered.

This topological winding upon adiabatic transport around a singular locus is reminiscent of what is referred to in quantum field theory as a diabolical locus~\cite{hsinBerryPhaseQuantum2020}, also referred to as a locus of ``unnecessary criticality''~\cite{biAdventureTopologicalPhase2019,jianGenericUnnecessaryQuantum2020,prakashMultiversalityUnnecessaryCriticality2023,prakashClassicalOriginsLandauincompatible2025,zhangDiracSpinLiquid2025}. 
A diabolic locus is a manifold in parameters space along which the ground state is gapless, contained inside a single gapped phase, rather than on a phase boundary. 
They can be characterized by higher Berry phase topological invariants in the space of couplings, which classify generalized Thouless pumping cycles that transport symmetry charges across a system~\cite{kapustinHigherdimensionalGeneralizationsThouless2020,kapustinHigherdimensionalGeneralizationsBerry2020,wenFlowHigherBerry2023}. 
By a bulk-boundary correspondence they generally imply the existence of boundary phase transitions in the phase diagram ~\cite{wenFlowHigherBerry2023,prakashChargePumpsBoundary2025}.

The $T_1$-degenerate locus shares some analogous, albeit fully classical, behaviors.
First, it is ``critical'' in the sense of having an enhanced ground state degeneracy, in particular with an accidental U(1) symmetry of the ground state manifold. 
Second, as explained above it acts as a source of topological winding of the canting angle, somewhat analogous to the charge pumping action of diabolical loci.
Third, the phenomenon occurs here due to a level repulsion between two eigenvalues of the interaction matrix with the same irreducible character, thanks to their symmetry-allowed mixing term in \cref{eq:T1_quadratic_form}.
Whereas tuning to a phase boundary requires tuning only a single parameter, making the two $T_1$ irreps degenerate, i.e. setting $J_{T_{1+}} = J_{T_{1-}}$ in \cref{tab:irrep_energies}, requires tuning two independent parameters, making the locus one dimension lower than a phase boundary. 
Such a level repulsion is also at the root of quantum diabolical loci~\cite{hsinBerryPhaseQuantum2020}.
While those studied examples generally have the critical surface either completely contained inside a single phase or terminating within the phase, here we have a locus which bridges between two phase boundaries.

This phenomenon should be quite common in the phase diagrams of quadratic corner-sharing-cluster Hamiltonians (i.e. those on line graph lattices~\cite{henleyCoulombPhaseFrustrated2010}), and we sketch here the basic idea.
Performing a symmetry decomposition for all the spins in a cluster into irreducible representations will result in $d$ distinct irreps $I$, each with multiplicity $n_I$.  
The total number of symmetry-allowed tuning parameters $D$ should be equal to the total number of distinct copies of irreps, $D=\sum_{I=1}^d n_I$, meaning that the model space is topologically a $(D-1)$-sphere.
There should be a single $(D-1)$-dimensional phase for each distinct irrep appearing in the decomposition---if $n_I>1$ then all duplicates of an irrep are allowed to couple and can be continuously rotated into each other, resulting in a single continuous phase. 
For $n_I$ copies of an irrep one must perform an SO($n_I$) rotation in the irrep eigenspace to decouple them, meaning that there will be $n_I(n_I-1)/2$ ``canting angles'' parameterizing elements of $\mathrm{SO}(n_I)$.
Note that, because the irrep eigenvectors are only defined up to an overall sign, we actually only need an element $\mathrm{PSO}(n_I)\simeq \mathrm{SO}(n_I)/\mathbbm{Z}_2$.
At a generic point in the phase one copy of the irrep has the lowest energy, and the canting angles vary continuously as the parameters are tuned. 
One then expects critical loci within the phase where multiple copies of the irrep are degenerate.
There must be $(D-2)$-dimensional loci along which two copies of the irrep are degenerate---that is, a phase boundary is a codimension-1 surface while a degenerate irrep locus is codimension-2.
We expect there to be one such codimension-2 locus for each of the $n_I(n_I-1)/2$ pairs of copies of the irrep.
Each such locus is singular in the sense that there is a topological winding number for closed 1-dimensional paths in parameter space which link with it. 
These loci may further intersect each other in lower-dimensional regions (see for example~\cite{chungGeometricallyFrustratedQuadrupoles2025}).

In the case of the pyrochlore studied here with zero single ion anisotropy we have $D=4$, $d_l=1$, and $n_I=2$, meaning that a 1-parameter family of Hamiltonians carries a $\pi_1(\mathrm{PSO}(2)) \simeq \mathbbm{Z}$ winding number which counts how many times $2\theta_c$ winds around the $T_1$-degenerate locus.
We can in principle define a Berry connection in the parameter space whose holonomy measures the winding, such as
\begin{equation}
    a_\mu \propto \langle \psi_{\bm{0},T_{1-},\alpha} \vert \frac{\partial}{\partial {J_\mu}} \vert \psi_{\bm{0},T_{1-},\alpha}\rangle,
\end{equation}
where $\vert \psi_{\bm{0},T_{1-},\alpha} \rangle$ is one of the $\bm{q}=\bm{0}$ eigenvector of the interaction matrix corresponding to the lowest-energy $T_1$ irrep (the result is independent of $\alpha$), and $J_\mu$ indicates the $\mu$'th coupling parameter.\footnote{
    It is notable that a 1-parameter family of Hamiltonians entirely contained in the $T_1$ phase which links the diabolical locus must cross through the $T_{1,\text{ice}}$ surface where the system is a deconfined spin liquid. It is unclear if there is any direct relation between the existence of the canting cycle and the existence of a classical spin liquid on a surface which terminates on the diabolical locus. 
}
Outside the $T_1$ phase the line of degeneracy and the topological invariant associated to 1-dimensional families of Hamiltonians is well-defined, even though the $T_1$ irreps are not in the ground state.
We note that in Ref.~\cite{chungGeometricallyFrustratedQuadrupoles2025} we have reported more complex instances of diabolical loci in a model with two irreps which appear twice and one which appears three times. 
We leave further study of such topological winding cycles in the phase diagrams of classical spin models for future study.

\section{Flat Bands and Spin Liquids From Ground State Irrep Collisions}
\label{sec:flat_bands}

We now turn to tabulating the flat band degeneracies which arise at the collision points of multiple irreps, i.e. which appear by tuning degenerate band touchings at $\bm{q}=\bm{0}$.
Having zero-energy flat bands (measured relative to the minimum eigenvalues of $\Jmat$) is a prerequisite for a classical spin liquid with Heisenberg spins, and almost always a prerequisite for Ising spins as well.\footnote{ 
    The primary counter-example that we know of is the Ising antiferromagnet on the triangular lattice, which does not have flat bands but nonetheless has power-law decaying correlations at low temperature~\cite{wannierAntiferromagnetismTriangularIsing1950,stephensonIsingModelSpinCorrelations1970a}.
}
The presence of flat bands at zero energy implies that any spin configuration constructed from the eigenvectors of those flat bands is a ground state, generally meaning that the system has a massive ground state manifold. 
This ground state manifold is characterized by the local constraint on every tetrahedron that the spin configuration has no overlap with the gapped irreps. 
This may be formulated as an emergent Gauss law for an emergent tensor gauge field~\cite{henleyCoulombPhaseFrustrated2010,yanClassificationClassicalSpin2024,yanClassificationClassicalSpin2024a}, and the resulting phase is a classical deconfined liquid, a sort of condensate of a massive number of ground state configurations. 
Excitations are violations of the Gauss law coming from exciting locally the gapped irreps, and behave as charges of the emergent gauge theory.

This can be seen at the level of the band structure, where the excitations correspond to a set of quadratically dispersing bands which touch the flat bands at the zone center.\footnote{
    Models with further-neighbor interactions may realize cases where the quadratic band touchings occur away from the zone-center~\cite{bentonTopologicalRouteNew2021}, but these do not occur in the nearest-neighbor model considered here.
}
It is also possible for the quadratic band touchings to be not only at a point but along an extended line, or possibly along a plane. 
Even in the cases without zero-energy flat bands, zero-energy flat planes or flat lines can occur. 
In this section we consider all possible flat band degeneracies of the nearest-neighbor pyrochlore Hamiltonian by considering different combinations of irrep degeneracies.
We have tabulated all such cases as follows: \cref{tab:flat_bands_phases} lists flat degeneracies within each phases, \cref{tab:flat_bands_phase_boundaries} lists them for the phase boundaries, and \cref{tab:flat_bands_triple_lines} lists them for the triply degenerate lines and points.
In this section we discuss how these enhanced degeneracies may or may not give rise to spin liquids.

\subsection{Ground State Selection vs. Spin Liquidity}

Having a set of flat bands suggests a massive ground state degeneracy and classical spin liquidity, but it does not guarantee the presence of a classical spin liquid, owing to the hard spin length constraint. 
The pyrochlore phase diagram hosts a plethora of interesting flat bands, but it is unlikely that these will all realize stable spin liquids. 
Heuristically, there is common folklore that the fewer the number of flat bands (relative to the total number of bands) the more likely the system is to fail to realize a spin liquid. 
Since we now have at hand a large family of models hosting a variety of flat degeneracies listed in \cref{tab:flat_bands_triple_lines,tab:flat_bands_phase_boundaries} it is worthwhile to distinguish the mechanisms that may preempt classical spin liquidity and lead to more traditional ground state selections which are likely to play a role in the pyrochlore phase diagram.

Classical spin configurations can be viewed as living inside the linear space $\smash{V=\mathbbm{R}^{3N_{\text{spin}}}}$ for $\smash{N_{\text{spin}}\propto L^3}$ spins with three spin components each, where $L$ is the linear dimension of the system.\footnote{
    For example, the FCC primitive cell contains four spins, and we can construct an $L\times L \times L$ periodic system with $N_{\text{spin}}=4L^3$ spins.
    }
The physical classical spin configuration space is a non-linear subspace defined by $\vert\bm{S}_i\vert^2 = 1$ for every spin, denoted $\mathscr{M}$, which is topologically a product of 2-spheres. 
The interaction matrix defines a quadratic form on $\mathscr{V}$, which then induces an energy function $H:\mathscr{M}\to\mathbbm{R}$, the classical Hamiltonian, on the non-linear physical configuration space $\mathscr{M}$. 
Let $\mathscr{F}$ denote the null space of the interaction matrix (spanned by the zero-energy eigenvectors, assuming the minimum eigenvalue is set to zero).
The ground state manifold $\mathscr{G}$ of the system is determined by the intersection
\begin{equation}
    \mathscr{G} = \mathscr{F} \cap \mathscr{M}.
\end{equation}
Of course actually determining the structure of this ground state manifold is generally a hard  problem~\cite{luttingerTheoryDipoleInteraction1946,litvinLuttingerTiszaMethod1974}.\footnote{
    It may happen that the intersection is empty, $\mathscr{G}=\varnothing$. Then the true ground state manifold is the locus in $\mathscr{M}$ where $H$ takes its minimum value, which is greater than the minimum eigenvalue of the interaction matrix. 
    Examples with Ising spins are easy to construct, where $\mathscr{M}$ is a discrete space---a product of 0-spheres, corresponding to the corners of a hypercube. 
    In particular this happens in all Hamiltonians of the form $\smash{\cramped{H=\sum_c \big(\sum_{i \in c}S_i^z \,\big)^2}}$ with an odd number of spins per cluster $c$, since the constraint $\sum_{i\in c} S_i^z = 0$ cannot be satisfied.
    This happens in the nearest-neighbor Ising antiferromagnet on the kagome~\cite{willsModelLocalizedHighly2002} and triangular~\cite{wannierAntiferromagnetismTriangularIsing1950} lattices. 
    We are not aware of any examples where this occurs with continuous spins---for corner-sharing cluster Hamiltonians this would mean that some ground states on a single cluster cannot be constructed from a single irrep. 
    }

\subsubsection{Symmetry Breaking}

The most common situation is that $\mathscr{F}$ is 1-, 2-, or 3-dimensional, corresponding to one $\bq=\bm{0}$ irrep having the lowest energy. 
In that case $\mathscr{G}$ usually consists of a discrete set of states related by the discrete space group symmetries, of which one is spontaneously chosen below a critical temperature, breaking the symmetry.
All spin wave excitations are then gapped as the ground state is a stable minimum of the Hamiltonian. 
We can expand this function to quadratic order in small perturbations around a point in $\mathscr{G}$ by parameterizing $\mathscr{M}$ locally in a small patch.
For each spin $\bm{S}_i$ define a local orthonormal basis $\{\tilde{\bm x}_i,\tilde{\bm y}_i,\tilde{\bm z}_i\}$ such that in the ground state all spins point along $\tilde{\bm z}_i$. Then small deviations away from this configuration can be parameterized by $S_i^{\tilde x}$ and $S_i^{\tilde y}$ along with the constraint 
\begin{equation}
    \cramped{S_i^{\tilde z} = \sqrt{1 - (S_i^{\tilde x})^2 - (S_i^{\tilde y})^2}}.
\end{equation}
Taylor expanding the Hamiltonian to quadratic order,
\begin{equation}
    H = E_0 + 
    \sum_{ij} 
    \sum_{\tilde{a},\tilde{b}\in\{\tilde{x},\tilde{y}\}}
    S_i^{\tilde a} \Hessian_{ij}^{\tilde{a}\tilde{b}} S_j^{\tilde b} 
    + 
    \mathcal{O}[(S_i^{\tilde a})^4],
\end{equation}
we obtain the Hessian $\Hessian$. 
If the ground state manifold is discrete, then the Hessian has all positive eigenvalues. A convenient formula for the Hessian is to let $\xyprojector$ be the rectangular matrix that projects into the space spanned by the $\tilde{\bm{x}}_i$ and $\tilde{\bm{y}}_i$ spin components, then
\begin{equation}
    \Hessian = \xyprojector\Jmat\xyprojector + \tilde{h}_0,
    \label{eq:Hessian}
\end{equation}
where we have defined the matrix
\begin{equation}
    [\tilde{h}_0]_{ij}^{\tilde{a}\tilde{b}} 
    \equiv 
    \left(
    - 
    \sum_{k} 
    \sum_{\tilde{c},\tilde{d}} 
    \tilde{z}_i^{\tilde{c}} \Jmat_{ij} \tilde{z}_j^{\tilde{d}}
    \right)
    \delta_{ij} \xyprojector^{\tilde{a}\tilde{b}}.
    \label{eq:on_site_energy}
\end{equation}
The first term in \cref{eq:Hessian} measures the energy cost coming from the transverse (to the ordering axis) deformation of the spin configuration, while the second term defined in \cref{eq:on_site_energy} measures the energy cost of reducing the longitudinal component (along the ordering axis) of the spin configuration.

\subsubsection{Order by Disorder}

An exception occurs in the pyrochlore $E$ phase ($\dim\mathscr{F} = 2$), \cref{fig:ground_states}(b,c), which spontaneously breaks SO(2) spin rotation symmetry without breaking lattice symmetries.
The ground state manifold is a 1-dimensional circle, $\mathscr{G}\simeq S^1$, parameterized by continuous rotation of spins about the local easy-axis.
In this case, a set of discrete ground states, either $\psi_2$ or $\psi_3$ in \cref{fig:ground_states}(a,b), are selected at finite temperature due to the asymmetry of small fluctuations about different points in $\mathscr{G}$~\cite{moessnerLowtemperaturePropertiesClassical1998,wongGroundStatePhase2013,yanTheoryMultiplephaseCompetition2017,hallasExperimentalInsightsGroundState2018,rauFrustratedQuantumRareEarth2019,chernPyrochloreAntiferromagnetAntisymmetric2010,javanparastOrderbydisorderCriticalityXY2015,javanparastFluctuationDrivenSelectionCriticality2015}, which is a standard example of order-by-disorder~\cite{henleyOrderingDueDisorder1989}.
While the ground state manifold does not break lattice symmetries, the spectrum of the Hessian is sensitive to the fact that different points on $\mathscr{G}$ are symmetry-inequivalent---the ground state has an \emph{accidental} SO(2) symmetry that is not a symmetry of the Hamiltonian.
Thus the energy costs of small deformations away from different points on $\mathcal{G}$ are inequivalent. 
The $\psi_2$ and $\psi_3$ configurations have higher symmetry than generic points on $\mathscr{G}$, with spins lying either in mirror planes ($\psi_2$) or along $\pi$ rotation axes ($\psi_3$), and are selected due to enhanced finite-temperature fluctuations. 
Thus the $E$ phase is actually further split into two phases---when $J_{z\pm} = 0$, $J_{\pm\pm}>0$ selects $\psi_2$ and $J_{\pm\pm}<0$ selects $\psi_3$, while the $J_{z\pm}\neq 0$ shape of the phase boundaries is significantly more complicated~\cite{yanTheoryMultiplephaseCompetition2017,wongGroundStatePhase2013,rauFrustratedQuantumRareEarth2019}.

While this XY selection due to accidental SO(2) degeneracy has been well-known for a long time, it was only recently realized that the same mechanism occurs in the collinear ferromagnet ground state~\cite{hickeyOrderbydisorderQuantumZeropoint2025}.
This has an accidental SO(3) symmetry in the ground state, from which high-symmetry [001], [110], and [111] directions are spontaneously chosen at finite temperature. 
The authors of Ref.~\cite{hickeyOrderbydisorderQuantumZeropoint2025} studied the Heisenberg-plus-DM model (cf. \cref{fig:phase_diagram_HDM}), but it would be very interesting to extend their results to the entire $\Tpar\oplus\Tperp$ constant-canting surface ($\alpha=1/3$ in \cref{fig:canting_foliation}) in the $T_1$ phase. 
For $J_{z\pm}>0$ the ground state is a collinear ferromagnet, while for $J_{z\pm}<0$ the ground state is made of linear combinations of the $\Tperp$ configurations from \cref{fig:ground_states}(f), which also has an accidental SO(3) symmetry. 
Furthermore, the $J_{z\pm}$-dual of this constant-canting surface ($\alpha = -1/3$) will also exhibit an accidental SO(3) symmetry in the $T_1$ phase. 

Assuming one can parameterize all of $\mathscr{G}$, this sort of order by disorder can be diagnosed from the Hessian spectrum at different points in $\mathscr{G}$. 
The system generally selects points with the softest fluctuations out of $\mathscr{G}$ (fluctuations within $\mathscr{G}$ always cost zero energy). 
This is equivalent to performing a low-temperature expansion of the free energy, which induces an entropic potential landscape on the ground state manifold~\cite{yanTheoryMultiplephaseCompetition2017,hickeyOrderbydisorderQuantumZeropoint2025}.

\subsubsection{Order by Singularity}

Note that despite the standard terminology ``ground state manifold'', $\mathscr{G}$ need not be a manifold in the mathematical sense. 
For example, the XY antiferromagnet on a single tetrahedron has a ground state space $\mathscr{G}$ consisting of three tori which touch pairwise along three circles~\cite{khatuaEffectiveTheoriesQuantum2019}. The dimension $\dim \mathscr{G}$ of the ground state subspace is well-defined locally, but there are singular subspaces where $\mathscr{G}$ intersects itself where the dimension is ill-defined. 
Refs~\cite{khatuaEffectiveTheoriesQuantum2019,srinivasanOrderSingularityKitaev2020,khatuaStateSelectionFrustrated2021} have proposed that these singular points generate ``order-by-singularity'', i.e. despite the degeneracy of the ground state manifold the system selects the singular points on $\mathscr{G}$. 
Heuristically, one may view this mechanism as a sort of ``inverse gimbal lock'', where the system gets stuck near singular points where it has enhanced degrees of freedom.

If $\mathscr{G}$ is continuous then one expects there to be gapless modes in the spin wave spectrum about any point in $\mathscr{G}$, and these should be enhanced at the intersection points. Ref.~\cite{noculakClassicalQuantumPhases2023} found that this mechanism selects the $\psi_3$ states when $E$ and $\Tperp$ are degenerate (cf. \cref{fig:phase_diagram_HDM}), because the ground state manifold locally has the form of two circles which intersect at the $\psi_3$ configuration.
The result is that the spin wave spectrum about the $\psi_3$ state has two gapless modes, whereas at a generic point in $\mathscr{G}$ it only has one.
Similarly, Ref.~\cite{yanTheoryMultiplephaseCompetition2017} studied the $E \oplus T_2$ phase boundary and found that singularities in the ground state manifold select the $\psi_2$ state. See also \cref{foot:obd}.

\subsubsection{Sub-extensive Flat Bands}

It can occur that the dimension of $\mathscr{F}$ is large but subextensive, e.g. when the interaction matrix has zero-energy flat lines or flat planes. 
This is actually quite a common feature throughout the pyrochlore phase, as tabulated in \cref{tab:flat_bands_triple_lines,tab:flat_bands_phase_boundaries,tab:flat_bands_phases}.
For example, the entire $T_2$ phase along with the $T_{1,\text{planar}}$ portion of the $T_{1}$ phase exhibit flat lines along $(hhh)$.
Furthermore, the $A_2 \oplus E \oplus T_1$ triple line and the $T_1 \oplus T_2$ phase boundary exhibit flat planes in the high-symmetry $\{hhl\}$ reciprocal space planes.

Due to the subextensive degeneracy, $\mathscr{G}$ can be at most a subextensive ground state manifold.
We believe it is most likely that $\mathscr{G}$ will be a discrete space, however. 
For example, models with a $T_2$ or $T_{1,\text{planar}}$ ground state host flat lines, but we know that these phases do not host $\bm{q}\neq\bm{0}$ ground states, i.e. $\mathscr{F}$ and $\mathscr{M}$ only intersect in the $\bm{q}=\bm{0}$ subspace.\footnote{
    The flat lines can, however, have significant impacts on the nature of the finite-temperature ordering transition, because the corresponding flat modes will have to be included in a Ginzburg-Landau description of the phase transition~\cite{cepasDegeneracyStrongFluctuationinduced2005,scheieDynamicalScalingSignature2022}.
}
On the other hand, Ref.~\cite{franciniHigherRankSpinLiquids2025} has reported that the $T_1\oplus T_2$ phase boundary exhibits a spin nematic ground state rather than either $T_1$ or $T_2$ order, which may be a common occurrence when flat planes are present.

One example where $\mathscr{G}$ is connected and has subextensive dimension are spiral spin liquids~\cite{bergmanOrderbydisorderSpiralSpinliquid2007,niggemannClassicalSpiralSpin2019,yaoGenericSpiralSpin2021,gaoLineGraphApproachSpiral2022,yanLowenergyStructureSpiral2022}. 
These are models where the subextensive flat bands form a closed surface in the Brillouin zone with non-trivial homotopy. 
In some models~\cite{niggemannClassicalSpiralSpin2019} it is known that each point on the zero-energy flat band corresponds to a ground state, so $\mathscr{G}$ has the same dimension as the zero-energy locus. 
These generally require fine-tuned further-neighbor interactions that bring bands down at wavevectors away from the zone center or boundaries~\cite{ballaAffineLatticeConstruction2019}.
However, we note that it is difficult to differentiate the effects of order-by-disorder and order-by-singularity, and they likely often occur in tandem, i.e. the effective entropic potential which selects a ground state exhibits minima at the singular points.

\subsubsection{Extensive Flat Bands and Spin Liquids}

If the interaction matrix has $\cramped{n_{\text{flat}}>0}$ zero-energy flat bands out of $n_{\text{tot}}$ total bands,\footnote{
    Note that $n_{\text{tot}}$ is three times the number of spins per primitive unit cell, $n_{\text{tot}}=12$ for the pyrochlore lattice.} 
then $\mathscr{F}$ is a $(n_{\text{flat}}/n_{\text{tot}})N_{\text{spin}}$-dimensional linear subspace of $\mathscr{V}$ spanned by the flat band eigenvectors, opening the possibility that the (local) dimension of $\mathscr{G}$ is extensive. 
Having extensive zero-energy flat modes does not necessarily guarantee that $\mathscr{G}$ is extensive, however. 
There is general folklore that if the number of flat bands is too low (compared to the total number of bands) the system will order rather than realize a stable spin liquid, though a proof that there is a lower bound on the required number is lacking. 
As a rough rule of thumb, we can use an elementary formula from linear algebra: the intersection of a $d$-dimensional and $d'$-dimensional linear subspace in a $D$-dimensional vector space has dimension $d+d'-D$.\footnote{
    We thank Tim Henke for pointing this out to us.
    } 
``Blindly'' applying this formula, we have
\begin{equation}
    \dim \mathscr{G} = \dim (\mathscr{F}\cap \mathscr{M}) \sim \dim \mathscr{F} + \dim \mathscr{M} - \dim \mathscr{V}.
\end{equation}
We use a $\sim$ to emphasize that the formula does not strictly apply because $\mathscr{M}$ is not a linear subspace. 
Since $\mathscr{M}$ is a product of 2-spheres it has dimension $\dim\mathscr{M}=(2/3)\dim\mathscr{V}$, which implies that for $\mathscr{G}$ to have dimension greater than zero we should have
\begin{equation}
    \dim\mathscr{F} \gtrsim \dim\mathscr{V}/3 \quad (\dim\mathscr{G}>0).
\end{equation} 
This suggests that at least a third of the bands should be flat to have a continuous ground state degeneracy. 
This is, of course, not a proof, since $\mathscr{M}$ is a non-linear subspace, and we expect many counter examples exist.\footnote{
    One straightforward possibility is that not all bands ``participate'' in the spin liquid ground state, in which case we should count the ratio only in terms of those that do participate. For example, in the spin ice region of the phase diagram there are only two flat bands of twelve, suggesting a ratio $1/6 < 1/3$; however, since $J_{z\pm} = 0$ the local-$xy$ degrees of freedom are decoupled from the local-$z$, and we can separate the band structure into four $z$ bands and eight $xy$ bands; only the $z$ bands contribute to the ground state, thus the actual relevant ratio is $1/2 > 1/3$. 
}
Nevertheless, we believe that it should serve well as a general rule of thumb for a crossover between stability and instability of classical spin liquids.

\begin{figure*}[t]
    \centering
    \includegraphics[width=\textwidth]{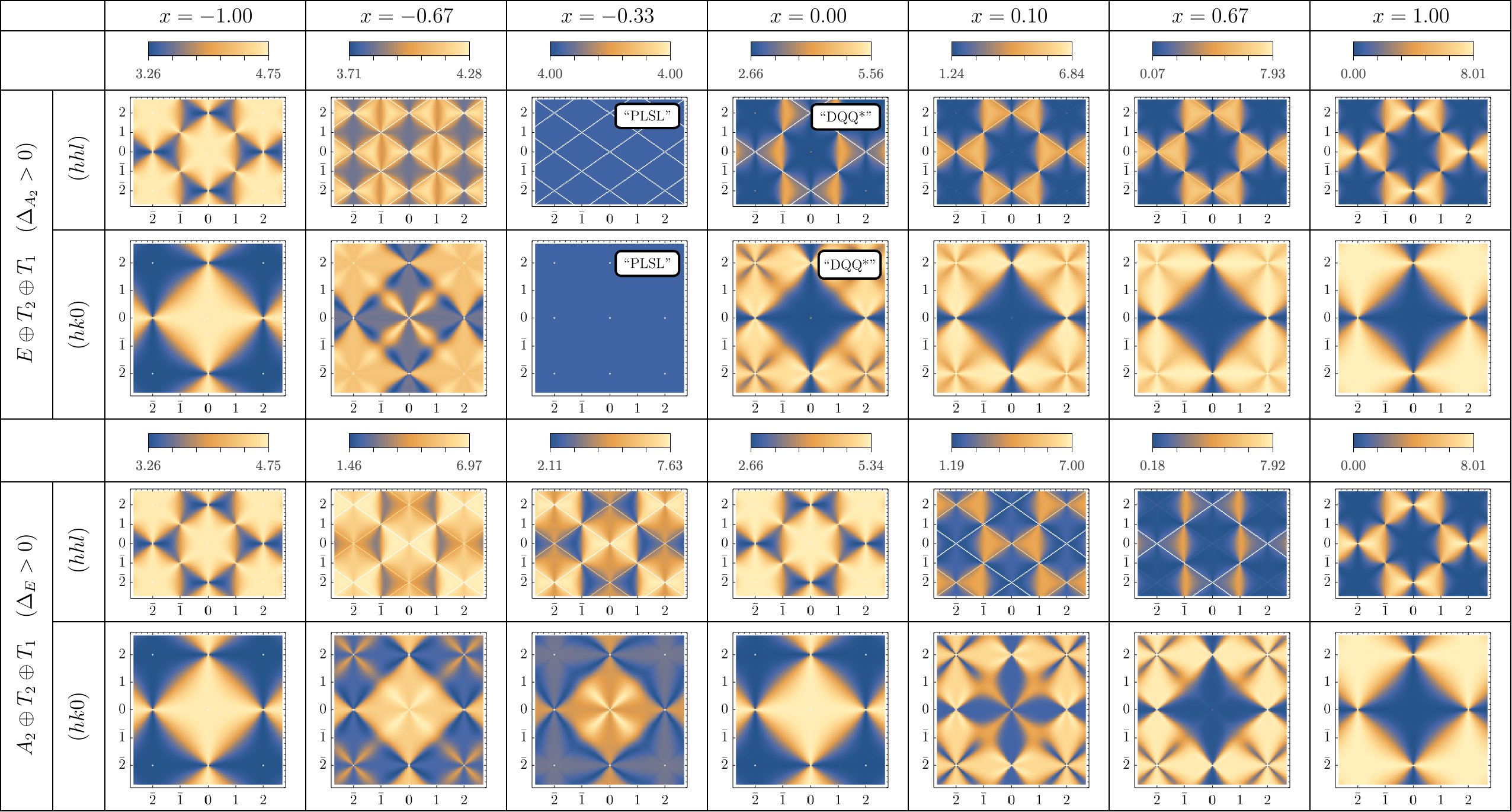}
	\caption{Evolution of the zero-temperature SCGA structure factors (\cref{apx:scga}), computed from the projector to the flat bands, for two of the four 3-fold degenerate lines---$E\oplus T_2\oplus T_1$ (top two rows) and $A_2\oplus T_2\oplus T_1$ (bottom two rows)---in the cubic $(hk0)$ and $(hhl)$ planes of reciprocal space.
    At the right end is the structure factor for the HAFM, while at the left end is the structure factor of the \HAFMdual, while the middle column is the $J_{z\pm} = 0$ case. 
    All cases show flat lines along $(hhh)$, visible in the $(hhl)$ plane, and fourfold pinch points, visible in the $(hk0)$ plane.
    Note that the color scale starts from the minimum intensity and goes to the maximum intensity of the flat band; higher intensities from the singular band touching lines and points appear white. 
    The $\Delta_{A_2}>0$ triple line was first discussed in Ref.~\cite{bentonSpinliquidPinchlineSingularities2016}, but only one point on the line---the ``pinch-line spin liquid'' (PLSL), corresponding to $J_3 < 0$ and all other couplings zero---was studied.
    The $J_{z\pm}=0$ point on this triple line is the $J_{\pm\pm}$-dual (denoted with an asterisk) of the ``DQQ'' model from Ref.~\cite{lozano-gomezCompetingGaugeFields2024}.
    The $\Delta_E>0$ triple line was recently studied in detail in Ref.~\cite{franciniHigherRankSpinLiquids2025}.
    }
    \label{fig:structure_factors_triple_lines}
\end{figure*}

There are a few possible scenarios which can then in principle distinguish:
\begin{enumerate}[leftmargin=*]
    \item The intersection is zero-dimensional, $\dim\mathscr{G}=0$. 
    Then $\mathscr{G}$ is a discrete space, and at each point in $\mathscr{G}$ the Hessian is positive-definite.
    A generic spin configuration is orthogonal to $\mathscr{F}$.
    We can distinguish the following cases\footnote{We use the notation $\pi_0(X)$ is the zero'th homotopy group of $X$, and $\vert\pi_0(X)\vert$ is the number of disconnected components of $X$. }
    \begin{enumerate}
        \item The number of discrete ground states, $\vert \pi_0 (\mathscr{G}) \vert$, is of order one. The system orders.
        \item The ground state entropy is subextensive, such that $\log \vert \pi_0 (\mathscr{G}) \vert \propto L^k$ with $0<k<3$. The system may potentially undergo dimensional reduction without ordering~\cite{yanTheoryMultiplephaseCompetition2017}. 
        \item The ground state entropy is extensive, such that $\log \vert \pi_0 (\mathscr{G}) \vert\propto L^3$. We further differentiate:
        \begin{enumerate}
            \item The ground state is a trivial paramagnet;
            \item the ground state is a symmetry-broken~\cite{chung2formU1Spin2025,hallenThermodynamicsFractalDynamics2024} or fragmented~\cite{brooks-bartlettMagneticMomentFragmentationMonopole2014,lhotelFragmentationFrustratedMagnets2020} spin liquid;
            \item the ground state is a symmetric spin liquid, as in the $T_{1,\text{ice}}$ region of the pyrochlore phase diagram.
        \end{enumerate}
    \end{enumerate}
    \item The intersection is sub-extensive, i.e. $\cramped{\dim \mathscr{G} \propto L^k}$ with $\cramped{0< k<3}$.
    Then $\mathscr{G}$ is parameterized locally by a subextensive number of continuous compact parameters.
    It may have one or more connected components.
    A generic spin configuration has a sub-extensive projection into~$\mathscr{F}$.
    \item The intersection is extensive, $\dim \mathscr{G} \propto L^3$, and is parameterized locally by an extensive number of parameters. 
    It may have one or more connected components.
    A generic spin configuration has non-zero overlap with $\mathscr{F}$ of order $(\dim\mathscr{G})/L^3$. 
    We can further distinguish the following possibilities: 
    \begin{enumerate}
        \item The system exhibits symmetry breaking, either
        \begin{enumerate}
            \item order-by-singularity selects a ground state;
            \item order-by-disorder selects a ground state;
            \item nematic but liquid-like order~\cite{taillefumierCompetingSpinLiquids2017,franciniHigherRankSpinLiquids2025}.
        \end{enumerate}
        \item The system exhibits no symmetry breaking, either
        \begin{enumerate}
            \item a liquid-to-liquid crossover~\cite{lozano-gomezCompetingGaugeFields2024};
            \item the ground state is a fully symmetric spin liquid, e.g. the pyrochlore HAFM~\cite{moessnerLowtemperaturePropertiesClassical1998,henleyCoulombPhaseFrustrated2010}.
        \end{enumerate}
    \end{enumerate}
\end{enumerate}

Case 1(a) is likely a common occurrence whenever the number of flat bands is less than a third of the total number of bands, following the rule of thumb. 
Case 1(b) and case 2 both imply there is an obstruction to consistently ``gluing together'' single-tetrahedron ground states in a corner-sharing fashion to build ground states of the entire system, i.e. what Ref.~\cite{yanTheoryMultiplephaseCompetition2017} calls the ``Lego-brick'' construction. 
In case 2 $\mathscr{G}$ is connected, but we expect that the subextensive dimension will result in either order-by-singularity or order-by-disorder ground state selection.
Unfortunately we do not known any explicit examples where this case occurs. 
Lastly, case 3(a) has an extensive ground state manifold yet a symmetry broken ground state is selected. 
We are not aware of examples of possibilities (i) or (ii), but believe one of these may occur on the $\Delta_E > 0$ line, where  Ref.~\cite{franciniHigherRankSpinLiquids2025} recently reported selection of the $A_2$ all-in-all-out configurations when $J_{z\pm}$ is small. We discuss this further in \cref{sec:stacked_flat_bands}.

If the system avoids these ordering mechanisms, this leaves cases 1(c) and 3(b). Case 1(c)(i), a trivial paramagnetic ground state, can occur for example in an antiferromagnetic Potts model with a sufficiently large number of states, where the system is trivially disordered to zero temperature~\cite{dengFiniteTemperaturePhaseTransition2011}. Case 1(c)(ii) means that the system exhibits an extensive but unsaturated ordered moment, while continuing to fluctuate as a spin liquid to zero temperature. 
Case 3(b)(i) was recently reported in Ref.~\cite{lozano-gomezCompetingGaugeFields2024} which we discuss briefly in the next section. 
Finally, cases 1(c)(iii) and 3(b)(ii) correspond to discrete and continuous symmetric spin liquids, respectively, many examples of which are known.

\subsection{History of Identified Pyrochlore Spin Liquids}
\label{sec:pyro_spin_liquids}

\begin{figure*}[t]
    \includegraphics[width=\textwidth]{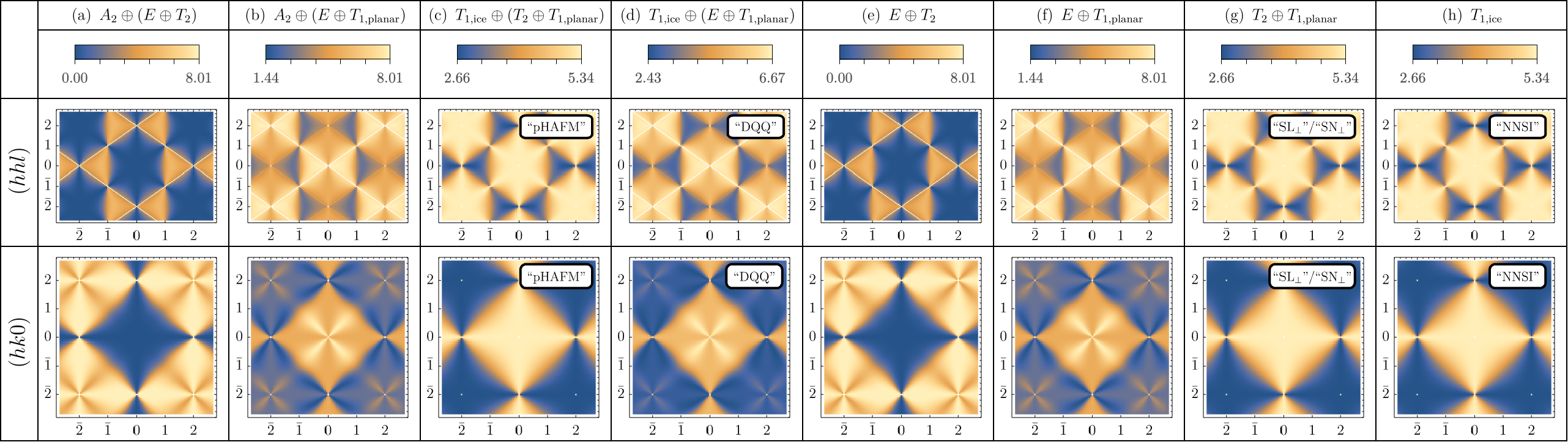}
	\caption{Spin structure factor for flat band cases not on the two lines shown in \cref{fig:structure_factors_triple_lines}. (a) On the $\Delta_{T_{-}}>0$ line there are two flat bands with pinch line singularities. Because the ground state does not involve a $T_1$ irrep, the ground state physics is independent of position along this line, and the structure factor does not evolve along it. (b) The $\Delta_{T_2}>0$ line does not have flat bands (only flat planes) except when $J_{z\pm}=0$. Note that in both cases (a) and (b) the $A_2$ phase is accidentally degenerate and does not contribute to the flat bands. (c) The isolated pseudo-HAFM triple point~\cite{taillefumierCompetingSpinLiquids2017} contains decoupled contributions from $T_{1,\text{ice}}$ and the easy-plane $T_1$ and $T_2$ irreps. (d) The isolated ``DQQ'' triple point~\cite{lozano-gomezCompetingGaugeFields2024} contains decoupled contributions from the $T_{1,\text{ice}}$ and the easy-plane $T_1$ and $E$ irreps. (e) The easy-plane $E$ and $T_2$ irreps combine to form non-trivial flat bands on this phase boundary. (f) The easy-plane $E$ and $T_1$ irreps combine to form non-trivial flat bands on this phase boundary in the $J_{z\pm}=0$ plane. (g) The easy-plane $T_2$ and $T_1$ irreps combine to form non-trivial flat bands on this phase boundary in the $J_{z\pm}=0$ plane. This region was studied in Ref.~\cite{taillefumierCompetingSpinLiquids2017} and found to be a nematic spin liquid. (h) The structure factor in the nearest-neighbor spin ice (NNSI) surface in the $T_1$ phase.
    Note that (a) is equivalent to (e) plus extra intensity at zone center due to the extra decoupled $A_2$ irrep; (b) is equivalent to (f) for the same reason; (c) is equivalent to (g) plus (h); and (d) is equivalent to (f) plus (h). 
    }
    \label{fig:structure_factors_special}
\end{figure*}

Previous studies have looked at a variety of spin liquid models on the pyrochlore lattice emerging at the triple-intersection of phases. 
The earliest recognized was the HAFM, which has long been a canonical example of a classical spin liquid~\cite{moessnerLowtemperaturePropertiesClassical1998,canalsPyrochloreAntiferromagnetThreeDimensional1998,canalsSpinliquidPhasePyrochlore2001,isakovDipolarSpinCorrelations2004,henleyPowerlawSpinCorrelations2005,conlonAbsentPinchPoints2010,henleyCoulombPhaseFrustrated2010}.
Referring to the first row of \cref{tab:flat_bands_triple_lines}, it has six flat bands with three quadratically dispersing bands touching at $\bm{q}=\bm{0}$.
At low temperature each tetrahedron is constrained to have zero net spin, i.e. $T_{1\parallel}$ is gapped (cf. \cref{eq:HAFM}). 
The low-temperature $T \to 0^+$ spin-spin correlations ($T$ being temperature) $\langle S_i S_j\rangle$ are $1/r^3$ power-law decaying~\cite{henleyPowerlawSpinCorrelations2005,henleyCoulombPhaseFrustrated2010}. 
Their Fourier transform, the spin structure factor,
\begin{equation}
    S(\bq) = \frac{1}{L^3}\sum_{i,j} \langle S_i S_j\rangle e^{-i\bq\cdot(\bm{r}_j - \bm{r}_i)},
    \label{eq:structure_factor}
\end{equation}
exhibits characteristic ``pinch-point'' correlations, i.e. a diffusive profile with bow-tie singularities located at zone centers due to the long-range correlations arising from a local constraint~\cite{henleyPowerlawSpinCorrelations2005,henleyCoulombPhaseFrustrated2010,conlonAbsentPinchPoints2010}.
The HAFM structure factor is shown in the right-most column of \cref{fig:structure_factors_triple_lines}.
As we have discussed throughout this paper, there is an additional dual HAFM point with negative $J_{z\pm}$, whose physics is exactly the same and whose ground states are in bijection with those of the HAFM by a $\pi$ rotation about the local easy-axis.
The corresponding structure factor is shown in the left-most column of \cref{fig:structure_factors_triple_lines}.
Similarly, the $T_{1,\text{ice}}$ ground state, occurring only when $J_{z\pm}=0$, is the spin ice spin liquid, with spins aligned along their easy-axes satisfying the two-in-two-out ice rules in the ground state~\cite{moessnerSpinIceCoulomb2021,castelnovoSpinIceFractionalization2012}, whose structure factor is shown in \cref{fig:structure_factors_special}(h). 
As mentioned in the introduction, these are the two limiting cases of completely isotropic and completely Ising spins, respectively. They both inherit their frustration from the flat bands of the pyrochlore lattice adjacency matrix, with Hamiltonians expressed as 
\begin{equation}
    H_{\text{HAFM}} \propto \sum_{ij} A_{ij} \bm{S}_i \cdot \bm{S}_j , 
    \quad 
    H_{\text{SI}} \propto \sum_{ij} A_{ij} S_i^z S_j^z,
\end{equation}
where $A_{ij}$ is the adjacency matrix of the pyrochlore lattice, $A_{ij} = +1$ for nearest-neighbors and zero otherwise. 
This matrix has two flat bands at the base of its spectrum.

It was not until 2016 that a new spin liquid would be identified with non-trivial anisotropies intermediate between the HAFM and spin ice limits. 
It was identified with the model $J_3<0$ and $J_1=J_2=J_4=0$ in terms of the global basis parameters in \cref{eq:Jijab}, and dubbed the ``pinch line spin liquid'' (PLSL)~\cite{bentonSpinliquidPinchlineSingularities2016}. 
In fact, Ref.~\cite{bentonSpinliquidPinchlineSingularities2016} discussed this model as a particular example of a family of models characterized by gapping $A_2$ and $T_{1+}$, i.e. the $\Delta_{A_2}>0$ line in \cref{fig:full-phase-diagram}, though the authors did not parameterize the entirely family.  
The interest new feature of this model was that it exhibits \emph{pinch line} singularities rather than pinch points, i.e. singular correlations everywhere along the high-symmetry $(hhh)$ line in reciprocal space. 
It corresponds to the second row of \cref{tab:flat_bands_triple_lines}, hosting four flat bands and a singe flat line along the high-symmetry $(hhh)$ lines in reciprocal space plus three quadratically dispersing bands at the zone center.
From the rest of the table it is clear that pinch lines are actually not unique to this model or this triple line---indeed they appear at all the triple intersections of phases.

In 2017 the ``pseudo''-HAFM (pHAFM) point was pointed out in Ref.~\cite{taillefumierCompetingSpinLiquids2017}, which was studying the XXZ family (i.e. with $J_{\pm\pm}=J_{z\pm}=0$).
The pHAFM is just the HAFM in the local basis, i.e. the isotropic limit of the XXZ family when $J_{\pm}=-J_{zz}/2$, $J_{zz}>0$.
The physics of this model is identical to the HAFM, since they are mapped to each other by a local on-site mapping of each spin from its local frame to the global frame. 
We have marked it in \cref{fig:full-phase-diagram}, it is the isolated triple point occurring on the $T_2\oplus T_1$ phase boundary where it is pierced by the $T_1$-degenerate locus.
It has degenerate irreps $T_2\oplus T_{1,\text{ice}}\oplus T_{1,\text{planar}}$ in \cref{tab:flat_bands_triple_lines}.

Ref.~\cite{taillefumierCompetingSpinLiquids2017} also studied the $T_2 \oplus T_{1,\text{planar}}$ line, which is also contained in the XXZ family.
This is the intersection of the $T_1\oplus T_2$ phase boundary with the $J_{z\pm}=0$ plane, starting at the pHAFM and extending to the $A_2$ phase.
According to \cref{tab:flat_bands_phase_boundaries} it hosts four flat bands with two quadratically dispersing bands at the zone center. 
This line contains a special point: the anti-ferromagnetic XY model,
\begin{equation}
    H_{XY} \propto \sum_{ij} A_{ij} (S_i^+ S_j^- + S_i^- S_j^+),
    \label{eq:H_XY}
\end{equation}
which was originally studied in \cite{moessnerLowtemperaturePropertiesClassical1998}, where it was found to undergo an order-by-disorder transition at low temperature.
However, the ground state may still be viewed as a type of spin liquid---the global U(1) symmetry is spontaneously broken, but once a direction in the local XY planes is chosen the system retains a two-in-two-out ice-type degeneracy.
Schematically, the ground state manifold is $\mathscr{G}_{XY} = \mathscr{G}_{\text{ice}}\times$U(1), with the U(1) degeneracy spontaneously broken.
Ref.~\cite{taillefumierCompetingSpinLiquids2017} found this behavior persists along the entire $T_2 \oplus T_{1,\text{planar}}$ line when weak $J_{zz}$ interactions are turned on: the system exhibits a finite-temperature spin liquid regime (dubbed ``SL$_{\perp}$''), with a phase transition to a nematic phase at low temperature with broken U(1) symmetry (dubbed ``SN$_{\perp}$'').

In 2020 another non-trivial spin liquid was then found, the so-called ``rank-2 U(1)'' spin liquid~\cite{yanRank2U1Spin2020}. 
This was inspired by a surge of interest in gauge theories with a symmetric rank-2 tensor electric fields, which are prototypical fractonic theories~\cite{pretkoGeneralizedElectromagnetismSubdimensional2017,pretkoSubdimensionalParticleStructure2017,premPinchPointSingularities2018,pretkoFractonPhasesMatter2020,nandkishoreFractons2019} whose excitations have highly constrained mobility due to local conservation of dipole and quadrupole moments.
Ref.~\cite{bentonSpinliquidPinchlineSingularities2016} first showed how to organize the irreps into a tensor, and Ref.~\cite{yanRank2U1Spin2020} exploited this structure by recognizing that a low-energy symmetric trace-free tensor field could be obtained by gapping the $A_2$ (trace component) and $T_2$ (anti-symmetric components) on each tetrahedron, along with the vector $T_{1\parallel}$.
This leaves only $T_{1\perp}\oplus E$ in the ground state, which together form the desired symmetric trace-free tensor on each tetrahedron---the quadrupole moment. 
This occurs along a one-dimensional locus in the $T_{1}\oplus E$ phase boundary along which $J_3 = 0$, so that the decoupled $T_1$ irreps are $T_{1\parallel}$ and $T_{1\perp}$.
This line is precisely traced by restricting the Hamiltonian \cref{eq:H_named_interactions} to only Heisenberg and DM interactions, which is shown in \cref{fig:phase_diagram_HDM} by the green circle (the inset shows how this family runs along the $T_1\oplus E$ phase boundary).
However, just being on this phase boundary is insufficient degeneracy for flat bands (c.f \cref{tab:flat_bands_phase_boundaries}), and the system selects the $E$ phase ground states~\cite{noculakClassicalQuantumPhases2023}. 
Ref.~\cite{yanRank2U1Spin2020} remedied this ordering by introducing breathing anisotropy to construct the Hamiltonian 
\begin{equation}
    H_{\text{R2U1}} = \sum_{t \in A} H_t^{E\oplus T_{1,\perp}} + \sum_{t\in B} H_t^{\text{HAFM}},
\end{equation}
where the first term selects the $E\oplus T_{1,\perp}$ ground states on the $A$ tetrahedra and the second term selects the  $(E\oplus T_{1,\perp}) \oplus A_1 \oplus T_2$ ground states on the $B$ tetrahedra.
This relaxes some of the constraints on the $B$ tetrahedra by allowing more fluctuating degrees of freedom in the ground state, and gives the correct number of flat bands and long-wavelength Gauss law. 
The rank-2 U(1) spin liquid is then obtained at finite temperatures, though the system still orders at sufficiently low temperature~\cite{sadouneHumanmachineCollaborationOrdering2025}.

In 2023 Ref.~\cite{lozano-gomezCompetingGaugeFields2024} studied another spin liquid, which was called the ``dipolar-quadrupolar-quadrupolar'' (``DQQ'') model. 
This is the second isolated triple point other than the pHAFM, sitting at the intersection of the $T_1\oplus E$ phase boundary and the $T_1$-degenerate line in \cref{fig:full-phase-diagram}, i.e. with degenerate irreps $E\oplus T_{1,\text{ice}}\oplus T_{1,\text{planar}}$ in \cref{tab:flat_bands_triple_lines}. 
This model was found to exhibit a classical liquid-to-liquid crossover, where the spin ice states are entropically selected from the ground state manifold at very low temperatures. 
It also has a $J_{\pm\pm}$-dual model, which is the point on the $\Delta_{A_2}>0$ line at $J_{z\pm}=0$. 

Recently, Ref.~\cite{franciniHigherRankSpinLiquids2025} studied the line with $\Delta_E>0$, and found that for small $J_{z\pm}$ it exhibits a ground state selection of $A_2$ order, while for larger $J_{z\pm}$ (closer to the HAFM) it exhibits a nematic ordering similar to that on the $\cramped{T_2\oplus T_{1,\text{planar}}}$ line. It was also reported that the nematic symmetry breaking occurs ubiquitously on the $T_1\oplus T_2$ phase boundary, which may be a consequence of flat planes (rather than flat bands), cf.~\cref{tab:flat_bands_phase_boundaries}.

Finally, Ref.~\cite{lozano-gomezAtlasClassicalPyrochlore2024}, which appeared concurrently with this work, performed an exhaustive search of all degenerate combinations of irreps and determined the corresponding Gauss laws and stability for pyrochlore spin liquids. 
They identify the \HAFMdual$\,$ model and show that the $\Delta_{A_2}>0$ line is a stable spin liquid family, but otherwise have found that there are no additional stable classical spin liquids with nearest-neighbor interactions.

Before moving on, we note that further-neighbor interactions can stabilize entirely novel spin liquids, in particular the author has shown in Ref.~\cite{chung2formU1Spin2025} the existence of a \emph{2-form U(1)} classical spin liquid with Ising spins on the pyrochlore lattice, by generalizing the zero-divergence constraint of spin ice to a zero-curl constraint. 
Its isotropic extension generalizes the HAFM spin liquid, and the possibility of anisotropic generalization of 2-form U(1) spin liquids remains an open possibility.

\begin{table}
    \centering
    \includegraphics[width=\columnwidth]{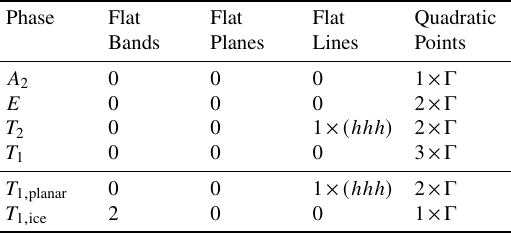}
    \\
    \vspace{-2.5ex}
    \caption{Flat degeneracies within each phase (cf.~\cref{fig:ground_states}). For the $T_1$ phase additional degeneracies occur when $J_{z\pm}=0$. $T_2$ and $T_{1,\text{planar}}$ have flat lines along the $(hhh)$ directions, while $T_{1,\text{ice}}$ hosts two flat bands corresponding to the spin ice ground states.
    Note that the sum of numbers in each row equals the dimension of the irrep in the first column---the number of zero modes at the $\Gamma$ ($\bm{q}=\bm{0}$) point.
    }
    \label{tab:flat_bands_phases}
    \includegraphics[width=\columnwidth]{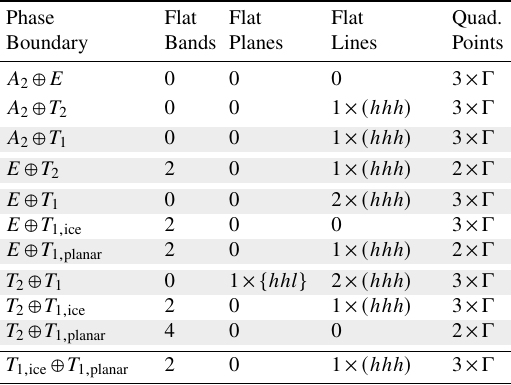}
    \\
    \vspace{-2.5ex}
    \caption{Flat degeneracies on the phase boundaries. Rows highlighted in gray indicate that the flat degeneracies are more than the sum of parts in \cref{tab:flat_bands_phases}. The $E\oplus T_2$ boundary carries two flat bands, while the $E\oplus T_1$ and $T_2\oplus T_1$ phase boundaries carry non-trivial flat bands when $T_1 = T_{1,\text{planar}}$. 
    }
    \label{tab:flat_bands_phase_boundaries}
    \includegraphics[width=\columnwidth]{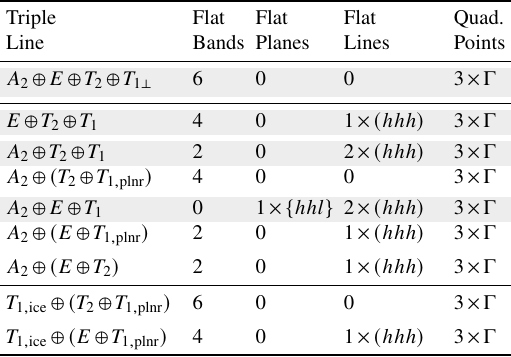}
    \\
    \vspace{-2.5ex}
    \caption{Flat band degeneracies on the triply-degenerate cases listed in \cref{tab:high_degeneracies} and depicted in \cref{fig:full-phase-diagram}. The HAFM points have the highest degeneracy, which is partially lifted along the four triply-degenerate lines. 
    Note that combining easy-axis with easy-plane irreps always results in a trivial stacking of degeneracies, indicated with parentheses on non-highlighted lines.
    }
    \label{tab:flat_bands_triple_lines}
\end{table}

\subsection{Irrep Collisions and Catalog of Pyrochlore Flat Bands}

With the full structure of the phase diagram now plainly exposed in \cref{fig:full-phase-diagram}, we can systematically account for all possible collisions (degenerate combinations) of ground state irreps, how they relate to each other in the phase diagram, and how they generate flat bands. 
Starting with the triple lines, these can be parameterized by a single parameter $x\in[-1,1]$, such that 
\begin{equation}
    \sign(J_{z\pm})= \sign(x) 
    \quad 
    \text{and}
    \quad 
    \frac{\Delta_{I'}}{\Delta_{T_{1+}}}=\max\left(\frac{\Delta_{I'}}{\Delta_{T_{1+}}}\right) (1-\vert x \vert),
    \label{eq:triple_line_parameterization}
\end{equation}
where the maximum is given by the right hand side of the inequalities in \cref{tab:high_degeneracies}.
Note that for $I'=T_{1-}$ the line is parameterized in \cref{tab:high_degeneracies} according to the splitting of the $T_1$ energies, but for the $\Delta_{T_{1-}}\geq 0$ portion it can be parameterized as in \cref{eq:triple_line_parameterization} with $0\leq \Delta_{T_{1-}}/\Delta_{T_{1+}}\leq 1/2$. 
For all four lines this parameterization yields the HAFM at $x=1$ and the \smash{\HAFMdual} at $x=-1$.
The flat band degeneracies on these lines are listed in \cref{tab:flat_bands_triple_lines}.

For each of the lines we have computed the zero-temperature spin structure factor,\footnote{
    While it would be more experimentally relevant to present neutron scattering cross sections, these would require a choice of $g$-tensor which relates the pseudo-spin to the magnetic moment of the ion, which is a material-dependent quantity that must be determined on a case-by-case basis. The spin structure factor corresponds to an isotropic $g$-tensor. 
} \cref{eq:structure_factor}, in the self-consistent Gaussian approximation (SCGA)~\cite{garaninClassicalSpinLiquid1999,canalsClassicalSpinLiquid2002,conlonAbsentPinchPoints2010} in the limit of zero temperature, which is proportional to the projector $P_{\mathscr{F}}$ to the flat band subspace (cf. \cref{apx:scga}),
\begin{equation}
    S(\bq)_{T\to 0^+} 
    = \frac{n_{\text{tot}}}{3n_{\text{flat}}}
    \sum_{\mu,\nu} \sum_\alpha 
    [P_{\mathscr{F}}(\bq)]_{\mu\nu}^{\alpha\alpha} \quad (\text{SCGA}).
\end{equation}
The touching of the dispersive bands with the flat bands give rise to characteristic pinch singularities in these structure factors. 
\Cref{fig:structure_factors_triple_lines} shows the evolution of the structure factor along the $\Delta_{A_2}>0$ and $\Delta_{E}>0$ lines, both exhibiting a combination of pinch lines and pinch points. 
The triple line with $\cramped{\Delta_{T_{1-}}>0}$ has no dependence on the canting angle because both $T_1$ irreps are gapped along this line, meaning that the structure factor is the same everywhere on this line, shown in \cref{fig:structure_factors_special}(a). 
The last triple line, with $\Delta_{T_2}>0$ has no flat bands, only flat planes, except when $\cramped{J_{z\pm}=0}$, shown in \cref{fig:structure_factors_special}(b). 
The remaining two triple points, the pHAFM and ``DQQ'' models, are shown in \cref{fig:structure_factors_special}(c,d).
Various phase boundaries also exhibit flat band degeneracies, which are shown in \cref{fig:structure_factors_special}(e,f,g). 
Lastly, the structure factors for spin ice, i.e. the  $T_{1,\text{ice}}$ ground state, are shown in \cref{fig:structure_factors_special}(h).

\subsubsection{Trivially Stacked Band Touchings and Easy-Axis Selection}
\label{sec:stacked_flat_bands}

Simply tuning additional irreps to be degenerate does not necessarily enhance the flat band degeneracies, meaning that the long-wavelength band structure at these points is just a trivial ``stacking'' of the band structure of the separate irreps, i.e. an accidental degeneracy. 
We should therefore ask which combinations of irreps \emph{enhance} the degeneracies beyond the individual components. 
Another way to state this is that, when different irreps are made degenerate, their combined orthonormalized eigenvectors can wind topologically around the reciprocal space loci where the flat bands touch the dispersive bands, characterized by a set of topological invariants~\cite{yanClassificationClassicalSpin2024,yanClassificationClassicalSpin2024a,fangClassificationClassicalSpin2024}; when we add an additional irrep degeneracy, do these topological invariants change?
If so, then the new irrep eigenvectors became ``intertwined'' with the old ones, and correspondingly there are new degeneracies in the flat band spectrum.
To study this systematically, we provide the degeneracies within each phase in \cref{tab:flat_bands_phases}, and on each phase boundary in \cref{tab:flat_bands_phase_boundaries}. 

We begin by looking at the easy-axis irreps $A_2$ and $T_{1,\text{ice}}$, checking the two irrep combinations in \cref{tab:flat_bands_phase_boundaries} against the corresponding degeneracies in \cref{tab:flat_bands_phases}. 
These cases are not highlighted because they do not exhibit intertwinement: they do not contain any additional degeneracies beyond their individual components.
For example, the degeneracies on the $A_2 \oplus T_2$ phase boundary are simply the sum of the degeneracies of $A_2$ plus those of $T_2$, and similarly for $A_2\oplus E$ and the three combinations involving $T_{1,\text{ice}}$. 
Tuning these irreps to become degenerate does not lead to any intertwining, they are trivially stacked and the eigenvectors do not mix. 
In the cases with $T_{1,\text{ice}}$ the reason is clear: these combinations involve an easy-axis and easy-plane irrep with $J_{z\pm}=0$, meaning that the easy-axis and easy-plane components are not coupled in the Hamiltonian and cannot mix. 
The cases with $A_2$ are similar, with a stacked easy-axis and easy-plane irrep, and such stacking remains trivial when $J_{z\pm} \neq 0$.
This pattern continues in the triple degeneracies listed in \cref{tab:flat_bands_triple_lines}: stacking an easy-axis irrep with easy-plane irreps is always trivial.

We can also see the trivial stacking in the corresponding structure factors shown in \cref{fig:structure_factors_special}.
For example, the $\cramped{A_2\oplus E \oplus T_2}$ line in \cref{tab:flat_bands_triple_lines} has the same flat degeneracies of the $\cramped{E\oplus T_2}$ phase boundary in \cref{tab:flat_bands_phase_boundaries} plus one extra zone-center quadratic band touching coming from the $A_2$ irrep. 
Correspondingly, the SCGA structure factor for this line shown in \cref{fig:structure_factors_special}(a) is exactly the same as that of the $E\oplus T_2$ phase boundary in \cref{fig:structure_factors_special}(e) (plus some extra zone-center intensity from the $A_2$ degeneracy not visible in the figure).
The same occurs for the $\cramped{A_2\oplus E \oplus T_{1,\text{planar}}}$ combination in \cref{fig:structure_factors_special}(b) compared to the $\cramped{E\oplus T_{1,\text{planar}}}$ combination in \cref{fig:structure_factors_special}(f).
Similarly, the two combinations with $T_{1,\text{ice}}$, (b) and (c), are equivalent to adding the same combination without $T_{1,\text{ice}}$, (g)~and~(f) respectively, with the $T_{1,\text{ice}}$ structure factor in~(h). 

Such trivial stacking does not necessarily rule out interesting physics arising due to the enhanced degeneracy. 
For example, the $E \oplus T_{1,\text{planar}}\oplus T_{1,\text{ice}}$ triple point (cf. \cref{fig:structure_factors_special}(d)) was studied in detail in Ref.~\cite{lozano-gomezCompetingGaugeFields2024} (dubbed the ``DQQ'' model), and exhibits a finite-temperature liquid-to-liquid crossover. 
At higher temperatures it develops correlations matching the SCGA, while at very low temperature it selects the spin ice manifold without undergoing a phase transition. 
The reason is because the degeneracy with the easy-plane configurations implies that if the spins are aligned along the easy-axis their transverse fluctuations are greatly enhanced, meaning that the Hessian about the easy-axis configurations will contain gapless easy-plane modes on each tetrahedron. 
The Hessian for a spin ice state has an extensive number of zero modes, while for a generic ground state it does not~\cite{lozano-gomezCompetingGaugeFields2024}.

We believe that this mechanism should be universal: selection of the easy-axis configurations will be preferred for all cases in which the ground state irreps split into decoupled easy-axis and easy-plane components.
From \cref{eq:Hessian} it clear that if $\tilde{\bm{z}}_i$ are along the easy-axis then the Hessian will inherit all of the zero-energy easy-plane normal modes of the interaction matrix.
In particular, this implies that cases that are trivially stacked with $A_2$ will select the all-in-all-out ground state, an example of order-by-singularity.  
In this sense, the liquid-to-liquid crossover of the DQQ model studied in Ref.~\cite{lozano-gomezCompetingGaugeFields2024} is an example of a selection of the ice manifold due to singularities in the ground state manifold, a ``easy-axis-spin-liquid-by-singularity''.

\subsubsection{Intertwined Flat Bands}

On the other hand, combinations of two easy-plane irreps always yields an enhanced degeneracy, highlighted gray in \cref{tab:flat_bands_phase_boundaries}. 
For example, $T_2 \oplus T_{1,\text{planar}}$ has four flat bands despite neither the $T_2$ phase nor the $T_{1,\text{planar}}$ region of the $T_1$ phase having any flat bands. 
This implies that the combined orthonormalized irrep eigenvectors have a non-trivial winding around the zone center.
We refer to such cases as ``intertwined'' flat bands, which are \emph{more than the sum of their parts}. 
The complete list of phase boundaries with intertwined extensive flat bands are precisely the three combinations of the easy-plane ground states: $E \oplus T_2$, $E\oplus T_{1,\text{planar}}$, and $T_2\oplus T_{1,\text{planar}}$. 
The three boundaries of the $T_1$ phase generically contain intertwined but subextensive flat bands: $A_2 \oplus T_1$ and $E\oplus T_1$ have intertwined flat lines while $T_2\oplus T_1$ has intertwined flat planes. 
Finally, moving to the triple lines and points in \cref{tab:flat_bands_triple_lines}, the $\Delta_{T_{1-}}>0$ has no intrinsically new flat bands, it is a trivial stacking of the $E\oplus T_2$ phase boundary flat bands with the $A_2$ irrep. 
The $\Delta_{T_2}>0$ triple line has no flat bands but has intertwined flat planes.
The remaining two triple lines, $\Delta_{A_2}>0$ and $\Delta_{E}>0$, both have intertwined flat bands, and the HAFM points have intertwined flat bands.

\subsection{Charge-Flux Relations Between Irreps, Band Touchings, and Gauss Laws}
\label{sec:charge-flux-intertwined}

We have now looked at the possible degenerate combinations of irreps listed in \cref{tab:flat_bands_phase_boundaries,tab:flat_bands_triple_lines} and seen that many of them yield flat bands, which are prerequisite to spin liquid ground states. 
At the level of the SCGA, all of these flat band models are spin liquids with a well-defined long-wavelength theory characterized by a field satisfying a Gauss law constraint~\cite{henleyCoulombPhaseFrustrated2010,yanClassificationClassicalSpin2024,yanClassificationClassicalSpin2024a}. 
In direct space, the flat band eigenvectors can in principle be used to construct compact localized deformations of ground state spin configurations which preserve the local constraint that all excited irreps are suppressed~\cite{bergmanBandTouchingRealspace2008}.
We now look to unpack further the underlying mechanism by which the degenerate irreps become intertwined to produce these flat bands and Gauss laws. 
The irreps are not entirely independent of each other, instead some act as fluxes and others act as charges of the emergent gauge theory describing these spin liquids.\footnote{
    From the gauge theory perspective, the ``fluxes'' we are discussing correspond to the electric field, not magnetic (gauge) fluxes.
    }

\subsubsection{Easy-Axis Irreps}

\begin{figure}
    \centering
    \includegraphics[width=\columnwidth]{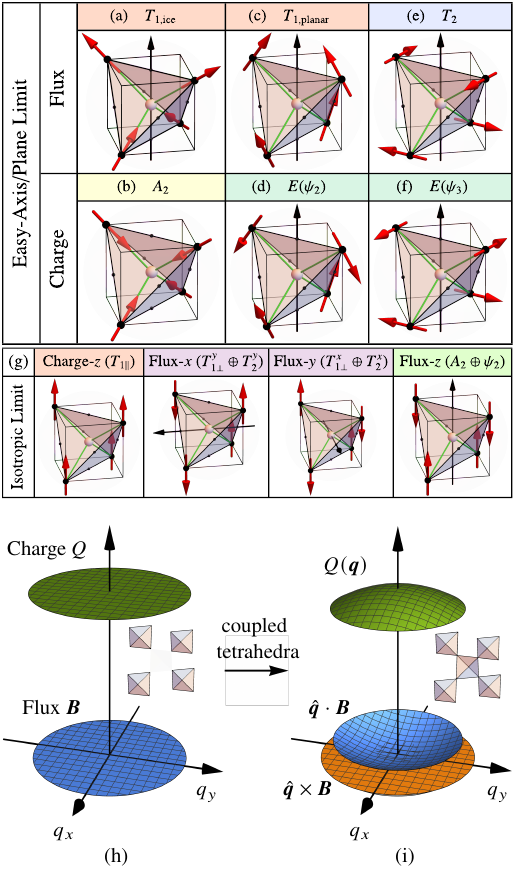}
    \caption{Illustration of the flux-charge relation between irreps. Each 3-component vector irrep acts as a flux, and its corresponding charge is obtained by reversing two spins. In the long-wavelength limit, gapping one of the charges will force the corresponding flux to have zero divergence. (a-f) In the easy-axis/easy-plane limit $J_{z\pm}=J_{\pm\pm}=0$, the local $x,y,z$ components are decoupled into three independent charge-flux combinations: (a,b) The $T_{1,\text{ice}}$ irrep correspond to 2-in-2-out states with zero lattice divergence, while the $A_2$ irrep corresponds to all-in-all-out states with maximal lattice divergence~\cite{henleyPowerlawSpinCorrelations2005,conlonAbsentPinchPoints2010,isakovDipolarSpinCorrelations2004}; (c,d) Flipping two spins from the $T_{1,\text{planar}}$ irrep results in the a $\psi_2$ configurationl; (e,f) Flipping two spins from a $T_2$ configuration results in a $\psi_3$ configuration. 
    (g) In the isotropic limit the global $x,y,z$ components decouple, the three $\smash{\Tpar}$ irrep components $\mathcal{D}$ act as the charges, and the fluxes correspond to the columns of $\mathcal{B}$. 
    (h,i) Illustration of how gapping the charges yields a zero-divergence constraint on the fluxes at the level of bands. 
    (h) In the limit where the $A$ tetrahedra are decoupled (zero coupling on the $B$ tetrahedra) the irreps label a set of flat bands, one for the charge and three for the flux, shown on the left. 
    (i) Turning on the coupling between tetrahedra, the necessity that the charge is zero on the $B$ tetrahedra restricts the allowed configurations on the surrounding $A$ tetrahedra, resulting in the longitudinal (along $\hat{\bm{q}}$) component of the coarse-grained flux becoming dispersive. The transverse (orthogonal to $\hat{\bm{q}}$) components remain flat.
    }
    \label{fig:flux-charge}
\end{figure}

Consider first the limit of the pure-$J_{zz}$ Hamiltonian with all other couplings zero. 
This corresponds to the maximally anisotropic limit where all spins lie along their easy-axes, with corresponding irrep decomposition $A_2 \oplus T_{1,\text{ice}}$.
All other irreps, corresponding to easy-plane modes, do not appear in the Hamiltonian, their energies are $J_I =0$.
Those absent easy-plane irreps cannot be the ground state, therefore we must have that $J_{A_2}<0$ and $J_{T_{1,\text{ice}}}>0$ or vice-versa.
To see how these irreps are coupled, it is instructive to write the Hamiltonian in this limit in terms of the pyrochlore adjacency matrix, 
\begin{align}
    H_{J_{zz}} &= \frac{J_{zz}}{2} \sum_{ij} A_{ij} S_i^z S_j^z \quad (\text{local $z$})
    \nonumber
    \\
    &= \text{const.} + J_{A_2} \sum_t \vert m_{A_2}\vert^2,
\end{align}
where $A_{ij}=+1$ if $i$ and $j$ are nearest-neighbor sites and zero otherwise. On a single tetrahedron, this has the form 
\begin{equation}
    A_{ij} = \begin{pmatrix}
        0 & 1 & 1 & 1 \\ 1 & 0 & 1 & 1 \\ 1 & 1 & 0 & 1 \\ 1  & 1 & 1 & 0
    \end{pmatrix} \quad (i,j\in t),
\end{equation}
which has a single eigenvalue $+3$ corresponding to the $A_2$ irrep and a triply degenerate eigenvalue $-1$ corresponding to the $T_{1,\text{ice}}$ irrep, whose associated eigenvectors are
\begin{align}
   A_2:& \qquad (+1,+1,+1,+1) \\
   T_{1,\text{ice}}:& \quad \begin{cases}
       (+1,+1,-1,-1)\\
       (+1,-1,+1,-1)\\
       (-1,+1,+1,-1)
   \end{cases}
   \label{eq:charge_flux}
\end{align}
Notice that the three $T_{1,\text{ice}}$ eigenvectors differ from the $A_2$ eigenvector by reversing two spins. 
Correspondingly the $A_2$ ground states are all-in-all-out, \cref{fig:flux-charge}(b), while the $T_{1,\text{ice}}$ ground states are 2-in-2-out, \cref{fig:flux-charge}(a).
If we think of each $S_i^z$ as representing a flux going in or out of the tetrahedron,\footnote{More precisely, these fluxes should be thought of as being along the edges of the parent diamond lattice.} then 2-in-2-out states may be viewed as having ``zero divergence'', while the all-in-all-out states have maximal divergence. 
Thus if we associate the $T_{1,\text{ice}}$ eigenmodes to zero-divergence fluxes, the $A_2$ eigenmode is the corresponding charge which sources a non-zero divergence.
This illustrates the charge-flux interrelation of these two irreps, which underlies the emergent gauge theory description of spin ice~\cite{henleyPowerlawSpinCorrelations2005,castelnovoMagneticMonopolesSpin2008,castelnovoSpinIceFractionalization2012,moessnerSpinIceCoulomb2021}.
Another way to see this is that there is a sum rule relating these irreps,
\begin{equation}
    \vert m_{A_2}\vert^2 + \vert \bm{m}_{T_{1,\text{ice}}}\vert^2 = \sum_i (S_i^z)^2.
    \label{eq:sum_rule_easy_axis}
\end{equation}
If all spins lie along the easy-axis in the ground state then the right hand side is a positive constant and either the $A_2$ or $T_{1,\text{ice}}$ is maximized, implying the other is minimized.
Therefore it is not possible to have both the $T_{1,\text{ice}}$ and $A_2$ eigenmodes degenerate in the ground state, because raising the energy of one lowers the energy of the other and vice-versa.
We see this in \cref{fig:full-phase-diagram}, and also in the fact that the last triple combination in \cref{tab:high_degeneracies} has no solution.

This flux-charge relation has a dramatic consequence in the long-wavelength expansion of the Hamiltonian. 
The band structure of the interaction matrix $\mathscr{J}$ has a single and triple degeneracy at the zone center corresponding to the $\bm{q}=\bm{0}$ single-tetrahedron irrep eigenmodes. 
Away from the zone center these bands disperse and need not remain degenerate.
The single-tetrahedron modes $m_{A_2}$ and $\bm{m}_{T_{1,\text{ice}}}$ vary from tetrahedron to tetrahedron, thus we can give a long-wavelength description in which the single-tetrahedron modes become coarse-grained fields---$A_2$ becomes a scalar field $Q$ and $T_{1,\text{ice}}$ a vector field $\bm{B}$.
Enforcing that $Q = 0$ then further forces the divergence of $\bm{B}$ to be zero~\cite{isakovDipolarSpinCorrelations2004,conlonAbsentPinchPoints2010,henleyCoulombPhaseFrustrated2010} due to the relation of the $A_2$ and $T_{1,\text{ice}}$ irreps.
That is, the coarse-grained low-energy theory satisfies the Gauss law constraint
\begin{equation}
    Q_{A_2}=0 \quad \Rightarrow 
    \quad 
    \bm{q}\cdot\bm{B}_{T_{1,\text{ice}}} = 0 
    \quad 
    (\nabla\cdot\bm{B}_{T_{1,\text{ice}}} = 0).
    \label{eq:gauss_ice}
\end{equation}
To see where this extra constraint arises from, consider the fact that the pyrochlore lattice can be divided into two sets of tetrahedra, ``$A$'' and ``$B$'', such that an $A$ tetrahedron is surrounded by four $B$ tetrahedra and vice-versa (see \cref{fig:tetrahedron_vectors}(a)). 
If we turn off the couplings on the $B$ tetrahedra then the bands of the interaction matrix would be dispersionless (flat) with degeneracies determined by the single-tetrahedron irreps, \cref{fig:flux-charge}(h), corresponding to completely decoupled $A$ tetrahedra which only have to satisfy their local constraint but can otherwise be in arbitrary ground states. 
Turning on the $B$ tetrahedra couplings, we realize that we cannot choose the $A$ tetrahedra ground states completely arbitrarily, because we have to ensure that the $B$ tetrahedra are also in a ground state. 
Thus some of the flat bands from the decoupled limit must be lifted in energy, but in such a way that the $\bm{q}=\bm{0}$ degeneracy is preserved. 
The result is a set of flat bands plus a set of quadratically dispersing bands which touch at the zone center, \cref{fig:flux-charge}(i).
The details of this band touching, i.e. which eigenvectors become dispersive and which do not, correspond to the long-wavelength Gauss law for the fluxes when the corresponding charges are gapped: the longitudinal components of the flux field (those that became dispersive) are suppressed, while the transverse modes (those forming the residual flat bands) remain free to fluctuate.\footnote{
    Note that this ``atomic limit'' argument only works because the tetrahedra centers form a bipartite lattice~\cite{henleyCoulombPhaseFrustrated2010}, for the non-bipartite case it is possible that the flat bands become completely gapped~\cite{rehnFractionalizedZ2Classical2017}.
}

In the opposite case, where the charge irrep has the lowest energy and the flux irrep is gapped, then the constraint on the scalar charge field must be a zero-gradient constraint, $\nabla Q = 0$.
Taking \cref{fig:flux-charge}(i) and flipping it upside-down (reversing the signs of the couplings) results in a band structure with a unique minimum and a quadratic dispersion, corresponding to a long-wavelength expansion of the form $\vert \bm{q}\vert^2 \vert Q\vert^2$. 

\subsubsection{Easy-Plane Irreps}

We can extend the easy-axis $A_2$-$T_{1,\text{ice}}$ relation to the easy-plane irreps $E$, $T_2$, and $T_{1,\text{planar}}$. 
Consider the isotropic easy-plane XY model in \cref{eq:H_XY}, written as
\begin{align}
    H_{XY} &= -J_{\pm} \sum_{ij} A_{ij} (S_i^x S_j^x + S_i^y S_j^y) \quad (\text{local $x,y$}),
    \nonumber
    \\
    &= \text{const}. - J_E \sum_{t} \vert \bm{m}_E\vert^2.
    \label{eq:H_XY2}
\end{align}
It follows from the first line that the local $x$ and $y$ modes are decoupled, and the corresponding eigenvectors are again given by \cref{eq:charge_flux}, one for the $x$ components and one for the $y$.
From the second line it is clear that the pair of charge modes must correspond to the pair of $E$ modes, leaving the three-dimensional $T_2$ and $T_{1,\text{planar}}$ irreps for the two flux modes.
Indeed it is easy to see that reversing two spins in a $T_{1,\text{planar}}$ state yields a $\psi_2$ configuration (\cref{fig:flux-charge}(c,d)), while reversing two spins in a $T_2$ state yields a $\psi_3$ configurations (\cref{fig:flux-charge}(e,f)).
Indeed they satisfy the sum rule
\begin{equation}
    \vert \bm{m}_E\vert^2 + \vert \bm{m}_{T_2}\vert^2 + \vert \bm{m}_{T_{1,\text{planar}}}\vert^2 = \sum_i [(S_i^x)^2 + (S_i^y)^2],
    \label{eq:sum_rule_easy_plane}
\end{equation}
so raising the energy of the charge $E$ irrep with $J_{\pm} < 0$ lowers the energy of the two fluxes, which should both satisfy a zero-divergence condition in the long-wavelength coarse-grained limit.\footnote{
    Due to the symmetry of the lattice we cannot split the energy of the $\psi_2$ and $\psi_3$ charges since they are two components of a single irreducible representation. 
}
\begin{equation}
    Q_{\psi_2} =Q_{\psi_3} = 0 
    \quad
    \Rightarrow 
    \quad
    \bm{q}\cdot\bm{B}_{T_{1,\text{planar}}}
    = 
    \bm{q}\cdot \bm{B}_{T_2} 
    = 
    0.
    \label{eq:gauss_T1_T2}
\end{equation}
While this might suggest that the ground state behaves as two decoupled copies of a Coulomb phase, there is a U(1) symmetry which rotates the two fluxes into each other. 
This symmetry is spontaneously broken~\cite{moessnerLowtemperaturePropertiesClassical1998,taillefumierCompetingSpinLiquids2017}, thus the low-energy theory is effectively that of a single Coulomb phase plus a Goldstone mode, a ``nematic spin liquid''~\cite{taillefumierCompetingSpinLiquids2017}.

\subsubsection{Easy-Axis Plus Easy-Plane}

Combining the analysis for the pure-$J_{zz}$ and pure-$J_{\pm}$ models to the XXZ model, we note that the fine-tuned point when $J_{zz} = -2J_{\pm} >0$ corresponds to the isolated $T_{1,\text{ice}}\oplus T_{1,\text{planar}}\oplus T_2$ triple point, where the ground state is described by three Gauss laws, \cref{eq:gauss_ice,eq:gauss_T1_T2}, which is precisely the pHAFM. 
The Hamiltonian can be expressed using the three-component diagonal components of the multipole tensor defined in \cref{eq:tensor_Q}, as 
\begin{equation}
    H_{\text{pHAFM}} = J_{E/A_2} \sum_t \vert \bm{\mathcal{Q}}_t\vert^2,
\end{equation}
where $J_E = J_{A_2}$ at this point. 
For $J_{E/A_2} > 0$ the ground state is described by a trio of divergence-free vector fields, \cref{eq:gauss_ice,eq:gauss_T1_T2}, with an (unbroken) SO(3) symmetry rotating between them. 
For $J_{E/A_2}<0$ the ground state becomes the pseudo-\emph{ferro}magnet (pHFM), which is a special point on the $A_2\oplus E$ phase boundary with a spontaneously broken global SO(3) symmetry, see \cref{sec:triple_ferro}.

On the other hand there is the $A_2 \oplus T_2 \oplus T_{1,\text{planar}}$ triple point when $J_{zz}<0$. 
However, as discussed above  (\cref{sec:stacked_flat_bands}), this is one example of a trivially stacked band degeneracy.
It does not yield a more complicated Gauss law.
Instead it combines the two Gauss laws \cref{eq:gauss_T1_T2} with the constraint $\nabla Q_{A_2}=0$, and these irrep fields do not intertwine. 
The result is that the $T_2\oplus T_{1,\text{planar}}$ and $A_2$ behaviors compete and, as argued above, the easy-axis order wins~\cite{franciniHigherRankSpinLiquids2025}.

\subsubsection{Isotropic Limit}

Next consider the isotropic case, i.e. the Heisenberg model, which we can write as a sum of three (classically) decoupled Hamiltonians
\begin{align}
    H_{J_{\text{Heis}}} &= \frac{J_{\text{Heis}}}{2} \!\!\sum_{\alpha \in \{x,y,z\}} \left(\sum_{ij}A_{ij} S_i^\alpha S_j^\alpha\right) \quad (\text{global $x,y,z$}),
    \nonumber
    \\
    &= \text{const.} + J_{\Tpar}\sum_t \vert \bm{\mathcal{D}}_t\vert^2,
\end{align}
where $\bm{\mathcal{D}}_t$ is the net dipole moment of a tetrahedron (\cref{eq:tensor_M}).
We can immediately infer that the three charges are the three components of the net-moment $\Tpar$ irrep and the ground state should be described by a trio of divergence-free vector fields, one for each spin component, with an SO(3) symmetry rotating between them. 
The $z$-component of the charge is shown in the first panel of \cref{fig:flux-charge}(g), corresponding to a collinear state.
The three corresponding flux components are obtained by flipping two spins, shown in the last three panels of \cref{fig:flux-charge}.
Because the canting angle is non-zero, the fluxes are mixtures of irreps---the $x$ and $y$ components are mixtures of $\Tperp$ and $T_2$ while the $z$ component is a mix of $A_2$ and the $\psi_2$ component of the $E$ irrep.
Notably, however, these flux components corresponding precisely to the last column of the multipole tensor $\bm{\mathcal{B}}$, \cref{eq:tensor_B}. 
Indeed each column of the tensor corresponds to one of the three vector fluxes, and we can define $\bm{\mathcal{B}}\equiv(\bm{B}_x,\bm{B}_y,\bm{B}_z)$.
In fact these are precisely the coarse-grained fluxes which were first defined and used in Refs.~\cite{isakovDipolarSpinCorrelations2004,conlonAbsentPinchPoints2010}, combined into a single rank-2 tensor.
Each of these fluxes satisfies a coarse-grained zero-divergence constraint at low temperature when $J_{\text{Heis}}>0$, 
\begin{equation}
    \bm{\mathcal{D}} = \bm{0} 
    \quad
    \Rightarrow 
    \quad
    \bm{q}\cdot\bm{B}_\beta = 0 
    \quad 
    (\beta = x,y,z),
\end{equation}
which can also be expressed as a single rank-2 tensor Gauss law,
\begin{equation}
    \partial_\alpha \mathcal{B}^{\alpha\beta} = 0,
    \label{eq:tensor_gauss_law}
\end{equation}
where all nine components of $\mathcal{B}$ are allowed to fluctuate. This describes three degenerate sets of two flat bands and a quadratic band touching, and the low-energy spin liquid behaves as three copies of a Coulomb phase with an SO(3) symmetry rotating between them~\cite{henleyCoulombPhaseFrustrated2010,isakovDipolarSpinCorrelations2004}.

\subsubsection{Anisotropy, Tensor Gauss Laws, and Pinch Lines}

So far we have discussed the easy-axis, easy-plane, and  Heisenberg limits. 
Each of these is ``isotropic'', in the sense that the Hamiltonian decouples into independent $x,y,z$ components, each related by the adjacency matrix, and the corresponding Gauss laws follow from the adjacency matrix band structure.
The result is a set of decoupled Coulomb phases with a symmetry rotating between them.
We can then consider perturbing from these limits with bond-dependent anisotropies to tune towards the wide variety of other flat band cases presented in \cref{tab:flat_bands_phase_boundaries,tab:flat_bands_triple_lines}.
This will yield new types of spin liquids which are not described by simple divergence-free vector fields, instead the coarse-grained fields are multipole tensor fields.  
Generally, the long-wavelength theory must be in terms of the coarse-grained version of the vector $\bm{\mathcal{D}}$ and the tensor $\bm{\mathcal{B}}$, which will satisfy certain constraints on their components.

For example, starting from the easy-plane limit, \cref{eq:H_XY2}, adding $J_{\pm\pm}\neq 0$ splits the energies of the $T_2$ and $T_{1,\text{planar}}$ irreps.
This makes it possible to tune the $E$ irrep to be degenerate with either the $T_{1,\text{planar}}$ or $T_2$ irrep, something which is not possible for $A_2$ and $T_{1,\text{ice}}$, because \cref{eq:sum_rule_easy_plane} involves three irreps, any two of which can be made degenerate.\footnote{
    Interestingly, the charge-flux relation between the easy-plane modes nicely explains why near the $T_{1,\text{planar}}$ phase boundary the $\psi_3$ state is selected via order by disorder, and similarly near the $T_2$ phase boundary $\psi_2$ is selected~\cite{rauFrustratedQuantumRareEarth2019}.
    For example, near the $T_2\oplus E$ phase boundary, $\bm{m}_{T_{1,\text{planar}}}$ is minimized, which stiffens the $\psi_3$ fluctuations and makes the $\psi_2$ mode fluctuations softer, driving the order-by-disorder selection.
    \label{foot:obd}
}
Referring to \cref{tab:flat_bands_phase_boundaries}, on the $T_2\oplus E$ phase boundary the band structure has two flat bands, a three-fold quadratic band touching at zone center, and flat nodal lines along the high-symmetry $(hhh)$ directions.\footnote{
    The same holds for $T_{1,\text{planar}}$ by the $J_{\pm\pm}$ duality when $J_{z\pm}=0$.
    } 
The coarse-grained fields will have the form of the tensor $\mathcal{B}$, \cref{eq:tensor_B}, but with $m_{A_2} = m_{\Tperp}^\alpha=0$, i.e. the long-wavelength field theory is of a trace-free tensor with diagonal and anti-symmetric components. 
The Gauss law is then given by the tensor expression \cref{eq:tensor_gauss_law}, but with the additional constraints that the trace and symmetric tensor components must be zero. 
The existence of the flat lines is not obvious from inspecting the Gauss law, as they arise from the additional constraints on the tensor components.

Another important example is the rank-2 U(1) spin liquid reported in Ref.~\cite{yanRank2U1Spin2020}. 
In retrospect, we can easily see why the model they consider has a Gauss law described by a rank-2 symmetric trace free tensor---the ground state irreps on each tetrahedron are $E\oplus T_{1,\perp}$ $(J_3 = 0)$, which are precisely the components of the quadrupole moment (\cref{sec:irreps_symmetries_multipoles}). 
This naturally gives the right tensor structure for the low-energy long-wavelength expansion.
The system remains over-constrained, without flat bands, but adding breathing anisotropy to relax the constraints on the B tetrahedra yields the desired band structure.\footnote{
    Another way to say this is that with breathing anisotropy all irreps are doubled into even (\textit{gerade}) and odd (\textit{ungerade}) copies, with the even copies controlling the band structure at the zone center and the odd copies controlling the band structure at the zone boundary. Putting couplings on the B tetrahedron to be those of the HAFM, i.e. bringing the $A_2$ and $T_2$ irreps to the ground state on the $B$ tetrahedra, brings down some of the bands at the zone boundary to be degenerate with those at the zone center, resulting in flat bands. 
}
This idea was applied by the author in Ref.~\cite{chungGeometricallyFrustratedQuadrupoles2025} to construct a rank-3 U(1) spin liquid by putting irreps corresponding to an octupole moment into the ground state.

On the other hand, starting from the isotropic Heisenberg point and inducing spin canting with the $J_3$ interaction (mixing of the two $T_1$ irreps), for example moving along the $\Delta_{A_2}>0$ line, will cause the vector and symmetric-off-diagonal tensor components to mix due to canting.
This results in more complex constrained tensor Gauss laws, which generically exhibit nodal lines touching the flat bands~\cite{bentonSpinliquidPinchlineSingularities2016,franciniHigherRankSpinLiquids2025,lozano-gomezCompetingGaugeFields2024}.
Ref.~\cite{lozano-gomezAtlasClassicalPyrochlore2024}, which appeared concurrently with this work, gives a detailed prescription for how to derive the inter-tetrahedron constraints and tensor Gauss laws and has done so for all degenerate irrep combinations in the pyrochlore lattice.
It remains unclear at this point how to fully account for the emergence of these more complicated tensor Gauss laws directly from the multipole tensors.

\begin{figure}[t]
    \includegraphics[width=\columnwidth]{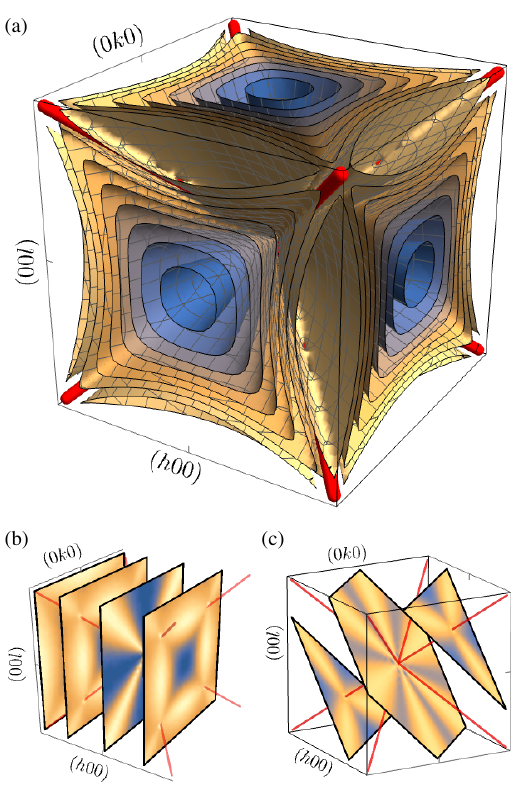}
	\caption{Three-dimensional structure of a fourfold pinch point. Here we have plotted the $\bq=\bm{0}$ pinch points from the $\Delta_E>0$ line with $x=0.1$, but all fourfold pinch points visible the in the various structure factors have the same qualitative structure. (a) equal-intensity contours of the structure factor for a cube centered at the zone center with $-0.1\leq h,k,l\leq 0.1$. 
    Three pinch line singularities run along the $(hhh)$ directions of reciprocal space, i.e. the diagonals of a cube (red lines), meeting at the zone center. 
    (b) Cutting through this in a cubic plane such as $(h0l)$ reveals a four-fold pinch structure. Slicing away from the zone center the pinch lines are visible as three-fold pinch point singularities. Cutting it along a cube diagonal plane like $(hhl)$ reveals the pinch lines, cf. \cref{fig:structure_factors_triple_lines,fig:structure_factors_special} (not shown here). 
    (c) Cutting along a plane orthogonal to $(hhh)$ through the zone center shows a six-fold pinch. Cutting orthogonal to the $(hhh)$ line away from the zone center shows three-fold pinches.
    }
    \label{fig:fourfold_pinchpoint3d}
\end{figure}

\subsection{Pinch Lines, Multi-fold Pinch Points, and Fractons}

We close with a brief discussion about the relation between pinch lines, multi-fold pinch points, and fractons. 
Fourfold pinch points arise from the Gauss law of a rank-2 symmetry tensor gauge theory, which describes fracton topological order~\cite{pretkoSubdimensionalParticleStructure2017,premPinchPointSingularities2018,yanRank2U1Spin2020,nandkishoreFractons2019,pretkoFractonPhasesMatter2020}.
There are two basic types: the scalar charge theory has Gauss law $\partial_\alpha\partial_\beta E_{\alpha\beta} = \rho = 0$, where $E_{\alpha\beta}$ is the rank-2 tensor electric field; the vector charge theory has Gauss law $\partial_\alpha E_{\alpha\beta} = \rho_\beta = 0$. 
In a spin liquid the scalar Gauss law would arise due to a \emph{quartic} band touching in the spectrum of the interaction matrix, corresponding to a long-wavelength expansion of the form $(q_\alpha q_\beta E_{\alpha\beta})^2$, which does occur in the model we consider here. 
All of the Gauss laws occurring for the spin liquids occurring in the pyrochlore model considered here are vector charge theories with quadratic dispersions.  
None of them are of the symmetric tensor type, instead they generally have more complicated Gauss law which mix symmetric tensor fields, anti-symmetric tensor fields, and vector fields together~\cite{lozano-gomezAtlasClassicalPyrochlore2024}. 
As can be seen in \cref{fig:structure_factors_triple_lines,fig:structure_factors_special}, we have a plethora of four-fold pinch points arising from these complex Gauss laws.
Here we wish to emphasize, that all of these fourfold pinch points arise from nodal (pinch) lines, rather than from the quadratic band touching of a rank-2 symmetric tensor gauge theory.

The full three-dimensional structure of the pinch singularity at the zone center is shown in \cref{fig:fourfold_pinchpoint3d}, which is qualitatively identical for all the fourfold pinches observable in \cref{fig:structure_factors_triple_lines,fig:structure_factors_special}.
The nodal lines run along the three cube diagonals, colored red in \cref{fig:fourfold_pinchpoint3d}(a), and intersect at the zone center.
\Cref{fig:fourfold_pinchpoint3d}(a) shows equal-intensity contours of the structure factor, which is singular along the pinch lines and goes to a minimum along the cubic $(h00)$ axes. 
Slicing through this in a cubic plane such as $(h0l)$, illustrated in \cref{fig:fourfold_pinchpoint3d}(b), reveals a fourfold pinch point at the zone center, while cutting along such a cubic plane away from the zone center reveals four pinch points where this plane intersects the pinch lines. 
On the other hand, cutting along a plane orthogonal to the $(hhh)$ pinch lines reveals a six-fold pinch pattern at the zone center, or a three-fold pinch pattern away from the zone center, as illustrated in \cref{fig:fourfold_pinchpoint3d}(c)
The three-fold pattern arises due to high intensity on the three cube edges which merge together along the cube diagonals, which can be seen in \cref{fig:fourfold_pinchpoint3d}(a).

Thus the four-fold pinch points of these spin liquids seen in \cref{fig:structure_factors_triple_lines,fig:structure_factors_special} actually reflect the cutting of these nodal lines in high-symmetry planes.
They are distinct from those originally envisaged in the symmetric tensor gauge theory, which would arise at rotationally-symmetric quadratic or quadratic band touching point~\cite{pretkoSubdimensionalParticleStructure2017,premPinchPointSingularities2018,yanRank2U1Spin2020}. 
It is important to ask the question of whether these multi-fold pinch points and their parent nodal lines are indications that these are fracton spin liquids or not. 
On the one hand, the presence of nodal lines in 3D is indicative that the system satisfies a Gauss law in \emph{every plane} orthogonal to the nodal line direction in direct space~\cite{davierPinchLineSpin2025}. 
This is strongly suggestive of fractonic behavior, as most fracton field theories are built on a foliated space by coupling gauge theories in neighboring leaves of the foliation~\cite{slagleFoliatedQuantumField2021}.
Nodal lines spin liquids are clearly much more common than rotationally-symmetric tensor spin liquids, which is sensible since the nodal lines break rotation symmetry and are only allowed in a crystalline environment. 
Crystalline environments have a natural foliation structure, so it would not be surprising if nodal line spin liquids are indeed fractonic. 
Unfortunately, the classical spin liquids discussed here are defined as statistical mechanical systems of continuous spins, and do not have well-defined quantized excitations which can exhibit fractonic behavior such as restricted mobility~\cite{pretkoSubdimensionalParticleStructure2017}. 
A clearer answer to the relation between nodal lines, multi-fold pinch points, and fractons in classical spin liquids will require the study of such spin liquids with Ising spins, i.e. fractonic analogs of classical spin ice, with well-defined quantized quasiparticle excitations (see \cite{sternFractonSpinLiquid2026} for an example).

\begin{figure*}[t]
    \centering
    \includegraphics[width=\textwidth]{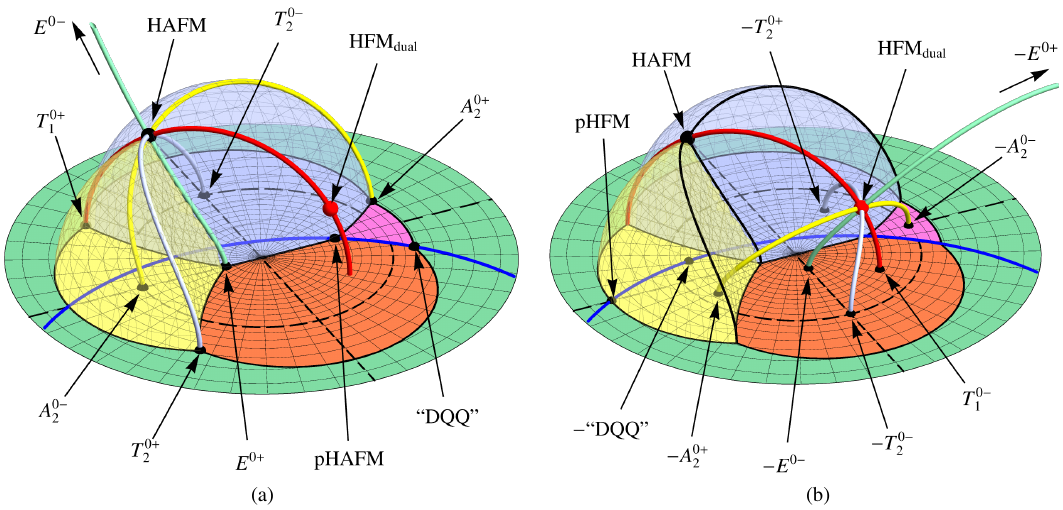}
    \caption{(a) The $J_{z\pm}>0$ phase diagram showing all four triple lines from \cref{tab:high_degeneracies} including the $\Delta_{I'}<0$ portions where the triple degeneracy occurs in the irreps above the ground state (red region on the left of the plots in \cref{fig:3-fold-lines-params}). Colors are the same as \cref{fig:full-phase-diagram,fig:phase_diagram_HDM}. Each is a circle which winds around the line of ``diabolical'' $T_1$ locus (blue line), and all four intersect at the two HAFM points. Points where these lines intersect the $J_{z\pm}=0$ plane (left and right ends of plots in \cref{fig:3-fold-lines-params}) are labeled $(I')^{0\pm}$, where $\pm$ denotes the sign of $\Delta_{I'}$: $+$ indicates this irrep is gapped and $-$ indicates it is the ground state. 
    (b) The ``ferro'' versions of the three triple lines and two isolated triple points, with reversed signs of all couplings $H \to - H$. The Heisenberg-plus-$J_{z\pm}$ family (red circle, cf. \cref{fig:phase_diagram_HDM}) with both $\Delta_{T_{1\pm}}\neq 0$ maps to itself, while the other three are mapped to new lines with triple degeneracies above the ground state. All four intersect at the two HFM points. The intersections with $J_{z\pm}=0$ are indicated with a minus sign compared to the same labels in (a). Note that HFM$\,\equiv-$HAFM, pHFM$\,\equiv-$pHAFM, and $T_1^{0-}\equiv -T_1^{0+}$.}
    \label{fig:triple_lines_AFM_FM}
\end{figure*}

\subsection{Quantum Spin Liquids}

In the future it will be interesting to study degenerate intersections of phases in more detail, and to determine whether any of them gives rise to a stable quantum spin liquid. 
Unfortunately in three dimensions there are few tools available to determine this. 
The only example where we have a meaningful level of analytical control is quantum spin ice, i.e. perturbing around the classical spin liquid point $J_{zz}>0$ with all other couplings zero. 
There, we can perform a direct mapping from the spin Hamiltonian to a U(1) lattice gauge theory Hamiltonian~\cite{hermelePyrochlorePhotonsU12004}, where the $S_i^z$ operators map to electric field operators and the spin raising and lowering operators $S_i^{\pm}$ map to electric field raising and lowering operations $\exp(\pm i A),$ where $A$ is the lattice vector potential. 
This is only possible because the transverse fluctuations can be treated perturbatively on top of a classical Ising limit. 
This method---treating the classical ground state manifold as electric field eigenstates and spin-flipping terms as Wilson string operators---cannot be applied to study, say, the quantum HAFM, because the isotropic nature of the classical ground state means there is no preferred way to split spin operators into longitudinal and transverse components. 
Indeed the ground state of the quantum HAFM has been debated extensively, with the general view being that it exhibits some form of order, symmetry breaking, or dimensional reduction~\cite{canalsClassicalSpinLiquid2002,iqbalQuantumClassicalPhases2019,hagymasiPossibleInversionSymmetry2021,astrakhantsevBrokenSymmetryGroundStates2021,pohleGroundStateS12023,schaferAbundanceHardHexagonCrystals2023}.
Similarly, the various tensor spin liquids arising from flat bands in the model discussed in this paper do not have a perturbative limit---all three spin components are used in the definition of the electric field, and all three also act as raising and lowering operators for each other---though there is some evidence that these quantum models do have some fractonic properties~\cite{hanRealizationFractonicQuantum2022}.

It will, in the long term, be interesting to determine whether the various degenerate loci in the pyrochlore phase diagram host quantum spin liquids.
In particular, it will be interesting to understand the quantum ground states along the four triply degenerate lines and how they merge together at the quantum HAFM point.
Previous studies have classified various types of quantum spin liquid mean field ground state wavefunctions compatible with the symmetries of the pyrochlore lattice~\cite{liuCompetingOrdersPyrochlore2019,liuSymmetricU1Z22021,desrochersCompetingU1Z22022,chernCompetingQuantumSpin2022,desrochersSymmetryFractionalizationGauge2023}.
Ref.~\cite{liuCompetingOrdersPyrochlore2019} emphasized how these are closely related to the competition of various intertwined irrep orders. 
Beyond mean field analysis, numerical methods are required. 
Unfortunately Quantum Monte Carlo algorithms are not generic enough to explore the entire phase diagram efficiently, and other powerful techniques like DMRG have extremely limited applicability in three dimensions. 
Relatively recently, pseudo-fermion functional renormalization (PFFRG) schemes have been developed which work well in three dimensions---see Ref.~\cite{mullerPseudofermionFunctionalRenormalization2024} for a review---and applied to the pyrochlore lattice~\cite{iqbalFunctionalRenormalizationGroup2016,iqbalQuantumClassicalPhases2019,niggemannQuantumEffectsUnconventional2023,noculakClassicalQuantumPhases2023,chernPseudofermionFunctionalRenormalization2024,hagymasiPhaseDiagramAntiferromagnetic2024,gresistaQuantumEffectsPyrochlore2025}.
These methods can detect certain ordering instabilities and predict two-point correlation function, but involve uncontrolled approximations and so must be used in combinations with other numerical and analytical methods to approach the question of whether the quantum ground state avoids ordering and becomes a spin liquid.

\section{Triple Degeneracies in Excited States}
\label{sec:triple_ferro}

While we have discussed so far the cataloging of cases which have large degeneracies in the ground state and the resulting zero-energy flat bands, these high degeneracies can still be present even when they are not at zero energy.
Such excited state degeneracies do not come into play in the ground state determination but may leave interesting imprints on the excitations above the ground state, e.g. in the spin wave (magnon) spectrum. 
As such, we close by briefly discussing the loci in the phase diagram where triple irrep collisions occur above zero energy. 
The effects of these enhanced degeneracies on the excitation spectrum within each phase will be an interesting topic for future study.

As mentioned previously, the triply-degenerate lines parameterized in \cref{tab:high_degeneracies} can be continued to negative $\Delta_{I'}$ and form complete circles in the phase diagram, as is shown in \cref{fig:3-fold-lines-params}. 
All four such lines wind non-trivially around the degenerate $T_1$ locus. 
They are shown in \cref{fig:triple_lines_AFM_FM}(a) in the $J_{z\pm}>0$ half of the phase diagram, along with the two triple points.
All four cycles intersect at the two fourfold-degenerate HAFM points. 
Each line, labeled by a single irrep $I'$ which is tuned while the others are kept degenerate, intersects the $J_{z\pm}=0$ plane at two points, one where $\Delta_{I'}>0$ and one where $\Delta_{I'}<0$, which we have marked and labeled as $(I')^{0\pm}$ depending on the sign of $\Delta_{I'}$. 
The $\Delta_E<0$ line is not shown in its entirety as the stereographic projection maps the $E^{0-}$ point far from the origin. 
The $(I')^{0-}$ points are in a sense ``centers'' of their respective phases: they are the points where all three neighboring phases have equal and maximal gap, implying that these points are equidistant from all three phase boundaries.
This can be seen in \cref{fig:3-fold-lines-params}(c), where these points correspond to the left-most limit of each plot: the triply-degenerate irreps have $J_{I}=0$ at these points while $\Delta_{I'}<0$ reaches its minimum value.
The line where both $T_1$ irreps are tuned, corresponding to the Heisenberg-plus-$J_{z\pm}$ model discussed in \cref{sec:two_families}, is slightly different.
It cannot be parameterized by $\Delta_{T_{1-}}/\Delta_{T_{1+}}$ because both go negative on part of the path, so we have parameterized it instead by the splitting of the two $T_1$ irreps. 
This line passes through two special points in the $T_1$ phase, namely the Heisenberg ferromagnet and its dual. 
These may respectively be thought of as two different ``centers'' of the $T_1$ phase, because in \cref{fig:phase_diagram_HDM} one can see that the $T_1$ phase appears to be the union of two roughly-spherical lobes centered on these two points.

These are not all of the triply degenerate cases, however, only the ones that are obtained from the four-fold degenerate HAFM by lifting the degeneracy of one irrep. 
The remaining ones are obtained by flipping the signs of all the couplings, i.e. changing the sign of the Hamiltonian $H\to -H$.
This produces another set of triple lines which all intersect at the Heisenberg \emph{ferro}magnet (HFM), shown in \cref{fig:triple_lines_AFM_FM}(b) along with two additional triple points, the pseudo-HFM and the ferromagnetic version of the DQQ (also appearing in \cref{fig:phase_diagram_HDM}).
We have indicated the locations of the intersections of these lines with the $J_{z\pm}=0$ plane, labeled the same as in \cref{fig:triple_lines_AFM_FM}(a) but with a minus sign to indicate these are the ``ferro'' counterparts.
The $I'=T_{1}$ line (red), corresponding to the Heisenberg-plus-$J_{z\pm}$ model, is special because this family of Hamiltonians is mapped to itself under $H\to -H$. 
Since this line connects the HAFM and HFM points, it connects all of the triply-degenerate models together.

\section{Conclusion}

In this paper we have considered in detail the structure of the classical phase diagram of rare-earth pyrochlore magnetic insulators with nearest-neighbor anisotropic interactions, described by the Hamiltonian \cref{eq:H_named_interactions,eq:H_local} with zero single-ion anisotropy. 

In \cref{sec:symmetry_classification} we gave an intuitive derivation of the organization of spins into tensor order parameters, corresponding to the multipole moments of a single tetrahedron. 
We directly related the various tensor components to irreducible representations (irreps) of the tetrahedral symmetry group $T_d$, listed in \cref{tab:multipoles}, each of which corresponds to a ground state shown in \cref{fig:ground_states}. 
The presence of two copies of the $T_1$ irrep results in the canting cycle shown in \cref{fig:T1_mixing}.
We described how different irreps are coupled together, where the 3-components vector irreps serve as ``fluxes'' with corresponding ``charges'' given by the three components of $A_2\oplus E$: raising the energy of a charge lowers the energy of its corresponding fluxes and vice-versa, and gapping a charge forces a zero-divergence condition on the corresponding flux. 

In \cref{sec:model_space_parameterization} we parameterized the model space---the space of equivalence classes of models differing by affine rescalings, topologically a 3-sphere---in terms of the relative irrep energies.
This allowed us to perform an exhaustive search of all possible ground state degeneracies, which uncovered four triply-degenerate lines where three phases meet, listed in \cref{tab:high_degeneracies}, all of which meet at the Heisenberg anti-ferromagnet point and its $J_{z\pm}$-dual, where all four phases become degenerate. 
Two isolated triple points also occur when $J_{z\pm}=0$.

In \cref{sec:mapping_the_phase_diagram} we visualized the structure of the phase diagram by using a stereographic projection from the unit 3-sphere in the space of the four local couplings to $\mathbbm{R}^3$. 
Since the phase diagram is reflection-symmetric about $J_{z\pm}=0$, we first mapped this stereographically in \cref{fig:jzpm0}, which yields the phase diagram of non-Kramers pyrochlore magnets (for which $J_{z\pm}=0$). 
We then mapped the phase diagram for $J_{z\pm}\neq 0$ (allowed for Kramers doublets) in \cref{fig:full-phase-diagram}, which exposes all of the phase boundaries, the four triply-degenerate lines, and shows how they merge at the two four-fold degenerate HAFM points. 
We also noted the locus along which the two $T_1$ irreps are degenerate, which is a circle lying in the $J_{z\pm}=0$ plane. 
This locus pierces the $T_1 \oplus T_2$ and $T_1 \oplus E$ phase boundaries, giving rise to the two isolated triple points in the phase diagram. 

In \cref{sec:canting_cycles} we discussed the $T_1$-degenerate locus in more detail. Going around any path in parameter space that links it the canting angle winds by $\pi$. In \cref{fig:canting_foliation} we showed how the phase diagram is foliated into 2-spheres on which the canting angle is constant, all of which intersection along the $T_1$-degenerate locus. 
This is analogous to ``diabolical loci'' which gives the phase diagram a topological structure not evident from energetic considerations alone. 
Adiabatically transporting the system around this locus within the $T_1$ phase will cause the spins to perform a half-rotation around the canting cycle illustrated in \cref{fig:T1_mixing}.
There is a topological invariant associated to 1-parameter families of Hamiltonians which identifies how many times they wind around the locus.
We briefly described a general framework for such loci in frustrated magnets made of corner-sharing clusters in terms of homotopy groups of PSO($n$). 

In \cref{sec:flat_bands} we considered how the enhanced degeneracies obtained by collision of ground state irreps at classical multi-critical points can give rise to flat bands, and how these may lead to spin liquids. 
We provided a general discussion of the connection between flat bands and spin liquids, and discussed various ground state selection mechanisms that may co-opt the formation of a stable classical spin liquid. 
We provided an overview of known spin liquids on the pyrochlore lattice obtained from the anisotropic Hamiltonian, and in \cref{tab:flat_bands_phases,tab:flat_bands_phase_boundaries,tab:flat_bands_triple_lines} provided a catalog of all flat band degeneracies in the phase diagram, and identified which combinations of irreps yield flat band degeneracies which are more than the sum of their parts (i.e. intertwined) versus trivially stacked.
In \cref{fig:structure_factors_triple_lines,fig:structure_factors_special} we provided reference structure factors computed in the self-consistent Gaussian approximation for various interesting flat band cases, and described the relation between fourfold pinch points and pinch lines observed ubiquitously in these models: \cref{fig:fourfold_pinchpoint3d} demonstrates the 3D structure of the pinched correlations at the intersection of pinch lines, which when cut produce fourfold pinch points. 

Lastly, in \cref{sec:triple_ferro} we identified all loci in the phase diagram where three irreps are degenerate above the ground state, including both the continuation of the triple phase degeneracies into each phase along with their ferro-counterparts which all intersect at the Heisenberg ferromagnetic points.
This is summarized in \cref{fig:triple_lines_AFM_FM}.

Beyond the issue of whether any of the flat band cases we have cataloged host stable quantum spin liquids, in the future it will be of interest to extend the analysis here to study the broader phase diagram of breathing pyrochlores, which can have different couplings on the two symmetry-inequivalent sets of tetrahedra and may host additional spin liquids~\cite{essafiFlatBandsDirac2017,ghoshBreathingChromiumSpinels2019,chernCompetingQuantumSpin2022}.
Furthermore, there may be surprising physics to be gleaned within the ordered phases of the pyrochlore in the spectrum of (magnon) excitations, as evidenced by the extended $T_1$-degenerate locus and the various lines of high degeneracy in the band structure above zero energy. 
Lastly, it will be interesting to study in more detail the canting cycle in the pyrochlore and other frustrated magnets, to explore its potential connections to diabolical loci, symmetry pumps, and boundary symmetry breaking, and whether it has experimentally observable consequences. 

Beyond the pyrochlore Hamiltonian considered here, the methods used in this paper are applicable to other models. 
For example, in Ref.~\cite{chungGeometricallyFrustratedQuadrupoles2025} we have applied the same methods to study the equivalent of \cref{eq:H_local} with spins (dipole operators) replaced by quadrupoles, which increases the number of degrees of freedom, and for which one can similarly identify diabolical loci and spin liquids arising from the collision of irreps.
In general, flat band classical spin liquids arise in Hamiltonians which enforce local constraints, sometimes called ``constrainer Hamiltonians''~\cite{yanClassificationClassicalSpin2024,yanClassificationClassicalSpin2024a,bentonTopologicalRouteNew2021,henleyCoulombPhaseFrustrated2010}.
In the pyrochlore model considered here, the local constraint is that ground states have zero overlap with the gapped irreps on each tetrahedron. 
In general constrainer Hamiltonians, the local constraints involve a constellation of dipole operators---spins---and it is natural that different constraints correspond to extinction of certain multipole moments of the spin distribution, and correspondingly to the collision of ground state irreps. 
It will be interesting to extend this to further-neighbor interactions and other lattices, and to eventually perform a complete symmetry classification of constrainer Hamiltonians.

\begin{acknowledgments}
We acknowledge helpful discussions with Michel Gingras, Johannes Reuther, Chris Hooley, and Daniel Lozano-G\'omez. This work was in part supported by the Deutsche Forschungsgemeinschaft  under the cluster of excellence ct.qmat (EXC-2147, project number 390858490).
\end{acknowledgments}

\appendix

\section{Coordinate Conventions}
\label{apx:conventions}

\begin{figure}[t]
    \centering
    \includegraphics[width=0.85\columnwidth]{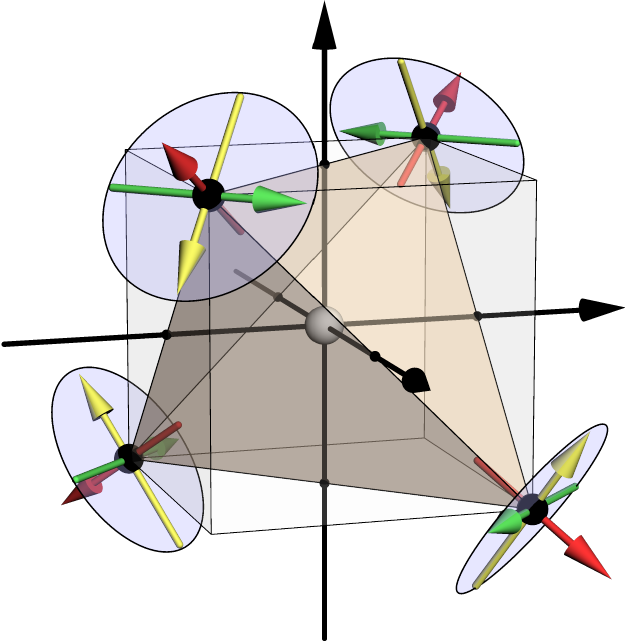}
    \caption{(a) A single tetrahedron inscribed in a cube showing the local orthonormal basis vectors defined in \cref{eq:local_z,eq:local_xy}. Red vectors are $\uvec{z}_\mu$ along the local three-fold easy-axis. Yellow vectors are the $\uvec{x}_\mu$ which lie in mirror planes. Green vectors are the $\uvec{y}_\mu$ which lie along local $C_2$ axes which exchange neighboring tetrahedra. Note that the $\psi_2$ ($\psi_3$) configurations in \cref{fig:ground_states} correspond to all spin lying along $\uvec{x}_\mu$ ($\uvec{y}_\mu$). Black arrows through the tetrahedron center denote the cubic [001] directions.
    }
    \label{fig:tetrahedron_SM_local-basis}
\end{figure}

Although not crucial for the results presented in this paper, for completeness we provide a set of coordinates for the local bases and parameterizations discussed throughout the paper. 
Defining the cubic axes of the pyrochlore crystal structure to be along the three Cartesian axes $\uvec{e}_x,\uvec{e}_y,\uvec{e}_z$, the FCC primitive vectors can be given by
\begin{equation}
    \bm{a}_1 = a_0 \frac{\uvec{e}_x+\uvec{e}_y}{2}, \quad
    \bm{a}_2 = a_0 \frac{\uvec{e}_y+\uvec{e}_z}{2}, \quad 
    \bm{a}_3 = a_0 \frac{\uvec{e}_z+\uvec{e}_x}{2},
\end{equation}
where $a_0$ is the cubic conventional cell lattice constant. 
The four FCC sublattice vectors, corresponding to the four corners of a tetrahedron, can then be given by 
\begin{equation}
    \bm{c}_1 = \bm{0}, \quad 
    \bm{c}_2 = \bm{a}_1/2, \quad 
    \bm{c}_3 = \bm{a}_2/2, \quad 
    \bm{c}_4 = \bm{a}_3/2.
\end{equation}
The centers of the tetrahedra form a bipartite diamond lattice with sublattice positions
\begin{equation}
    \bm{\delta}_{\pm} = \pm \frac{1}{4}\sum_{\mu=1}^4 \bm{c}_{\mu},
\end{equation}
where we use $\mu$ to index the FCC sublattices of the pyrochlore lattice. 
The four sublattice easy-axis vectors can be then be specified by normalizing the displacement vector from the center of a tetrahedron to each corner,
\begin{equation}
    \uvec{z}_\mu = \frac{\bm{c}_\mu - \bm{\delta}_+}{\vert\bm{c}_\mu - \bm{\delta}_+\vert}.
    \label{eq:local_z}
\end{equation}
Using the notation 
\begin{equation}
    [abc] \equiv a \uvec{e}_x + b \uvec{e}_y + c \uvec{e}_z
\end{equation}
along with $\bar{a} \equiv -a$, we can provide bases for the local frames at each corner of the tetrahedron (up to normalization),
\begin{align}
    \uvec{x}_1 &\propto [11\bar{2}] \quad
    \uvec{x}_2 \propto [\bar{1}\bar{1}\bar{2}] \quad
    \uvec{x}_3 \propto [\bar{1}12] \quad
    \uvec{x}_4 \propto [1\bar{1}2] 
    \nonumber
    \\[2ex]
    \uvec{y}_1 &\propto [1\bar{1}0]  \quad
    \uvec{y}_2 \propto [\bar{1}10]  \quad
    \uvec{y}_3 \propto [\bar{1}\bar{1}0]  \quad
    \uvec{y}_4 \propto [110] 
    \nonumber
    \\[2ex]
    \uvec{z}_1 &\propto [\bar{1}\bar{1}\bar{1}]  \quad
    \uvec{z}_2 \propto [11\bar{1}]  \quad
    \uvec{z}_3 \propto [1\bar{1}1]  \quad
    \uvec{z}_4 \propto [\bar{1}11] .
    \label{eq:local_xy}
\end{align}
These are shown in \cref{fig:tetrahedron_SM_local-basis}. The $\uvec{x}_\mu$ vectors have been chosen to lie in mirror planes, while the $\uvec{y}_\mu$ vectors lie along a $C_2$ rotation axis.
With these convention, referring to \cref{fig:ground_states} and \cref{fig:T1_mixing}, $\psi_2$ has all spins along $\uvec{x}_\mu$; $\psi_3$ has all spins along $\uvec{y}_\mu$, $T_2$ has two spins along $\uvec{y}_\mu$ and two along $-\uvec{y}_\mu$; $T_{1,\text{planar}}$ has two spins along $\uvec{x}_\mu$ and two along $-\uvec{x}_\mu$; $A_2$ has all spins along $\uvec{z}_\mu$; and $T_{1,\text{ice}}$ has two spins along $\uvec{z}_\mu$ and two along $-\uvec{z}_\mu$.

\section{Fourier Transforms}

\begin{figure}
    \centering
    \includegraphics[width=0.9\linewidth]{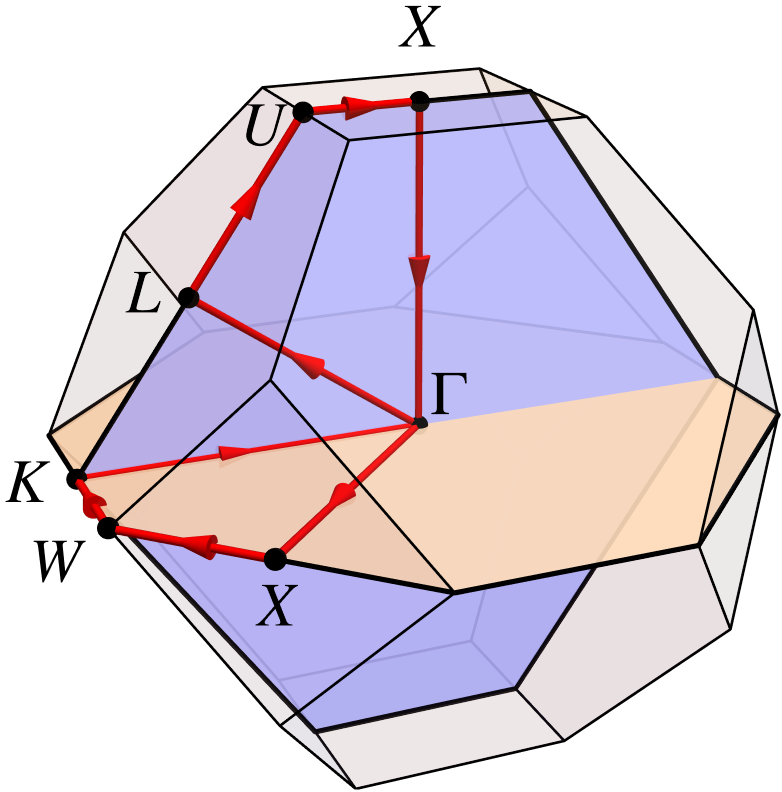}
    \caption{
    The first Brillouin zone of the FCC lattice, with high symmetry points and lines marked. The orange surface indicates the $(hk0)$ plane and the blue surface indicates the $(hhl)$ plane. The red path corresponds to the line cuts shown in \cref{fig:bands}. The labels correspond to reciprocal space Miller indices, \cref{eq:apx_hkl}: $\Gamma = \{000\}$, $X = \{001\}$, $L = \{\tfrac{1}{2}\tfrac{1}{2}\tfrac{1}{2}\}$, $K=\{\tfrac{3}{4}\tfrac{3}{4}0\}$, $U=\{\tfrac{1}{4}\tfrac{1}{4}1\}$, $W = \{\tfrac{3}{4}\tfrac{3}{4}\tfrac{1}{4}\}$.
    }
    \label{fig:FBZ_path}
\end{figure}

Due to translation invariance the interaction matrix in \cref{eq:H_generic} is diagonal in the Fourier basis.
We take the system to consist of $L^3$ unit cells with periodic boundaries, in which case
\begin{align}
    \Jmat_{\mu\nu}^{\alpha\beta}(\bq,\bq') 
    &\equiv \frac{1}{L^3} \sum_{i\in \mu}\sum_{j\in\nu} J_{ij}^{\alpha\beta} e^{i\bq\cdot\bm{r}_j} e^{-i\bq'\cdot\bm{r}_i} 
    \nonumber
    \\
    &= \delta_{\bq,\bq'} \sum_{j\in\nu} J_{ij}^{\alpha\beta} e^{i\bq\cdot(\bm{r}_j-\bm{r}_j)} \quad (i\in\mu) 
    \nonumber 
    \\
    &\equiv \delta_{\bq,\bq'} J_{\mu\nu}^{\alpha\beta}(\bq),
\end{align}
where the sums are restricted to sites in a single sublattice.
For nearest-neighbor interactions it is useful to define a modified adjacency matrix of the pyrochlore lattice,
\begin{equation}
    \mathcal{A}_{ij} = \begin{cases}
        1 &\quad \text{$i,j$ are nearest-neighbors,}
        \\
        1/2 &\quad i=j,
        \\
        0 &\quad \text{otherwise,}
    \end{cases}
    \label{eq:adjacency_matrix}
\end{equation}
whose Fourier transform is
\begin{equation}
    \mathcal{A}_{\mu\nu}(\bq) = (2-\delta_{\mu\nu})\cos[\bq\cdot(\bm{c}_\mu-\bm{c}_\nu)].
\end{equation}
The Fourier transformed nearest-neighbor interaction matrix is then
\begin{equation}
    \Jmat_{\mu\nu}^{\alpha\beta}(\bq) = J_{\mu\nu}^{\alpha\beta} \, \mathcal{A}_{\mu\nu}(\bq),
\end{equation}
where the $12 \times 12$ matrix $J_{\mu\nu}^{\alpha\beta}$ is given in \cref{eq:J14_full_matrix}.
The on-site term in \cref{eq:adjacency_matrix} is added to allow for the on-site single-ion anisotropy interaction, which is double counted in the unrestricted double sum in \cref{eq:H_generic}.
For reciprocal space we use Miller indices relative to the cubic conventional unit cell, denoted with round brackets
\begin{equation}
    \bq = \frac{2\pi}{a_0} (h \uvec{e}_x + k \uvec{e}_y + l \uvec{e}_z) \equiv (hkl).
    \label{eq:apx_hkl}
\end{equation}
\Cref{fig:FBZ_path} shows the first Brillouin zone of the FCC lattice with high-symmetry points, lines, and planes marked.

\section{Mapping Between Parameterizations}
\label{apx:parameter_maps}

The number of free parameters in the interaction matrix $\Jmat$ in \cref{eq:H_generic} is restricted by symmetry. A given space group $G$ acts on the spin configuration $S_i^\alpha$ by permutation of the site index $i$ and pseudovector rotation of the spin components index $\alpha$. 
For nearest-neighbor interactions in the (non-breathing) pyrochlore lattice, the interaction matrix is the same on every tetrahedron, so the symmetry constraints can be obtained from just looking at the symmetry group of the tetrahedron, $T_d$, with action defined by the representation $\rho$ in \cref{eq:rep_matrix_pi_R}.
The allowed spin-spin couplings can be deduced from the fact that the Hamiltonian on a tetrahedron reduces to a sum over the irrep order parameters squared, or more generally by solving the equations 
\begin{equation}
    [\Jmat, \rho(g)]=0\quad \forall g \in G \quad \text{and} \quad \Jmat = \Jmat^{\mathrm{T}}.
\end{equation}
\subsection{Global Frame \texorpdfstring{$J_1$-$J_4$}{J1-J4}}
With the conventions in \cref{apx:conventions}, the symmetry-allowed matrix elements in \cref{eq:H_generic} for a single tetrahedron, denoted $\Jmat_{\mu\nu}^{\alpha\beta}$ where $\mu,\nu$ index sublattices (tetrahedron corners), take the form
\begin{equation}
\begin{pmatrix*}[c]
    \begingroup 
    \setlength\arraycolsep{5pt}
    \begin{pmatrix*}[c]
     d & d & d \\
     d & d & d \\
     d & d & d \\
    \end{pmatrix*} 
    \endgroup
    & 
    {\begin{pmatrix*}[c]
     J_1 & J_3 & \overline{J}_4 \\
     J_3 & J_1 & \overline{J}_4 \\
     J_4 & J_4 & J_2 \\
    \end{pmatrix*}} 
    & 
    \begin{pmatrix*}[c]
     J_1 & \overline{J}_4 & J_3 \\
     J_4 & J_2 & J_4 \\
     J_3 & \overline{J}_4 & J_1 \\
    \end{pmatrix*} 
    & 
    \begin{pmatrix*}[c]
     J_2 & J_4 & J_4 \\
     \overline{J}_4 & J_1 & J_3 \\
     \overline{J}_4 & J_3 & J_1 \\
    \end{pmatrix*} 
    \\[1ex]
    \begin{pmatrix*}[c]
     J_1 & J_3 & J_4 \\
     J_3 & J_1 & J_4 \\
     \overline{J}_4 & \overline{J}_4 & J_2 \\
    \end{pmatrix*} 
    & 
    \begingroup 
    \setlength\arraycolsep{5pt}
    \begin{pmatrix*}[c]
     d & d & \overline{d} \\
     d & d & \overline{d} \\
     \overline{d} & \overline{d} & d \\
    \end{pmatrix*} 
    \endgroup
    & 
    \begin{pmatrix*}[c]
     J_2 & J_4 & \overline{J}_4 \\
     \overline{J}_4 & J_1 & \overline{J}_3 \\
     J_4 & \overline{J}_3 & J_1 \\
    \end{pmatrix*} 
    & 
    \begin{pmatrix*}[c]
     J_1 & \overline{J}_4 & \overline{J}_3 \\
     J_4 & J_2 & \overline{J}_4 \\
     \overline{J}_3 & J_4 & J_1 \\
    \end{pmatrix*} 
    \\[1ex]
    \begin{pmatrix*}[c]
     J_1 & J_4 & J_3 \\
     \overline{J}_4 & J_2 & \overline{J}_4 \\
     J_3 & J_4 & J_1 \\
    \end{pmatrix*} 
    & 
    \begin{pmatrix*}[c]
     J_2 & \overline{J}_4 & J_4 \\
     J_4 & J_1 & \overline{J}_3 \\
     \overline{J}_4 & \overline{J}_3 & J_1 \\
    \end{pmatrix*} 
    & 
    \begingroup 
    \setlength\arraycolsep{5pt}
    \begin{pmatrix*}[c]
     d & \overline{d} & d \\
     \overline{d} & d & \overline{d} \\
     d & \overline{d} & d \\
    \end{pmatrix*}
    \endgroup
    & 
    \begin{pmatrix*}[c]
     J_1 & \overline{J}_3 & \overline{J}_4 \\
     \overline{J}_3 & J_1 & J_4 \\
     J_4 & \overline{J}_4 & J_2 \\
    \end{pmatrix*} 
    \\
    \begin{pmatrix*}[c]
     J_2 & \overline{J}_4 & \overline{J}_4 \\
     J_4 & J_1 & J_3 \\
     J_4 & J_3 & J_1 \\
    \end{pmatrix*} 
    & 
    \begin{pmatrix*}[c]
     J_1 & J_4 & \overline{J}_3 \\
     \overline{J}_4 & J_2 & J_4 \\
     \overline{J}_3 & \overline{J}_4 & J_1 \\
    \end{pmatrix*} 
    & 
    \begin{pmatrix*}[c]
     J_1 & \overline{J}_3 & J_4 \\
     \overline{J}_3 & J_1 & \overline{J}_4 \\
     \overline{J}_4 & J_4 & J_2 \\
    \end{pmatrix*} 
    & 
    \begingroup 
    \setlength\arraycolsep{5pt}
    \begin{pmatrix*}[c]
     d & \overline{d} & \overline{d} \\
     \overline{d} & d & d \\
     \overline{d} & d & d \\
    \end{pmatrix*} 
    \endgroup
\end{pmatrix*},
\label{eq:J14_full_matrix}
\end{equation}
where for compactness we have used a bar to indicate a negative, $\overline{J}_4 = -J_4$ and $\overline{d}=-d$, and defined $d=3J_{\text{SIA}}$ (cf. \cref{eq:H_named_interactions}). 
\Cref{eq:Jijab} showed only the matrix elements for $\mu=1$ and $\nu=2$.

\subsection{Global Frame Coordinate-Free}
In \cref{eq:H_named_interactions} we provided another version of the Hamiltonian with various named interactions: Heisenberg, pseudo-dipolar, Kitaev-like, and Dzyaloshinskii-Moriya (DM).
Their definitions are provided by \cref{fig:tetrahedron_vectors}.
They are directly related to the global-basis parameters $J_1$ through $J_4$. 
It suffices to compare the interaction matrix $J_{\mu\nu}^{\alpha\beta}$ in \cref{eq:J14_full_matrix} for a single pair $\mu\neq\nu$. 
With the conventions for the vectors given in \cref{fig:tetrahedron_vectors} and \cref{apx:conventions}, we have the identification for sublattices $\mu=1$ and $\nu=2$
\begin{equation}
    \begin{pmatrix*}[c]
         J_1 & J_3 & -J_4 \\
         J_3 & J_1 & -J_4 \\
         J_4 & J_4 & J_2 
    \end{pmatrix*}
    =
    \begin{pmatrix*}[c]
         J_{\text{Heis}} + J_{\text{PD}}/2 & J_{\text{PD}}/2 & J_{\text{DM}}/\sqrt{2} \\
         J_{\text{PD}}/2 & J_{\text{Heis}} + J_{\text{PD}}/2 & J_{\text{DM}}/\sqrt{2} \\
         -J_{\text{DM}}/\sqrt{2} & -J_{\text{DM}}/\sqrt{2} & J_{\text{Heis}} + J_{\text{K}}
    \end{pmatrix*} .
\end{equation}

\subsection{Local Frame}

To obtain the local basis \cref{eq:H_local}, let $R_\mu$ denote the SO(3) matrices rotating from the global Cartesian basis $\{\uvec{e}_x,\uvec{e}_y,\uvec{e}_z\}$ to the local basis on each sublattice defined by \cref{eq:local_z} and \cref{eq:local_xy}, and let $Q$ denote the non-unitary transformation
\begin{equation}
    Q = \begin{pmatrix}
        1 & i & 0 \\ 1 & -i & 0 \\ 0 & 0 & 1
    \end{pmatrix} : (S^x_\mu,S^y_\mu,S^z_\mu)\mapsto (S^+_\mu, S^-_\mu, S^z_\mu),
\end{equation}
let $\mathcal{R}$ denote the block-diagonal matrix $\diag(R_1,R_2,R_3,R_4)$ and let $\mathscr{Q}$ denote the block-diagonal matrix $\diag(Q,Q,Q,Q)$.
Then the local couplings in \cref{eq:H_local} are obtained by transforming \cref{eq:J14_full_matrix} as\footnote{
 Note that we are taking the Hamiltonian to be defined by the complex sesquilinear inner product, $H=\bm{S}^\dagger \cdot  \mathcal{J} \cdot \bm{S}/2$. Because of this, the $S^+_i S^-_j$ terms appears as diagonal components in $\Jmat_{\text{local}}$, while the $S_i^+ S_j^+$ terms appear as off-diagonal components. Note further that $\mathcal{J}_{\text{local}}$ and $\mathcal{J}_{\text{global}}$ do not have the same eigenvalues as $\mathscr{Q}$ is not unitary.
}
\begin{equation}
    \Jmat_{\text{local}} = (\mathscr{Q}^{-1})^\dagger \mathcal{R}\,\Jmat_{\text{global}}\, \mathcal{R}^{-1} \mathscr{Q}^{-1},
\end{equation}
where $\scriptstyle{\dagger}$ denotes the conjugate transpose. 
Note that the complex phase factors appearing in \cref{eq:H_local} depend on the convention for the local bases. 
With the choice of basis from \cref{apx:conventions}, they are
\begin{equation}
    \left(
\begin{array}{ccc|ccc|ccc|ccc}
 0 & 0 & 0 & 1 & 1 & 1 & 1 & -\omega  & \omega^2 & 1 & \omega^2 & -\omega  \\
 0 & 0 & 0 & 1 & 1 & 1 & \omega^2 & 1 & -\omega  & -\omega  & 1 & \omega^2 \\
 0 & 0 & 0 & 1 & 1 & 1 & -\omega  & \omega^2 & 1 & \omega^2 & -\omega  & 1 \Bstrut
 \\ 
 \hline
 \Tstrut
 1 & 1 & 1 & 0 & 0 & 0 & 1 & \omega^2 & -\omega  & 1 & -\omega  & \omega^2 \\
 1 & 1 & 1 & 0 & 0 & 0 & -\omega  & 1 & \omega^2 & \omega^2 & 1 & -\omega  \\
 1 & 1 & 1 & 0 & 0 & 0 & \omega^2 & -\omega  & 1 & -\omega  & \omega^2 & 1  \Bstrut
 \\ 
 \hline
 \Tstrut
 1 & -\omega  & \omega^2 & 1 & \omega^2 & -\omega  & 0 & 0 & 0 & 1 & 1 & 1 \\
 \omega^2 & 1 & -\omega  & -\omega  & 1 & \omega^2 & 0 & 0 & 0 & 1 & 1 & 1 \\
 -\omega  & \omega^2 & 1 & \omega^2 & -\omega  & 1 & 0 & 0 & 0 & 1 & 1 & 1  \Bstrut
 \\ 
 \hline
 \Tstrut
 1 & \omega^2 & -\omega  & 1 & -\omega  & \omega^2 & 1 & 1 & 1 & 0 & 0 & 0 \\
 -\omega  & 1 & \omega^2 & \omega^2 & 1 & -\omega  & 1 & 1 & 1 & 0 & 0 & 0 \\
 \omega^2 & -\omega  & 1 & -\omega  & \omega^2 & 1 & 1 & 1 & 1 & 0 & 0 & 0 \\
\end{array}
\right)
\end{equation}
where $\omega=(-1)^{1/3}=e^{i\pi/3}$.
The relation between the global and local coupling parameters are given in \cref{eq:JloctoJ14,eq:J14toJloc}.

\section{SCGA Structure Factors}
\label{apx:scga}

In the self-consistent Gaussian approximation (SCGA), the spin length constraint is relaxed to a Gaussian distribution whose variance is enforced by a Lagrange multiplier $\lambda$. Given a quadratic spin Hamiltonian $H$ of the form \cref{eq:H_generic}, we define the SCGA Hamiltonian
\begin{equation}
    H_{\text{SCGA}} = H + T\sum_i \lambda_i \vert\bm{S}_i\vert^2,
\end{equation}
where $T$ is temperature. The partition function is then simply a multivariate Gaussian integral,
\begin{equation}
    Z_{\text{SCGA}} = \int \prod_{i,\alpha}\mathrm{d}S_i^\alpha \, \exp(-\sum_{ij}\sum_{\alpha\beta} S_i^\alpha (\underbrace{\lambda_i \delta_{ij} \delta_{\alpha\beta}}_{\Lambda_{ij}^{\alpha\beta}} + \beta \Jmat_{ij}^{\alpha\beta})S_j^\beta),
\end{equation}
where $\beta=1/T$ is the inverse temperature and $\delta$ is the Kronecker delta.
Correlation functions are then simply given by 
\begin{equation}
    \langle S_i^\alpha S_j^\beta\rangle \equiv \mathcal{G}_{ij}^{\alpha\beta} = ([\Lambda + \beta\Jmat]^{-1})_{ij}^{\alpha\beta}.
\end{equation}
The Lagrange multipliers are solved for self-consistently via the constraint
\begin{equation}
    1 = \langle \vert \bm{S}_i\vert^2 \rangle = \sum_\alpha ([\Lambda + \beta\Jmat]^{-1})_{ii}^{\alpha\alpha}.
\end{equation}
Since every site is symmetry-equivalent in the pyrochlore lattice, the Lagrange multipliers are all the same, $\lambda_i \equiv \lambda$, and the self-consistency condition can be solved by averaging over all sites,
\begin{equation}
    1 = \frac{1}{N_{\text{spins}}} \Tr [\lambda \mathbbm{1} + \beta\Jmat]^{-1},
\end{equation}
where $\mathbbm{1}$ is the  $12L^3$-dimensional identity matrix. 
We assume that the minimum eigenvalue of $\Jmat$ has been shifted to zero, let $\mathscr{F}$ denote the kernel of $\Jmat$, and let $N_0=\dim \mathscr{F}$ denote the number of zero eigenvalues of $\Jmat$ (if not, then let $v$ denote the minimum eigenvalue and redefine $\Jmat\to \Jmat - v \mathbbm{1}$ and $\Lambda \to \Lambda + \beta v \mathbbm{1}$).
In the limit of high temperature $\beta\to 0$ the solution is $\lambda = 3$ (the number of spin components).
In the low-temperature limit $\beta\to \infty$ all positive eigenvalues of $\Jmat$ are suppressed in the inverse, and the solution is $\lambda = N_0/N_{\text{spins}}$.
If $N_0$ is subextensive then $N_0 / N_{\text{spins}} \to 0$ in the thermodynamic limit $N_\text{spins}\to\infty$, meaning that $\lambda$ vanishes at some critical temperature $T_c$, indicating a phase transition as the stiffness of the zero-energy modes vanishes, and a breakdown of the SCGA below $T_c$.
If the number of zero eigenvalues is extensive, i.e. $\Jmat$ has zero-energy flat bands, with $N_0 = 3N_{\text{spins}}f$ for $0<f=n_{\text{flat}}/n_{\text{tot}}<1$, then the solution is $\lambda = 3f$. 
The zero-temperature correlation functions are then given by the projector $P_{\mathscr{F}}$ to the kernel of $\Jmat$,
\begin{equation}
    \lim_{T\to 0} \mathcal{G} = \frac{1}{3f} P_{\mathscr{F}}.
\end{equation}
We use this formula to calculate the structure factor plots in \cref{fig:structure_factors_triple_lines,fig:structure_factors_special}.

\clearpage


\begin{thebibliography}{151}%
\makeatletter
\providecommand \@ifxundefined [1]{%
 \@ifx{#1\undefined}
}%
\providecommand \@ifnum [1]{%
 \ifnum #1\expandafter \@firstoftwo
 \else \expandafter \@secondoftwo
 \fi
}%
\providecommand \@ifx [1]{%
 \ifx #1\expandafter \@firstoftwo
 \else \expandafter \@secondoftwo
 \fi
}%
\providecommand \natexlab [1]{#1}%
\providecommand \enquote  [1]{``#1''}%
\providecommand \bibnamefont  [1]{#1}%
\providecommand \bibfnamefont [1]{#1}%
\providecommand \citenamefont [1]{#1}%
\providecommand \href@noop [0]{\@secondoftwo}%
\providecommand \href [0]{\begingroup \@sanitize@url \@href}%
\providecommand \@href[1]{\@@startlink{#1}\@@href}%
\providecommand \@@href[1]{\endgroup#1\@@endlink}%
\providecommand \@sanitize@url [0]{\catcode `\\12\catcode `\$12\catcode `\&12\catcode `\#12\catcode `\^12\catcode `\_12\catcode `\%12\relax}%
\providecommand \@@startlink[1]{}%
\providecommand \@@endlink[0]{}%
\providecommand \url  [0]{\begingroup\@sanitize@url \@url }%
\providecommand \@url [1]{\endgroup\@href {#1}{\urlprefix }}%
\providecommand \urlprefix  [0]{URL }%
\providecommand \Eprint [0]{\href }%
\providecommand \doibase [0]{https://doi.org/}%
\providecommand \selectlanguage [0]{\@gobble}%
\providecommand \bibinfo  [0]{\@secondoftwo}%
\providecommand \bibfield  [0]{\@secondoftwo}%
\providecommand \translation [1]{[#1]}%
\providecommand \BibitemOpen [0]{}%
\providecommand \bibitemStop [0]{}%
\providecommand \bibitemNoStop [0]{.\EOS\space}%
\providecommand \EOS [0]{\spacefactor3000\relax}%
\providecommand \BibitemShut  [1]{\csname bibitem#1\endcsname}%
\let\auto@bib@innerbib\@empty
\bibitem [{\citenamefont {Senthil}\ \emph {et~al.}(2004{\natexlab{a}})\citenamefont {Senthil}, \citenamefont {Vishwanath}, \citenamefont {Balents}, \citenamefont {{Sachdev, Subir}},\ and\ \citenamefont {Fisher}}]{senthilDeconfinedQuantumCritical2004}%
  \BibitemOpen
  \bibfield  {author} {\bibinfo {author} {\bibfnamefont {T.}~\bibnamefont {Senthil}}, \bibinfo {author} {\bibfnamefont {A.}~\bibnamefont {Vishwanath}}, \bibinfo {author} {\bibfnamefont {L.}~\bibnamefont {Balents}}, \bibinfo {author} {\bibnamefont {{Sachdev, Subir}}},\ and\ \bibinfo {author} {\bibfnamefont {M.~P.~A.}\ \bibnamefont {Fisher}},\ }\bibfield  {title} {\bibinfo {title} {Deconfined {{Quantum Critical Points}}},\ }\href {https://doi.org/10.1126/science.1091806} {\bibfield  {journal} {\bibinfo  {journal} {Science}\ }\textbf {\bibinfo {volume} {303}},\ \bibinfo {pages} {1490} (\bibinfo {year} {2004}{\natexlab{a}})}\BibitemShut {NoStop}%
\bibitem [{\citenamefont {Senthil}\ \emph {et~al.}(2004{\natexlab{b}})\citenamefont {Senthil}, \citenamefont {Balents}, \citenamefont {Sachdev}, \citenamefont {Vishwanath},\ and\ \citenamefont {Fisher}}]{senthilQuantumCriticalityLandauGinzburgWilson2004}%
  \BibitemOpen
  \bibfield  {author} {\bibinfo {author} {\bibfnamefont {T.}~\bibnamefont {Senthil}}, \bibinfo {author} {\bibfnamefont {L.}~\bibnamefont {Balents}}, \bibinfo {author} {\bibfnamefont {S.}~\bibnamefont {Sachdev}}, \bibinfo {author} {\bibfnamefont {A.}~\bibnamefont {Vishwanath}},\ and\ \bibinfo {author} {\bibfnamefont {M.~P.~A.}\ \bibnamefont {Fisher}},\ }\bibfield  {title} {\bibinfo {title} {Quantum criticality beyond the {{Landau-Ginzburg-Wilson}} paradigm},\ }\href {https://doi.org/10.1103/PhysRevB.70.144407} {\bibfield  {journal} {\bibinfo  {journal} {Phys. Rev. B}\ }\textbf {\bibinfo {volume} {70}},\ \bibinfo {pages} {144407} (\bibinfo {year} {2004}{\natexlab{b}})}\BibitemShut {NoStop}%
\bibitem [{\citenamefont {Hermele}\ \emph {et~al.}(2005)\citenamefont {Hermele}, \citenamefont {Senthil},\ and\ \citenamefont {Fisher}}]{hermeleAlgebraicSpinLiquid2005}%
  \BibitemOpen
  \bibfield  {author} {\bibinfo {author} {\bibfnamefont {M.}~\bibnamefont {Hermele}}, \bibinfo {author} {\bibfnamefont {T.}~\bibnamefont {Senthil}},\ and\ \bibinfo {author} {\bibfnamefont {M.~P.~A.}\ \bibnamefont {Fisher}},\ }\bibfield  {title} {\bibinfo {title} {Algebraic spin liquid as the mother of many competing orders},\ }\href {https://doi.org/10.1103/PhysRevB.72.104404} {\bibfield  {journal} {\bibinfo  {journal} {Phys. Rev. B}\ }\textbf {\bibinfo {volume} {72}},\ \bibinfo {pages} {104404} (\bibinfo {year} {2005})}\BibitemShut {NoStop}%
\bibitem [{\citenamefont {Balents}\ \emph {et~al.}(2005)\citenamefont {Balents}, \citenamefont {Bartosch}, \citenamefont {Burkov}, \citenamefont {Sachdev},\ and\ \citenamefont {Sengupta}}]{balentsCompetingOrdersNonLandauGinzburgWilson2005}%
  \BibitemOpen
  \bibfield  {author} {\bibinfo {author} {\bibfnamefont {L.}~\bibnamefont {Balents}}, \bibinfo {author} {\bibfnamefont {L.}~\bibnamefont {Bartosch}}, \bibinfo {author} {\bibfnamefont {A.}~\bibnamefont {Burkov}}, \bibinfo {author} {\bibfnamefont {S.}~\bibnamefont {Sachdev}},\ and\ \bibinfo {author} {\bibfnamefont {K.}~\bibnamefont {Sengupta}},\ }\bibfield  {title} {\bibinfo {title} {Competing orders and non-{{Landau-Ginzburg-Wilson}} criticality in ({{Bose}}) {{Mott}} transitions},\ }\href {https://doi.org/10.1143/PTPS.160.314} {\bibfield  {journal} {\bibinfo  {journal} {Progress of Theoretical Physics Supplement}\ }\textbf {\bibinfo {volume} {160}},\ \bibinfo {pages} {314} (\bibinfo {year} {2005})}\BibitemShut {NoStop}%
\bibitem [{\citenamefont {Senthil}\ and\ \citenamefont {Fisher}(2006)}]{senthilCompetingOrdersNonlinear2006}%
  \BibitemOpen
  \bibfield  {author} {\bibinfo {author} {\bibfnamefont {T.}~\bibnamefont {Senthil}}\ and\ \bibinfo {author} {\bibfnamefont {M.~P.~A.}\ \bibnamefont {Fisher}},\ }\bibfield  {title} {\bibinfo {title} {Competing orders, nonlinear sigma models, and topological terms in quantum magnets},\ }\href {https://doi.org/10.1103/PhysRevB.74.064405} {\bibfield  {journal} {\bibinfo  {journal} {Phys. Rev. B}\ }\textbf {\bibinfo {volume} {74}},\ \bibinfo {pages} {064405} (\bibinfo {year} {2006})}\BibitemShut {NoStop}%
\bibitem [{\citenamefont {Wang}\ \emph {et~al.}(2017)\citenamefont {Wang}, \citenamefont {Nahum}, \citenamefont {Metlitski}, \citenamefont {Xu},\ and\ \citenamefont {Senthil}}]{wangDeconfinedQuantumCritical2017}%
  \BibitemOpen
  \bibfield  {author} {\bibinfo {author} {\bibfnamefont {C.}~\bibnamefont {Wang}}, \bibinfo {author} {\bibfnamefont {A.}~\bibnamefont {Nahum}}, \bibinfo {author} {\bibfnamefont {M.~A.}\ \bibnamefont {Metlitski}}, \bibinfo {author} {\bibfnamefont {C.}~\bibnamefont {Xu}},\ and\ \bibinfo {author} {\bibfnamefont {T.}~\bibnamefont {Senthil}},\ }\bibfield  {title} {\bibinfo {title} {Deconfined {{Quantum Critical Points}}: {{Symmetries}} and {{Dualities}}},\ }\href {https://doi.org/10.1103/PhysRevX.7.031051} {\bibfield  {journal} {\bibinfo  {journal} {Phys. Rev. X}\ }\textbf {\bibinfo {volume} {7}},\ \bibinfo {pages} {031051} (\bibinfo {year} {2017})}\BibitemShut {NoStop}%
\bibitem [{\citenamefont {Song}\ \emph {et~al.}(2019)\citenamefont {Song}, \citenamefont {Wang}, \citenamefont {Vishwanath},\ and\ \citenamefont {He}}]{songUnifyingDescriptionCompeting2019}%
  \BibitemOpen
  \bibfield  {author} {\bibinfo {author} {\bibfnamefont {X.-Y.}\ \bibnamefont {Song}}, \bibinfo {author} {\bibfnamefont {C.}~\bibnamefont {Wang}}, \bibinfo {author} {\bibfnamefont {A.}~\bibnamefont {Vishwanath}},\ and\ \bibinfo {author} {\bibfnamefont {Y.-C.}\ \bibnamefont {He}},\ }\bibfield  {title} {\bibinfo {title} {Unifying description of competing orders in two-dimensional quantum magnets},\ }\href {https://doi.org/10.1038/s41467-019-11727-3} {\bibfield  {journal} {\bibinfo  {journal} {Nature Communications}\ }\textbf {\bibinfo {volume} {10}},\ \bibinfo {pages} {4254} (\bibinfo {year} {2019})}\BibitemShut {NoStop}%
\bibitem [{\citenamefont {Liu}\ \emph {et~al.}(2019)\citenamefont {Liu}, \citenamefont {Hal{\'a}sz},\ and\ \citenamefont {Balents}}]{liuCompetingOrdersPyrochlore2019}%
  \BibitemOpen
  \bibfield  {author} {\bibinfo {author} {\bibfnamefont {C.}~\bibnamefont {Liu}}, \bibinfo {author} {\bibfnamefont {G.~B.}\ \bibnamefont {Hal{\'a}sz}},\ and\ \bibinfo {author} {\bibfnamefont {L.}~\bibnamefont {Balents}},\ }\bibfield  {title} {\bibinfo {title} {Competing orders in pyrochlore magnets from a {{$\mathbb{Z}$}}{$_2$} spin liquid perspective},\ }\href {https://doi.org/10.1103/PhysRevB.100.075125} {\bibfield  {journal} {\bibinfo  {journal} {Phys. Rev. B}\ }\textbf {\bibinfo {volume} {100}},\ \bibinfo {pages} {075125} (\bibinfo {year} {2019})}\BibitemShut {NoStop}%
\bibitem [{\citenamefont {Zou}\ \emph {et~al.}(2021)\citenamefont {Zou}, \citenamefont {He},\ and\ \citenamefont {Wang}}]{zouStiefelLiquidsPossible2021}%
  \BibitemOpen
  \bibfield  {author} {\bibinfo {author} {\bibfnamefont {L.}~\bibnamefont {Zou}}, \bibinfo {author} {\bibfnamefont {Y.-C.}\ \bibnamefont {He}},\ and\ \bibinfo {author} {\bibfnamefont {C.}~\bibnamefont {Wang}},\ }\bibfield  {title} {\bibinfo {title} {Stiefel {{Liquids}}: {{Possible Non-Lagrangian Quantum Criticality}} from {{Intertwined Orders}}},\ }\href {https://doi.org/10.1103/PhysRevX.11.031043} {\bibfield  {journal} {\bibinfo  {journal} {Phys. Rev. X}\ }\textbf {\bibinfo {volume} {11}},\ \bibinfo {pages} {031043} (\bibinfo {year} {2021})}\BibitemShut {NoStop}%
\bibitem [{\citenamefont {Lacroix}\ \emph {et~al.}(2011)\citenamefont {Lacroix}, \citenamefont {Mendels},\ and\ \citenamefont {Mila}}]{lacroixIntroductionFrustratedMagnetism2011}%
  \BibitemOpen
  \bibinfo {editor} {\bibfnamefont {C.}~\bibnamefont {Lacroix}}, \bibinfo {editor} {\bibfnamefont {P.}~\bibnamefont {Mendels}},\ and\ \bibinfo {editor} {\bibfnamefont {F.}~\bibnamefont {Mila}},\ eds.,\ \href {https://doi.org/10.1007/978-3-642-10589-0} {\emph {\bibinfo {title} {Introduction to {{Frustrated Magnetism}}: {{Materials}}, {{Experiments}}, {{Theory}}}}},\ \bibinfo {series} {Springer {{Series}} in {{Solid-State Sciences}}}, Vol.\ \bibinfo {volume} {164}\ (\bibinfo  {publisher} {Springer},\ \bibinfo {address} {Berlin, Heidelberg},\ \bibinfo {year} {2011})\BibitemShut {NoStop}%
\bibitem [{\citenamefont {Diep}(2020)}]{diepFrustratedSpinSystems2020}%
  \BibitemOpen
  \bibinfo {editor} {\bibfnamefont {H.~T.}\ \bibnamefont {Diep}},\ ed.,\ \href@noop {} {\emph {\bibinfo {title} {Frustrated Spin Systems}}},\ \bibinfo {edition} {3rd}\ ed.\ (\bibinfo  {publisher} {World Scientific Publishing Co. Pte. Ltd.},\ \bibinfo {address} {Singapore},\ \bibinfo {year} {2020})\BibitemShut {NoStop}%
\bibitem [{\citenamefont {Balents}(2010)}]{balentsSpinLiquidsFrustrated2010}%
  \BibitemOpen
  \bibfield  {author} {\bibinfo {author} {\bibfnamefont {L.}~\bibnamefont {Balents}},\ }\bibfield  {title} {\bibinfo {title} {Spin liquids in frustrated magnets},\ }\href {https://doi.org/10.1038/nature08917} {\bibfield  {journal} {\bibinfo  {journal} {Nature}\ }\textbf {\bibinfo {volume} {464}},\ \bibinfo {pages} {199} (\bibinfo {year} {2010})}\BibitemShut {NoStop}%
\bibitem [{\citenamefont {Lhuillier}\ and\ \citenamefont {Misguich}(2011)}]{lhuillierIntroductionQuantumSpin2011}%
  \BibitemOpen
  \bibfield  {author} {\bibinfo {author} {\bibfnamefont {C.}~\bibnamefont {Lhuillier}}\ and\ \bibinfo {author} {\bibfnamefont {G.}~\bibnamefont {Misguich}},\ }\bibfield  {title} {\bibinfo {title} {Introduction to {{Quantum Spin Liquids}}},\ }in\ \href {https://doi.org/10.1007/978-3-642-10589-0_2} {\emph {\bibinfo {booktitle} {Introduction to {{Frustrated Magnetism}}: {{Materials}}, {{Experiments}}, {{Theory}}}}},\ \bibinfo {series and number} {Springer {{Series}} in {{Solid-State Sciences}}},\ \bibinfo {editor} {edited by\ \bibinfo {editor} {\bibfnamefont {C.}~\bibnamefont {Lacroix}}, \bibinfo {editor} {\bibfnamefont {P.}~\bibnamefont {Mendels}},\ and\ \bibinfo {editor} {\bibfnamefont {F.}~\bibnamefont {Mila}}}\ (\bibinfo  {publisher} {Springer},\ \bibinfo {address} {Berlin, Heidelberg},\ \bibinfo {year} {2011})\ pp.\ \bibinfo {pages} {23--41}\BibitemShut {NoStop}%
\bibitem [{\citenamefont {Savary}\ and\ \citenamefont {Balents}(2016)}]{savaryQuantumSpinLiquids2016}%
  \BibitemOpen
  \bibfield  {author} {\bibinfo {author} {\bibfnamefont {L.}~\bibnamefont {Savary}}\ and\ \bibinfo {author} {\bibfnamefont {L.}~\bibnamefont {Balents}},\ }\bibfield  {title} {\bibinfo {title} {Quantum spin liquids: A review},\ }\href {https://doi.org/10.1088/0034-4885/80/1/016502} {\bibfield  {journal} {\bibinfo  {journal} {Rep. Prog. Phys.}\ }\textbf {\bibinfo {volume} {80}},\ \bibinfo {pages} {016502} (\bibinfo {year} {2016})}\BibitemShut {NoStop}%
\bibitem [{\citenamefont {Zhou}\ \emph {et~al.}(2017)\citenamefont {Zhou}, \citenamefont {Kanoda},\ and\ \citenamefont {Ng}}]{zhouQuantumSpinLiquid2017}%
  \BibitemOpen
  \bibfield  {author} {\bibinfo {author} {\bibfnamefont {Y.}~\bibnamefont {Zhou}}, \bibinfo {author} {\bibfnamefont {K.}~\bibnamefont {Kanoda}},\ and\ \bibinfo {author} {\bibfnamefont {T.-K.}\ \bibnamefont {Ng}},\ }\bibfield  {title} {\bibinfo {title} {Quantum spin liquid states},\ }\href {https://doi.org/10.1103/RevModPhys.89.025003} {\bibfield  {journal} {\bibinfo  {journal} {Rev. Mod. Phys.}\ }\textbf {\bibinfo {volume} {89}},\ \bibinfo {pages} {025003} (\bibinfo {year} {2017})}\BibitemShut {NoStop}%
\bibitem [{\citenamefont {Wen}\ \emph {et~al.}(2019)\citenamefont {Wen}, \citenamefont {Yu}, \citenamefont {Li}, \citenamefont {Yu},\ and\ \citenamefont {Li}}]{wenExperimentalIdentificationQuantum2019}%
  \BibitemOpen
  \bibfield  {author} {\bibinfo {author} {\bibfnamefont {J.}~\bibnamefont {Wen}}, \bibinfo {author} {\bibfnamefont {S.-L.}\ \bibnamefont {Yu}}, \bibinfo {author} {\bibfnamefont {S.}~\bibnamefont {Li}}, \bibinfo {author} {\bibfnamefont {W.}~\bibnamefont {Yu}},\ and\ \bibinfo {author} {\bibfnamefont {J.-X.}\ \bibnamefont {Li}},\ }\bibfield  {title} {\bibinfo {title} {Experimental identification of quantum spin liquids},\ }\href {https://doi.org/10.1038/s41535-019-0151-6} {\bibfield  {journal} {\bibinfo  {journal} {npj Quantum Mater.}\ }\textbf {\bibinfo {volume} {4}},\ \bibinfo {pages} {1} (\bibinfo {year} {2019})}\BibitemShut {NoStop}%
\bibitem [{\citenamefont {Knolle}\ and\ \citenamefont {Moessner}(2019)}]{knolleFieldGuideSpin2019}%
  \BibitemOpen
  \bibfield  {author} {\bibinfo {author} {\bibfnamefont {J.}~\bibnamefont {Knolle}}\ and\ \bibinfo {author} {\bibfnamefont {R.}~\bibnamefont {Moessner}},\ }\bibfield  {title} {\bibinfo {title} {A field guide to spin liquids},\ }\href {https://doi.org/10.1146/annurev-conmatphys-031218-013401} {\bibfield  {journal} {\bibinfo  {journal} {Annual Review of Condensed Matter Physics}\ }\textbf {\bibinfo {volume} {10}},\ \bibinfo {pages} {451} (\bibinfo {year} {2019})}\BibitemShut {NoStop}%
\bibitem [{\citenamefont {Broholm}\ \emph {et~al.}(2020)\citenamefont {Broholm}, \citenamefont {Cava}, \citenamefont {Kivelson}, \citenamefont {Nocera}, \citenamefont {Norman},\ and\ \citenamefont {Senthil}}]{broholmQuantumSpinLiquids2020}%
  \BibitemOpen
  \bibfield  {author} {\bibinfo {author} {\bibfnamefont {C.}~\bibnamefont {Broholm}}, \bibinfo {author} {\bibfnamefont {R.~J.}\ \bibnamefont {Cava}}, \bibinfo {author} {\bibfnamefont {S.~A.}\ \bibnamefont {Kivelson}}, \bibinfo {author} {\bibfnamefont {D.~G.}\ \bibnamefont {Nocera}}, \bibinfo {author} {\bibfnamefont {M.~R.}\ \bibnamefont {Norman}},\ and\ \bibinfo {author} {\bibfnamefont {T.}~\bibnamefont {Senthil}},\ }\bibfield  {title} {\bibinfo {title} {Quantum spin liquids},\ }\href {https://doi.org/10.1126/science.aay0668} {\bibfield  {journal} {\bibinfo  {journal} {Science}\ }\textbf {\bibinfo {volume} {367}},\ \bibinfo {pages} {eaay0668} (\bibinfo {year} {2020})}\BibitemShut {NoStop}%
\bibitem [{\citenamefont {Subramanian}\ \emph {et~al.}(1983)\citenamefont {Subramanian}, \citenamefont {Aravamudan},\ and\ \citenamefont {Subba~Rao}}]{subramanianOxidePyrochloresReview1983}%
  \BibitemOpen
  \bibfield  {author} {\bibinfo {author} {\bibfnamefont {M.~A.}\ \bibnamefont {Subramanian}}, \bibinfo {author} {\bibfnamefont {G.}~\bibnamefont {Aravamudan}},\ and\ \bibinfo {author} {\bibfnamefont {G.~V.}\ \bibnamefont {Subba~Rao}},\ }\bibfield  {title} {\bibinfo {title} {Oxide pyrochlores --- {{A}} review},\ }\href {https://doi.org/10.1016/0079-6786(83)90001-8} {\bibfield  {journal} {\bibinfo  {journal} {Progress in Solid State Chemistry}\ }\textbf {\bibinfo {volume} {15}},\ \bibinfo {pages} {55} (\bibinfo {year} {1983})}\BibitemShut {NoStop}%
\bibitem [{\citenamefont {Gardner}\ \emph {et~al.}(2010)\citenamefont {Gardner}, \citenamefont {Gingras},\ and\ \citenamefont {Greedan}}]{gardnerMagneticPyrochloreOxides2010}%
  \BibitemOpen
  \bibfield  {author} {\bibinfo {author} {\bibfnamefont {J.~S.}\ \bibnamefont {Gardner}}, \bibinfo {author} {\bibfnamefont {M.~J.~P.}\ \bibnamefont {Gingras}},\ and\ \bibinfo {author} {\bibfnamefont {J.~E.}\ \bibnamefont {Greedan}},\ }\bibfield  {title} {\bibinfo {title} {Magnetic pyrochlore oxides},\ }\href {https://doi.org/10.1103/RevModPhys.82.53} {\bibfield  {journal} {\bibinfo  {journal} {Rev. Mod. Phys.}\ }\textbf {\bibinfo {volume} {82}},\ \bibinfo {pages} {53} (\bibinfo {year} {2010})}\BibitemShut {NoStop}%
\bibitem [{\citenamefont {Lee}\ \emph {et~al.}(2010)\citenamefont {Lee}, \citenamefont {Takagi}, \citenamefont {Louca}, \citenamefont {Matsuda}, \citenamefont {Ji}, \citenamefont {Ueda}, \citenamefont {Ueda}, \citenamefont {Katsufuji}, \citenamefont {Chung}, \citenamefont {Park}, \citenamefont {Cheong},\ and\ \citenamefont {Broholm}}]{leeFrustratedMagnetismCooperative2010}%
  \BibitemOpen
  \bibfield  {author} {\bibinfo {author} {\bibfnamefont {S.-H.}\ \bibnamefont {Lee}}, \bibinfo {author} {\bibfnamefont {H.}~\bibnamefont {Takagi}}, \bibinfo {author} {\bibfnamefont {D.}~\bibnamefont {Louca}}, \bibinfo {author} {\bibfnamefont {M.}~\bibnamefont {Matsuda}}, \bibinfo {author} {\bibfnamefont {S.}~\bibnamefont {Ji}}, \bibinfo {author} {\bibfnamefont {H.}~\bibnamefont {Ueda}}, \bibinfo {author} {\bibfnamefont {Y.}~\bibnamefont {Ueda}}, \bibinfo {author} {\bibfnamefont {T.}~\bibnamefont {Katsufuji}}, \bibinfo {author} {\bibfnamefont {J.-H.}\ \bibnamefont {Chung}}, \bibinfo {author} {\bibfnamefont {S.}~\bibnamefont {Park}}, \bibinfo {author} {\bibfnamefont {S.-W.}\ \bibnamefont {Cheong}},\ and\ \bibinfo {author} {\bibfnamefont {C.}~\bibnamefont {Broholm}},\ }\bibfield  {title} {\bibinfo {title} {Frustrated {{Magnetism}} and {{Cooperative Phase Transitions}} in {{Spinels}}},\ }\href {https://doi.org/10.1143/JPSJ.79.011004} {\bibfield  {journal} {\bibinfo  {journal} {J. Phys. Soc. Jpn.}\ }\textbf
  {\bibinfo {volume} {79}},\ \bibinfo {pages} {011004} (\bibinfo {year} {2010})}\BibitemShut {NoStop}%
\bibitem [{\citenamefont {Wiebe}\ and\ \citenamefont {Hallas}(2015)}]{wiebeFrustrationPressureExotic2015}%
  \BibitemOpen
  \bibfield  {author} {\bibinfo {author} {\bibfnamefont {C.~R.}\ \bibnamefont {Wiebe}}\ and\ \bibinfo {author} {\bibfnamefont {A.~M.}\ \bibnamefont {Hallas}},\ }\bibfield  {title} {\bibinfo {title} {Frustration under pressure: {{Exotic}} magnetism in new pyrochlore oxides},\ }\href {https://doi.org/10.1063/1.4916020} {\bibfield  {journal} {\bibinfo  {journal} {APL Materials}\ }\textbf {\bibinfo {volume} {3}},\ \bibinfo {pages} {041519} (\bibinfo {year} {2015})}\BibitemShut {NoStop}%
\bibitem [{\citenamefont {Ghosh}\ \emph {et~al.}(2019)\citenamefont {Ghosh}, \citenamefont {Iqbal}, \citenamefont {M{\"u}ller}, \citenamefont {Ponnaganti}, \citenamefont {Thomale}, \citenamefont {Narayanan}, \citenamefont {Reuther}, \citenamefont {Gingras},\ and\ \citenamefont {Jeschke}}]{ghoshBreathingChromiumSpinels2019}%
  \BibitemOpen
  \bibfield  {author} {\bibinfo {author} {\bibfnamefont {P.}~\bibnamefont {Ghosh}}, \bibinfo {author} {\bibfnamefont {Y.}~\bibnamefont {Iqbal}}, \bibinfo {author} {\bibfnamefont {T.}~\bibnamefont {M{\"u}ller}}, \bibinfo {author} {\bibfnamefont {R.~T.}\ \bibnamefont {Ponnaganti}}, \bibinfo {author} {\bibfnamefont {R.}~\bibnamefont {Thomale}}, \bibinfo {author} {\bibfnamefont {R.}~\bibnamefont {Narayanan}}, \bibinfo {author} {\bibfnamefont {J.}~\bibnamefont {Reuther}}, \bibinfo {author} {\bibfnamefont {M.~J.~P.}\ \bibnamefont {Gingras}},\ and\ \bibinfo {author} {\bibfnamefont {H.~O.}\ \bibnamefont {Jeschke}},\ }\bibfield  {title} {\bibinfo {title} {Breathing chromium spinels: A showcase for a variety of pyrochlore {{Heisenberg Hamiltonians}}},\ }\href {https://doi.org/10.1038/s41535-019-0202-z} {\bibfield  {journal} {\bibinfo  {journal} {npj Quantum Materials}\ }\textbf {\bibinfo {volume} {4}},\ \bibinfo {pages} {63} (\bibinfo {year} {2019})}\BibitemShut {NoStop}%
\bibitem [{\citenamefont {{Reig-i-Plessis}}\ and\ \citenamefont {Hallas}(2021)}]{reig-i-plessisFrustratedMagnetismFluoride2021}%
  \BibitemOpen
  \bibfield  {author} {\bibinfo {author} {\bibfnamefont {D.}~\bibnamefont {{Reig-i-Plessis}}}\ and\ \bibinfo {author} {\bibfnamefont {A.~M.}\ \bibnamefont {Hallas}},\ }\bibfield  {title} {\bibinfo {title} {Frustrated magnetism in fluoride and chalcogenide pyrochlore lattice materials},\ }\href {https://doi.org/10.1103/PhysRevMaterials.5.030301} {\bibfield  {journal} {\bibinfo  {journal} {Phys. Rev. Materials}\ }\textbf {\bibinfo {volume} {5}},\ \bibinfo {pages} {030301} (\bibinfo {year} {2021})}\BibitemShut {NoStop}%
\bibitem [{\citenamefont {Rau}\ and\ \citenamefont {Gingras}(2019)}]{rauFrustratedQuantumRareEarth2019}%
  \BibitemOpen
  \bibfield  {author} {\bibinfo {author} {\bibfnamefont {J.~G.}\ \bibnamefont {Rau}}\ and\ \bibinfo {author} {\bibfnamefont {M.~J.}\ \bibnamefont {Gingras}},\ }\bibfield  {title} {\bibinfo {title} {Frustrated {{Quantum Rare-Earth Pyrochlores}}},\ }\href {https://doi.org/10.1146/annurev-conmatphys-022317-110520} {\bibfield  {journal} {\bibinfo  {journal} {Annual Review of Condensed Matter Physics}\ }\textbf {\bibinfo {volume} {10}},\ \bibinfo {pages} {357} (\bibinfo {year} {2019})}\BibitemShut {NoStop}%
\bibitem [{\citenamefont {Moessner}\ and\ \citenamefont {Chalker}(1998)}]{moessnerLowtemperaturePropertiesClassical1998}%
  \BibitemOpen
  \bibfield  {author} {\bibinfo {author} {\bibfnamefont {R.}~\bibnamefont {Moessner}}\ and\ \bibinfo {author} {\bibfnamefont {J.~T.}\ \bibnamefont {Chalker}},\ }\bibfield  {title} {\bibinfo {title} {Low-temperature properties of classical geometrically frustrated antiferromagnets},\ }\href {https://doi.org/10.1103/PhysRevB.58.12049} {\bibfield  {journal} {\bibinfo  {journal} {Phys. Rev. B}\ }\textbf {\bibinfo {volume} {58}},\ \bibinfo {pages} {12049} (\bibinfo {year} {1998})}\BibitemShut {NoStop}%
\bibitem [{\citenamefont {Henley}(2005)}]{henleyPowerlawSpinCorrelations2005}%
  \BibitemOpen
  \bibfield  {author} {\bibinfo {author} {\bibfnamefont {C.~L.}\ \bibnamefont {Henley}},\ }\bibfield  {title} {\bibinfo {title} {Power-law spin correlations in pyrochlore antiferromagnets},\ }\href {https://doi.org/10.1103/PhysRevB.71.014424} {\bibfield  {journal} {\bibinfo  {journal} {Phys. Rev. B}\ }\textbf {\bibinfo {volume} {71}},\ \bibinfo {pages} {014424} (\bibinfo {year} {2005})}\BibitemShut {NoStop}%
\bibitem [{\citenamefont {Henley}(2010)}]{henleyCoulombPhaseFrustrated2010}%
  \BibitemOpen
  \bibfield  {author} {\bibinfo {author} {\bibfnamefont {C.~L.}\ \bibnamefont {Henley}},\ }\bibfield  {title} {\bibinfo {title} {The ``{{Coulomb Phase}}'' in {{Frustrated Systems}}},\ }\href {https://doi.org/10.1146/annurev-conmatphys-070909-104138} {\bibfield  {journal} {\bibinfo  {journal} {Annual Review of Condensed Matter Physics}\ }\textbf {\bibinfo {volume} {1}},\ \bibinfo {pages} {179} (\bibinfo {year} {2010})}\BibitemShut {NoStop}%
\bibitem [{\citenamefont {Udagawa}\ and\ \citenamefont {Jaubert}(2021)}]{udagawaSpinIce2021}%
  \BibitemOpen
  \bibinfo {editor} {\bibfnamefont {M.}~\bibnamefont {Udagawa}}\ and\ \bibinfo {editor} {\bibfnamefont {L.}~\bibnamefont {Jaubert}},\ eds.,\ \href {https://doi.org/10.1007/978-3-030-70860-3} {\emph {\bibinfo {title} {Spin {{Ice}}}}},\ \bibinfo {series} {Springer {{Series}} in {{Solid-State Sciences}}}, Vol.\ \bibinfo {volume} {197}\ (\bibinfo  {publisher} {Springer International Publishing},\ \bibinfo {address} {Cham},\ \bibinfo {year} {2021})\BibitemShut {NoStop}%
\bibitem [{\citenamefont {Castelnovo}\ \emph {et~al.}(2012)\citenamefont {Castelnovo}, \citenamefont {Moessner},\ and\ \citenamefont {Sondhi}}]{castelnovoSpinIceFractionalization2012}%
  \BibitemOpen
  \bibfield  {author} {\bibinfo {author} {\bibfnamefont {C.}~\bibnamefont {Castelnovo}}, \bibinfo {author} {\bibfnamefont {R.}~\bibnamefont {Moessner}},\ and\ \bibinfo {author} {\bibfnamefont {S.}~\bibnamefont {Sondhi}},\ }\bibfield  {title} {\bibinfo {title} {Spin {{Ice}}, {{Fractionalization}}, and {{Topological Order}}},\ }\href {https://doi.org/10.1146/annurev-conmatphys-020911-125058} {\bibfield  {journal} {\bibinfo  {journal} {Annual Review of Condensed Matter Physics}\ }\textbf {\bibinfo {volume} {3}},\ \bibinfo {pages} {35} (\bibinfo {year} {2012})}\BibitemShut {NoStop}%
\bibitem [{\citenamefont {Hermele}\ \emph {et~al.}(2004)\citenamefont {Hermele}, \citenamefont {Fisher},\ and\ \citenamefont {Balents}}]{hermelePyrochlorePhotonsU12004}%
  \BibitemOpen
  \bibfield  {author} {\bibinfo {author} {\bibfnamefont {M.}~\bibnamefont {Hermele}}, \bibinfo {author} {\bibfnamefont {M.~P.~A.}\ \bibnamefont {Fisher}},\ and\ \bibinfo {author} {\bibfnamefont {L.}~\bibnamefont {Balents}},\ }\bibfield  {title} {\bibinfo {title} {Pyrochlore photons: {{The U}}(1) spin liquid in a {{S}}=1/2 three-dimensional frustrated magnet},\ }\href {https://doi.org/10.1103/PhysRevB.69.064404} {\bibfield  {journal} {\bibinfo  {journal} {Phys. Rev. B}\ }\textbf {\bibinfo {volume} {69}},\ \bibinfo {pages} {064404} (\bibinfo {year} {2004})}\BibitemShut {NoStop}%
\bibitem [{\citenamefont {Benton}\ \emph {et~al.}(2012)\citenamefont {Benton}, \citenamefont {Sikora},\ and\ \citenamefont {Shannon}}]{bentonSeeingLightExperimental2012}%
  \BibitemOpen
  \bibfield  {author} {\bibinfo {author} {\bibfnamefont {O.}~\bibnamefont {Benton}}, \bibinfo {author} {\bibfnamefont {O.}~\bibnamefont {Sikora}},\ and\ \bibinfo {author} {\bibfnamefont {N.}~\bibnamefont {Shannon}},\ }\bibfield  {title} {\bibinfo {title} {Seeing the light: {{Experimental}} signatures of emergent electromagnetism in a quantum spin ice},\ }\href {https://doi.org/10.1103/PhysRevB.86.075154} {\bibfield  {journal} {\bibinfo  {journal} {Phys. Rev. B}\ }\textbf {\bibinfo {volume} {86}},\ \bibinfo {pages} {075154} (\bibinfo {year} {2012})}\BibitemShut {NoStop}%
\bibitem [{\citenamefont {Savary}\ and\ \citenamefont {Balents}(2012)}]{savaryCoulombicQuantumLiquids2012}%
  \BibitemOpen
  \bibfield  {author} {\bibinfo {author} {\bibfnamefont {L.}~\bibnamefont {Savary}}\ and\ \bibinfo {author} {\bibfnamefont {L.}~\bibnamefont {Balents}},\ }\bibfield  {title} {\bibinfo {title} {Coulombic {{Quantum Liquids}} in {{Spin-1}}/2 {{Pyrochlores}}},\ }\href {https://doi.org/10.1103/PhysRevLett.108.037202} {\bibfield  {journal} {\bibinfo  {journal} {Phys. Rev. Lett.}\ }\textbf {\bibinfo {volume} {108}},\ \bibinfo {pages} {037202} (\bibinfo {year} {2012})}\BibitemShut {NoStop}%
\bibitem [{\citenamefont {Gingras}\ and\ \citenamefont {McClarty}(2014)}]{gingrasQuantumSpinIce2014}%
  \BibitemOpen
  \bibfield  {author} {\bibinfo {author} {\bibfnamefont {M.~J.~P.}\ \bibnamefont {Gingras}}\ and\ \bibinfo {author} {\bibfnamefont {P.~A.}\ \bibnamefont {McClarty}},\ }\bibfield  {title} {\bibinfo {title} {Quantum spin ice: A search for gapless quantum spin liquids in pyrochlore magnets},\ }\href {https://doi.org/10.1088/0034-4885/77/5/056501} {\bibfield  {journal} {\bibinfo  {journal} {Rep. Prog. Phys.}\ }\textbf {\bibinfo {volume} {77}},\ \bibinfo {pages} {056501} (\bibinfo {year} {2014})}\BibitemShut {NoStop}%
\bibitem [{\citenamefont {Wen}(2002)}]{wenQuantumOrdersSymmetric2002}%
  \BibitemOpen
  \bibfield  {author} {\bibinfo {author} {\bibfnamefont {X.-G.}\ \bibnamefont {Wen}},\ }\bibfield  {title} {\bibinfo {title} {Quantum orders and symmetric spin liquids},\ }\href {https://doi.org/10.1103/PhysRevB.65.165113} {\bibfield  {journal} {\bibinfo  {journal} {Phys. Rev. B}\ }\textbf {\bibinfo {volume} {65}},\ \bibinfo {pages} {165113} (\bibinfo {year} {2002})}\BibitemShut {NoStop}%
\bibitem [{\citenamefont {Smith}\ \emph {et~al.}(2022)\citenamefont {Smith}, \citenamefont {Benton}, \citenamefont {Yahne}, \citenamefont {Placke}, \citenamefont {Sch{\"a}fer}, \citenamefont {Gaudet}, \citenamefont {Dudemaine}, \citenamefont {Fitterman}, \citenamefont {Beare}, \citenamefont {Wildes}, \citenamefont {Bhattacharya}, \citenamefont {DeLazzer}, \citenamefont {Buhariwalla}, \citenamefont {Butch}, \citenamefont {Movshovich}, \citenamefont {Garrett}, \citenamefont {Marjerrison}, \citenamefont {Clancy}, \citenamefont {Kermarrec}, \citenamefont {Luke}, \citenamefont {Bianchi}, \citenamefont {Ross},\ and\ \citenamefont {Gaulin}}]{smithCaseU1piQuantum2022}%
  \BibitemOpen
  \bibfield  {author} {\bibinfo {author} {\bibfnamefont {E.~M.}\ \bibnamefont {Smith}}, \bibinfo {author} {\bibfnamefont {O.}~\bibnamefont {Benton}}, \bibinfo {author} {\bibfnamefont {D.~R.}\ \bibnamefont {Yahne}}, \bibinfo {author} {\bibfnamefont {B.}~\bibnamefont {Placke}}, \bibinfo {author} {\bibfnamefont {R.}~\bibnamefont {Sch{\"a}fer}}, \bibinfo {author} {\bibfnamefont {J.}~\bibnamefont {Gaudet}}, \bibinfo {author} {\bibfnamefont {J.}~\bibnamefont {Dudemaine}}, \bibinfo {author} {\bibfnamefont {A.}~\bibnamefont {Fitterman}}, \bibinfo {author} {\bibfnamefont {J.}~\bibnamefont {Beare}}, \bibinfo {author} {\bibfnamefont {A.~R.}\ \bibnamefont {Wildes}}, \bibinfo {author} {\bibfnamefont {S.}~\bibnamefont {Bhattacharya}}, \bibinfo {author} {\bibfnamefont {T.}~\bibnamefont {DeLazzer}}, \bibinfo {author} {\bibfnamefont {C.~R.~C.}\ \bibnamefont {Buhariwalla}}, \bibinfo {author} {\bibfnamefont {N.~P.}\ \bibnamefont {Butch}}, \bibinfo {author} {\bibfnamefont {R.}~\bibnamefont {Movshovich}}, \bibinfo {author}
  {\bibfnamefont {J.~D.}\ \bibnamefont {Garrett}}, \bibinfo {author} {\bibfnamefont {C.~A.}\ \bibnamefont {Marjerrison}}, \bibinfo {author} {\bibfnamefont {J.~P.}\ \bibnamefont {Clancy}}, \bibinfo {author} {\bibfnamefont {E.}~\bibnamefont {Kermarrec}}, \bibinfo {author} {\bibfnamefont {G.~M.}\ \bibnamefont {Luke}}, \bibinfo {author} {\bibfnamefont {A.~D.}\ \bibnamefont {Bianchi}}, \bibinfo {author} {\bibfnamefont {K.~A.}\ \bibnamefont {Ross}},\ and\ \bibinfo {author} {\bibfnamefont {B.~D.}\ \bibnamefont {Gaulin}},\ }\bibfield  {title} {\bibinfo {title} {Case for a {{U}}(1)-pi {{Quantum Spin Liquid Ground State}} in the {{Dipole-Octupole Pyrochlore Ce}}{$_{2}$}{{Zr}}{$_{2}$}{{O}}{$_{7}$}},\ }\href {https://doi.org/10.1103/PhysRevX.12.021015} {\bibfield  {journal} {\bibinfo  {journal} {Phys. Rev. X}\ }\textbf {\bibinfo {volume} {12}},\ \bibinfo {pages} {021015} (\bibinfo {year} {2022})}\BibitemShut {NoStop}%
\bibitem [{\citenamefont {Por{\'e}e}\ \emph {et~al.}(2025)\citenamefont {Por{\'e}e}, \citenamefont {Yan}, \citenamefont {Desrochers}, \citenamefont {Petit}, \citenamefont {Lhotel}, \citenamefont {Appel}, \citenamefont {Ollivier}, \citenamefont {Kim}, \citenamefont {Nevidomskyy},\ and\ \citenamefont {Sibille}}]{poreeEvidenceFractionalMatter2025}%
  \BibitemOpen
  \bibfield  {author} {\bibinfo {author} {\bibfnamefont {V.}~\bibnamefont {Por{\'e}e}}, \bibinfo {author} {\bibfnamefont {H.}~\bibnamefont {Yan}}, \bibinfo {author} {\bibfnamefont {F.}~\bibnamefont {Desrochers}}, \bibinfo {author} {\bibfnamefont {S.}~\bibnamefont {Petit}}, \bibinfo {author} {\bibfnamefont {E.}~\bibnamefont {Lhotel}}, \bibinfo {author} {\bibfnamefont {M.}~\bibnamefont {Appel}}, \bibinfo {author} {\bibfnamefont {J.}~\bibnamefont {Ollivier}}, \bibinfo {author} {\bibfnamefont {Y.~B.}\ \bibnamefont {Kim}}, \bibinfo {author} {\bibfnamefont {A.~H.}\ \bibnamefont {Nevidomskyy}},\ and\ \bibinfo {author} {\bibfnamefont {R.}~\bibnamefont {Sibille}},\ }\bibfield  {title} {\bibinfo {title} {Evidence for fractional matter coupled to an emergent gauge field in a quantum spin ice},\ }\href {https://doi.org/10.1038/s41567-024-02711-w} {\bibfield  {journal} {\bibinfo  {journal} {Nat. Phys.}\ }\textbf {\bibinfo {volume} {21}},\ \bibinfo {pages} {83} (\bibinfo {year} {2025})}\BibitemShut {NoStop}%
\bibitem [{\citenamefont {Gao}\ \emph {et~al.}(2025)\citenamefont {Gao}, \citenamefont {Desrochers}, \citenamefont {Tam}, \citenamefont {Kirschbaum}, \citenamefont {Steffens}, \citenamefont {Hiess}, \citenamefont {Nguyen}, \citenamefont {Su}, \citenamefont {Cheong}, \citenamefont {Paschen}, \citenamefont {Kim},\ and\ \citenamefont {Dai}}]{gaoNeutronScatteringThermodynamic2025}%
  \BibitemOpen
  \bibfield  {author} {\bibinfo {author} {\bibfnamefont {B.}~\bibnamefont {Gao}}, \bibinfo {author} {\bibfnamefont {F.}~\bibnamefont {Desrochers}}, \bibinfo {author} {\bibfnamefont {D.~W.}\ \bibnamefont {Tam}}, \bibinfo {author} {\bibfnamefont {D.~M.}\ \bibnamefont {Kirschbaum}}, \bibinfo {author} {\bibfnamefont {P.}~\bibnamefont {Steffens}}, \bibinfo {author} {\bibfnamefont {A.}~\bibnamefont {Hiess}}, \bibinfo {author} {\bibfnamefont {D.~H.}\ \bibnamefont {Nguyen}}, \bibinfo {author} {\bibfnamefont {Y.}~\bibnamefont {Su}}, \bibinfo {author} {\bibfnamefont {S.-W.}\ \bibnamefont {Cheong}}, \bibinfo {author} {\bibfnamefont {S.}~\bibnamefont {Paschen}}, \bibinfo {author} {\bibfnamefont {Y.~B.}\ \bibnamefont {Kim}},\ and\ \bibinfo {author} {\bibfnamefont {P.}~\bibnamefont {Dai}},\ }\bibfield  {title} {\bibinfo {title} {Neutron scattering and thermodynamic evidence for emergent photons and fractionalization in a pyrochlore spin ice},\ }\bibfield  {journal} {\bibinfo  {journal} {Nat. Phys.}\ }\href
  {https://doi.org/10.1038/s41567-025-02922-9} {10.1038/s41567-025-02922-9} (\bibinfo {year} {2025})\BibitemShut {NoStop}%
\bibitem [{\citenamefont {Smith}\ \emph {et~al.}(2025)\citenamefont {Smith}, \citenamefont {Lhotel}, \citenamefont {Petit},\ and\ \citenamefont {Gaulin}}]{smithExperimentalInsightsQuantum2025}%
  \BibitemOpen
  \bibfield  {author} {\bibinfo {author} {\bibfnamefont {E.~M.}\ \bibnamefont {Smith}}, \bibinfo {author} {\bibfnamefont {E.}~\bibnamefont {Lhotel}}, \bibinfo {author} {\bibfnamefont {S.}~\bibnamefont {Petit}},\ and\ \bibinfo {author} {\bibfnamefont {B.~D.}\ \bibnamefont {Gaulin}},\ }\bibfield  {title} {\bibinfo {title} {Experimental {{Insights}} into {{Quantum Spin Ice Physics}} in {{Dipole}}--{{Octupole Pyrochlore Magnets}}},\ }\href {https://doi.org/10.1146/annurev-conmatphys-041124-015101} {\bibfield  {journal} {\bibinfo  {journal} {Annual Review of Condensed Matter Physics}\ }\textbf {\bibinfo {volume} {16}},\ \bibinfo {pages} {387} (\bibinfo {year} {2025})}\BibitemShut {NoStop}%
\bibitem [{\citenamefont {Bertin}\ \emph {et~al.}(2012)\citenamefont {Bertin}, \citenamefont {Chapuis}, \citenamefont {de~R{\'e}otier},\ and\ \citenamefont {Yaouanc}}]{bertinCrystalElectricField2012}%
  \BibitemOpen
  \bibfield  {author} {\bibinfo {author} {\bibfnamefont {A.}~\bibnamefont {Bertin}}, \bibinfo {author} {\bibfnamefont {Y.}~\bibnamefont {Chapuis}}, \bibinfo {author} {\bibfnamefont {P.~D.}\ \bibnamefont {de~R{\'e}otier}},\ and\ \bibinfo {author} {\bibfnamefont {A.}~\bibnamefont {Yaouanc}},\ }\bibfield  {title} {\bibinfo {title} {Crystal electric field in the {{R}}{$_{2}$}{{Ti}}{$_{2}$}{{O}}{$_7$} pyrochlore compounds},\ }\href {https://doi.org/10.1088/0953-8984/24/25/256003} {\bibfield  {journal} {\bibinfo  {journal} {J. Phys.: Condens. Matter}\ }\textbf {\bibinfo {volume} {24}},\ \bibinfo {pages} {256003} (\bibinfo {year} {2012})}\BibitemShut {NoStop}%
\bibitem [{\citenamefont {Rau}\ \emph {et~al.}(2016)\citenamefont {Rau}, \citenamefont {Lee},\ and\ \citenamefont {Kee}}]{rauSpinOrbitPhysicsGiving2016}%
  \BibitemOpen
  \bibfield  {author} {\bibinfo {author} {\bibfnamefont {J.~G.}\ \bibnamefont {Rau}}, \bibinfo {author} {\bibfnamefont {E.~K.-H.}\ \bibnamefont {Lee}},\ and\ \bibinfo {author} {\bibfnamefont {H.-Y.}\ \bibnamefont {Kee}},\ }\bibfield  {title} {\bibinfo {title} {Spin-{{Orbit Physics Giving Rise}} to {{Novel Phases}} in {{Correlated Systems}}: {{Iridates}} and {{Related Materials}}},\ }\href {https://doi.org/10.1146/annurev-conmatphys-031115-011319} {\bibfield  {journal} {\bibinfo  {journal} {Annual Review of Condensed Matter Physics}\ }\textbf {\bibinfo {volume} {7}},\ \bibinfo {pages} {195} (\bibinfo {year} {2016})}\BibitemShut {NoStop}%
\bibitem [{\citenamefont {Essafi}\ \emph {et~al.}(2017{\natexlab{a}})\citenamefont {Essafi}, \citenamefont {Benton},\ and\ \citenamefont {Jaubert}}]{essafiGenericNearestneighborKagome2017}%
  \BibitemOpen
  \bibfield  {author} {\bibinfo {author} {\bibfnamefont {K.}~\bibnamefont {Essafi}}, \bibinfo {author} {\bibfnamefont {O.}~\bibnamefont {Benton}},\ and\ \bibinfo {author} {\bibfnamefont {L.~D.~C.}\ \bibnamefont {Jaubert}},\ }\bibfield  {title} {\bibinfo {title} {Generic nearest-neighbor kagome model: {{XYZ}} and {{Dzyaloshinskii-Moriya}} couplings with comparison to the pyrochlore-lattice case},\ }\href {https://doi.org/10.1103/PhysRevB.96.205126} {\bibfield  {journal} {\bibinfo  {journal} {Phys. Rev. B}\ }\textbf {\bibinfo {volume} {96}},\ \bibinfo {pages} {205126} (\bibinfo {year} {2017}{\natexlab{a}})}\BibitemShut {NoStop}%
\bibitem [{\citenamefont {Maksimov}\ \emph {et~al.}(2019)\citenamefont {Maksimov}, \citenamefont {Zhu}, \citenamefont {White},\ and\ \citenamefont {Chernyshev}}]{maksimovAnisotropicExchangeMagnetsTriangular2019}%
  \BibitemOpen
  \bibfield  {author} {\bibinfo {author} {\bibfnamefont {P.~A.}\ \bibnamefont {Maksimov}}, \bibinfo {author} {\bibfnamefont {Z.}~\bibnamefont {Zhu}}, \bibinfo {author} {\bibfnamefont {S.~R.}\ \bibnamefont {White}},\ and\ \bibinfo {author} {\bibfnamefont {A.~L.}\ \bibnamefont {Chernyshev}},\ }\bibfield  {title} {\bibinfo {title} {Anisotropic-{{Exchange Magnets}} on a {{Triangular Lattice}}: {{Spin Waves}}, {{Accidental Degeneracies}}, and {{Dual Spin Liquids}}},\ }\href {https://doi.org/10.1103/PhysRevX.9.021017} {\bibfield  {journal} {\bibinfo  {journal} {Physical Review X}\ }\textbf {\bibinfo {volume} {9}},\ \bibinfo {pages} {021017} (\bibinfo {year} {2019})}\BibitemShut {NoStop}%
\bibitem [{\citenamefont {Kitaev}(2006)}]{kitaevAnyonsExactlySolved2006}%
  \BibitemOpen
  \bibfield  {author} {\bibinfo {author} {\bibfnamefont {A.}~\bibnamefont {Kitaev}},\ }\bibfield  {title} {\bibinfo {title} {Anyons in an exactly solved model and beyond},\ }\href {https://doi.org/10.1016/j.aop.2005.10.005} {\bibfield  {journal} {\bibinfo  {journal} {Annals of Physics}\ }\bibinfo {series} {January {{Special Issue}}},\ \textbf {\bibinfo {volume} {321}},\ \bibinfo {pages} {2} (\bibinfo {year} {2006})}\BibitemShut {NoStop}%
\bibitem [{\citenamefont {Hermanns}\ \emph {et~al.}(2018)\citenamefont {Hermanns}, \citenamefont {Kimchi},\ and\ \citenamefont {Knolle}}]{hermannsPhysicsKitaevModel2018}%
  \BibitemOpen
  \bibfield  {author} {\bibinfo {author} {\bibfnamefont {M.}~\bibnamefont {Hermanns}}, \bibinfo {author} {\bibfnamefont {I.}~\bibnamefont {Kimchi}},\ and\ \bibinfo {author} {\bibfnamefont {J.}~\bibnamefont {Knolle}},\ }\bibfield  {title} {\bibinfo {title} {Physics of the {{Kitaev Model}}: {{Fractionalization}}, {{Dynamic Correlations}}, and {{Material Connections}}},\ }\href {https://doi.org/10.1146/annurev-conmatphys-033117-053934} {\bibfield  {journal} {\bibinfo  {journal} {Annual Review of Condensed Matter Physics}\ }\textbf {\bibinfo {volume} {9}},\ \bibinfo {pages} {17} (\bibinfo {year} {2018})}\BibitemShut {NoStop}%
\bibitem [{\citenamefont {Takagi}\ \emph {et~al.}(2019)\citenamefont {Takagi}, \citenamefont {Takayama}, \citenamefont {Jackeli}, \citenamefont {Khaliullin},\ and\ \citenamefont {Nagler}}]{takagiConceptRealizationKitaev2019}%
  \BibitemOpen
  \bibfield  {author} {\bibinfo {author} {\bibfnamefont {H.}~\bibnamefont {Takagi}}, \bibinfo {author} {\bibfnamefont {T.}~\bibnamefont {Takayama}}, \bibinfo {author} {\bibfnamefont {G.}~\bibnamefont {Jackeli}}, \bibinfo {author} {\bibfnamefont {G.}~\bibnamefont {Khaliullin}},\ and\ \bibinfo {author} {\bibfnamefont {S.~E.}\ \bibnamefont {Nagler}},\ }\bibfield  {title} {\bibinfo {title} {Concept and realization of {{Kitaev}} quantum spin liquids},\ }\href {https://doi.org/10.1038/s42254-019-0038-2} {\bibfield  {journal} {\bibinfo  {journal} {Nat Rev Phys}\ }\textbf {\bibinfo {volume} {1}},\ \bibinfo {pages} {264} (\bibinfo {year} {2019})}\BibitemShut {NoStop}%
\bibitem [{\citenamefont {Taillefumier}\ \emph {et~al.}(2017)\citenamefont {Taillefumier}, \citenamefont {Benton}, \citenamefont {Yan}, \citenamefont {Jaubert},\ and\ \citenamefont {Shannon}}]{taillefumierCompetingSpinLiquids2017}%
  \BibitemOpen
  \bibfield  {author} {\bibinfo {author} {\bibfnamefont {M.}~\bibnamefont {Taillefumier}}, \bibinfo {author} {\bibfnamefont {O.}~\bibnamefont {Benton}}, \bibinfo {author} {\bibfnamefont {H.}~\bibnamefont {Yan}}, \bibinfo {author} {\bibfnamefont {L.~D.~C.}\ \bibnamefont {Jaubert}},\ and\ \bibinfo {author} {\bibfnamefont {N.}~\bibnamefont {Shannon}},\ }\bibfield  {title} {\bibinfo {title} {Competing {{Spin Liquids}} and {{Hidden Spin-Nematic Order}} in {{Spin Ice}} with {{Frustrated Transverse Exchange}}},\ }\href {https://doi.org/10.1103/PhysRevX.7.041057} {\bibfield  {journal} {\bibinfo  {journal} {Phys. Rev. X}\ }\textbf {\bibinfo {volume} {7}},\ \bibinfo {pages} {041057} (\bibinfo {year} {2017})}\BibitemShut {NoStop}%
\bibitem [{\citenamefont {Francini}\ \emph {et~al.}(2025{\natexlab{a}})\citenamefont {Francini}, \citenamefont {Janssen},\ and\ \citenamefont {{Lozano-G{\'o}mez}}}]{franciniHigherRankSpinLiquids2025}%
  \BibitemOpen
  \bibfield  {author} {\bibinfo {author} {\bibfnamefont {N.}~\bibnamefont {Francini}}, \bibinfo {author} {\bibfnamefont {L.}~\bibnamefont {Janssen}},\ and\ \bibinfo {author} {\bibfnamefont {D.}~\bibnamefont {{Lozano-G{\'o}mez}}},\ }\bibfield  {title} {\bibinfo {title} {Higher-rank spin liquids and spin nematics from competing orders in pyrochlore magnets},\ }\href {https://doi.org/10.1103/PhysRevB.111.085140} {\bibfield  {journal} {\bibinfo  {journal} {Phys. Rev. B}\ }\textbf {\bibinfo {volume} {111}},\ \bibinfo {pages} {085140} (\bibinfo {year} {2025}{\natexlab{a}})}\BibitemShut {NoStop}%
\bibitem [{\citenamefont {Benton}\ \emph {et~al.}(2016)\citenamefont {Benton}, \citenamefont {Jaubert}, \citenamefont {Yan},\ and\ \citenamefont {Shannon}}]{bentonSpinliquidPinchlineSingularities2016}%
  \BibitemOpen
  \bibfield  {author} {\bibinfo {author} {\bibfnamefont {O.}~\bibnamefont {Benton}}, \bibinfo {author} {\bibfnamefont {L.~D.~C.}\ \bibnamefont {Jaubert}}, \bibinfo {author} {\bibfnamefont {H.}~\bibnamefont {Yan}},\ and\ \bibinfo {author} {\bibfnamefont {N.}~\bibnamefont {Shannon}},\ }\bibfield  {title} {\bibinfo {title} {A spin-liquid with pinch-line singularities on the pyrochlore lattice},\ }\href {https://doi.org/10.1038/ncomms11572} {\bibfield  {journal} {\bibinfo  {journal} {Nature Communications}\ }\textbf {\bibinfo {volume} {7}},\ \bibinfo {pages} {11572} (\bibinfo {year} {2016})}\BibitemShut {NoStop}%
\bibitem [{\citenamefont {Yan}\ \emph {et~al.}(2020)\citenamefont {Yan}, \citenamefont {Benton}, \citenamefont {Jaubert},\ and\ \citenamefont {Shannon}}]{yanRank2U1Spin2020}%
  \BibitemOpen
  \bibfield  {author} {\bibinfo {author} {\bibfnamefont {H.}~\bibnamefont {Yan}}, \bibinfo {author} {\bibfnamefont {O.}~\bibnamefont {Benton}}, \bibinfo {author} {\bibfnamefont {L.~D.~C.}\ \bibnamefont {Jaubert}},\ and\ \bibinfo {author} {\bibfnamefont {N.}~\bibnamefont {Shannon}},\ }\bibfield  {title} {\bibinfo {title} {Rank-2 {{U}}(1) {{Spin Liquid}} on the {{Breathing Pyrochlore Lattice}}},\ }\href {https://doi.org/10.1103/PhysRevLett.124.127203} {\bibfield  {journal} {\bibinfo  {journal} {Phys. Rev. Lett.}\ }\textbf {\bibinfo {volume} {124}},\ \bibinfo {pages} {127203} (\bibinfo {year} {2020})}\BibitemShut {NoStop}%
\bibitem [{\citenamefont {{Lozano-G{\'o}mez}}\ \emph {et~al.}(2024{\natexlab{a}})\citenamefont {{Lozano-G{\'o}mez}}, \citenamefont {Noculak}, \citenamefont {Oitmaa}, \citenamefont {Singh}, \citenamefont {Iqbal}, \citenamefont {Reuther},\ and\ \citenamefont {Gingras}}]{lozano-gomezCompetingGaugeFields2024}%
  \BibitemOpen
  \bibfield  {author} {\bibinfo {author} {\bibfnamefont {D.}~\bibnamefont {{Lozano-G{\'o}mez}}}, \bibinfo {author} {\bibfnamefont {V.}~\bibnamefont {Noculak}}, \bibinfo {author} {\bibfnamefont {J.}~\bibnamefont {Oitmaa}}, \bibinfo {author} {\bibfnamefont {R.~R.~P.}\ \bibnamefont {Singh}}, \bibinfo {author} {\bibfnamefont {Y.}~\bibnamefont {Iqbal}}, \bibinfo {author} {\bibfnamefont {J.}~\bibnamefont {Reuther}},\ and\ \bibinfo {author} {\bibfnamefont {M.~J.~P.}\ \bibnamefont {Gingras}},\ }\bibfield  {title} {\bibinfo {title} {Competing gauge fields and entropically driven spin liquid to spin liquid transition in non-{{Kramers}} pyrochlores},\ }\href {https://doi.org/10.1073/pnas.2403487121} {\bibfield  {journal} {\bibinfo  {journal} {Proceedings of the National Academy of Sciences}\ }\textbf {\bibinfo {volume} {121}},\ \bibinfo {pages} {e2403487121} (\bibinfo {year} {2024}{\natexlab{a}})}\BibitemShut {NoStop}%
\bibitem [{\citenamefont {Pretko}(2017{\natexlab{a}})}]{pretkoSubdimensionalParticleStructure2017}%
  \BibitemOpen
  \bibfield  {author} {\bibinfo {author} {\bibfnamefont {M.}~\bibnamefont {Pretko}},\ }\bibfield  {title} {\bibinfo {title} {Subdimensional particle structure of higher rank {{U}}(1) spin liquids},\ }\href {https://doi.org/10.1103/PhysRevB.95.115139} {\bibfield  {journal} {\bibinfo  {journal} {Phys. Rev. B}\ }\textbf {\bibinfo {volume} {95}},\ \bibinfo {pages} {115139} (\bibinfo {year} {2017}{\natexlab{a}})}\BibitemShut {NoStop}%
\bibitem [{\citenamefont {Prem}\ \emph {et~al.}(2018)\citenamefont {Prem}, \citenamefont {Vijay}, \citenamefont {Chou}, \citenamefont {Pretko},\ and\ \citenamefont {Nandkishore}}]{premPinchPointSingularities2018}%
  \BibitemOpen
  \bibfield  {author} {\bibinfo {author} {\bibfnamefont {A.}~\bibnamefont {Prem}}, \bibinfo {author} {\bibfnamefont {S.}~\bibnamefont {Vijay}}, \bibinfo {author} {\bibfnamefont {Y.-Z.}\ \bibnamefont {Chou}}, \bibinfo {author} {\bibfnamefont {M.}~\bibnamefont {Pretko}},\ and\ \bibinfo {author} {\bibfnamefont {R.~M.}\ \bibnamefont {Nandkishore}},\ }\bibfield  {title} {\bibinfo {title} {Pinch point singularities of tensor spin liquids},\ }\href {https://doi.org/10.1103/PhysRevB.98.165140} {\bibfield  {journal} {\bibinfo  {journal} {Phys. Rev. B}\ }\textbf {\bibinfo {volume} {98}},\ \bibinfo {pages} {165140} (\bibinfo {year} {2018})}\BibitemShut {NoStop}%
\bibitem [{\citenamefont {Benton}\ and\ \citenamefont {Moessner}(2021)}]{bentonTopologicalRouteNew2021}%
  \BibitemOpen
  \bibfield  {author} {\bibinfo {author} {\bibfnamefont {O.}~\bibnamefont {Benton}}\ and\ \bibinfo {author} {\bibfnamefont {R.}~\bibnamefont {Moessner}},\ }\bibfield  {title} {\bibinfo {title} {Topological {{Route}} to {{New}} and {{Unusual Coulomb Spin Liquids}}},\ }\href {https://doi.org/10.1103/PhysRevLett.127.107202} {\bibfield  {journal} {\bibinfo  {journal} {Phys. Rev. Lett.}\ }\textbf {\bibinfo {volume} {127}},\ \bibinfo {pages} {107202} (\bibinfo {year} {2021})}\BibitemShut {NoStop}%
\bibitem [{\citenamefont {Yan}\ \emph {et~al.}(2024{\natexlab{a}})\citenamefont {Yan}, \citenamefont {Benton}, \citenamefont {Moessner},\ and\ \citenamefont {Nevidomskyy}}]{yanClassificationClassicalSpin2024}%
  \BibitemOpen
  \bibfield  {author} {\bibinfo {author} {\bibfnamefont {H.}~\bibnamefont {Yan}}, \bibinfo {author} {\bibfnamefont {O.}~\bibnamefont {Benton}}, \bibinfo {author} {\bibfnamefont {R.}~\bibnamefont {Moessner}},\ and\ \bibinfo {author} {\bibfnamefont {A.~H.}\ \bibnamefont {Nevidomskyy}},\ }\bibfield  {title} {\bibinfo {title} {Classification of classical spin liquids: {{Typology}} and resulting landscape},\ }\href {https://doi.org/10.1103/PhysRevB.110.L020402} {\bibfield  {journal} {\bibinfo  {journal} {Phys. Rev. B}\ }\textbf {\bibinfo {volume} {110}},\ \bibinfo {pages} {L020402} (\bibinfo {year} {2024}{\natexlab{a}})}\BibitemShut {NoStop}%
\bibitem [{\citenamefont {Yan}\ \emph {et~al.}(2024{\natexlab{b}})\citenamefont {Yan}, \citenamefont {Benton}, \citenamefont {Nevidomskyy},\ and\ \citenamefont {Moessner}}]{yanClassificationClassicalSpin2024a}%
  \BibitemOpen
  \bibfield  {author} {\bibinfo {author} {\bibfnamefont {H.}~\bibnamefont {Yan}}, \bibinfo {author} {\bibfnamefont {O.}~\bibnamefont {Benton}}, \bibinfo {author} {\bibfnamefont {A.~H.}\ \bibnamefont {Nevidomskyy}},\ and\ \bibinfo {author} {\bibfnamefont {R.}~\bibnamefont {Moessner}},\ }\bibfield  {title} {\bibinfo {title} {Classification of classical spin liquids: {{Detailed}} formalism and suite of examples},\ }\href {https://doi.org/10.1103/PhysRevB.109.174421} {\bibfield  {journal} {\bibinfo  {journal} {Phys. Rev. B}\ }\textbf {\bibinfo {volume} {109}},\ \bibinfo {pages} {174421} (\bibinfo {year} {2024}{\natexlab{b}})}\BibitemShut {NoStop}%
\bibitem [{\citenamefont {Fang}\ \emph {et~al.}(2024)\citenamefont {Fang}, \citenamefont {Cano}, \citenamefont {Nevidomskyy},\ and\ \citenamefont {Yan}}]{fangClassificationClassicalSpin2024}%
  \BibitemOpen
  \bibfield  {author} {\bibinfo {author} {\bibfnamefont {Y.}~\bibnamefont {Fang}}, \bibinfo {author} {\bibfnamefont {J.}~\bibnamefont {Cano}}, \bibinfo {author} {\bibfnamefont {A.~H.}\ \bibnamefont {Nevidomskyy}},\ and\ \bibinfo {author} {\bibfnamefont {H.}~\bibnamefont {Yan}},\ }\bibfield  {title} {\bibinfo {title} {Classification of classical spin liquids: {{Topological}} quantum chemistry and crystalline symmetry},\ }\href {https://doi.org/10.1103/PhysRevB.110.054421} {\bibfield  {journal} {\bibinfo  {journal} {Phys. Rev. B}\ }\textbf {\bibinfo {volume} {110}},\ \bibinfo {pages} {054421} (\bibinfo {year} {2024})}\BibitemShut {NoStop}%
\bibitem [{\citenamefont {Nutakki}\ \emph {et~al.}(2023)\citenamefont {Nutakki}, \citenamefont {Jaubert},\ and\ \citenamefont {Pollet}}]{nutakkiClassicalHeisenbergModel2023}%
  \BibitemOpen
  \bibfield  {author} {\bibinfo {author} {\bibfnamefont {R.~P.}\ \bibnamefont {Nutakki}}, \bibinfo {author} {\bibfnamefont {L.~D.~C.}\ \bibnamefont {Jaubert}},\ and\ \bibinfo {author} {\bibfnamefont {L.}~\bibnamefont {Pollet}},\ }\bibfield  {title} {\bibinfo {title} {The classical {{Heisenberg}} model on the centred pyrochlore lattice},\ }\href {https://doi.org/10.21468/SciPostPhys.15.2.040} {\bibfield  {journal} {\bibinfo  {journal} {SciPost Physics}\ }\textbf {\bibinfo {volume} {15}},\ \bibinfo {pages} {040} (\bibinfo {year} {2023})}\BibitemShut {NoStop}%
\bibitem [{\citenamefont {Wong}\ \emph {et~al.}(2013)\citenamefont {Wong}, \citenamefont {Hao},\ and\ \citenamefont {Gingras}}]{wongGroundStatePhase2013}%
  \BibitemOpen
  \bibfield  {author} {\bibinfo {author} {\bibfnamefont {A.~W.~C.}\ \bibnamefont {Wong}}, \bibinfo {author} {\bibfnamefont {Z.}~\bibnamefont {Hao}},\ and\ \bibinfo {author} {\bibfnamefont {M.~J.~P.}\ \bibnamefont {Gingras}},\ }\bibfield  {title} {\bibinfo {title} {Ground state phase diagram of generic {{XY}} pyrochlore magnets with quantum fluctuations},\ }\href {https://doi.org/10.1103/PhysRevB.88.144402} {\bibfield  {journal} {\bibinfo  {journal} {Phys. Rev. B}\ }\textbf {\bibinfo {volume} {88}},\ \bibinfo {pages} {144402} (\bibinfo {year} {2013})}\BibitemShut {NoStop}%
\bibitem [{\citenamefont {Javanparast}\ \emph {et~al.}(2015{\natexlab{a}})\citenamefont {Javanparast}, \citenamefont {Day}, \citenamefont {Hao},\ and\ \citenamefont {Gingras}}]{javanparastOrderbydisorderCriticalityXY2015}%
  \BibitemOpen
  \bibfield  {author} {\bibinfo {author} {\bibfnamefont {B.}~\bibnamefont {Javanparast}}, \bibinfo {author} {\bibfnamefont {A.~G.~R.}\ \bibnamefont {Day}}, \bibinfo {author} {\bibfnamefont {Z.}~\bibnamefont {Hao}},\ and\ \bibinfo {author} {\bibfnamefont {M.~J.~P.}\ \bibnamefont {Gingras}},\ }\bibfield  {title} {\bibinfo {title} {Order-by-disorder near criticality in {{XY}} pyrochlore magnets},\ }\href {https://doi.org/10.1103/PhysRevB.91.174424} {\bibfield  {journal} {\bibinfo  {journal} {Phys. Rev. B}\ }\textbf {\bibinfo {volume} {91}},\ \bibinfo {pages} {174424} (\bibinfo {year} {2015}{\natexlab{a}})}\BibitemShut {NoStop}%
\bibitem [{\citenamefont {Yan}\ \emph {et~al.}(2017)\citenamefont {Yan}, \citenamefont {Benton}, \citenamefont {Jaubert},\ and\ \citenamefont {Shannon}}]{yanTheoryMultiplephaseCompetition2017}%
  \BibitemOpen
  \bibfield  {author} {\bibinfo {author} {\bibfnamefont {H.}~\bibnamefont {Yan}}, \bibinfo {author} {\bibfnamefont {O.}~\bibnamefont {Benton}}, \bibinfo {author} {\bibfnamefont {L.}~\bibnamefont {Jaubert}},\ and\ \bibinfo {author} {\bibfnamefont {N.}~\bibnamefont {Shannon}},\ }\bibfield  {title} {\bibinfo {title} {Theory of multiple-phase competition in pyrochlore magnets with anisotropic exchange with application to {{Yb}}{$_{2}$}{{Ti}}{$_{2}$}{{O}}{$_7$}, {{Er}}{$_{2}$}{{Ti}}{$_{2}$}{{O}}{$_7$}, {{Er}}{$_{2}$}{{Sn}}{$_{2}$}{{O}}{$_{7}$}},\ }\href {https://doi.org/10.1103/PhysRevB.95.094422} {\bibfield  {journal} {\bibinfo  {journal} {Phys. Rev. B}\ }\textbf {\bibinfo {volume} {95}},\ \bibinfo {pages} {094422} (\bibinfo {year} {2017})}\BibitemShut {NoStop}%
\bibitem [{\citenamefont {Wei}\ and\ \citenamefont {Curnoe}(2023)}]{weiExactDiagonalization16site2023}%
  \BibitemOpen
  \bibfield  {author} {\bibinfo {author} {\bibfnamefont {C.}~\bibnamefont {Wei}}\ and\ \bibinfo {author} {\bibfnamefont {S.~H.}\ \bibnamefont {Curnoe}},\ }\bibfield  {title} {\bibinfo {title} {Exact diagonalization for a 16-site spin-1/2 pyrochlore cluster},\ }\href {https://doi.org/10.1088/1361-648X/acccc8} {\bibfield  {journal} {\bibinfo  {journal} {J. Phys.: Condens. Matter}\ }\textbf {\bibinfo {volume} {35}},\ \bibinfo {pages} {295802} (\bibinfo {year} {2023})}\BibitemShut {NoStop}%
\bibitem [{\citenamefont {Noculak}\ \emph {et~al.}(2023)\citenamefont {Noculak}, \citenamefont {{Lozano-G{\'o}mez}}, \citenamefont {Oitmaa}, \citenamefont {Singh}, \citenamefont {Iqbal}, \citenamefont {Gingras},\ and\ \citenamefont {Reuther}}]{noculakClassicalQuantumPhases2023}%
  \BibitemOpen
  \bibfield  {author} {\bibinfo {author} {\bibfnamefont {V.}~\bibnamefont {Noculak}}, \bibinfo {author} {\bibfnamefont {D.}~\bibnamefont {{Lozano-G{\'o}mez}}}, \bibinfo {author} {\bibfnamefont {J.}~\bibnamefont {Oitmaa}}, \bibinfo {author} {\bibfnamefont {R.~R.~P.}\ \bibnamefont {Singh}}, \bibinfo {author} {\bibfnamefont {Y.}~\bibnamefont {Iqbal}}, \bibinfo {author} {\bibfnamefont {M.~J.~P.}\ \bibnamefont {Gingras}},\ and\ \bibinfo {author} {\bibfnamefont {J.}~\bibnamefont {Reuther}},\ }\bibfield  {title} {\bibinfo {title} {Classical and quantum phases of the pyrochlore {{S}}=1/2 magnet with {{Heisenberg}} and {{Dzyaloshinskii-Moriya}} interactions},\ }\href {https://doi.org/10.1103/PhysRevB.107.214414} {\bibfield  {journal} {\bibinfo  {journal} {Phys. Rev. B}\ }\textbf {\bibinfo {volume} {107}},\ \bibinfo {pages} {214414} (\bibinfo {year} {2023})}\BibitemShut {NoStop}%
\bibitem [{\citenamefont {Hsin}\ \emph {et~al.}(2020)\citenamefont {Hsin}, \citenamefont {Kapustin},\ and\ \citenamefont {Thorngren}}]{hsinBerryPhaseQuantum2020}%
  \BibitemOpen
  \bibfield  {author} {\bibinfo {author} {\bibfnamefont {P.-S.}\ \bibnamefont {Hsin}}, \bibinfo {author} {\bibfnamefont {A.}~\bibnamefont {Kapustin}},\ and\ \bibinfo {author} {\bibfnamefont {R.}~\bibnamefont {Thorngren}},\ }\bibfield  {title} {\bibinfo {title} {Berry phase in quantum field theory: {{Diabolical}} points and boundary phenomena},\ }\href {https://doi.org/10.1103/PhysRevB.102.245113} {\bibfield  {journal} {\bibinfo  {journal} {Phys. Rev. B}\ }\textbf {\bibinfo {volume} {102}},\ \bibinfo {pages} {245113} (\bibinfo {year} {2020})}\BibitemShut {NoStop}%
\bibitem [{\citenamefont {Bi}\ and\ \citenamefont {Senthil}(2019)}]{biAdventureTopologicalPhase2019}%
  \BibitemOpen
  \bibfield  {author} {\bibinfo {author} {\bibfnamefont {Z.}~\bibnamefont {Bi}}\ and\ \bibinfo {author} {\bibfnamefont {T.}~\bibnamefont {Senthil}},\ }\bibfield  {title} {\bibinfo {title} {Adventure in {{Topological Phase Transitions}} in \$3+1\$-{{D}}: {{Non-Abelian Deconfined Quantum Criticalities}} and a {{Possible Duality}}},\ }\href {https://doi.org/10.1103/PhysRevX.9.021034} {\bibfield  {journal} {\bibinfo  {journal} {Phys. Rev. X}\ }\textbf {\bibinfo {volume} {9}},\ \bibinfo {pages} {021034} (\bibinfo {year} {2019})}\BibitemShut {NoStop}%
\bibitem [{\citenamefont {Jian}\ and\ \citenamefont {Xu}(2020)}]{jianGenericUnnecessaryQuantum2020}%
  \BibitemOpen
  \bibfield  {author} {\bibinfo {author} {\bibfnamefont {C.-M.}\ \bibnamefont {Jian}}\ and\ \bibinfo {author} {\bibfnamefont {C.}~\bibnamefont {Xu}},\ }\bibfield  {title} {\bibinfo {title} {Generic ``unnecessary'' quantum critical points with minimal degrees of freedom},\ }\href {https://doi.org/10.1103/PhysRevB.101.035118} {\bibfield  {journal} {\bibinfo  {journal} {Phys. Rev. B}\ }\textbf {\bibinfo {volume} {101}},\ \bibinfo {pages} {035118} (\bibinfo {year} {2020})}\BibitemShut {NoStop}%
\bibitem [{\citenamefont {Prakash}\ \emph {et~al.}(2023)\citenamefont {Prakash}, \citenamefont {Fava},\ and\ \citenamefont {Parameswaran}}]{prakashMultiversalityUnnecessaryCriticality2023}%
  \BibitemOpen
  \bibfield  {author} {\bibinfo {author} {\bibfnamefont {A.}~\bibnamefont {Prakash}}, \bibinfo {author} {\bibfnamefont {M.}~\bibnamefont {Fava}},\ and\ \bibinfo {author} {\bibfnamefont {S.~A.}\ \bibnamefont {Parameswaran}},\ }\bibfield  {title} {\bibinfo {title} {Multiversality and {{Unnecessary Criticality}} in {{One Dimension}}},\ }\href {https://doi.org/10.1103/PhysRevLett.130.256401} {\bibfield  {journal} {\bibinfo  {journal} {Phys. Rev. Lett.}\ }\textbf {\bibinfo {volume} {130}},\ \bibinfo {pages} {256401} (\bibinfo {year} {2023})}\BibitemShut {NoStop}%
\bibitem [{\citenamefont {Prakash}\ and\ \citenamefont {Jones}(2025)}]{prakashClassicalOriginsLandauincompatible2025}%
  \BibitemOpen
  \bibfield  {author} {\bibinfo {author} {\bibfnamefont {A.}~\bibnamefont {Prakash}}\ and\ \bibinfo {author} {\bibfnamefont {N.~G.}\ \bibnamefont {Jones}},\ }\bibfield  {title} {\bibinfo {title} {Classical {{Origins}} of {{Landau-Incompatible Transitions}}},\ }\href {https://doi.org/10.1103/PhysRevLett.134.097103} {\bibfield  {journal} {\bibinfo  {journal} {Phys. Rev. Lett.}\ }\textbf {\bibinfo {volume} {134}},\ \bibinfo {pages} {097103} (\bibinfo {year} {2025})}\BibitemShut {NoStop}%
\bibitem [{\citenamefont {Rau}\ and\ \citenamefont {Gingras}(2018)}]{rauFrustrationAnisotropicExchange2018}%
  \BibitemOpen
  \bibfield  {author} {\bibinfo {author} {\bibfnamefont {J.~G.}\ \bibnamefont {Rau}}\ and\ \bibinfo {author} {\bibfnamefont {M.~J.~P.}\ \bibnamefont {Gingras}},\ }\bibfield  {title} {\bibinfo {title} {Frustration and anisotropic exchange in ytterbium magnets with edge-shared octahedra},\ }\href {https://doi.org/10.1103/PhysRevB.98.054408} {\bibfield  {journal} {\bibinfo  {journal} {Physical Review B}\ }\textbf {\bibinfo {volume} {98}},\ \bibinfo {pages} {054408} (\bibinfo {year} {2018})}\BibitemShut {NoStop}%
\bibitem [{\citenamefont {Huang}\ \emph {et~al.}(2014)\citenamefont {Huang}, \citenamefont {Chen},\ and\ \citenamefont {Hermele}}]{huangQuantumSpinIces2014}%
  \BibitemOpen
  \bibfield  {author} {\bibinfo {author} {\bibfnamefont {Y.-P.}\ \bibnamefont {Huang}}, \bibinfo {author} {\bibfnamefont {G.}~\bibnamefont {Chen}},\ and\ \bibinfo {author} {\bibfnamefont {M.}~\bibnamefont {Hermele}},\ }\bibfield  {title} {\bibinfo {title} {Quantum {{Spin Ices}} and {{Topological Phases}} from {{Dipolar-Octupolar Doublets}} on the {{Pyrochlore Lattice}}},\ }\href {https://doi.org/10.1103/PhysRevLett.112.167203} {\bibfield  {journal} {\bibinfo  {journal} {Phys. Rev. Lett.}\ }\textbf {\bibinfo {volume} {112}},\ \bibinfo {pages} {167203} (\bibinfo {year} {2014})}\BibitemShut {NoStop}%
\bibitem [{\citenamefont {Ross}\ \emph {et~al.}(2011)\citenamefont {Ross}, \citenamefont {Savary}, \citenamefont {Gaulin},\ and\ \citenamefont {Balents}}]{rossQuantumExcitationsQuantum2011}%
  \BibitemOpen
  \bibfield  {author} {\bibinfo {author} {\bibfnamefont {K.~A.}\ \bibnamefont {Ross}}, \bibinfo {author} {\bibfnamefont {L.}~\bibnamefont {Savary}}, \bibinfo {author} {\bibfnamefont {B.~D.}\ \bibnamefont {Gaulin}},\ and\ \bibinfo {author} {\bibfnamefont {L.}~\bibnamefont {Balents}},\ }\bibfield  {title} {\bibinfo {title} {Quantum {{Excitations}} in {{Quantum Spin Ice}}},\ }\href {https://doi.org/10.1103/PhysRevX.1.021002} {\bibfield  {journal} {\bibinfo  {journal} {Phys. Rev. X}\ }\textbf {\bibinfo {volume} {1}},\ \bibinfo {pages} {021002} (\bibinfo {year} {2011})}\BibitemShut {NoStop}%
\bibitem [{\citenamefont {Remund}\ \emph {et~al.}(2022)\citenamefont {Remund}, \citenamefont {Pohle}, \citenamefont {Akagi}, \citenamefont {Romh{\'a}nyi},\ and\ \citenamefont {Shannon}}]{remundSemiclassicalSimulationSpin12022}%
  \BibitemOpen
  \bibfield  {author} {\bibinfo {author} {\bibfnamefont {K.}~\bibnamefont {Remund}}, \bibinfo {author} {\bibfnamefont {R.}~\bibnamefont {Pohle}}, \bibinfo {author} {\bibfnamefont {Y.}~\bibnamefont {Akagi}}, \bibinfo {author} {\bibfnamefont {J.}~\bibnamefont {Romh{\'a}nyi}},\ and\ \bibinfo {author} {\bibfnamefont {N.}~\bibnamefont {Shannon}},\ }\bibfield  {title} {\bibinfo {title} {Semi-classical simulation of spin-1 magnets},\ }\href {https://doi.org/10.1103/PhysRevResearch.4.033106} {\bibfield  {journal} {\bibinfo  {journal} {Phys. Rev. Res.}\ }\textbf {\bibinfo {volume} {4}},\ \bibinfo {pages} {033106} (\bibinfo {year} {2022})}\BibitemShut {NoStop}%
\bibitem [{\citenamefont {Zhang}\ and\ \citenamefont {Batista}(2021)}]{zhangClassicalSpinDynamics2021}%
  \BibitemOpen
  \bibfield  {author} {\bibinfo {author} {\bibfnamefont {H.}~\bibnamefont {Zhang}}\ and\ \bibinfo {author} {\bibfnamefont {C.~D.}\ \bibnamefont {Batista}},\ }\bibfield  {title} {\bibinfo {title} {Classical spin dynamics based on {{SU}}({{N}}) coherent states},\ }\href {https://doi.org/10.1103/PhysRevB.104.104409} {\bibfield  {journal} {\bibinfo  {journal} {Physical Review B}\ }\textbf {\bibinfo {volume} {104}},\ \bibinfo {pages} {104409} (\bibinfo {year} {2021})}\BibitemShut {NoStop}%
\bibitem [{\citenamefont {Chung}\ \emph {et~al.}(2025)\citenamefont {Chung}, \citenamefont {Petit}, \citenamefont {Robert},\ and\ \citenamefont {McClarty}}]{chungGeometricallyFrustratedQuadrupoles2025}%
  \BibitemOpen
  \bibfield  {author} {\bibinfo {author} {\bibfnamefont {K.~T.~K.}\ \bibnamefont {Chung}}, \bibinfo {author} {\bibfnamefont {S.}~\bibnamefont {Petit}}, \bibinfo {author} {\bibfnamefont {J.}~\bibnamefont {Robert}},\ and\ \bibinfo {author} {\bibfnamefont {P.}~\bibnamefont {McClarty}},\ }\href@noop {} {\bibinfo {title} {Geometrically {{Frustrated Quadrupoles}} on the {{Pyrochlore Lattice}} and {{Generalized Spin Liquids}}}} (\bibinfo {year} {2025}),\ \Eprint {https://arxiv.org/abs/2506.19908} {arXiv:2506.19908} \BibitemShut {NoStop}%
\bibitem [{\citenamefont {Zhang}\ \emph {et~al.}(2019)\citenamefont {Zhang}, \citenamefont {Changlani}, \citenamefont {Plumb}, \citenamefont {Tchernyshyov},\ and\ \citenamefont {Moessner}}]{zhangDynamicalStructureFactor2019}%
  \BibitemOpen
  \bibfield  {author} {\bibinfo {author} {\bibfnamefont {S.}~\bibnamefont {Zhang}}, \bibinfo {author} {\bibfnamefont {H.~J.}\ \bibnamefont {Changlani}}, \bibinfo {author} {\bibfnamefont {K.~W.}\ \bibnamefont {Plumb}}, \bibinfo {author} {\bibfnamefont {O.}~\bibnamefont {Tchernyshyov}},\ and\ \bibinfo {author} {\bibfnamefont {R.}~\bibnamefont {Moessner}},\ }\bibfield  {title} {\bibinfo {title} {Dynamical {{Structure Factor}} of the {{Three-Dimensional Quantum Spin Liquid Candidate NaCaNi}}{$_{2}$}{{F}}{$_{7}$}},\ }\href {https://doi.org/10.1103/PhysRevLett.122.167203} {\bibfield  {journal} {\bibinfo  {journal} {Phys. Rev. Lett.}\ }\textbf {\bibinfo {volume} {122}},\ \bibinfo {pages} {167203} (\bibinfo {year} {2019})}\BibitemShut {NoStop}%
\bibitem [{\citenamefont {Plumb}\ \emph {et~al.}(2019)\citenamefont {Plumb}, \citenamefont {Changlani}, \citenamefont {Scheie}, \citenamefont {Zhang}, \citenamefont {Krizan}, \citenamefont {{Rodriguez-Rivera}}, \citenamefont {Qiu}, \citenamefont {Winn}, \citenamefont {Cava},\ and\ \citenamefont {Broholm}}]{plumbContinuumQuantumFluctuations2019}%
  \BibitemOpen
  \bibfield  {author} {\bibinfo {author} {\bibfnamefont {K.~W.}\ \bibnamefont {Plumb}}, \bibinfo {author} {\bibfnamefont {H.~J.}\ \bibnamefont {Changlani}}, \bibinfo {author} {\bibfnamefont {A.}~\bibnamefont {Scheie}}, \bibinfo {author} {\bibfnamefont {S.}~\bibnamefont {Zhang}}, \bibinfo {author} {\bibfnamefont {J.~W.}\ \bibnamefont {Krizan}}, \bibinfo {author} {\bibfnamefont {J.~A.}\ \bibnamefont {{Rodriguez-Rivera}}}, \bibinfo {author} {\bibfnamefont {Y.}~\bibnamefont {Qiu}}, \bibinfo {author} {\bibfnamefont {B.}~\bibnamefont {Winn}}, \bibinfo {author} {\bibfnamefont {R.~J.}\ \bibnamefont {Cava}},\ and\ \bibinfo {author} {\bibfnamefont {C.~L.}\ \bibnamefont {Broholm}},\ }\bibfield  {title} {\bibinfo {title} {Continuum of quantum fluctuations in a three-dimensional {{S}} = 1 {{Heisenberg}} magnet},\ }\href {https://doi.org/10.1038/s41567-018-0317-3} {\bibfield  {journal} {\bibinfo  {journal} {Nature Physics}\ }\textbf {\bibinfo {volume} {15}},\ \bibinfo {pages} {54} (\bibinfo {year} {2019})}\BibitemShut
  {NoStop}%
\bibitem [{\citenamefont {Hagym{\'a}si}\ \emph {et~al.}(2024)\citenamefont {Hagym{\'a}si}, \citenamefont {Niggemann},\ and\ \citenamefont {Reuther}}]{hagymasiPhaseDiagramAntiferromagnetic2024}%
  \BibitemOpen
  \bibfield  {author} {\bibinfo {author} {\bibfnamefont {I.}~\bibnamefont {Hagym{\'a}si}}, \bibinfo {author} {\bibfnamefont {N.}~\bibnamefont {Niggemann}},\ and\ \bibinfo {author} {\bibfnamefont {J.}~\bibnamefont {Reuther}},\ }\bibfield  {title} {\bibinfo {title} {Phase diagram of the antiferromagnetic {{J1-J2}} spin-1 pyrochlore {{Heisenberg}} model},\ }\href {https://doi.org/10.1103/PhysRevB.110.224416} {\bibfield  {journal} {\bibinfo  {journal} {Phys. Rev. B}\ }\textbf {\bibinfo {volume} {110}},\ \bibinfo {pages} {224416} (\bibinfo {year} {2024})}\BibitemShut {NoStop}%
\bibitem [{\citenamefont {Luttinger}\ and\ \citenamefont {Tisza}(1946)}]{luttingerTheoryDipoleInteraction1946}%
  \BibitemOpen
  \bibfield  {author} {\bibinfo {author} {\bibfnamefont {J.~M.}\ \bibnamefont {Luttinger}}\ and\ \bibinfo {author} {\bibfnamefont {L.}~\bibnamefont {Tisza}},\ }\bibfield  {title} {\bibinfo {title} {Theory of {{Dipole Interaction}} in {{Crystals}}},\ }\href {https://doi.org/10.1103/PhysRev.70.954} {\bibfield  {journal} {\bibinfo  {journal} {Phys. Rev.}\ }\textbf {\bibinfo {volume} {70}},\ \bibinfo {pages} {954} (\bibinfo {year} {1946})}\BibitemShut {NoStop}%
\bibitem [{\citenamefont {Litvin}(1974)}]{litvinLuttingerTiszaMethod1974}%
  \BibitemOpen
  \bibfield  {author} {\bibinfo {author} {\bibfnamefont {D.~B.}\ \bibnamefont {Litvin}},\ }\bibfield  {title} {\bibinfo {title} {The {{Luttinger-Tisza}} method},\ }\href {https://doi.org/10.1016/0031-8914(74)90257-2} {\bibfield  {journal} {\bibinfo  {journal} {Physica}\ }\textbf {\bibinfo {volume} {77}},\ \bibinfo {pages} {205} (\bibinfo {year} {1974})}\BibitemShut {NoStop}%
\bibitem [{\citenamefont {Davier}\ \emph {et~al.}(2023)\citenamefont {Davier}, \citenamefont {G{\'o}mez~Albarrac{\'i}n}, \citenamefont {Rosales},\ and\ \citenamefont {Pujol}}]{davierCombinedApproachAnalyze2023}%
  \BibitemOpen
  \bibfield  {author} {\bibinfo {author} {\bibfnamefont {N.}~\bibnamefont {Davier}}, \bibinfo {author} {\bibfnamefont {F.~A.}\ \bibnamefont {G{\'o}mez~Albarrac{\'i}n}}, \bibinfo {author} {\bibfnamefont {H.~D.}\ \bibnamefont {Rosales}},\ and\ \bibinfo {author} {\bibfnamefont {P.}~\bibnamefont {Pujol}},\ }\bibfield  {title} {\bibinfo {title} {Combined approach to analyze and classify families of classical spin liquids},\ }\href {https://doi.org/10.1103/PhysRevB.108.054408} {\bibfield  {journal} {\bibinfo  {journal} {Phys. Rev. B}\ }\textbf {\bibinfo {volume} {108}},\ \bibinfo {pages} {054408} (\bibinfo {year} {2023})}\BibitemShut {NoStop}%
\bibitem [{\citenamefont {Isakov}\ \emph {et~al.}(2004)\citenamefont {Isakov}, \citenamefont {Gregor}, \citenamefont {Moessner},\ and\ \citenamefont {Sondhi}}]{isakovDipolarSpinCorrelations2004}%
  \BibitemOpen
  \bibfield  {author} {\bibinfo {author} {\bibfnamefont {S.~V.}\ \bibnamefont {Isakov}}, \bibinfo {author} {\bibfnamefont {K.}~\bibnamefont {Gregor}}, \bibinfo {author} {\bibfnamefont {R.}~\bibnamefont {Moessner}},\ and\ \bibinfo {author} {\bibfnamefont {S.~L.}\ \bibnamefont {Sondhi}},\ }\bibfield  {title} {\bibinfo {title} {Dipolar spin correlations in classical pyrochlore magnets},\ }\href {https://doi.org/10.1103/PhysRevLett.93.167204} {\bibfield  {journal} {\bibinfo  {journal} {Phys. Rev. Lett.}\ }\textbf {\bibinfo {volume} {93}},\ \bibinfo {pages} {167204} (\bibinfo {year} {2004})}\BibitemShut {NoStop}%
\bibitem [{\citenamefont {Pandey}\ \emph {et~al.}(2020)\citenamefont {Pandey}, \citenamefont {Moessner},\ and\ \citenamefont {Castelnovo}}]{pandeyAnalyticalTheoryPyrochlore2020}%
  \BibitemOpen
  \bibfield  {author} {\bibinfo {author} {\bibfnamefont {A.}~\bibnamefont {Pandey}}, \bibinfo {author} {\bibfnamefont {R.}~\bibnamefont {Moessner}},\ and\ \bibinfo {author} {\bibfnamefont {C.}~\bibnamefont {Castelnovo}},\ }\bibfield  {title} {\bibinfo {title} {Analytical theory of pyrochlore cooperative paramagnets},\ }\href {https://doi.org/10.1103/PhysRevB.101.115107} {\bibfield  {journal} {\bibinfo  {journal} {Phys. Rev. B}\ }\textbf {\bibinfo {volume} {101}},\ \bibinfo {pages} {115107} (\bibinfo {year} {2020})}\BibitemShut {NoStop}%
\bibitem [{\citenamefont {Castelnovo}\ \emph {et~al.}(2008)\citenamefont {Castelnovo}, \citenamefont {Moessner},\ and\ \citenamefont {Sondhi}}]{castelnovoMagneticMonopolesSpin2008}%
  \BibitemOpen
  \bibfield  {author} {\bibinfo {author} {\bibfnamefont {C.}~\bibnamefont {Castelnovo}}, \bibinfo {author} {\bibfnamefont {R.}~\bibnamefont {Moessner}},\ and\ \bibinfo {author} {\bibfnamefont {S.~L.}\ \bibnamefont {Sondhi}},\ }\bibfield  {title} {\bibinfo {title} {Magnetic monopoles in spin ice},\ }\href {https://doi.org/10.1038/nature06433} {\bibfield  {journal} {\bibinfo  {journal} {Nature}\ }\textbf {\bibinfo {volume} {451}},\ \bibinfo {pages} {42} (\bibinfo {year} {2008})}\BibitemShut {NoStop}%
\bibitem [{\citenamefont {Palmer}\ and\ \citenamefont {Chalker}(2000)}]{palmerOrderInducedDipolar2000}%
  \BibitemOpen
  \bibfield  {author} {\bibinfo {author} {\bibfnamefont {S.~E.}\ \bibnamefont {Palmer}}\ and\ \bibinfo {author} {\bibfnamefont {J.~T.}\ \bibnamefont {Chalker}},\ }\bibfield  {title} {\bibinfo {title} {Order induced by dipolar interactions in a geometrically frustrated antiferromagnet},\ }\href {https://doi.org/10.1103/PhysRevB.62.488} {\bibfield  {journal} {\bibinfo  {journal} {Phys. Rev. B}\ }\textbf {\bibinfo {volume} {62}},\ \bibinfo {pages} {488} (\bibinfo {year} {2000})}\BibitemShut {NoStop}%
\bibitem [{\citenamefont {Francini}\ \emph {et~al.}(2025{\natexlab{b}})\citenamefont {Francini}, \citenamefont {Schmidt}, \citenamefont {Janssen},\ and\ \citenamefont {{Lozano-G{\'o}mez}}}]{franciniExactNematicMixed2025}%
  \BibitemOpen
  \bibfield  {author} {\bibinfo {author} {\bibfnamefont {N.}~\bibnamefont {Francini}}, \bibinfo {author} {\bibfnamefont {L.}~\bibnamefont {Schmidt}}, \bibinfo {author} {\bibfnamefont {L.}~\bibnamefont {Janssen}},\ and\ \bibinfo {author} {\bibfnamefont {D.}~\bibnamefont {{Lozano-G{\'o}mez}}},\ }\href@noop {} {\bibinfo {title} {Exact nematic and mixed magnetic phases driven by competing orders on the pyrochlore lattice}} (\bibinfo {year} {2025}{\natexlab{b}}),\ \Eprint {https://arxiv.org/abs/2510.23704} {arXiv:2510.23704} \BibitemShut {NoStop}%
\bibitem [{\citenamefont {Scheie}\ \emph {et~al.}(2022)\citenamefont {Scheie}, \citenamefont {Benton}, \citenamefont {Taillefumier}, \citenamefont {Jaubert}, \citenamefont {Sala}, \citenamefont {Jalarvo}, \citenamefont {Koohpayeh},\ and\ \citenamefont {Shannon}}]{scheieDynamicalScalingSignature2022}%
  \BibitemOpen
  \bibfield  {author} {\bibinfo {author} {\bibfnamefont {A.}~\bibnamefont {Scheie}}, \bibinfo {author} {\bibfnamefont {O.}~\bibnamefont {Benton}}, \bibinfo {author} {\bibfnamefont {M.}~\bibnamefont {Taillefumier}}, \bibinfo {author} {\bibfnamefont {L.~D.~C.}\ \bibnamefont {Jaubert}}, \bibinfo {author} {\bibfnamefont {G.}~\bibnamefont {Sala}}, \bibinfo {author} {\bibfnamefont {N.}~\bibnamefont {Jalarvo}}, \bibinfo {author} {\bibfnamefont {S.~M.}\ \bibnamefont {Koohpayeh}},\ and\ \bibinfo {author} {\bibfnamefont {N.}~\bibnamefont {Shannon}},\ }\bibfield  {title} {\bibinfo {title} {Dynamical {{Scaling}} as a {{Signature}} of {{Multiple Phase Competition}} in {{Yb}}{$_{2}$}{{Ti}}{$_{2}$}{{O}}{$_{7}$}},\ }\href {https://doi.org/10.1103/PhysRevLett.129.217202} {\bibfield  {journal} {\bibinfo  {journal} {Phys. Rev. Lett.}\ }\textbf {\bibinfo {volume} {129}},\ \bibinfo {pages} {217202} (\bibinfo {year} {2022})}\BibitemShut {NoStop}%
\bibitem [{\citenamefont {Reimers}(1992)}]{reimersAbsenceLongrangeOrder1992}%
  \BibitemOpen
  \bibfield  {author} {\bibinfo {author} {\bibfnamefont {J.~N.}\ \bibnamefont {Reimers}},\ }\bibfield  {title} {\bibinfo {title} {Absence of long-range order in a three-dimensional geometrically frustrated antiferromagnet},\ }\href {https://doi.org/10.1103/PhysRevB.45.7287} {\bibfield  {journal} {\bibinfo  {journal} {Phys. Rev. B}\ }\textbf {\bibinfo {volume} {45}},\ \bibinfo {pages} {7287} (\bibinfo {year} {1992})}\BibitemShut {NoStop}%
\bibitem [{\citenamefont {Canals}\ and\ \citenamefont {Lacroix}(1998)}]{canalsPyrochloreAntiferromagnetThreeDimensional1998}%
  \BibitemOpen
  \bibfield  {author} {\bibinfo {author} {\bibfnamefont {B.}~\bibnamefont {Canals}}\ and\ \bibinfo {author} {\bibfnamefont {C.}~\bibnamefont {Lacroix}},\ }\bibfield  {title} {\bibinfo {title} {Pyrochlore {{Antiferromagnet}}: {{A Three-Dimensional Quantum Spin Liquid}}},\ }\href {https://doi.org/10.1103/PhysRevLett.80.2933} {\bibfield  {journal} {\bibinfo  {journal} {Phys. Rev. Lett.}\ }\textbf {\bibinfo {volume} {80}},\ \bibinfo {pages} {2933} (\bibinfo {year} {1998})}\BibitemShut {NoStop}%
\bibitem [{\citenamefont {Canals}\ and\ \citenamefont {Garanin}(2001)}]{canalsSpinliquidPhasePyrochlore2001}%
  \BibitemOpen
  \bibfield  {author} {\bibinfo {author} {\bibfnamefont {B.}~\bibnamefont {Canals}}\ and\ \bibinfo {author} {\bibfnamefont {D.~A.}\ \bibnamefont {Garanin}},\ }\bibfield  {title} {\bibinfo {title} {Spin-liquid phase in the pyrochlore anti-ferromagnet},\ }\href {https://doi.org/10.1139/p01-101} {\bibfield  {journal} {\bibinfo  {journal} {Can. J. Phys.}\ }\textbf {\bibinfo {volume} {79}},\ \bibinfo {pages} {1323} (\bibinfo {year} {2001})}\BibitemShut {NoStop}%
\bibitem [{\citenamefont {Conlon}\ and\ \citenamefont {Chalker}(2010)}]{conlonAbsentPinchPoints2010}%
  \BibitemOpen
  \bibfield  {author} {\bibinfo {author} {\bibfnamefont {P.~H.}\ \bibnamefont {Conlon}}\ and\ \bibinfo {author} {\bibfnamefont {J.~T.}\ \bibnamefont {Chalker}},\ }\bibfield  {title} {\bibinfo {title} {Absent pinch points and emergent clusters: {{Further}} neighbor interactions in the pyrochlore {{Heisenberg}} antiferromagnet},\ }\href {https://doi.org/10.1103/PhysRevB.81.224413} {\bibfield  {journal} {\bibinfo  {journal} {Phys. Rev. B}\ }\textbf {\bibinfo {volume} {81}},\ \bibinfo {pages} {224413} (\bibinfo {year} {2010})}\BibitemShut {NoStop}%
\bibitem [{\citenamefont {Iqbal}\ \emph {et~al.}(2019)\citenamefont {Iqbal}, \citenamefont {M{\"u}ller}, \citenamefont {Ghosh}, \citenamefont {Gingras}, \citenamefont {Jeschke}, \citenamefont {Rachel}, \citenamefont {Reuther},\ and\ \citenamefont {Thomale}}]{iqbalQuantumClassicalPhases2019}%
  \BibitemOpen
  \bibfield  {author} {\bibinfo {author} {\bibfnamefont {Y.}~\bibnamefont {Iqbal}}, \bibinfo {author} {\bibfnamefont {T.}~\bibnamefont {M{\"u}ller}}, \bibinfo {author} {\bibfnamefont {P.}~\bibnamefont {Ghosh}}, \bibinfo {author} {\bibfnamefont {M.~J.~P.}\ \bibnamefont {Gingras}}, \bibinfo {author} {\bibfnamefont {H.~O.}\ \bibnamefont {Jeschke}}, \bibinfo {author} {\bibfnamefont {S.}~\bibnamefont {Rachel}}, \bibinfo {author} {\bibfnamefont {J.}~\bibnamefont {Reuther}},\ and\ \bibinfo {author} {\bibfnamefont {R.}~\bibnamefont {Thomale}},\ }\bibfield  {title} {\bibinfo {title} {Quantum and {{Classical Phases}} of the {{Pyrochlore Heisenberg Model}} with {{Competing Interactions}}},\ }\href {https://doi.org/10.1103/PhysRevX.9.011005} {\bibfield  {journal} {\bibinfo  {journal} {Phys. Rev. X}\ }\textbf {\bibinfo {volume} {9}},\ \bibinfo {pages} {11005} (\bibinfo {year} {2019})}\BibitemShut {NoStop}%
\bibitem [{\citenamefont {Berry}\ and\ \citenamefont {Wilkinson}(1997)}]{berryDiabolicalPointsSpectra1997}%
  \BibitemOpen
  \bibfield  {author} {\bibinfo {author} {\bibfnamefont {M.~V.}\ \bibnamefont {Berry}}\ and\ \bibinfo {author} {\bibfnamefont {M.}~\bibnamefont {Wilkinson}},\ }\bibfield  {title} {\bibinfo {title} {Diabolical points in the spectra of triangles},\ }\href {https://doi.org/10.1098/rspa.1984.0022} {\bibfield  {journal} {\bibinfo  {journal} {Proceedings of the Royal Society of London. A. Mathematical and Physical Sciences}\ }\textbf {\bibinfo {volume} {392}},\ \bibinfo {pages} {15} (\bibinfo {year} {1997})}\BibitemShut {NoStop}%
\bibitem [{\citenamefont {Elhajal}\ \emph {et~al.}(2004)\citenamefont {Elhajal}, \citenamefont {Canals},\ and\ \citenamefont {Lacroix}}]{elhajalOrderingPyrochloreCompounds2004}%
  \BibitemOpen
  \bibfield  {author} {\bibinfo {author} {\bibfnamefont {M.}~\bibnamefont {Elhajal}}, \bibinfo {author} {\bibfnamefont {B.}~\bibnamefont {Canals}},\ and\ \bibinfo {author} {\bibfnamefont {C.}~\bibnamefont {Lacroix}},\ }\bibfield  {title} {\bibinfo {title} {Ordering in pyrochlore compounds due to {{Dzyaloshinsky}}--{{Moriya}} interactions: The case of {{Cu4O3}}},\ }\href {https://doi.org/10.1088/0953-8984/16/11/049} {\bibfield  {journal} {\bibinfo  {journal} {J. Phys.: Condens. Matter}\ }\textbf {\bibinfo {volume} {16}},\ \bibinfo {pages} {S917} (\bibinfo {year} {2004})}\BibitemShut {NoStop}%
\bibitem [{\citenamefont {Elhajal}\ \emph {et~al.}(2005)\citenamefont {Elhajal}, \citenamefont {Canals}, \citenamefont {Sunyer},\ and\ \citenamefont {Lacroix}}]{elhajalOrderingPyrochloreAntiferromagnet2005}%
  \BibitemOpen
  \bibfield  {author} {\bibinfo {author} {\bibfnamefont {M.}~\bibnamefont {Elhajal}}, \bibinfo {author} {\bibfnamefont {B.}~\bibnamefont {Canals}}, \bibinfo {author} {\bibfnamefont {R.}~\bibnamefont {Sunyer}},\ and\ \bibinfo {author} {\bibfnamefont {C.}~\bibnamefont {Lacroix}},\ }\bibfield  {title} {\bibinfo {title} {Ordering in the pyrochlore antiferromagnet due to {{Dzyaloshinsky-Moriya}} interactions},\ }\href {https://doi.org/10.1103/PhysRevB.71.094420} {\bibfield  {journal} {\bibinfo  {journal} {Phys. Rev. B}\ }\textbf {\bibinfo {volume} {71}},\ \bibinfo {pages} {094420} (\bibinfo {year} {2005})}\BibitemShut {NoStop}%
\bibitem [{\citenamefont {Canals}\ \emph {et~al.}(2008)\citenamefont {Canals}, \citenamefont {Elhajal},\ and\ \citenamefont {Lacroix}}]{canalsIsinglikeOrderDisorder2008}%
  \BibitemOpen
  \bibfield  {author} {\bibinfo {author} {\bibfnamefont {B.}~\bibnamefont {Canals}}, \bibinfo {author} {\bibfnamefont {M.}~\bibnamefont {Elhajal}},\ and\ \bibinfo {author} {\bibfnamefont {C.}~\bibnamefont {Lacroix}},\ }\bibfield  {title} {\bibinfo {title} {Ising-like order by disorder in the pyrochlore antiferromagnet with {{Dzyaloshinskii-Moriya}} interactions},\ }\href {https://doi.org/10.1103/PhysRevB.78.214431} {\bibfield  {journal} {\bibinfo  {journal} {Phys. Rev. B}\ }\textbf {\bibinfo {volume} {78}},\ \bibinfo {pages} {214431} (\bibinfo {year} {2008})}\BibitemShut {NoStop}%
\bibitem [{\citenamefont {Hickey}\ \emph {et~al.}(2025)\citenamefont {Hickey}, \citenamefont {{Lozano-G{\'o}mez}},\ and\ \citenamefont {Gingras}}]{hickeyOrderbydisorderQuantumZeropoint2025}%
  \BibitemOpen
  \bibfield  {author} {\bibinfo {author} {\bibfnamefont {A.}~\bibnamefont {Hickey}}, \bibinfo {author} {\bibfnamefont {D.}~\bibnamefont {{Lozano-G{\'o}mez}}},\ and\ \bibinfo {author} {\bibfnamefont {M.~J.~P.}\ \bibnamefont {Gingras}},\ }\bibfield  {title} {\bibinfo {title} {Order-by-disorder without quantum zero-point fluctuations in the pyrochlore {{Heisenberg}} ferromagnet with {{Dzyaloshinskii-Moriya}} interactions},\ }\href {https://doi.org/10.1103/PhysRevB.111.184434} {\bibfield  {journal} {\bibinfo  {journal} {Phys. Rev. B}\ }\textbf {\bibinfo {volume} {111}},\ \bibinfo {pages} {184434} (\bibinfo {year} {2025})}\BibitemShut {NoStop}%
\bibitem [{\citenamefont {Zhang}\ \emph {et~al.}(2025)\citenamefont {Zhang}, \citenamefont {Song},\ and\ \citenamefont {Senthil}}]{zhangDiracSpinLiquid2025}%
  \BibitemOpen
  \bibfield  {author} {\bibinfo {author} {\bibfnamefont {Y.}~\bibnamefont {Zhang}}, \bibinfo {author} {\bibfnamefont {X.-Y.}\ \bibnamefont {Song}},\ and\ \bibinfo {author} {\bibfnamefont {T.}~\bibnamefont {Senthil}},\ }\bibfield  {title} {\bibinfo {title} {Dirac spin liquid as an ``unnecessary'' quantum critical point on square lattice antiferromagnets},\ }\href {https://doi.org/10.21468/SciPostPhysCore.8.1.024} {\bibfield  {journal} {\bibinfo  {journal} {SciPost Physics Core}\ }\textbf {\bibinfo {volume} {8}},\ \bibinfo {pages} {024} (\bibinfo {year} {2025})}\BibitemShut {NoStop}%
\bibitem [{\citenamefont {Kapustin}\ and\ \citenamefont {Spodyneiko}(2020{\natexlab{a}})}]{kapustinHigherdimensionalGeneralizationsThouless2020}%
  \BibitemOpen
  \bibfield  {author} {\bibinfo {author} {\bibfnamefont {A.}~\bibnamefont {Kapustin}}\ and\ \bibinfo {author} {\bibfnamefont {L.}~\bibnamefont {Spodyneiko}},\ }\href@noop {} {\bibinfo {title} {Higher-dimensional generalizations of the {{Thouless}} charge pump}} (\bibinfo {year} {2020}{\natexlab{a}}),\ \Eprint {https://arxiv.org/abs/2003.09519} {arXiv:2003.09519} \BibitemShut {NoStop}%
\bibitem [{\citenamefont {Kapustin}\ and\ \citenamefont {Spodyneiko}(2020{\natexlab{b}})}]{kapustinHigherdimensionalGeneralizationsBerry2020}%
  \BibitemOpen
  \bibfield  {author} {\bibinfo {author} {\bibfnamefont {A.}~\bibnamefont {Kapustin}}\ and\ \bibinfo {author} {\bibfnamefont {L.}~\bibnamefont {Spodyneiko}},\ }\bibfield  {title} {\bibinfo {title} {Higher-dimensional generalizations of {{Berry}} curvature},\ }\href {https://doi.org/10.1103/PhysRevB.101.235130} {\bibfield  {journal} {\bibinfo  {journal} {Phys. Rev. B}\ }\textbf {\bibinfo {volume} {101}},\ \bibinfo {pages} {235130} (\bibinfo {year} {2020}{\natexlab{b}})}\BibitemShut {NoStop}%
\bibitem [{\citenamefont {Wen}\ \emph {et~al.}(2023)\citenamefont {Wen}, \citenamefont {Qi}, \citenamefont {Beaudry}, \citenamefont {Moreno}, \citenamefont {Pflaum}, \citenamefont {Spiegel}, \citenamefont {Vishwanath},\ and\ \citenamefont {Hermele}}]{wenFlowHigherBerry2023}%
  \BibitemOpen
  \bibfield  {author} {\bibinfo {author} {\bibfnamefont {X.}~\bibnamefont {Wen}}, \bibinfo {author} {\bibfnamefont {M.}~\bibnamefont {Qi}}, \bibinfo {author} {\bibfnamefont {A.}~\bibnamefont {Beaudry}}, \bibinfo {author} {\bibfnamefont {J.}~\bibnamefont {Moreno}}, \bibinfo {author} {\bibfnamefont {M.~J.}\ \bibnamefont {Pflaum}}, \bibinfo {author} {\bibfnamefont {D.}~\bibnamefont {Spiegel}}, \bibinfo {author} {\bibfnamefont {A.}~\bibnamefont {Vishwanath}},\ and\ \bibinfo {author} {\bibfnamefont {M.}~\bibnamefont {Hermele}},\ }\bibfield  {title} {\bibinfo {title} {Flow of higher {{Berry}} curvature and bulk-boundary correspondence in parametrized quantum systems},\ }\href {https://doi.org/10.1103/PhysRevB.108.125147} {\bibfield  {journal} {\bibinfo  {journal} {Phys. Rev. B}\ }\textbf {\bibinfo {volume} {108}},\ \bibinfo {pages} {125147} (\bibinfo {year} {2023})}\BibitemShut {NoStop}%
\bibitem [{\citenamefont {Prakash}\ and\ \citenamefont {Parameswaran}(2025)}]{prakashChargePumpsBoundary2025}%
  \BibitemOpen
  \bibfield  {author} {\bibinfo {author} {\bibfnamefont {A.}~\bibnamefont {Prakash}}\ and\ \bibinfo {author} {\bibfnamefont {S.~A.}\ \bibnamefont {Parameswaran}},\ }\bibfield  {title} {\bibinfo {title} {Charge pumps, boundary modes, and the necessity of unnecessary criticality},\ }\href {https://doi.org/10.1103/kztj-jcyc} {\bibfield  {journal} {\bibinfo  {journal} {Physical Review B}\ }\textbf {\bibinfo {volume} {112}},\ \bibinfo {pages} {L241117} (\bibinfo {year} {2025})}\BibitemShut {NoStop}%
\bibitem [{\citenamefont {Wannier}(1950)}]{wannierAntiferromagnetismTriangularIsing1950}%
  \BibitemOpen
  \bibfield  {author} {\bibinfo {author} {\bibfnamefont {G.~H.}\ \bibnamefont {Wannier}},\ }\bibfield  {title} {\bibinfo {title} {Antiferromagnetism. {{The Triangular Ising Net}}},\ }\href {https://doi.org/10.1103/PhysRev.79.357} {\bibfield  {journal} {\bibinfo  {journal} {Phys. Rev.}\ }\textbf {\bibinfo {volume} {79}},\ \bibinfo {pages} {357} (\bibinfo {year} {1950})}\BibitemShut {NoStop}%
\bibitem [{\citenamefont {Stephenson}(1970)}]{stephensonIsingModelSpinCorrelations1970a}%
  \BibitemOpen
  \bibfield  {author} {\bibinfo {author} {\bibfnamefont {J.}~\bibnamefont {Stephenson}},\ }\bibfield  {title} {\bibinfo {title} {Ising-{{Model Spin Correlations}} on the {{Triangular Lattice}}. {{III}}. {{Isotropic Antiferromagnetic Lattice}}},\ }\href {https://doi.org/10.1063/1.1665154} {\bibfield  {journal} {\bibinfo  {journal} {Journal of Mathematical Physics}\ }\textbf {\bibinfo {volume} {11}},\ \bibinfo {pages} {413} (\bibinfo {year} {1970})}\BibitemShut {NoStop}%
\bibitem [{\citenamefont {Wills}\ \emph {et~al.}(2002)\citenamefont {Wills}, \citenamefont {Ballou},\ and\ \citenamefont {Lacroix}}]{willsModelLocalizedHighly2002}%
  \BibitemOpen
  \bibfield  {author} {\bibinfo {author} {\bibfnamefont {A.~S.}\ \bibnamefont {Wills}}, \bibinfo {author} {\bibfnamefont {R.}~\bibnamefont {Ballou}},\ and\ \bibinfo {author} {\bibfnamefont {C.}~\bibnamefont {Lacroix}},\ }\bibfield  {title} {\bibinfo {title} {Model of localized highly frustrated ferromagnetism: {{The}} kagome spin ice},\ }\href {https://doi.org/10.1103/PhysRevB.66.144407} {\bibfield  {journal} {\bibinfo  {journal} {Phys. Rev. B}\ }\textbf {\bibinfo {volume} {66}},\ \bibinfo {pages} {144407} (\bibinfo {year} {2002})}\BibitemShut {NoStop}%
\bibitem [{\citenamefont {Hallas}\ \emph {et~al.}(2018)\citenamefont {Hallas}, \citenamefont {Gaudet},\ and\ \citenamefont {Gaulin}}]{hallasExperimentalInsightsGroundState2018}%
  \BibitemOpen
  \bibfield  {author} {\bibinfo {author} {\bibfnamefont {A.~M.}\ \bibnamefont {Hallas}}, \bibinfo {author} {\bibfnamefont {J.}~\bibnamefont {Gaudet}},\ and\ \bibinfo {author} {\bibfnamefont {B.~D.}\ \bibnamefont {Gaulin}},\ }\bibfield  {title} {\bibinfo {title} {Experimental {{Insights}} into {{Ground-State Selection}} of {{Quantum XY Pyrochlores}}},\ }\href {https://doi.org/10.1146/annurev-conmatphys-031016-025218} {\bibfield  {journal} {\bibinfo  {journal} {Annual Review of Condensed Matter Physics}\ }\textbf {\bibinfo {volume} {9}},\ \bibinfo {pages} {105} (\bibinfo {year} {2018})}\BibitemShut {NoStop}%
\bibitem [{\citenamefont {Chern}(2010)}]{chernPyrochloreAntiferromagnetAntisymmetric2010}%
  \BibitemOpen
  \bibfield  {author} {\bibinfo {author} {\bibfnamefont {G.-W.}\ \bibnamefont {Chern}},\ }\href@noop {} {\bibinfo {title} {Pyrochlore antiferromagnet with antisymmetric exchange interactions: Critical behavior and order from disorder}} (\bibinfo {year} {2010}),\ \Eprint {https://arxiv.org/abs/1008.3038} {arXiv:1008.3038} \BibitemShut {NoStop}%
\bibitem [{\citenamefont {Javanparast}\ \emph {et~al.}(2015{\natexlab{b}})\citenamefont {Javanparast}, \citenamefont {Hao}, \citenamefont {Enjalran},\ and\ \citenamefont {Gingras}}]{javanparastFluctuationDrivenSelectionCriticality2015}%
  \BibitemOpen
  \bibfield  {author} {\bibinfo {author} {\bibfnamefont {B.}~\bibnamefont {Javanparast}}, \bibinfo {author} {\bibfnamefont {Z.}~\bibnamefont {Hao}}, \bibinfo {author} {\bibfnamefont {M.}~\bibnamefont {Enjalran}},\ and\ \bibinfo {author} {\bibfnamefont {M.~J.~P.}\ \bibnamefont {Gingras}},\ }\bibfield  {title} {\bibinfo {title} {Fluctuation-{{Driven Selection}} at {{Criticality}} in a {{Frustrated Magnetic System}}: {{The Case}} of {{Multiple-k Partial Order}} on the {{Pyrochlore Lattice}}},\ }\href {https://doi.org/10.1103/PhysRevLett.114.130601} {\bibfield  {journal} {\bibinfo  {journal} {Phys. Rev. Lett.}\ }\textbf {\bibinfo {volume} {114}},\ \bibinfo {pages} {130601} (\bibinfo {year} {2015}{\natexlab{b}})}\BibitemShut {NoStop}%
\bibitem [{\citenamefont {Henley}(1989)}]{henleyOrderingDueDisorder1989}%
  \BibitemOpen
  \bibfield  {author} {\bibinfo {author} {\bibfnamefont {C.~L.}\ \bibnamefont {Henley}},\ }\bibfield  {title} {\bibinfo {title} {Ordering due to disorder in a frustrated vector antiferromagnet},\ }\href {https://doi.org/10.1103/PhysRevLett.62.2056} {\bibfield  {journal} {\bibinfo  {journal} {Phys. Rev. Lett.}\ }\textbf {\bibinfo {volume} {62}},\ \bibinfo {pages} {2056} (\bibinfo {year} {1989})}\BibitemShut {NoStop}%
\bibitem [{\citenamefont {Khatua}\ \emph {et~al.}(2019)\citenamefont {Khatua}, \citenamefont {Sen},\ and\ \citenamefont {Ganesh}}]{khatuaEffectiveTheoriesQuantum2019}%
  \BibitemOpen
  \bibfield  {author} {\bibinfo {author} {\bibfnamefont {S.}~\bibnamefont {Khatua}}, \bibinfo {author} {\bibfnamefont {D.}~\bibnamefont {Sen}},\ and\ \bibinfo {author} {\bibfnamefont {R.}~\bibnamefont {Ganesh}},\ }\bibfield  {title} {\bibinfo {title} {Effective theories for quantum spin clusters: {{Geometric}} phases and state selection by singularity},\ }\href {https://doi.org/10.1103/PhysRevB.100.134411} {\bibfield  {journal} {\bibinfo  {journal} {Phys. Rev. B}\ }\textbf {\bibinfo {volume} {100}},\ \bibinfo {pages} {134411} (\bibinfo {year} {2019})}\BibitemShut {NoStop}%
\bibitem [{\citenamefont {Srinivasan}\ \emph {et~al.}(2020)\citenamefont {Srinivasan}, \citenamefont {Khatua}, \citenamefont {Baskaran},\ and\ \citenamefont {Ganesh}}]{srinivasanOrderSingularityKitaev2020}%
  \BibitemOpen
  \bibfield  {author} {\bibinfo {author} {\bibfnamefont {S.}~\bibnamefont {Srinivasan}}, \bibinfo {author} {\bibfnamefont {S.}~\bibnamefont {Khatua}}, \bibinfo {author} {\bibfnamefont {G.}~\bibnamefont {Baskaran}},\ and\ \bibinfo {author} {\bibfnamefont {R.}~\bibnamefont {Ganesh}},\ }\bibfield  {title} {\bibinfo {title} {Order by singularity in {{Kitaev}} clusters},\ }\href {https://doi.org/10.1103/PhysRevResearch.2.023212} {\bibfield  {journal} {\bibinfo  {journal} {Phys. Rev. Research}\ }\textbf {\bibinfo {volume} {2}},\ \bibinfo {pages} {023212} (\bibinfo {year} {2020})}\BibitemShut {NoStop}%
\bibitem [{\citenamefont {Khatua}\ \emph {et~al.}(2021)\citenamefont {Khatua}, \citenamefont {Srinivasan},\ and\ \citenamefont {Ganesh}}]{khatuaStateSelectionFrustrated2021}%
  \BibitemOpen
  \bibfield  {author} {\bibinfo {author} {\bibfnamefont {S.}~\bibnamefont {Khatua}}, \bibinfo {author} {\bibfnamefont {S.}~\bibnamefont {Srinivasan}},\ and\ \bibinfo {author} {\bibfnamefont {R.}~\bibnamefont {Ganesh}},\ }\bibfield  {title} {\bibinfo {title} {State selection in frustrated magnets},\ }\href {https://doi.org/10.1103/PhysRevB.103.174412} {\bibfield  {journal} {\bibinfo  {journal} {Phys. Rev. B}\ }\textbf {\bibinfo {volume} {103}},\ \bibinfo {pages} {174412} (\bibinfo {year} {2021})}\BibitemShut {NoStop}%
\bibitem [{\citenamefont {C{\'e}pas}\ \emph {et~al.}(2005)\citenamefont {C{\'e}pas}, \citenamefont {Young},\ and\ \citenamefont {Shastry}}]{cepasDegeneracyStrongFluctuationinduced2005}%
  \BibitemOpen
  \bibfield  {author} {\bibinfo {author} {\bibfnamefont {O.}~\bibnamefont {C{\'e}pas}}, \bibinfo {author} {\bibfnamefont {A.~P.}\ \bibnamefont {Young}},\ and\ \bibinfo {author} {\bibfnamefont {B.~S.}\ \bibnamefont {Shastry}},\ }\bibfield  {title} {\bibinfo {title} {Degeneracy and strong fluctuation-induced first-order phase transition in the dipolar pyrochlore antiferromagnet},\ }\href {https://doi.org/10.1103/PhysRevB.72.184408} {\bibfield  {journal} {\bibinfo  {journal} {Phys. Rev. B}\ }\textbf {\bibinfo {volume} {72}},\ \bibinfo {pages} {184408} (\bibinfo {year} {2005})}\BibitemShut {NoStop}%
\bibitem [{\citenamefont {Bergman}\ \emph {et~al.}(2007)\citenamefont {Bergman}, \citenamefont {Alicea}, \citenamefont {Gull}, \citenamefont {Trebst},\ and\ \citenamefont {Balents}}]{bergmanOrderbydisorderSpiralSpinliquid2007}%
  \BibitemOpen
  \bibfield  {author} {\bibinfo {author} {\bibfnamefont {D.}~\bibnamefont {Bergman}}, \bibinfo {author} {\bibfnamefont {J.}~\bibnamefont {Alicea}}, \bibinfo {author} {\bibfnamefont {E.}~\bibnamefont {Gull}}, \bibinfo {author} {\bibfnamefont {S.}~\bibnamefont {Trebst}},\ and\ \bibinfo {author} {\bibfnamefont {L.}~\bibnamefont {Balents}},\ }\bibfield  {title} {\bibinfo {title} {Order-by-disorder and spiral spin-liquid in frustrated diamond-lattice antiferromagnets},\ }\href {https://doi.org/10.1038/nphys622} {\bibfield  {journal} {\bibinfo  {journal} {Nature Phys}\ }\textbf {\bibinfo {volume} {3}},\ \bibinfo {pages} {487} (\bibinfo {year} {2007})}\BibitemShut {NoStop}%
\bibitem [{\citenamefont {Niggemann}\ \emph {et~al.}(2019)\citenamefont {Niggemann}, \citenamefont {Hering},\ and\ \citenamefont {Reuther}}]{niggemannClassicalSpiralSpin2019}%
  \BibitemOpen
  \bibfield  {author} {\bibinfo {author} {\bibfnamefont {N.}~\bibnamefont {Niggemann}}, \bibinfo {author} {\bibfnamefont {M.}~\bibnamefont {Hering}},\ and\ \bibinfo {author} {\bibfnamefont {J.}~\bibnamefont {Reuther}},\ }\bibfield  {title} {\bibinfo {title} {Classical spiral spin liquids as a possible route to quantum spin liquids},\ }\href {https://doi.org/10.1088/1361-648X/ab4480} {\bibfield  {journal} {\bibinfo  {journal} {J. Phys.: Condens. Matter}\ }\textbf {\bibinfo {volume} {32}},\ \bibinfo {pages} {024001} (\bibinfo {year} {2019})}\BibitemShut {NoStop}%
\bibitem [{\citenamefont {Yao}\ \emph {et~al.}(2021)\citenamefont {Yao}, \citenamefont {Liu}, \citenamefont {Huang}, \citenamefont {Wang},\ and\ \citenamefont {Chen}}]{yaoGenericSpiralSpin2021}%
  \BibitemOpen
  \bibfield  {author} {\bibinfo {author} {\bibfnamefont {X.-P.}\ \bibnamefont {Yao}}, \bibinfo {author} {\bibfnamefont {J.~Q.}\ \bibnamefont {Liu}}, \bibinfo {author} {\bibfnamefont {C.-J.}\ \bibnamefont {Huang}}, \bibinfo {author} {\bibfnamefont {X.}~\bibnamefont {Wang}},\ and\ \bibinfo {author} {\bibfnamefont {G.}~\bibnamefont {Chen}},\ }\bibfield  {title} {\bibinfo {title} {Generic spiral spin liquids},\ }\href {https://doi.org/10.1007/s11467-021-1074-9} {\bibfield  {journal} {\bibinfo  {journal} {Front. Phys.}\ }\textbf {\bibinfo {volume} {16}},\ \bibinfo {pages} {53303} (\bibinfo {year} {2021})}\BibitemShut {NoStop}%
\bibitem [{\citenamefont {Gao}\ \emph {et~al.}(2022)\citenamefont {Gao}, \citenamefont {Pokharel}, \citenamefont {May}, \citenamefont {Paddison}, \citenamefont {Pasco}, \citenamefont {Liu}, \citenamefont {Taddei}, \citenamefont {Calder}, \citenamefont {Mandrus}, \citenamefont {Stone},\ and\ \citenamefont {Christianson}}]{gaoLineGraphApproachSpiral2022}%
  \BibitemOpen
  \bibfield  {author} {\bibinfo {author} {\bibfnamefont {S.}~\bibnamefont {Gao}}, \bibinfo {author} {\bibfnamefont {G.}~\bibnamefont {Pokharel}}, \bibinfo {author} {\bibfnamefont {A.~F.}\ \bibnamefont {May}}, \bibinfo {author} {\bibfnamefont {J.~A.~M.}\ \bibnamefont {Paddison}}, \bibinfo {author} {\bibfnamefont {C.}~\bibnamefont {Pasco}}, \bibinfo {author} {\bibfnamefont {Y.}~\bibnamefont {Liu}}, \bibinfo {author} {\bibfnamefont {K.~M.}\ \bibnamefont {Taddei}}, \bibinfo {author} {\bibfnamefont {S.}~\bibnamefont {Calder}}, \bibinfo {author} {\bibfnamefont {D.~G.}\ \bibnamefont {Mandrus}}, \bibinfo {author} {\bibfnamefont {M.~B.}\ \bibnamefont {Stone}},\ and\ \bibinfo {author} {\bibfnamefont {A.~D.}\ \bibnamefont {Christianson}},\ }\bibfield  {title} {\bibinfo {title} {Line-{{Graph Approach}} to {{Spiral Spin Liquids}}},\ }\href {https://doi.org/10.1103/PhysRevLett.129.237202} {\bibfield  {journal} {\bibinfo  {journal} {Phys. Rev. Lett.}\ }\textbf {\bibinfo {volume} {129}},\ \bibinfo {pages} {237202} (\bibinfo
  {year} {2022})}\BibitemShut {NoStop}%
\bibitem [{\citenamefont {Yan}\ and\ \citenamefont {Reuther}(2022)}]{yanLowenergyStructureSpiral2022}%
  \BibitemOpen
  \bibfield  {author} {\bibinfo {author} {\bibfnamefont {H.}~\bibnamefont {Yan}}\ and\ \bibinfo {author} {\bibfnamefont {J.}~\bibnamefont {Reuther}},\ }\bibfield  {title} {\bibinfo {title} {Low-energy structure of spiral spin liquids},\ }\href {https://doi.org/10.1103/PhysRevResearch.4.023175} {\bibfield  {journal} {\bibinfo  {journal} {Phys. Rev. Res.}\ }\textbf {\bibinfo {volume} {4}},\ \bibinfo {pages} {023175} (\bibinfo {year} {2022})}\BibitemShut {NoStop}%
\bibitem [{\citenamefont {Balla}\ \emph {et~al.}(2019)\citenamefont {Balla}, \citenamefont {Iqbal},\ and\ \citenamefont {Penc}}]{ballaAffineLatticeConstruction2019}%
  \BibitemOpen
  \bibfield  {author} {\bibinfo {author} {\bibfnamefont {P.}~\bibnamefont {Balla}}, \bibinfo {author} {\bibfnamefont {Y.}~\bibnamefont {Iqbal}},\ and\ \bibinfo {author} {\bibfnamefont {K.}~\bibnamefont {Penc}},\ }\bibfield  {title} {\bibinfo {title} {Affine lattice construction of spiral surfaces in frustrated {{Heisenberg}} models},\ }\href {https://doi.org/10.1103/PhysRevB.100.140402} {\bibfield  {journal} {\bibinfo  {journal} {Phys. Rev. B}\ }\textbf {\bibinfo {volume} {100}},\ \bibinfo {pages} {140402} (\bibinfo {year} {2019})}\BibitemShut {NoStop}%
\bibitem [{\citenamefont {Chung}\ and\ \citenamefont {Gingras}(2025)}]{chung2formU1Spin2025}%
  \BibitemOpen
  \bibfield  {author} {\bibinfo {author} {\bibfnamefont {K.~T.~K.}\ \bibnamefont {Chung}}\ and\ \bibinfo {author} {\bibfnamefont {M.~J.~P.}\ \bibnamefont {Gingras}},\ }\bibfield  {title} {\bibinfo {title} {2-form {{U}}(1) spin liquids: {{A}} classical model and quantum aspects},\ }\href {https://doi.org/10.1103/PhysRevB.111.064417} {\bibfield  {journal} {\bibinfo  {journal} {Phys. Rev. B}\ }\textbf {\bibinfo {volume} {111}},\ \bibinfo {pages} {064417} (\bibinfo {year} {2025})}\BibitemShut {NoStop}%
\bibitem [{\citenamefont {Hall{\'e}n}\ \emph {et~al.}(2024)\citenamefont {Hall{\'e}n}, \citenamefont {Castelnovo},\ and\ \citenamefont {Moessner}}]{hallenThermodynamicsFractalDynamics2024}%
  \BibitemOpen
  \bibfield  {author} {\bibinfo {author} {\bibfnamefont {J.~N.}\ \bibnamefont {Hall{\'e}n}}, \bibinfo {author} {\bibfnamefont {C.}~\bibnamefont {Castelnovo}},\ and\ \bibinfo {author} {\bibfnamefont {R.}~\bibnamefont {Moessner}},\ }\bibfield  {title} {\bibinfo {title} {Thermodynamics and fractal dynamics of a nematic spin ice: {{A}} doubly frustrated pyrochlore {{Ising}} magnet},\ }\href {https://doi.org/10.1103/PhysRevB.109.014438} {\bibfield  {journal} {\bibinfo  {journal} {Phys. Rev. B}\ }\textbf {\bibinfo {volume} {109}},\ \bibinfo {pages} {014438} (\bibinfo {year} {2024})}\BibitemShut {NoStop}%
\bibitem [{\citenamefont {{Brooks-Bartlett}}\ \emph {et~al.}(2014)\citenamefont {{Brooks-Bartlett}}, \citenamefont {Banks}, \citenamefont {Jaubert}, \citenamefont {{Harman-Clarke}},\ and\ \citenamefont {Holdsworth}}]{brooks-bartlettMagneticMomentFragmentationMonopole2014}%
  \BibitemOpen
  \bibfield  {author} {\bibinfo {author} {\bibfnamefont {M.~E.}\ \bibnamefont {{Brooks-Bartlett}}}, \bibinfo {author} {\bibfnamefont {S.~T.}\ \bibnamefont {Banks}}, \bibinfo {author} {\bibfnamefont {L.~D.~C.}\ \bibnamefont {Jaubert}}, \bibinfo {author} {\bibfnamefont {A.}~\bibnamefont {{Harman-Clarke}}},\ and\ \bibinfo {author} {\bibfnamefont {P.~C.~W.}\ \bibnamefont {Holdsworth}},\ }\bibfield  {title} {\bibinfo {title} {Magnetic-{{Moment Fragmentation}} and {{Monopole Crystallization}}},\ }\href {https://doi.org/10.1103/PhysRevX.4.011007} {\bibfield  {journal} {\bibinfo  {journal} {Phys. Rev. X}\ }\textbf {\bibinfo {volume} {4}},\ \bibinfo {pages} {011007} (\bibinfo {year} {2014})}\BibitemShut {NoStop}%
\bibitem [{\citenamefont {Lhotel}\ \emph {et~al.}(2020)\citenamefont {Lhotel}, \citenamefont {Jaubert},\ and\ \citenamefont {Holdsworth}}]{lhotelFragmentationFrustratedMagnets2020}%
  \BibitemOpen
  \bibfield  {author} {\bibinfo {author} {\bibfnamefont {E.}~\bibnamefont {Lhotel}}, \bibinfo {author} {\bibfnamefont {L.~D.~C.}\ \bibnamefont {Jaubert}},\ and\ \bibinfo {author} {\bibfnamefont {P.~C.}\ \bibnamefont {Holdsworth}},\ }\bibfield  {title} {\bibinfo {title} {Fragmentation in {{Frustrated Magnets}}: {{A Review}}},\ }\href {https://doi.org/10.1007/s10909-020-02521-3} {\bibfield  {journal} {\bibinfo  {journal} {Journal of Low Temperature Physics}\ }\textbf {\bibinfo {volume} {201}},\ \bibinfo {pages} {710} (\bibinfo {year} {2020})}\BibitemShut {NoStop}%
\bibitem [{\citenamefont {Deng}\ \emph {et~al.}(2011)\citenamefont {Deng}, \citenamefont {Huang}, \citenamefont {Jacobsen}, \citenamefont {Salas},\ and\ \citenamefont {Sokal}}]{dengFiniteTemperaturePhaseTransition2011}%
  \BibitemOpen
  \bibfield  {author} {\bibinfo {author} {\bibfnamefont {Y.}~\bibnamefont {Deng}}, \bibinfo {author} {\bibfnamefont {Y.}~\bibnamefont {Huang}}, \bibinfo {author} {\bibfnamefont {J.~L.}\ \bibnamefont {Jacobsen}}, \bibinfo {author} {\bibfnamefont {J.}~\bibnamefont {Salas}},\ and\ \bibinfo {author} {\bibfnamefont {A.~D.}\ \bibnamefont {Sokal}},\ }\bibfield  {title} {\bibinfo {title} {Finite-{{Temperature Phase Transition}} in a {{Class}} of {{Four-State Potts Antiferromagnets}}},\ }\href {https://doi.org/10.1103/PhysRevLett.107.150601} {\bibfield  {journal} {\bibinfo  {journal} {Phys. Rev. Lett.}\ }\textbf {\bibinfo {volume} {107}},\ \bibinfo {pages} {150601} (\bibinfo {year} {2011})}\BibitemShut {NoStop}%
\bibitem [{\citenamefont {Moessner}(2021)}]{moessnerSpinIceCoulomb2021}%
  \BibitemOpen
  \bibfield  {author} {\bibinfo {author} {\bibfnamefont {R.}~\bibnamefont {Moessner}},\ }\bibfield  {title} {\bibinfo {title} {Spin {{Ice As}} a {{Coulomb Liquid}}: {{From Emergent Gauge Fields}} to {{Magnetic Monopoles}}},\ }in\ \href {https://doi.org/10.1007/978-3-030-70860-3_3} {\emph {\bibinfo {booktitle} {Spin {{Ice}}}}},\ \bibinfo {series and number} {Springer {{Series}} in {{Solid-State Sciences}}},\ \bibinfo {editor} {edited by\ \bibinfo {editor} {\bibfnamefont {M.}~\bibnamefont {Udagawa}}\ and\ \bibinfo {editor} {\bibfnamefont {L.}~\bibnamefont {Jaubert}}}\ (\bibinfo  {publisher} {Springer International Publishing},\ \bibinfo {address} {Cham},\ \bibinfo {year} {2021})\ pp.\ \bibinfo {pages} {37--70}\BibitemShut {NoStop}%
\bibitem [{\citenamefont {Pretko}(2017{\natexlab{b}})}]{pretkoGeneralizedElectromagnetismSubdimensional2017}%
  \BibitemOpen
  \bibfield  {author} {\bibinfo {author} {\bibfnamefont {M.}~\bibnamefont {Pretko}},\ }\bibfield  {title} {\bibinfo {title} {Generalized electromagnetism of subdimensional particles: {{A}} spin liquid story},\ }\href {https://doi.org/10.1103/PhysRevB.96.035119} {\bibfield  {journal} {\bibinfo  {journal} {Phys. Rev. B}\ }\textbf {\bibinfo {volume} {96}},\ \bibinfo {pages} {035119} (\bibinfo {year} {2017}{\natexlab{b}})}\BibitemShut {NoStop}%
\bibitem [{\citenamefont {Pretko}\ \emph {et~al.}(2020)\citenamefont {Pretko}, \citenamefont {Chen},\ and\ \citenamefont {You}}]{pretkoFractonPhasesMatter2020}%
  \BibitemOpen
  \bibfield  {author} {\bibinfo {author} {\bibfnamefont {M.}~\bibnamefont {Pretko}}, \bibinfo {author} {\bibfnamefont {X.}~\bibnamefont {Chen}},\ and\ \bibinfo {author} {\bibfnamefont {Y.}~\bibnamefont {You}},\ }\bibfield  {title} {\bibinfo {title} {Fracton phases of matter},\ }\href {https://doi.org/10.1142/S0217751X20300033} {\bibfield  {journal} {\bibinfo  {journal} {Int. J. Mod. Phys. A}\ }\textbf {\bibinfo {volume} {35}},\ \bibinfo {pages} {2030003} (\bibinfo {year} {2020})}\BibitemShut {NoStop}%
\bibitem [{\citenamefont {Nandkishore}\ and\ \citenamefont {Hermele}(2019)}]{nandkishoreFractons2019}%
  \BibitemOpen
  \bibfield  {author} {\bibinfo {author} {\bibfnamefont {R.~M.}\ \bibnamefont {Nandkishore}}\ and\ \bibinfo {author} {\bibfnamefont {M.}~\bibnamefont {Hermele}},\ }\bibfield  {title} {\bibinfo {title} {Fractons},\ }\href {https://doi.org/10.1146/annurev-conmatphys-031218-013604} {\bibfield  {journal} {\bibinfo  {journal} {Annual Review of Condensed Matter Physics}\ }\textbf {\bibinfo {volume} {10}},\ \bibinfo {pages} {295} (\bibinfo {year} {2019})}\BibitemShut {NoStop}%
\bibitem [{\citenamefont {Sadoune}\ \emph {et~al.}(2025)\citenamefont {Sadoune}, \citenamefont {Liu}, \citenamefont {Yan}, \citenamefont {Jaubert}, \citenamefont {Shannon},\ and\ \citenamefont {Pollet}}]{sadouneHumanmachineCollaborationOrdering2025}%
  \BibitemOpen
  \bibfield  {author} {\bibinfo {author} {\bibfnamefont {N.}~\bibnamefont {Sadoune}}, \bibinfo {author} {\bibfnamefont {K.}~\bibnamefont {Liu}}, \bibinfo {author} {\bibfnamefont {H.}~\bibnamefont {Yan}}, \bibinfo {author} {\bibfnamefont {L.~D.~C.}\ \bibnamefont {Jaubert}}, \bibinfo {author} {\bibfnamefont {N.}~\bibnamefont {Shannon}},\ and\ \bibinfo {author} {\bibfnamefont {L.}~\bibnamefont {Pollet}},\ }\bibfield  {title} {\bibinfo {title} {Human-machine collaboration: {{Ordering}} mechanism of rank-2 spin liquid on breathing pyrochlore lattice},\ }\href {https://doi.org/10.1103/c6z1-wh6l} {\bibfield  {journal} {\bibinfo  {journal} {Physical Review Research}\ }\textbf {\bibinfo {volume} {7}},\ \bibinfo {pages} {033061} (\bibinfo {year} {2025})}\BibitemShut {NoStop}%
\bibitem [{\citenamefont {{Lozano-G{\'o}mez}}\ \emph {et~al.}(2024{\natexlab{b}})\citenamefont {{Lozano-G{\'o}mez}}, \citenamefont {Benton}, \citenamefont {Gingras},\ and\ \citenamefont {Yan}}]{lozano-gomezAtlasClassicalPyrochlore2024}%
  \BibitemOpen
  \bibfield  {author} {\bibinfo {author} {\bibfnamefont {D.}~\bibnamefont {{Lozano-G{\'o}mez}}}, \bibinfo {author} {\bibfnamefont {O.}~\bibnamefont {Benton}}, \bibinfo {author} {\bibfnamefont {M.~J.~P.}\ \bibnamefont {Gingras}},\ and\ \bibinfo {author} {\bibfnamefont {H.}~\bibnamefont {Yan}},\ }\href@noop {} {\bibinfo {title} {An {{Atlas}} of {{Classical Pyrochlore Spin Liquids}}}} (\bibinfo {year} {2024}{\natexlab{b}}),\ \Eprint {https://arxiv.org/abs/2411.03547} {arXiv:2411.03547} \BibitemShut {NoStop}%
\bibitem [{\citenamefont {Garanin}\ and\ \citenamefont {Canals}(1999)}]{garaninClassicalSpinLiquid1999}%
  \BibitemOpen
  \bibfield  {author} {\bibinfo {author} {\bibfnamefont {D.~A.}\ \bibnamefont {Garanin}}\ and\ \bibinfo {author} {\bibfnamefont {B.}~\bibnamefont {Canals}},\ }\bibfield  {title} {\bibinfo {title} {Classical spin liquid: {{Exact}} solution for the infinite-component antiferromagnetic model on the kagome lattice},\ }\href {https://doi.org/10.1103/PhysRevB.59.443} {\bibfield  {journal} {\bibinfo  {journal} {Phys. Rev. B}\ }\textbf {\bibinfo {volume} {59}},\ \bibinfo {pages} {443} (\bibinfo {year} {1999})}\BibitemShut {NoStop}%
\bibitem [{\citenamefont {Canals}\ and\ \citenamefont {Garanin}(2002)}]{canalsClassicalSpinLiquid2002}%
  \BibitemOpen
  \bibfield  {author} {\bibinfo {author} {\bibfnamefont {B.}~\bibnamefont {Canals}}\ and\ \bibinfo {author} {\bibfnamefont {D.}~\bibnamefont {Garanin}},\ }\bibfield  {title} {\bibinfo {title} {Classical spin liquid properties of the infinite-component spin vector model on a fully frustrated two dimensional lattice},\ }\href {https://doi.org/10.1140/epjb/e20020112} {\bibfield  {journal} {\bibinfo  {journal} {Eur. Phys. J. B}\ }\textbf {\bibinfo {volume} {26}},\ \bibinfo {pages} {439} (\bibinfo {year} {2002})}\BibitemShut {NoStop}%
\bibitem [{\citenamefont {Bergman}\ \emph {et~al.}(2008)\citenamefont {Bergman}, \citenamefont {Wu},\ and\ \citenamefont {Balents}}]{bergmanBandTouchingRealspace2008}%
  \BibitemOpen
  \bibfield  {author} {\bibinfo {author} {\bibfnamefont {D.~L.}\ \bibnamefont {Bergman}}, \bibinfo {author} {\bibfnamefont {C.}~\bibnamefont {Wu}},\ and\ \bibinfo {author} {\bibfnamefont {L.}~\bibnamefont {Balents}},\ }\bibfield  {title} {\bibinfo {title} {Band touching from real-space topology in frustrated hopping models},\ }\href {https://doi.org/10.1103/PhysRevB.78.125104} {\bibfield  {journal} {\bibinfo  {journal} {Phys. Rev. B}\ }\textbf {\bibinfo {volume} {78}},\ \bibinfo {pages} {125104} (\bibinfo {year} {2008})}\BibitemShut {NoStop}%
\bibitem [{\citenamefont {Rehn}\ \emph {et~al.}(2017)\citenamefont {Rehn}, \citenamefont {Sen},\ and\ \citenamefont {Moessner}}]{rehnFractionalizedZ2Classical2017}%
  \BibitemOpen
  \bibfield  {author} {\bibinfo {author} {\bibfnamefont {J.}~\bibnamefont {Rehn}}, \bibinfo {author} {\bibfnamefont {A.}~\bibnamefont {Sen}},\ and\ \bibinfo {author} {\bibfnamefont {R.}~\bibnamefont {Moessner}},\ }\bibfield  {title} {\bibinfo {title} {Fractionalized {{Z2 Classical Heisenberg Spin Liquids}}},\ }\href {https://doi.org/10.1103/PhysRevLett.118.047201} {\bibfield  {journal} {\bibinfo  {journal} {Phys. Rev. Lett.}\ }\textbf {\bibinfo {volume} {118}},\ \bibinfo {pages} {047201} (\bibinfo {year} {2017})}\BibitemShut {NoStop}%
\bibitem [{\citenamefont {Davier}\ \emph {et~al.}(2025)\citenamefont {Davier}, \citenamefont {Albarrac{\'i}n}, \citenamefont {Rosales}, \citenamefont {Pujol},\ and\ \citenamefont {Jaubert}}]{davierPinchLineSpin2025}%
  \BibitemOpen
  \bibfield  {author} {\bibinfo {author} {\bibfnamefont {N.}~\bibnamefont {Davier}}, \bibinfo {author} {\bibfnamefont {F.~A.~G.}\ \bibnamefont {Albarrac{\'i}n}}, \bibinfo {author} {\bibfnamefont {H.~D.}\ \bibnamefont {Rosales}}, \bibinfo {author} {\bibfnamefont {P.}~\bibnamefont {Pujol}},\ and\ \bibinfo {author} {\bibfnamefont {L.~D.~C.}\ \bibnamefont {Jaubert}},\ }\bibfield  {title} {\bibinfo {title} {Pinch line spin liquids as layered {{Coulomb}} phases and applications to cubic models},\ }\href {https://doi.org/10.1103/tczh-wx28} {\bibfield  {journal} {\bibinfo  {journal} {Physical Review B}\ }\textbf {\bibinfo {volume} {112}},\ \bibinfo {pages} {104425} (\bibinfo {year} {2025})}\BibitemShut {NoStop}%
\bibitem [{\citenamefont {Slagle}(2021)}]{slagleFoliatedQuantumField2021}%
  \BibitemOpen
  \bibfield  {author} {\bibinfo {author} {\bibfnamefont {K.}~\bibnamefont {Slagle}},\ }\bibfield  {title} {\bibinfo {title} {Foliated {{Quantum Field Theory}} of {{Fracton Order}}},\ }\href {https://doi.org/10.1103/PhysRevLett.126.101603} {\bibfield  {journal} {\bibinfo  {journal} {Phys. Rev. Lett.}\ }\textbf {\bibinfo {volume} {126}},\ \bibinfo {pages} {101603} (\bibinfo {year} {2021})}\BibitemShut {NoStop}%
\bibitem [{\citenamefont {Stern}\ \emph {et~al.}(2026)\citenamefont {Stern}, \citenamefont {Burke}, \citenamefont {Gingras}, \citenamefont {Romh{\'a}nyi},\ and\ \citenamefont {Chung}}]{sternFractonSpinLiquid2026}%
  \BibitemOpen
  \bibfield  {author} {\bibinfo {author} {\bibfnamefont {M.}~\bibnamefont {Stern}}, \bibinfo {author} {\bibfnamefont {M.~D.}\ \bibnamefont {Burke}}, \bibinfo {author} {\bibfnamefont {M.~J.~P.}\ \bibnamefont {Gingras}}, \bibinfo {author} {\bibfnamefont {J.}~\bibnamefont {Romh{\'a}nyi}},\ and\ \bibinfo {author} {\bibfnamefont {K.~T.~K.}\ \bibnamefont {Chung}},\ }\href@noop {} {\bibinfo {title} {Fracton {{Spin Liquid}} and {{Exotic Frustrated Phases}} in {{Ising-like Octochlore Magnets}}}} (\bibinfo {year} {2026}),\ \Eprint {https://arxiv.org/abs/2603.12313} {arXiv:2603.12313} \BibitemShut {NoStop}%
\bibitem [{\citenamefont {Hagym{\'a}si}\ \emph {et~al.}(2021)\citenamefont {Hagym{\'a}si}, \citenamefont {Sch{\"a}fer}, \citenamefont {Moessner},\ and\ \citenamefont {Luitz}}]{hagymasiPossibleInversionSymmetry2021}%
  \BibitemOpen
  \bibfield  {author} {\bibinfo {author} {\bibfnamefont {I.}~\bibnamefont {Hagym{\'a}si}}, \bibinfo {author} {\bibfnamefont {R.}~\bibnamefont {Sch{\"a}fer}}, \bibinfo {author} {\bibfnamefont {R.}~\bibnamefont {Moessner}},\ and\ \bibinfo {author} {\bibfnamefont {D.~J.}\ \bibnamefont {Luitz}},\ }\bibfield  {title} {\bibinfo {title} {Possible {{Inversion Symmetry Breaking}} in the {{S}}=1/2 {{Pyrochlore Heisenberg Magnet}}},\ }\href {https://doi.org/10.1103/PhysRevLett.126.117204} {\bibfield  {journal} {\bibinfo  {journal} {Phys. Rev. Lett.}\ }\textbf {\bibinfo {volume} {126}},\ \bibinfo {pages} {117204} (\bibinfo {year} {2021})}\BibitemShut {NoStop}%
\bibitem [{\citenamefont {Astrakhantsev}\ \emph {et~al.}(2021)\citenamefont {Astrakhantsev}, \citenamefont {Westerhout}, \citenamefont {Tiwari}, \citenamefont {Choo}, \citenamefont {Chen}, \citenamefont {Fischer}, \citenamefont {Carleo},\ and\ \citenamefont {Neupert}}]{astrakhantsevBrokenSymmetryGroundStates2021}%
  \BibitemOpen
  \bibfield  {author} {\bibinfo {author} {\bibfnamefont {N.}~\bibnamefont {Astrakhantsev}}, \bibinfo {author} {\bibfnamefont {T.}~\bibnamefont {Westerhout}}, \bibinfo {author} {\bibfnamefont {A.}~\bibnamefont {Tiwari}}, \bibinfo {author} {\bibfnamefont {K.}~\bibnamefont {Choo}}, \bibinfo {author} {\bibfnamefont {A.}~\bibnamefont {Chen}}, \bibinfo {author} {\bibfnamefont {M.~H.}\ \bibnamefont {Fischer}}, \bibinfo {author} {\bibfnamefont {G.}~\bibnamefont {Carleo}},\ and\ \bibinfo {author} {\bibfnamefont {T.}~\bibnamefont {Neupert}},\ }\bibfield  {title} {\bibinfo {title} {Broken-{{Symmetry Ground States}} of the {{Heisenberg Model}} on the {{Pyrochlore Lattice}}},\ }\href {https://doi.org/10.1103/PhysRevX.11.041021} {\bibfield  {journal} {\bibinfo  {journal} {Phys. Rev. X}\ }\textbf {\bibinfo {volume} {11}},\ \bibinfo {pages} {041021} (\bibinfo {year} {2021})}\BibitemShut {NoStop}%
\bibitem [{\citenamefont {Pohle}\ \emph {et~al.}(2023)\citenamefont {Pohle}, \citenamefont {Yamaji},\ and\ \citenamefont {Imada}}]{pohleGroundStateS12023}%
  \BibitemOpen
  \bibfield  {author} {\bibinfo {author} {\bibfnamefont {R.}~\bibnamefont {Pohle}}, \bibinfo {author} {\bibfnamefont {Y.}~\bibnamefont {Yamaji}},\ and\ \bibinfo {author} {\bibfnamefont {M.}~\bibnamefont {Imada}},\ }\href@noop {} {\bibinfo {title} {Ground state of the {{S}}=1/2 pyrochlore {{Heisenberg}} antiferromagnet: {{A}} quantum spin liquid emergent from dimensional reduction}} (\bibinfo {year} {2023}),\ \Eprint {https://arxiv.org/abs/2311.11561} {arXiv:2311.11561} \BibitemShut {NoStop}%
\bibitem [{\citenamefont {Sch{\"a}fer}\ \emph {et~al.}(2023)\citenamefont {Sch{\"a}fer}, \citenamefont {Placke}, \citenamefont {Benton},\ and\ \citenamefont {Moessner}}]{schaferAbundanceHardHexagonCrystals2023}%
  \BibitemOpen
  \bibfield  {author} {\bibinfo {author} {\bibfnamefont {R.}~\bibnamefont {Sch{\"a}fer}}, \bibinfo {author} {\bibfnamefont {B.}~\bibnamefont {Placke}}, \bibinfo {author} {\bibfnamefont {O.}~\bibnamefont {Benton}},\ and\ \bibinfo {author} {\bibfnamefont {R.}~\bibnamefont {Moessner}},\ }\bibfield  {title} {\bibinfo {title} {Abundance of {{Hard-Hexagon Crystals}} in the {{Quantum Pyrochlore Antiferromagnet}}},\ }\href {https://doi.org/10.1103/PhysRevLett.131.096702} {\bibfield  {journal} {\bibinfo  {journal} {Phys. Rev. Lett.}\ }\textbf {\bibinfo {volume} {131}},\ \bibinfo {pages} {096702} (\bibinfo {year} {2023})}\BibitemShut {NoStop}%
\bibitem [{\citenamefont {Han}\ \emph {et~al.}(2022)\citenamefont {Han}, \citenamefont {Patri},\ and\ \citenamefont {Kim}}]{hanRealizationFractonicQuantum2022}%
  \BibitemOpen
  \bibfield  {author} {\bibinfo {author} {\bibfnamefont {S.}~\bibnamefont {Han}}, \bibinfo {author} {\bibfnamefont {A.~S.}\ \bibnamefont {Patri}},\ and\ \bibinfo {author} {\bibfnamefont {Y.~B.}\ \bibnamefont {Kim}},\ }\bibfield  {title} {\bibinfo {title} {Realization of fractonic quantum phases in the breathing pyrochlore lattice},\ }\href {https://doi.org/10.1103/PhysRevB.105.235120} {\bibfield  {journal} {\bibinfo  {journal} {Phys. Rev. B}\ }\textbf {\bibinfo {volume} {105}},\ \bibinfo {pages} {235120} (\bibinfo {year} {2022})}\BibitemShut {NoStop}%
\bibitem [{\citenamefont {Liu}\ \emph {et~al.}(2021)\citenamefont {Liu}, \citenamefont {Hal{\'a}sz},\ and\ \citenamefont {Balents}}]{liuSymmetricU1Z22021}%
  \BibitemOpen
  \bibfield  {author} {\bibinfo {author} {\bibfnamefont {C.}~\bibnamefont {Liu}}, \bibinfo {author} {\bibfnamefont {G.~B.}\ \bibnamefont {Hal{\'a}sz}},\ and\ \bibinfo {author} {\bibfnamefont {L.}~\bibnamefont {Balents}},\ }\bibfield  {title} {\bibinfo {title} {Symmetric {{U}}(1) and {{$\mathbb{Z}$}}{$_2$} spin liquids on the pyrochlore lattice},\ }\href {https://doi.org/10.1103/PhysRevB.104.054401} {\bibfield  {journal} {\bibinfo  {journal} {Phys. Rev. B}\ }\textbf {\bibinfo {volume} {104}},\ \bibinfo {pages} {054401} (\bibinfo {year} {2021})}\BibitemShut {NoStop}%
\bibitem [{\citenamefont {Desrochers}\ \emph {et~al.}(2022)\citenamefont {Desrochers}, \citenamefont {Chern},\ and\ \citenamefont {Kim}}]{desrochersCompetingU1Z22022}%
  \BibitemOpen
  \bibfield  {author} {\bibinfo {author} {\bibfnamefont {F.}~\bibnamefont {Desrochers}}, \bibinfo {author} {\bibfnamefont {L.~E.}\ \bibnamefont {Chern}},\ and\ \bibinfo {author} {\bibfnamefont {Y.~B.}\ \bibnamefont {Kim}},\ }\bibfield  {title} {\bibinfo {title} {Competing {{U}}(1) and {{Z2}} dipolar-octupolar quantum spin liquids on the pyrochlore lattice: {{Application}} to {{Ce}}{$_{2}$}{{Zr}}{$_{2}$}{{O}}{$_{7}$}},\ }\href {https://doi.org/10.1103/PhysRevB.105.035149} {\bibfield  {journal} {\bibinfo  {journal} {Phys. Rev. B}\ }\textbf {\bibinfo {volume} {105}},\ \bibinfo {pages} {035149} (\bibinfo {year} {2022})}\BibitemShut {NoStop}%
\bibitem [{\citenamefont {Chern}\ \emph {et~al.}(2022)\citenamefont {Chern}, \citenamefont {Kim},\ and\ \citenamefont {Castelnovo}}]{chernCompetingQuantumSpin2022}%
  \BibitemOpen
  \bibfield  {author} {\bibinfo {author} {\bibfnamefont {L.~E.}\ \bibnamefont {Chern}}, \bibinfo {author} {\bibfnamefont {Y.~B.}\ \bibnamefont {Kim}},\ and\ \bibinfo {author} {\bibfnamefont {C.}~\bibnamefont {Castelnovo}},\ }\bibfield  {title} {\bibinfo {title} {Competing quantum spin liquids, gauge fluctuations, and anisotropic interactions in a breathing pyrochlore lattice},\ }\href {https://doi.org/10.1103/PhysRevB.106.134402} {\bibfield  {journal} {\bibinfo  {journal} {Phys. Rev. B}\ }\textbf {\bibinfo {volume} {106}},\ \bibinfo {pages} {134402} (\bibinfo {year} {2022})}\BibitemShut {NoStop}%
\bibitem [{\citenamefont {Desrochers}\ \emph {et~al.}(2023)\citenamefont {Desrochers}, \citenamefont {Chern},\ and\ \citenamefont {Kim}}]{desrochersSymmetryFractionalizationGauge2023}%
  \BibitemOpen
  \bibfield  {author} {\bibinfo {author} {\bibfnamefont {F.}~\bibnamefont {Desrochers}}, \bibinfo {author} {\bibfnamefont {L.~E.}\ \bibnamefont {Chern}},\ and\ \bibinfo {author} {\bibfnamefont {Y.~B.}\ \bibnamefont {Kim}},\ }\bibfield  {title} {\bibinfo {title} {Symmetry fractionalization in the gauge mean-field theory of quantum spin ice},\ }\href {https://doi.org/10.1103/PhysRevB.107.064404} {\bibfield  {journal} {\bibinfo  {journal} {Phys. Rev. B}\ }\textbf {\bibinfo {volume} {107}},\ \bibinfo {pages} {064404} (\bibinfo {year} {2023})}\BibitemShut {NoStop}%
\bibitem [{\citenamefont {M{\"u}ller}\ \emph {et~al.}(2024)\citenamefont {M{\"u}ller}, \citenamefont {Kiese}, \citenamefont {Niggemann}, \citenamefont {Sbierski}, \citenamefont {Reuther}, \citenamefont {Trebst}, \citenamefont {Thomale},\ and\ \citenamefont {Iqbal}}]{mullerPseudofermionFunctionalRenormalization2024}%
  \BibitemOpen
  \bibfield  {author} {\bibinfo {author} {\bibfnamefont {T.}~\bibnamefont {M{\"u}ller}}, \bibinfo {author} {\bibfnamefont {D.}~\bibnamefont {Kiese}}, \bibinfo {author} {\bibfnamefont {N.}~\bibnamefont {Niggemann}}, \bibinfo {author} {\bibfnamefont {B.}~\bibnamefont {Sbierski}}, \bibinfo {author} {\bibfnamefont {J.}~\bibnamefont {Reuther}}, \bibinfo {author} {\bibfnamefont {S.}~\bibnamefont {Trebst}}, \bibinfo {author} {\bibfnamefont {R.}~\bibnamefont {Thomale}},\ and\ \bibinfo {author} {\bibfnamefont {Y.}~\bibnamefont {Iqbal}},\ }\bibfield  {title} {\bibinfo {title} {Pseudo-fermion functional renormalization group for spin models},\ }\href {https://doi.org/10.1088/1361-6633/ad208c} {\bibfield  {journal} {\bibinfo  {journal} {Rep. Prog. Phys.}\ }\textbf {\bibinfo {volume} {87}},\ \bibinfo {pages} {036501} (\bibinfo {year} {2024})}\BibitemShut {NoStop}%
\bibitem [{\citenamefont {Iqbal}\ \emph {et~al.}(2016)\citenamefont {Iqbal}, \citenamefont {Thomale}, \citenamefont {Parisen~Toldin}, \citenamefont {Rachel},\ and\ \citenamefont {Reuther}}]{iqbalFunctionalRenormalizationGroup2016}%
  \BibitemOpen
  \bibfield  {author} {\bibinfo {author} {\bibfnamefont {Y.}~\bibnamefont {Iqbal}}, \bibinfo {author} {\bibfnamefont {R.}~\bibnamefont {Thomale}}, \bibinfo {author} {\bibfnamefont {F.}~\bibnamefont {Parisen~Toldin}}, \bibinfo {author} {\bibfnamefont {S.}~\bibnamefont {Rachel}},\ and\ \bibinfo {author} {\bibfnamefont {J.}~\bibnamefont {Reuther}},\ }\bibfield  {title} {\bibinfo {title} {Functional renormalization group for three-dimensional quantum magnetism},\ }\href {https://doi.org/10.1103/PhysRevB.94.140408} {\bibfield  {journal} {\bibinfo  {journal} {Phys. Rev. B}\ }\textbf {\bibinfo {volume} {94}},\ \bibinfo {pages} {140408} (\bibinfo {year} {2016})}\BibitemShut {NoStop}%
\bibitem [{\citenamefont {Niggemann}\ \emph {et~al.}(2023)\citenamefont {Niggemann}, \citenamefont {Iqbal},\ and\ \citenamefont {Reuther}}]{niggemannQuantumEffectsUnconventional2023}%
  \BibitemOpen
  \bibfield  {author} {\bibinfo {author} {\bibfnamefont {N.}~\bibnamefont {Niggemann}}, \bibinfo {author} {\bibfnamefont {Y.}~\bibnamefont {Iqbal}},\ and\ \bibinfo {author} {\bibfnamefont {J.}~\bibnamefont {Reuther}},\ }\bibfield  {title} {\bibinfo {title} {Quantum {{Effects}} on {{Unconventional Pinch Point Singularities}}},\ }\href {https://doi.org/10.1103/PhysRevLett.130.196601} {\bibfield  {journal} {\bibinfo  {journal} {Phys. Rev. Lett.}\ }\textbf {\bibinfo {volume} {130}},\ \bibinfo {pages} {196601} (\bibinfo {year} {2023})}\BibitemShut {NoStop}%
\bibitem [{\citenamefont {Chern}\ \emph {et~al.}(2024)\citenamefont {Chern}, \citenamefont {Desrochers}, \citenamefont {Kim},\ and\ \citenamefont {Castelnovo}}]{chernPseudofermionFunctionalRenormalization2024}%
  \BibitemOpen
  \bibfield  {author} {\bibinfo {author} {\bibfnamefont {L.~E.}\ \bibnamefont {Chern}}, \bibinfo {author} {\bibfnamefont {F.}~\bibnamefont {Desrochers}}, \bibinfo {author} {\bibfnamefont {Y.~B.}\ \bibnamefont {Kim}},\ and\ \bibinfo {author} {\bibfnamefont {C.}~\bibnamefont {Castelnovo}},\ }\bibfield  {title} {\bibinfo {title} {Pseudofermion functional renormalization group study of dipolar-octupolar pyrochlore magnets},\ }\href {https://doi.org/10.1103/PhysRevB.109.184421} {\bibfield  {journal} {\bibinfo  {journal} {Phys. Rev. B}\ }\textbf {\bibinfo {volume} {109}},\ \bibinfo {pages} {184421} (\bibinfo {year} {2024})}\BibitemShut {NoStop}%
\bibitem [{\citenamefont {Gresista}\ \emph {et~al.}(2025)\citenamefont {Gresista}, \citenamefont {{Lozano-G{\'o}mez}}, \citenamefont {Vojta}, \citenamefont {Trebst},\ and\ \citenamefont {Iqbal}}]{gresistaQuantumEffectsPyrochlore2025}%
  \BibitemOpen
  \bibfield  {author} {\bibinfo {author} {\bibfnamefont {L.}~\bibnamefont {Gresista}}, \bibinfo {author} {\bibfnamefont {D.}~\bibnamefont {{Lozano-G{\'o}mez}}}, \bibinfo {author} {\bibfnamefont {M.}~\bibnamefont {Vojta}}, \bibinfo {author} {\bibfnamefont {S.}~\bibnamefont {Trebst}},\ and\ \bibinfo {author} {\bibfnamefont {Y.}~\bibnamefont {Iqbal}},\ }\bibfield  {title} {\bibinfo {title} {Quantum effects on pyrochlore higher-rank {{U}}(1) spin liquids: {{Pinch-line}} singularities, spin nematics, and connections to oxide materials},\ }\href {https://doi.org/10.1103/3y39-9ndv} {\bibfield  {journal} {\bibinfo  {journal} {Physical Review Research}\ }\textbf {\bibinfo {volume} {7}},\ \bibinfo {pages} {033109} (\bibinfo {year} {2025})}\BibitemShut {NoStop}%
\bibitem [{\citenamefont {Essafi}\ \emph {et~al.}(2017{\natexlab{b}})\citenamefont {Essafi}, \citenamefont {Jaubert},\ and\ \citenamefont {Udagawa}}]{essafiFlatBandsDirac2017}%
  \BibitemOpen
  \bibfield  {author} {\bibinfo {author} {\bibfnamefont {K.}~\bibnamefont {Essafi}}, \bibinfo {author} {\bibfnamefont {L.~D.~C.}\ \bibnamefont {Jaubert}},\ and\ \bibinfo {author} {\bibfnamefont {M.}~\bibnamefont {Udagawa}},\ }\bibfield  {title} {\bibinfo {title} {Flat bands and {{Dirac}} cones in breathing lattices},\ }\href {https://doi.org/10.1088/1361-648X/aa782f} {\bibfield  {journal} {\bibinfo  {journal} {J. Phys.: Condens. Matter}\ }\textbf {\bibinfo {volume} {29}},\ \bibinfo {pages} {315802} (\bibinfo {year} {2017}{\natexlab{b}})}\BibitemShut {NoStop}%
\end{thebibliography}
\end{document}